\documentclass[rmp,twocolumn,superscriptaddress,eqsecnum]{revtex4}
\usepackage{amsmath}
\usepackage{amsfonts}
\usepackage{amsthm}

\newtheorem*{thmnonum}{Theorem}
\newtheorem*{lemma}{Lemma}

\begin{document}
\title{Reference frames, superselection rules, and quantum information}
\author{Stephen D. Bartlett}
\affiliation{School of Physics, The University of Sydney,
  Sydney, New South Wales 2006, Australia}%
\author{Terry Rudolph}
\affiliation{Optics Section, Blackett Laboratory, Imperial College
  London, London SW7 2BW, United Kingdom}%
\affiliation{Institute for Mathematical Sciences, Imperial College
London, London SW7 2BW, United Kingdom}%
\author{Robert W. Spekkens}
\affiliation{Department of Applied Mathematics and Theoretical
Physics, University of Cambridge, Cambridge CB3 0WA, United Kingdom}%
\date{4 April 2007}

\begin{abstract}
Recently, there has been much interest in a new kind of
``unspeakable'' quantum information that stands to regular quantum
information in the same way that a direction in space or a moment in
time stands to a classical bit string: the former can only be
encoded using particular degrees of freedom while the latter are
indifferent to the physical nature of the information carriers. The
problem of correlating distant reference frames, of which aligning
Cartesian axes and synchronizing clocks are important instances, is
an example of a task that requires the exchange of unspeakable
information and for which it is interesting to determine the
fundamental quantum limit of efficiency. There have also been many
investigations into the information theory that is appropriate for
parties that lack reference frames or that lack correlation between
their reference frames, restrictions that result in global and local
superselection rules. In the presence of these, quantum unspeakable
information becomes a new kind of resource that can be manipulated,
depleted, quantified, etcetera. Methods have also been developed to
contend with these restrictions using relational encodings,
particularly in the context of computation, cryptography,
communication, and the manipulation of entanglement. This article
reviews the role of reference frames and superselection rules in the
theory of quantum information processing.
\end{abstract}
\maketitle

\tableofcontents

\section{Introduction -- Why consider reference frames in quantum
  information?}

Classical information theory is typically concerned with
\emph{fungible information}, that is, information for which the
means of encoding is not important. Shannon's coding theorems, for
instance, are indifferent to whether the two values ``0'' and ``1''
of a classical bit correspond to two values of magnetization on a
tape, two voltages on a transmission line, or two positions of a
bead on an abacus.  Most information-processing tasks of interest to
computer scientists and information theorists are of this sort,
whether they be communication tasks such as data compression,
cryptographic tasks such as key distribution, or computational tasks
such as factoring.  Nonetheless, there are many tasks that cannot be
achieved with fungible information but that are also aptly described
as ``information processing'' tasks.  Examples include the
synchronization of distant clocks, the alignment of distant
Cartesian frames, and the determination of one's global position.
Imagine for instance that Alice and Bob are in separate spaceships
with no shared Cartesian frame (in particular, no access to the
fixed stars).  There is clearly no way for Alice to \emph{describe}
a direction in space to Bob abstractly, that is, using nothing more
than a string of classical bits. Rather, she must send to Bob a
system that can point in some direction, a token of one of the axes
of her own Cartesian frame.  This token cannot be spherically
symmetric; it must have a degree of freedom that can encode
directional information. On the other hand, if she wishes to
synchronize her clock with Bob's by sending him a token system, she
will need to make use of a system that has a natural oscillation.
The information that is communicated in these sorts of tasks is said
to be \emph{nonfungible}.  These two sorts of information, fungible
and nonfungible, have also been referred to as \emph{speakable} and
\emph{unspeakable}~\cite{Per02b}.

The relatively young field of quantum information theory has been
primarily concerned with developing a quantum theory of speakable
information. Investigators have sought to determine the degree of
success with which various abstract information-processing tasks can
be achieved assuming that the systems used to implement these tasks
obey the laws of quantum theory.  Nonetheless, there has also been
progress in developing a quantum theory of \emph{unspeakable}
information, outlining, for instance, the success with which tasks
such as clock synchronization and Cartesian frame alignment can be
achieved in a quantum world.

That one must look to physics to answer questions of interest to
computer scientists is a fact that has not always been obvious.
(\textcite{Lan93} summarized this point in the slogan, ``Information
is physical''.) That one must look to physics to answers questions
about the processing of unspeakable information, on the other hand,
comes as no surprise. Nonetheless, the \emph{quantum} theory of
unspeakable information is only just beginning to be explored.

It is critical to note that when one has a system encoding
directional information, such as a spin-1/2 particle in a pure
state, the direction is not defined with respect to any purported
absolute Newtonian space, but rather with respect to another system,
for instance, a set of gyroscopes in the laboratory.  Similarly, a
system that contains phase information, such as a two-level atom in
a coherent superposition of ground and excited states, is not
defined relative to any purported absolute time, but rather relative
to a clock. We refer to the systems with respect to which
unspeakable/nonfungible information is defined, clocks, gyroscopes,
metre sticks and so forth, as \emph{reference frames}.  The tasks we
have highlighted thus far can all be described as the alignment of
reference frames. Nonfungible information is nonfungible precisely
because it can only be defined with respect to a particular type of
reference frame.

Even a quantum information theorist who is uninterested in tasks
such as clock synchronization and Cartesian frame alignment must
necessarily consider physical systems which make use of reference
frames. The reason is that although fungible information can be
encoded into \emph{any} degree of freedom, and thus defined with
respect to \emph{any} reference frame, it is still the case that
\emph{some} degree of freedom must be chosen, and consequently
\emph{some} reference frame is required.  For instance, if a
two-level atomic qubit is being used for some task, one still
requires a clock in the background in order to implement arbitrary
preparations and measurements on this qubit even if the task is to
perform abstract quantum information-processing rather than as a
means of distributing phase information.  In this example, one can
change the relative phase between the ground and excited states of a
two-level atom by a specified amount by turning on a static electric
field for a specific time interval, but this requires a suitably
precise clock as well as alignment of the field with the atomic
dipole moment.

It follows that to \emph{lack} a reference frame for a particular
degree of freedom has an impact on the success with which one can
perform certain quantum information processing tasks.  On several
occasions there has been considerable controversy over the
performance of certain tasks because this impact was ignored, or not
treated properly.  As we will see, the lack of a reference frame can
be treated within the quantum formalism as a form of
\emph{decoherence} -- quantum noise. As opposed to the typical
source of decoherence, which is due to correlation with an
environment to which one does not have access, this decoherence can
be viewed as resulting from correlation with a (possibly
hypothetical) reference frame to which one does not have access.
This is a powerful result, because if the lack of a reference frame
can be viewed as a form of decoherence, the now-standard techniques
of combating decoherence in quantum information theory (in
particular, the use of \emph{decoherence-free subsystems}) can be
applied.

As it turns out, the restriction of lacking a reference frame is
mathematically equivalent to that of so-called \emph{superselection
rules} -- postulated rules forbidding the preparation of quantum
states that exhibit coherence between eigenstates of certain
observables. Originally, superselection rules were introduced to
enforce additional constraints to quantum theory beyond the
well-studied constraints of selection rules (conservation laws).
They were considered to be axiomatic restrictions, applying to only
certain degrees of freedom.  For instance, a superselection rule for
electric charge asserts the impossibility of preparing a coherent
superposition of different charge eigenstates. As we shall see,
however, for superselection rules associated with compact symmetry
groups, the presence of appropriate reference frames can actually
allow for the preparation of such superposition states, thereby
obviating the superselection rules in practice. This shows that
there is an intimate connection between the restriction of lacking a
reference frame and that of a superselection rule.

As~\textcite{Sch03} has emphasized, interesting restrictions on
experimental operations yield interesting information theories. For
instance, the fact that classical channels and local operations are
a cheap resource compared to quantum channels leads us to study what
can be achieved with local operations and classical communication
(LOCC). The resulting information theory is the theory of
entanglement.  As another example, the relative ease with which one
can implement Gaussian operations in quantum optics leads one to
consider the information theory that results from the restriction to
these operations.  By comparing and contrasting the information
theories that result from various different restrictions we are led
to a much broader perspective on all of them.  In particular,
analogies between the resulting theories allow us to apply the
insights gained in the context of one to solve problems arising in
the context of another. In this sense, studying the restriction of a
superselection rule -- or equivalently, as we shall demonstrate, the
restriction of lacking a reference frame -- may yield lessons for
the rest of quantum information theory.

In some cases, it is difficult to imagine lacking a reference
frame. For example, Cartesian frames with precision on the order
of fractions of a degree and clocks with precision on the order of
fractions of a second are sufficiently ubiquitous that their
presence typically does not even warrant mention. However, these
same reference frames become quite difficult to prepare and
maintain if one requires very high precision or very good
stability. Furthermore, there are certain kinds of reference
frames that are difficult to prepare even if one requires only low
precision and poor stability.  For instance, a Bose-Einstein
condensate of alkali atoms can act as a reference frame for the
phase that is conjugate to atom number, and the reliable
preparation of these has only been achieved in the past decade. In
addition, it is straightforward to imagine two parties with
reference frames that are uncorrelated (such as the example of the
space-faring Alice and Bob provided earlier). In this case we say
that they lack a \emph{shared} reference frame. All of these facts
demonstrate that reference frames must be considered as
\emph{resources}.

Regardless of the degree of freedom in question, a reference frame
is always associated with some physical system.  As such, it may be
treated within the formalism of quantum mechanics.  In this case, we
speak of \emph{quantum reference frames}. Indeed, one can imagine an
extreme case wherein the only system in a party's possession that
plays the role of a reference frame (or plays the role of a shared
reference frame with another party) is of bounded size. For
instance, one can imagine a quantum clock consisting of an
oscillator with a small maximum number of excitations, or a quantum
gyroscope consisting of a handful of spin-1/2 systems. It is then
natural to ask how well such a bounded-size reference frame
approximates one that is of unbounded size.

The ability of a bounded reference frame to stand in for an
unbounded reference frame is analogous to the ability of an
entangled state to stand in for the possibility of implementing
non-local operations. Recall that the teleportation protocol permits
entanglement and classical communication to substitute for a
non-local operation. More generally, when one lacks the ability to
perform non-local operations (such as when qubits are remotely
separated), entanglement becomes a quantifiable resource. Similarly,
when one is subject to a superselection rule (i.e., when one lacks a
reference frame for some degree of freedom) bounded reference frames
become a quantifiable resource about which we can ask the same sorts
of questions as we do for entanglement. For instance, we may ask the
following questions: Which states are interconvertible?  How many
states of a standard form can be distilled from a given state and
how many are required to form a given state? How much of the
resource is required for a given task? How quickly is it used up?
etc.

Finally, because it is all too easy to forget about the presence of
reference frames, these are at the root of various conceptual
confusions. These include: the interpretation of quantum states
exhibiting coherence between number states in a single mode (a
subject of controversy in quantum optics, Bose-Einstein condensation
and superconductivity); the quantification of entanglement in
systems of bosons or fermions, or in situations when operations are
restricted; the efficiency with which frames may be aligned, clocks
synchronized, etc.; and the significance of superselection rules on
the possibility of implementing various quantum information
processing tasks.

In this article, we provide a review of the recent investigations
into these and related issues.  In Sec.~\ref{sec:Formal}, we
introduce the formalism for treating the lack of a general reference
frame in quantum theory, and show how this is equivalent to a
superselection rule. Sec.~\ref{sec:NoSRF} considers quantum
information processing without a shared reference frame.
Sec.~\ref{sec:QRFs} considers how to treat reference frames within
the quantum formalism, which provides the starting point for a
theory of distributing quantum reference frames -- the topic of
Sec.~\ref{sec:Aligning}. The effect of bounding the size of
reference frames for quantum information processing is considered in
Sec.~\ref{sec:BoundedRFs}. Finally, we provide an outlook to the
future of this field in Sec.~\ref{sec:Outlook}.

\section{Formalizing reference frames and superselection rules}
\label{sec:Formal}

\subsection{Reference frames in quantum theory}

Reference frames (RFs) are implicit in the definition of quantum
states.  For example, in the position representation of the
wavefunction of a quantum particle, $\psi(x)$, $x$ parameterizes the
position of the particle relative to a spatial reference frame. More
generally, the quantum state of a system is a description of the
system relative to a suitable reference frame.

Consider a quantum system with Hilbert space $\mathcal{H}$, prepared
in a state $|\psi_0\rangle$ relative to a reference frame. We can
now consider a transformation that changes this relation. Such a
transformation can be \emph{active}, changing the system such that
it subsequently holds a different relation to the reference frame,
or \emph{passive}, in which case the system is unchanged but is now
described relative to a new reference frame. In both situations, the
transformation can be represented by a unitary operator, $T(g)$,
where $g$ denotes the transformation; the transformed system is then
described by the state $T(g)|\psi_0\rangle$.  Note that these
operations can be composed, so that $T(g'g) \equiv T(g')T(g)$ is a
transformation if both $T(g')$ and $T(g)$ are, and this composition
is associative (i.e., $T(g''g')T(g) = T(g'')T(g'g)$).  Also, there
exists an \emph{inverse} transformation $T(g^{-1})$ to every
transformation $T(g)$, such that $T(g^{-1})T(g) = I$, the identity.
If this inverse is unique,\footnote{If the inverse is not unique,
then the RF is instead associated with a coset space.  This
situation occurs when the RF itself is invariant under some
transformations.  We consider an example of such an RF -- a
direction (as opposed to a full Cartesian frame) -- in
Sec.~\ref{sec:Aligning}.} the set of all transformations form a
group $G$. We use $g \in G$ to denote an abstract transformation
within the group, and say that $T$ is the \emph{unitary
representation} of this group on the quantum system (or
equivalently, on the Hilbert space $\mathcal{H}$).

In this review, we will often use two common examples of a reference
frame to illustrate the concepts and ideas we cover.  The first
example is a \emph{phase reference}, for which the relevant group of
transformations is U(1), the group of real numbers modulo $2\pi$
under addition. A representation of U(1) on a quantum system
determines how that system transforms under phase shifts. The second
example that we use extensively in this review is a \emph{Cartesian
frame} specifying three orthogonal spatial directions; the group of
transformations of orientation relative to a Cartesian frame is the
group of rotations SO(3).  An element $\Omega \in$ SO(3) can be
given, say, by a set of three Euler angles.  The representation of
SO(3) on a quantum system, then, determines how that system
transforms under rotations; for example, a spin-$j$ particle
transforms according to the unitary representation $R_{j}$ (a Wigner
rotation matrix). We will often extend the group of rotations SO(3)
to the group SU(2) to allow for spinor representations.

Because group theory provides a powerful mathematical tool for
analyzing the role of reference frames in quantum systems, we will
make frequent use of group theoretic techniques throughout this
review.  We present a short introduction to the relevant techniques
in this section, but the reader may wish to consult a standard group
theory text, such as~\textcite{FultonHarris} or
\textcite{Sternberg}, for further details.  Also, for an
introduction to the standard mathematical techniques of quantum
information, we direct the reader to~\textcite{Nie00}.

We begin by exploring an illustrative example.

\subsection{Lacking a phase reference implies a
photon-number superselection rule}
\label{subsec:PNSSR}

In this section, we investigate an explicit example of a reference
frame -- a phase reference -- and demonstrate that if one
\emph{lacks} a phase reference then the resulting quantum theory is
equivalent to one in which there is a superselection rule for photon
number.

In quantum optical experiments, states of an optical mode are always
referred to some phase reference.  Consider $K$ optical modes as
described by some party, Alice, relative to a phase reference in her
possession -- for example, a high intensity laser.  Let
$|n_1,\ldots,n_K\rangle$ be the Fock state basis for the Hilbert
space $\mathcal{H}^{(K)}$ describing these modes, with $n_i$ the
number of photons in the mode $i$, and $\hat{N}_i$ the number
operator for this mode.

Consider another party, Charlie, who has a different phase
reference.  Let $\phi$ be the angle that relates Charlie's phase
reference to Alice's.  Alice can perform an \emph{active}
transformation on her system of optical modes by allowing them to
evolve under a Hamiltonian proportional to $\hat{N}_{\rm tot} \equiv
\sum_{i=1}^K \hat{N}_i$, the total photon number operator.
Specifically, the unitary transformation $U(\phi) =
\exp(i\phi\hat{N}_{\rm tot})$ will actively advance her system by an
angle $\phi$.  Using the equivalence between the representations for
active and passive transformations, we thus conclude that states
prepared by Alice are represented by Charlie relative to his phase
reference by performing a passive transformation of $\phi$, using
the representation $U$ of U(1) on $K$ modes given by $U(\phi) =
\exp(i\phi\hat{N}_{\rm tot})$.  If $|\psi\rangle$ is the state
relative to Alice's phase reference, then this same state relative
to Charlie's phase reference is given by the transformed state
\begin{equation}\label{eq:GeneralStateindiffframe}
    U(\phi)|\psi\rangle = e^{i\phi\hat{N}_{\rm tot}}|\psi\rangle \,.
\end{equation}

For example, let Alice prepare the single-mode coherent state
\begin{equation}
  |\alpha\rangle \equiv \sum_{n=0}^\infty c_n |n\rangle\,, \quad c_n
  \equiv
  e^{-|\alpha|^2/2}\frac{\alpha^n}{\sqrt{n!}}\,,
\end{equation}
with $\alpha \in \mathbb{C}$; this state has a phase
$\arg(\alpha)$ relative to Alice's phase reference. Charlie would
describe this same state relative to his phase reference by a
coherent state with the same amplitude but with phase
$\arg(\alpha)+\phi$.  This passive transformation agrees with that
of Eq.~\eqref{eq:GeneralStateindiffframe} because
\begin{equation}\label{eq:CSindiffframe}
    |e^{i\phi}\alpha\rangle = e^{i\phi\hat{N}}|\alpha \rangle\,,
\end{equation}
where $\hat{N}$ is the number operator on this single mode.

As another example, let Alice prepare the two-mode state
$(|01\rangle + |10\rangle)/\sqrt{2}$.  Because this state is an
eigenstate of $\hat{N}_{\rm tot}$, the transformation $U(\phi)$
induces only an unobservable overall phase when acting on this
state.  Thus, Charlie also represents the state of the system as
$(|01\rangle + |10\rangle)/\sqrt{2}$ relative to his phase
reference.  This two-mode state is an example of an invariant state;
it is defined independently of any phase reference.

It will be useful for us to decompose the Hilbert space
$\mathcal{H}^{(K)}$ of $K$ modes into subspaces that transform in a
simple way under the action of the group U(1), as follows.  Defining
$\mathcal{H}_n$ to be the Hilbert space consisting of states of $K$
modes with precisely $n$ total photons, i.e., eigenspaces of
$\hat{N}_{\rm tot}$ with eigenvalue $n$, we can express the Hilbert
space $\mathcal{H}^{(K)}$ as a \emph{direct sum}
\begin{equation}
  \mathcal{H}^{(K)} = \bigoplus_{n=0}^{\infty}\mathcal{H}_{n}\,.
\end{equation}
Any state $|\psi_n\rangle \in \mathcal{H}_n$ transforms under phase
shifts, i.e., under the representation $U$ of U(1), as
\begin{equation}\label{eq:Uaction}
  U(\phi)|\psi_n\rangle = e^{in\phi}|\psi_n\rangle \,, \quad
  |\psi_n\rangle \in \mathcal{H}_n \,.
\end{equation}
Define $\Pi_n$ to be the projector onto $\mathcal{H}_n$.  Then an
arbitrary state $|\psi\rangle \in \mathcal{H}^{(K)}$ transforms as
\begin{equation}\label{eq:UactionOnEigenspaces}
    U(\phi)|\psi\rangle = \sum_n e^{in\phi}\Pi_n|\psi\rangle \,.
\end{equation}

Now consider the situation where Charlie has no knowledge of the
angle $\phi$ that relates his phase references to Alice's, i.e., the
laser serving as his phase reference is not phase-locked to
hers.\footnote{It should be noted that if the phase between Alice
and Charlie's references is changing in time in a known manner, then
the transformation relating their descriptions is still of the form
of Eq.~\eqref{eq:GeneralStateindiffframe} but with $\phi$ a function
of time.  Given that this function is known, the parties can
compensate for this effect.  However, a lack of knowledge of how
this phase is changing, for instance, an unknown drift, can
eventually lead to Alice and Charlie having no information about the
relative phase between their references.  In such a situation, the
timescale of the drift relative to the operations they perform is
critical; a slow drift may have negligible effect on a quick
protocol.} Let Alice prepare a quantum state $|\psi\rangle$ of $K$
modes relative to her phase reference. Given that $\phi$ is
completely unknown, one must average over its possible values to
obtain the state relative to Charlie. This averaging yields the
\emph{mixed} state
\begin{equation}
  \label{eq:LPNSSR1}
  \mathcal{U}\bigl[|\psi\rangle\langle\psi|\bigr]
  \equiv \int_0^{2\pi} \frac{{\rm d}\phi}{2\pi}\,
  U(\phi)|\psi\rangle\langle\psi| U(\phi)^\dag \,.
\end{equation}
Using Eq.~\eqref{eq:UactionOnEigenspaces} yields
\begin{align}
  \mathcal{U}\bigl[|\psi\rangle\langle\psi|\bigr]
  &= \int_0^{2\pi} \frac{{\rm d}\phi}{2\pi}\,\sum_{n,n'}
  e^{in\phi}\Pi_n|\psi\rangle\langle\psi|\Pi_{n'} e^{-in'\phi} \nonumber \\
  &= \sum_{n,n'} \Pi_n|\psi\rangle\langle\psi|\Pi_{n'}
  \Bigl( \int_0^{2\pi} \frac{{\rm d}\phi}{2\pi}\,
  e^{i(n-n')\phi} \Bigr) \nonumber \\
  &= \sum_{n} \Pi_n|\psi\rangle\langle\psi|\Pi_n \,.
\end{align}
Because this result applies to any state $|\psi\rangle$, we can
express the action of $\mathcal{U}$ on an arbitrary density operator
$\rho$ as
\begin{equation}
  \label{eq:LPNSSR2}
    \mathcal{U}[\rho] = \sum_{n} \Pi_n\rho\Pi_n \,.
\end{equation}
The map $\mathcal{U}$ removes all coherence between states of
differing total photon number on Alice's systems.  It follows in
particular that $\mathcal{U}[\rho]$ is invariant under phase shifts,
\begin{equation}
  \label{eq:phaseinvariance}
    [\mathcal{U}[\rho],U(\phi)] = 0\,,\quad \forall\ \phi \,.
\end{equation}

Thus, if states are described relative to Charlie's phase reference,
Alice faces a restriction in what she can prepare.  This restriction
is characterized by the quantum operation $\mathcal{U}$, which
ensures that Charlie will describe any state prepared by Alice as
block-diagonal in total photon number, or equivalently, as invariant
under phase shifts. We note in particular that the only pure states
that Alice can prepare are those which lie entirely within a single
eigenspace $\mathcal{H}_n$.

Now consider the related question for operations: If a unitary
operation $V$ is performed by Alice relative to her phase reference,
how is this operation described by Charlie relative to his phase
reference? Let $\sigma$ be the state of the system relative to
Charlie's phase reference.  To describe the action of $V$ on this
state if the angle $\phi$ that relates Charlie's phase reference to
Alice's is \emph{known}, Charlie could transform this state into
Alice's frame, then apply the unitary $V$, then transform back to
his frame; the resulting state is
\begin{equation}
    U(\phi)V U(\phi)^\dag \sigma U(\phi)V^\dag U(\phi)^\dag \,,
\end{equation}
relative to Charlie.  Thus, the operation is described by Charlie by
the unitary $V_{\phi} = U(\phi)V U(\phi)^\dag$. If the phase $\phi$
is \emph{unknown}, then Charlie would instead describe the operation
by an incoherent mixture of unitaries of this form, i.e., by the map
\begin{equation}
    \label{eq:U(1)supertwirling}
    \tilde{\mathcal{V}}[\sigma]
    \equiv \int_0^{2\pi} \frac{{\rm d}\phi}{2\pi}\,
    U(\phi)V U(\phi)^\dag \sigma U(\phi)V^\dag U(\phi)^\dag \,.
\end{equation}
A notable special case is if the system was prepared by Alice, so
that the state $\sigma$ relative to Charlie's RF is of the form
$\sigma = \mathcal{U}[\rho]$ as in Eq.~\eqref{eq:LPNSSR2}.  In this
case,
\begin{align}
    \label{eq:U(1)superblockdiagonal}
    \tilde{\mathcal{V}}[\sigma]
    &= \mathcal{U}[V \sigma V^{\dag}]\,,
\end{align}
so that $\tilde{\mathcal{V}}[\sigma]$ is also block-diagonal in
total photon number.  Thus, if operations are described relative to
Charlie's phase reference, then Alice experiences a restriction on
what operations she can perform.

We note that a restriction that requires states to be block-diagonal
in the eigenspaces of some operator is common in quantum theory:  it
is formally equivalent to a \emph{superselection rule}
(SSR)~\cite{Giu96}. Many superselection rules in non-relativistic
quantum theory, such as the superselection rule for
charge~\cite{Wic52}, are characterized by an inability to prepare
states with coherence between eigenspaces of some ``charge
operator'' corresponding to different eigenvalues.  Thus, we can
refer to the restriction described above as a superselection rule
for photon number~\cite{San03}. Alice cannot prepare, say, a
coherent state $|\alpha\rangle$ relative to Charlie's phase
reference, but she can prepare a phase-invariant state such as
$(|01\rangle + |10\rangle)/\sqrt{2}$.  In addition, she cannot
perform the unitary displacement operation that takes the vacuum
$|0\rangle$ to a coherent state $|\alpha\rangle$, but she can
perform any unitary operation on the two-dimensional subspace
spanned by $|01\rangle$ and $|10\rangle$.

We note that in the present context the SSR only restricts
preparations and operations \emph{by Alice} (or any party who does
not share Charlie's phase reference).  The SSR does not forbid
states with coherence between different total photon-number
eigenstates from existing within the theory, and in particular,
Charlie (or any party who \emph{does} share Charlie's phase
reference) experiences no such restriction on what states he can
prepare.  Thus, it makes sense within this context to consider what
manipulations Alice can perform under the restriction of an SSR on
general (possibly coherent) states.  For example, Alice can perform
the relative phase shift which takes the state
$(|0\rangle+|1\rangle)/\sqrt{2}$ to
$(|0\rangle-|1\rangle)/\sqrt{2}$.  Also, we note that Alice is able
to (incoherently) change the total photon number, i.e., she can
perform an operation that maps the vacuum $|0\rangle$ to the
single-photon state $|1\rangle$.  Thus, this restriction is not
equivalent to a conservation law for photon number.

\subsection{A general framework for reference frames and superselection rules}
\label{subsec:GeneralSSR}

In this section, we consider how to generalize the basic idea of the
previous section -- that lacking a reference frame leads to a
superselection rule -- beyond the case of a phase reference. We
present some formal mathematical tools, in particular, tools from
group theory and linear algebra, that we will use throughout this
review paper.

Suppose two parties, Alice and Charlie, are considering a single
quantum system described by a Hilbert space $\mathcal{H}$.  Let this
system transform via a group $G$ relative to some reference frame.
Throughout this review, we will consider both finite groups and
continuous (Lie) groups.  For the latter, we will restrict our
attention to Lie groups that (\textit{i}) are compact, so that they
possess a group-invariant (Haar) measure $\text{d}g$; and
(\textit{ii}) act on $\mathcal{H}$ via a \emph{unitary}
representation $T$, ensuring that they are completely
reducible~\cite{Sternberg}. Many of the techniques in this review
can be applied to other groups with some modification, but there are
many technical difficulties which are beyond the scope of this
review.

Let $g\in G$ be the group element relating Charlie's reference frame
to Alice's, i.e., the element in $G$ that describes the passive
transformation from Alice's to Charlie's reference frame.
Furthermore, suppose that $g$ is completely unknown, i.e., that
Alice's reference frame and Charlie's are uncorrelated.  It follows
that if Alice prepares a state $\rho$ on $\mathcal{H}$ relative to
her frame, the state of the system is represented relative to
Charlie's frame by the state\footnote{The invariant measure is
chosen using the maximum
  entropy principle: because Charlie has no prior knowledge about Alice's
  reference frame, he should assume a uniform measure over all
  possibilities.}
\begin{align}
  \tilde{\rho} &= \int_G \text{d}g\,
  T(g) \rho T^\dag(g) \nonumber \\
  &\equiv \mathcal{G}[\rho]  \, ,
  \label{eq:AveragedState}
\end{align}
with $T(g)$ a unitary representation of $g$ on $\mathcal{H}$, and
$\text{d}g$ the group-invariant (Haar) measure.\footnote{If the
group $G$ is instead a finite group, this expression is
$\mathcal{G}_{\rm finite}[\rho] \equiv |G|^{-1}\sum_{g\in G}
T(g)\rho T^\dag(g)$. In the following, we use the Lie group notation
exclusively; however, all results apply equally well to finite
groups.}  We call the operation $\mathcal{G}$ the ``$G$-twirling''
operation. If we choose to always represent preparations by Alice
relative to the reference frame of Charlie, then all states are of
the form $\tilde{\rho} = \mathcal{G}[\rho]$.

Any $\tilde{\rho}$ of this form satisfies
\begin{equation}
  \label{eq:StatesCommute}
  [\tilde{\rho},T(g)]=0\,, \quad \forall\ g\in G\,.
\end{equation}
and thus is said to be \emph{$G$-invariant}.  The proof follows from
the fact that $T(g)\tilde{\rho}T^{\dag}(g)=\int_G \text{d}g'\,
  T(gg') \rho T^\dag(gg')=\tilde{\rho}$.

Let $\mathcal{B}(\mathcal{H})$ denote the set of all bounded
operators on $\mathcal{H}$. Given that $\mathcal{B}(\mathcal{H})$
forms a Hilbert space under the Hilbert-Schmidt inner product
$(\sigma,\tau) = {\rm Tr}(\sigma^\dagger\tau)$, linear maps can be
regarded as operators acting on $\mathcal{B}(\mathcal{H})$. These
are called \textit{superoperators} to distinguish them from
operators acting on $\mathcal{H}$. It is useful to define the
superoperator $\mathcal{T}(g)$ by $\mathcal{T}(g)[\rho] = T(g)\rho
T^\dag(g)$, which is the unitary representation of $G$ on
$\mathcal{B}(\mathcal{H})$.  We may then express $\mathcal{G}$
simply as $\mathcal{G} = \int_G \text{d}g\, \mathcal{T}(g)$.

We now consider the representation of transformations. The most
general transformation upon a quantum system, i.e., the most general
quantum operation, is represented by a completely
positivity-preserving superoperator
$\mathcal{E}:\mathcal{B}(\mathcal{H})\rightarrow\mathcal{B}(\mathcal{H})$.
(See~\textcite{Nie00} for the definition and properties of these
superoperators.)  The question of interest to us is the following:
if an operation is represented by the superoperator $\mathcal{E}$
relative to Alice's frame, how is this same operation represented
relative to Charlie's frame? Generalizing the justification given
for Eq.~(\ref{eq:U(1)supertwirling}) in the case of a phase
reference, we conclude that relative to Charlie's frame the
operation is represented by the superoperator $\mathcal{\tilde{E}}$,
where
\begin{equation}
\label{eq:GInvariantOperationsaction}
  \mathcal{\tilde{E}}[\rho]= \int_G
  \text{d}g\,T(g)\mathcal{E}[T^{\dag}(g)\rho T(g)]T^{\dag}(g)\,,
\end{equation}
or, equivalently,
\begin{equation}
  \label{eq:GInvariantOperations}
  \mathcal{\tilde{E}}= \int_G
  \text{d}g\,\mathcal{T}(g)\circ \mathcal{E} \circ
  \mathcal{T}(g^{-1})\,,
\end{equation}
where
$\mathcal{A}\circ\mathcal{B}[\rho]=\mathcal{A}[\mathcal{B}[\rho]].$
Given that $\mathcal{T}(g)$ is a representation of $G$ on
$\mathcal{B}(\mathcal{H})$, Eq.~(\ref{eq:GInvariantOperations}) has
the form of Eq.~(\ref{eq:AveragedState}) except with operators
replaced by superoperators.  We therefore refer to the map taking
$\mathcal{E}$ to $\mathcal{\tilde{E}}$ as ``super-$G$-twirling''.
Any superoperator of the form of $\mathcal{\tilde{E}}$ satisfies
\begin{equation}
  \label{eq:OperationsCommute}
  [\mathcal{\tilde{E}},\mathcal{T}(g)]=0\,, \quad \forall\ g\in G\,,
\end{equation}
where
$[\mathcal{A},\mathcal{B}]=\mathcal{A}\circ\mathcal{B}-\mathcal{B}\circ\mathcal{A}$
is the superoperator commutator.  Thus, $\mathcal{\tilde{E}}$ is
invariant under the action of $G$; it is a $G$-invariant operation.

The superoperator $\mathcal{\tilde{E}}$ acts on a $G$-invariant
operator $\tilde{A}$ as
\begin{align}
  \mathcal{\tilde{E}}[\tilde{A}] &= \int_G
  \text{d}g\,\mathcal{T}(g)\circ \mathcal{E} \circ
  \mathcal{T}(g^{-1})[\mathcal{G}[\tilde{A}]] \nonumber \\
  &= \int_G \text{d}g\,\mathcal{T}(g)\circ \mathcal{E} \circ
  \mathcal{G}[\tilde{A}] \nonumber \\
  &= \mathcal{G}\circ \mathcal{E}\circ \mathcal{G}[\tilde{A}] \,,
\end{align}
where we have used the fact that $\tilde{A}=\mathcal{G}[\tilde{A}]$
and $\mathcal{T}(g^{-1})\circ\mathcal{G}=\mathcal{G}$.

Every completely positivity-preserving superoperator admits an
operator-sum decomposition of the form $\mathcal{E}[\rho]=\sum_k
A_k\rho A_k^{\dag}$ where the $A_k$ are called \emph{Kraus}
operators. Clearly, a sufficient condition for an operation
$\mathcal{E}$ to be a $G$-invariant operation is for all of its
Kraus operators $A_k$ to be $G$-invariant operators.  In general,
however, this is not a necessary condition. Note, in particular,
that if $V$ is a unitary operator that is $G$-invariant, then
$\mathcal{V}[\cdot] = V (\cdot) V^{\dag}$ is a $G$-invariant
superoperator and the associated unitary transformation can be
implemented without an RF. Nonetheless, there may exist
$G$-invariant superoperators arising from $G$-noninvariant unitary
operators.

Finally, we consider the representation of measurements. The most
general measurement on a quantum system is represented by a set of
completely positivity-preserving superoperators $\{\mathcal{E}_k\}$,
the sum of which is trace-preserving. The probability of outcome $k$
for the measurement is $p_k=\text{Tr}(\mathcal{E}_k[\rho])$ and upon
obtaining this outcome, $\rho$ is updated to
$\mathcal{E}_k[\rho]/p_k$.  The probability of outcome $k$ may also
be specified by $p_k=\text{Tr}(E_k \rho)$ where the set $\{ E_k \}$
is a positive operator valued measure (POVM) (defined by the
conditions $E_k\ge 0$ and $\sum_k E_k=I$).  The POVM $\{ E_k \}$
that is associated with a measurement is obtained from the set of
superoperators $\{ \mathcal{E}_k \}$ associated with it by
$E_k=\mathcal{E}^{\dag}_k[I]$, where the adjoint of a superoperator
is defined relative to the Hilbert-Schmidt inner product on the
operator space, ${\rm Tr}(\mathcal{E}^\dagger(\sigma)\tau) = {\rm
Tr}(\sigma\,\mathcal{E}(\tau))$.

Recalling how operations transform under a change of reference
frame, if a measurement is represented by the set of superoperators
$\{ \mathcal{E}_k \}$ relative to Alice's frame, then it is
represented by the set of superoperators $\{ \tilde{\mathcal{E}}_k
\}$ relative to Charlie's, where $\tilde{\mathcal{E}}_k$ is given by
Eq.~(\ref{eq:GInvariantOperations}). Taking the superoperator
adjoint of Eq.~(\ref{eq:GInvariantOperations}), and using the fact
that $E_k=\mathcal{E}^{\dag}_k[I]$, it follows that the POVM $\{ E_k
\}$ relative to Alice's frame is represented by the POVM $\{
\tilde{E}_k \}$ relative to Charlie's frame where
\begin{equation}
  \label{eq:GinvariantPOVM}
  \tilde{E}_k=\mathcal{G}[E_k]\,.
\end{equation}
It follows that
\begin{equation}
  \label{eq:GinvariantPOVM2}
  [\tilde{E}_k,T(g)]=0\,, \quad \forall\ g\in G\,,
\end{equation}
that is, the POVM $\{ \tilde{E}_k \}$ is G-invariant.

Thus, relative to Charlie's reference frame, the preparations,
operations and measurements that Alice can implement are represented
by states, superoperators and POVMs of the form of
\eqref{eq:AveragedState}, \eqref{eq:GInvariantOperations}, and
\eqref{eq:GinvariantPOVM}, respectively.  We now demonstrate that
this restriction has the same mathematical characterization as that
of a superselection rule for a (possibly non-Abelian) group $G$.

First, we note that the representation $T$ of the group $G$ allows
for a decomposition of the Hilbert space into \emph{charge sectors}
$\mathcal{H}_q$, labeled by an index $q$, as
\begin{equation}
    \label{eq:Hdecomp1}
    \mathcal{H} = \bigoplus_q \mathcal{H}_q \,,
\end{equation}
where each charge sector carries an inequivalent representation
$T_q$ of $G$.  In the U(1) phase reference example presented above,
the charge sectors corresponded to eigenspaces of total photon
number.  Each sector can be further decomposed into a tensor
product,
\begin{equation}
    \label{eq:Hdecomp2}
  \mathcal{H}_q = \mathcal{M}_q \otimes \mathcal{N}_q \,,
\end{equation}
of a subsystem $\mathcal{M}_q$ carrying an irreducible
representation (irrep) $T_q$ of $G$ and a subsystem $\mathcal{N}_q$
carrying a trivial representation of $G$.  (Recall that a
representation acts irreducibly on a space if there are no invariant
subspaces.)  Note that this tensor product does not correspond to
the standard tensor product obtained by combining multiple qubits:
it is \emph{virtual}~\cite{Zan01b}.  The spaces $\mathcal{M}_q$ and
$\mathcal{N}_q$ are therefore \emph{virtual subsystems}.  The
$\mathcal{M}_q$ and $\mathcal{N}_q$ are sometimes referred to as
\emph{gauge spaces} and \emph{multiplicity spaces}
respectively.\footnote{In high energy physics, the $\mathcal{M}_q$
are called colour spaces and the $\mathcal{N}_q$ are called flavour
spaces.} For the U(1) phase reference example, the subsystems
$\mathcal{M}_q$ are one-dimensional, and so the additional tensor
product structure within the irreps is not required; for a general
superselection rule corresponding to a non-Abelian group $G$,
however, they can be non-trivial.

Expressed in terms of this decomposition of the Hilbert space, the
map $\mathcal{G}$ takes a particularly simple form.  Because of the
broad utility of this form, we present it as a theorem.
\medskip

\begin{thmnonum}
The action of $\mathcal{G}$ in terms of the decomposition
\begin{equation}
    \label{eq:HdecompFull}
  \mathcal{H} = \bigoplus_q \mathcal{M}_q \otimes \mathcal{N}_q \,,
\end{equation}
is given by
\begin{equation}
  \label{eq:GeneralDecoherenceDfullDfree}
  \mathcal{G} = \sum_q (\mathcal{D}_{\mathcal{M}_q} \otimes
  \mathcal{I}_{\mathcal{N}_q}) \circ \mathcal{P}_q \,,
\end{equation}
where $\mathcal{P}_q$ is the superoperator associated with
projection into the charge sector $q$, that is,
$\mathcal{P}_q[\rho]= \Pi_q \rho \Pi_q$ with $\Pi_q$ the projection
onto $\mathcal{H}_q = \mathcal{M}_q \otimes \mathcal{N}_q$,
$\mathcal{D}_{\mathcal{M}}$ denotes the trace-preserving operation
that takes every operator on the Hilbert space $\mathcal{M}$ to a
constant times the identity operator on that space, and
$\mathcal{I}_{\mathcal{N}}$ denotes the identity map over operators
in the space $\mathcal{N}$.
\end{thmnonum}

We provide a short proof of this theorem at the end of this section.

Note that the operation $\mathcal{G}$ has the general form of
\emph{decoherence}.  Whereas decoherence typically describes
correlation with an environment to which one does not have access,
in this case the decoherence describes correlation to a reference
frame to which one does not have access.  Given that $\mathcal{G}$
acts as identity on subsystems $\mathcal{N}_q$, these subsystems are
called \emph{decoherence-free subsystems} (also known as
\emph{noiseless subsystems})~\cite{Zan97,Kni00}.  In stark contrast,
$\mathcal{G}$ acts as the completely depolarizing operation on the
subsystems $\mathcal{M}_q$; these are called \emph{decoherence-full
subsystems}~\cite{BRS04a}.

It follows, in particular, that a $G$-invariant operator
$\tilde{A}=\mathcal{G}(A)$ must have the form
\begin{equation}
  \label{eq:Ginvariantoperator}
  \tilde{A} = \bigoplus_q I_{\mathcal{M}_q} \otimes A_{\mathcal{N}_q}\,,
\end{equation}
where the $I_{\mathcal{M}_q}$ are identity operators on the
subsystems $\mathcal{M}_q$ and the $A_{\mathcal{N}_q}$ are arbitrary
operators on the subsystems $\mathcal{N}_q$.


We are now in a position to see how the restriction of lacking a
reference frame for the group $G$ is equivalent to the standard
notion of a superselection rule associated with this group.
Superselection rules are most commonly discussed in the context of
Abelian groups where they can be described simply as a restriction
of the physical states and observables to those that are
block-diagonal with respect to the inequivalent representations of
$G$~\cite{Giu96}. (Occasionally, this restriction is argued to hold
for the observables alone, but in this case every state that is not
restricted in this way is operationally indistinguishable from a
state that is, so one may as well assume this restriction for the
states also.) The standard notion of a superselection rule for an
arbitrary (possibly non-Abelian) group $G$ is a restriction of the
physical states and observables to those that commute with every
element of $G$~\cite{Giu96}. This restriction on states is precisely
what is asserted in Eq.~\eqref{eq:StatesCommute}, and the
restriction on observables is simply Eq.~\eqref{eq:GinvariantPOVM2}
applied to the special case of a projective measurement. The
restriction on transformations has traditionally only been
articulated for unitary transformations and asserts that only
$G$-invariant Hamiltonians are physical.  This is equivalent to
asserting that the unitary itself be $G$-invariant, and such
unitaries were identified above as the only ones that can be
achieved when lacking an RF for the group $G$.
Eq.~\eqref{eq:OperationsCommute} is a generalization of this
restriction to irreversible transformations. Thus, one can view the
restrictions of
Eqs.~\eqref{eq:StatesCommute},~\eqref{eq:OperationsCommute} and
\eqref{eq:GinvariantPOVM2} as the formalization of the restrictions
of a superselection rule associated with the group $G$ in the
language of quantum information theory. We shall say that the
restriction due to the lack of a reference frame for $G$ is
equivalent to a superselection rule associated with the group $G$.

We note that although the term ``superselection rule'' was initially
introduced to describe an \emph{axiomatic} restriction on quantum
states, observables, and operations~\cite{Wic52}, it has been
emphasized by \textcite{Aha67} that whether or not coherent
superpositions of a particular observable are possible is a
practical matter, depending on the availability of a suitable
reference system.  We return to this issue in Sec.~\ref{sec:QRFs}.

Finally, although we have thus far mentioned only the two limiting
possibilities for the correlations that might hold between Alice and
Charlie's reference frames -- completely correlated or completely
uncorrelated -- in general one might wish to consider the
intermediate scenario wherein they are partially correlated. To
model this, one replaces the uniform Haar measure appearing in
Eq.~(\ref{eq:AveragedState}) with the non-uniform measure that
characterizes Charlie's partial knowledge of the group element $g$
in order to obtain a \emph{weighted G-twirling} operation. Like
$G$-twirling, this operation is noiseless on the multiplicity
spaces, but unlike $G$-twirling, which is completely decohering on
the gauge spaces, the weighted $G$-twirling operation is only
partially decohering on these spaces.

\begin{proof}[Proof of Theorem 1]
Our proof, which follows~\textcite{Nie03}, will make use of two
central theorems of group representation theory known as Schur's
Lemmas. We state these lemmas here without proof.
\begin{lemma}[Schur's first]
If $T(g)$ is an irreducible representation of the group $G$ on the
Hilbert space $\mathcal{H}$, then any operator $A$ satisfying
$T(g)AT^{\dag}(g)=A$ for all $g\in G$ is a multiple of the identity
on $\mathcal{H}$.
\end{lemma}
\begin{lemma}[Schur's second]
If $T_1(g)$ and $T_2(g)$ are inequivalent representations of $G$,
then $T_1(g)AT_2^{\dag}(g)=A$ for all $g\in G$ implies $A=0$.
\end{lemma}
We begin by decomposing the representation $T(g)$ appearing in
Eq.~(\ref{eq:AveragedState}) into a sum of irreducible
representations, $T(g)=\oplus_{q,\lambda} T_{q,\lambda}(g)$ where
$q$ labels inequivalent irreps and $\lambda$ is a multiplicity
index. It follows that
\begin{equation}
    \label{eq:Proof1}
    \mathcal{G}[A]=\bigoplus_{q,q',\lambda,\lambda'} \int \mathrm{d}g\,
    T_{q,\lambda}(g) A T_{q',\lambda'}^\dag(g)\,.
\end{equation}
Define $A_{q,q',\lambda,\lambda'} =  \int \mathrm{d}g\,
T_{q,\lambda}(g) A T_{q',\lambda'}^\dag(g)$.  Because of the
invariance of the measure $\text{d}g$, it follows that
\begin{equation}
    T_{q,\lambda}(g) A_{q,q',\lambda,\lambda'} T_{q',\lambda'}^\dag(g)
    = A_{q,q',\lambda,\lambda'}\,, \quad \forall\ g \in G\,.
\end{equation}
Thus, by Schur's second lemma, $A_{q,q',\lambda,\lambda'}=0$ for
$q\ne q'$. Eq.~\eqref{eq:Proof1} can then be expressed as
\begin{equation}
  \mathcal{G}[A]=\bigoplus_{q,\lambda,\lambda'} \int \mathrm{d}g\,
  T_{q,\lambda}(g) A T_{q,\lambda'}^\dag(g)\,.
\end{equation}
Let $\Pi_{q,\lambda}$ be the projection of $\mathcal{H}$ onto the
carrier space of $T_{q,\lambda}$, and let $\Pi_q = \sum_\lambda
\Pi_{q,\lambda}$.  Then the above equation can be expressed as
\begin{equation}
  \mathcal{G}[A]=\sum_{q,\lambda,\lambda'} \int \mathrm{d}g\,
  T_{q,\lambda}(g) \Pi_q A \Pi_q T_{q,\lambda'}^\dag(g)\,,
\end{equation}
and thus we can express $\mathcal{G}$ as
\begin{equation}
  \mathcal{G} = \sum_q \mathcal{G}_q \circ \mathcal{P}_q \,,
\end{equation}
where $\mathcal{G}_q[A_q] = \sum_{\lambda,\lambda'} \int
\mathrm{d}g\, T_{q,\lambda}(g) A_q T_{q,\lambda'}^\dag(g)$ is a
superoperator on $\mathcal{H}_q$, and recall that $\mathcal{P}_q[A]
= \Pi_q A \Pi_q$.

We now determine the form of $\mathcal{G}_q$ in terms of the tensor
product structure $\mathcal{H}_q=\mathcal{M}_q\otimes
\mathcal{N}_q$.  The projector $\Pi_{q,\lambda}$ can be expressed in
terms of this tensor product as $\Pi_{q,\lambda} =
\Pi_{\mathcal{M}_q} \otimes \Pi_\lambda$, where
$\Pi_{\mathcal{M}_q}$ is the projector onto $\mathcal{M}_q$, and
$\Pi_\lambda$ is the rank-1 projector on $\mathcal{N}_q$ that
``picks out'' the representation $\lambda$ of $G$.  The rank-1
projectors $\Pi_\lambda$ form a basis for $\mathcal{N}_q$, so that
$\sum_\lambda \Pi_\lambda$ is the identity on $\mathcal{N}_q$.
Given that $T_q(g)$ acts nontrivially only on $\mathcal{H}_q$, we
can write $T_{q,\lambda}(g)= T_q(g)\otimes \Pi_{\lambda}$. It
follows that
\begin{align}
    \mathcal{G}[A] &= \sum_{q,\lambda,\lambda'} \int \mathrm{d}g\,
    \big(T_q(g) \otimes \Pi_\lambda\big)
    \Pi_q A \Pi_q \big( T_q^\dag(g) \otimes
    \Pi_{\lambda'}\big) \nonumber \\
    &= \sum_q \int \mathrm{d}g\,
    \big(T_q(g) \otimes \sum_\lambda \Pi_\lambda\big)
    \Pi_q A \Pi_q \big( T_q^\dag(g) \otimes \sum_{\lambda'}
    \Pi_{\lambda'}\big) \nonumber \\
    &=\sum_{q} (\mathcal{G}_{\mathcal{M}_q} \otimes
    \mathcal{I}_{\mathcal{N}_q})\circ \mathcal{P}_q[A] \,,
\end{align}
where the superoperator $\mathcal{G}_{\mathcal{M}_q}$ takes an
operator $B$ on $\mathcal{M}_q$ to $\mathcal{G}_{\mathcal{M}_q}[B]=
\int \mathrm{d}g\, T_{q}(g) B T_{q}^\dag(g)$.  By Schur's first
lemma, $\mathcal{G}_{\mathcal{M}_q}[B]$ is a multiple of identity on
$\mathcal{M}_q$. Therefore, because the map $\mathcal{G}$ is
trace-preserving, $\mathcal{G}_{\mathcal{M}_q}
=\mathcal{D}_{\mathcal{M}_q}$, the trace-preserving map that takes
every operator on $\mathcal{M}_q$ to a constant times the identity
on $\mathcal{M}_q$.
\end{proof}

\section{Quantum information without a shared reference frame}
\label{sec:NoSRF}

In implementing multi-partite cryptographic and communication tasks
using quantum systems, it is generally presumed, at least
implicitly, that all parties share perfect reference frames for all
relevant degrees of freedom. Moreover, one might think that in order
to achieve some or all of these tasks, they \emph{must} share such
reference frames; for instance, one might think that if they wish to
achieve quantum communication using the Fock space of an optical
mode, they must share a phase reference, and if they wish to do so
using spin-1/2 systems, they must share a reference frame for
spatial orientation. This impression is mistaken; quantum
information processing tasks can be achieved without first
establishing a shared reference frame by using entangled states of
multiple systems, that is, relational encodings.

A classical analogue is elucidating. If two parties do not share a
Cartesian frame, then they cannot communicate any classical
information to one another through encodings in the directional
degree of freedom of a system.  For instance, if Alice encodes
information into the orientation, relative to her frame, of a
physical arrow or gyroscope, Bob cannot access this information
because he can compare the system with his frame only.  Nonetheless,
they can still communicate by encoding information in the
\emph{relative} orientations of two or more such systems.  We shall
be concerned with the quantum analogue of such relational encodings.

The essential idea is to use the result, presented in
Sec.~\ref{sec:Formal}, that the effect of lacking a shared RF can be
expressed as a form of decoherence.  We then make use of the
techniques of decoherence-free subspaces and
subsystems~\cite{Kem01,Kni00,Zan97} to find quantum states that are
protected from the noise.  These techniques (and variants thereof)
can be interpreted as yielding relational encodings. They are in
fact ideally suited to the problem of overcoming the lack of a
shared RF because the existence of decoherence-free subspaces and
subsystems relies on there being nontrivial symmetries in the noise,
something that may not occur for a realistic noise model, but which
is guaranteed to occur in the present context. For instance, in
order to redescribe, relative to one RF, a qubit state that is
defined relative to a second, uncorrelated RF, one must apply to it
an unknown unitary. To redescribe, relative to this RF, \emph{many}
qubits that were all prepared relative to the same RF, one must
apply precisely the same unitary to each.

We begin in Sec.~\ref{subsec:NoSRFComm} by applying these techniques
and others to determine the efficiency with which classical and
quantum communication can be performed in the presence of such
noise.  The implications for quantum key distribution are discussed
in Sec.~\ref{subsec:QKD}. We also discuss the important issue of
sharing \emph{entanglement} between two parties who lack a shared
RF; we demonstrate in Sec.~\ref{subsec:Entanglement} that a rich
structure emerges in bipartite entanglement of pure states when this
restriction applies.  Finally, in Sec.~\ref{subsec:PrivateSRFs}, we
investigate the cryptographic power of \emph{private} shared RFs,
where it is assumed that it is an eavesdropper Eve who fails to have
a sample of Alice and Bob's RF.

\subsection{Communication without a shared reference frame}
\label{subsec:NoSRFComm}

\subsubsection{Communication using photons without a shared phase
reference} \label{subsubsec:NoSRFPhase}

Consider the following problem:  Alice wants to communicate some
amount of classical or quantum information to Bob using an optical
channel, i.e., using quantum states of some number of optical modes,
when they do not share a phase reference.  Using the formalism of
Sec.~\ref{subsec:PNSSR}, a state $\rho$ prepared by Alice is
represented by Bob as the (generally mixed) state $\mathcal{U}[\rho]
= \sum_{n} \Pi_n\rho\Pi_n$.  This problem thus takes the form of a
more standard one from quantum communication:  how to communicate
quantum or classical information through a noisy channel described
by a decoherence map $\mathcal{U}$.

The communication may be constrained in some additional way, such as
by a limit on the number of usable optical modes, or by an energy
limit that bounds the maximum number of photons that can be
transmitted, or both.  Because of these constraints, Alice and Bob
wish to use a communication protocol that makes optimal use of these
resources.

Let's first consider classical communication.  The simplest possible
problem is the one wherein Alice is restricted to sending at most
one photon to Bob, using a single optical mode.  Clearly, using such
a channel, Alice can communicate a single classical bit to Bob by
sending either a single photon $|1\rangle$ or no photon (the vacuum)
$|0\rangle$.\footnote{Such an encoding is not feasible in practice,
because all photon-based communication schemes rely on obtaining a
detector event (a ``click'') for every message. Specifically, the
detection of the vacuum cannot be discriminated from an event where
the photon is lost, or missed by the detector.} This protocol does
not rely on Alice and Bob sharing a phase reference, because both
the states $|0\rangle$ and $|1\rangle$ are invariant under the
superoperator $\mathcal{U}$. Generalizing this result, if Alice can
send at most $N$ photons in a single mode, she can communicate $N+1$
classical messages (equivalently, $\log_2(N+1)$ classical bits) to
Bob. With $K>1$ modes, one has to consider all the possible ways of
distributing $N$ photons among $K$ modes. The dimension of the
resulting Hilbert space is $(N+K)!/N!K!$, and specifies the number
of classical messages Alice can communicate using eigenstates of
photon number.

What about quantum communication?  Again, consider a situation
wherein Alice is restricted to sending at most a single photon to
Bob using a single optical mode. Any state $\rho_1$ Alice prepares
must then have support on the qubit Hilbert space spanned by
$\{|0\rangle,|1\rangle\}$, and any such state is represented by
Bob as $\mathcal{U}[\rho_1] = p_0|0\rangle\langle 0| +
p_1|1\rangle\langle 1|$ for $p_i = \langle i|\rho_1|i\rangle$,
i.e., as an \emph{incoherent mixture} of the zero- and one-photon
states. Any qubit state is completely depolarized according to
Bob.  Thus, quantum communication cannot be performed by using
only a single mode with at most one photon.  This negative result
is one of many disadvantages to this encoding of a qubit into
states spanned by $|0\rangle$ and $|1\rangle$, known as the
``single-rail'' encoding~\cite{Kok06}. Clearly, no quantum
communication can be performed using \emph{any} number of photons
in a single mode, because Bob represents all states prepared by
Alice as being diagonal in the photon number basis.

Now consider the case where Alice can make use of two modes in her
communication to Bob.  Noting that Bob will represent any
preparation by Alice as block-diagonal in the eigenspaces of
\emph{total} photon number, Alice should prepare states lying in
just one of these eigenspaces if she wishes to communicate quantum
information.  For example, the one-photon eigenspace of two modes
(labelled $a$ and $b$) is two-dimensional, and a general pure state
on this eigenspace has the form
\begin{equation}\label{eq:DualRail}
    |\psi\rangle_{n=1} = \alpha |1\rangle_a|0\rangle_b + \beta
    |0\rangle_a|1\rangle_b \,,
\end{equation}
for $\alpha,\beta\in\mathbb{C}$ satisfying $|\alpha|^2 + |\beta|^2 =
1$.  Any such state satisfies
$\mathcal{U}[|\psi\rangle_1\langle\psi|] =
|\psi\rangle_1\langle\psi|$; this two-dimensional subspace is a
\emph{decoherence-free subspace} of $\mathcal{U}$. Using states of
this form, Alice can communicate a single qubit to Bob without
requiring a shared phase reference. We note that this encoding is
the commonly-used ``dual-rail'' encoding of optical quantum
computing~\cite{Kok06}.  Evidently, to communicate quantum
information using at most $N$ photons in $M$ modes without a shared
phase reference, Alice and Bob should make use of the eigenspace of
total photon number $N'$ ($N'\le N$) that has the largest dimension.
This eigenspace is the one corresponding to $N'=N$, and has
dimension $(N+K-1)!/N!(K-1)!$.

Using multiple modes of the optical field raises addition issues
regarding the use of reference frames, depending on how these modes
are identified, and this can lead to a much richer structure.  For
example, in the dual-rail encoding of Eq.~\eqref{eq:DualRail}, the
modes $a$ and $b$ could represent different spatial or temporal
modes, in which case Alice and Bob would require a shared Cartesian
frame or a clock in order to identify these modes.  Another common
implementation for this encoding is for $a$ and $b$ to represent the
two polarization modes of the single photon (for example, horizontal
and vertical polarization) -- a so-called ``polarization
encoding''~\cite{Kok06}. For Alice and Bob to share quantum
information using such an encoding, although they do not need to
share a phase reference, they \emph{do} need to share a reference
frame for polarization, i.e., to agree on an axis for their
polarizing materials that are used to prepare, manipulate, and
measure such states.  The efficiencies of general schemes for
transmitting quantum information via the polarization and phase of
optical modes when parties do not share a reference frame for
polarization have been fully characterized~\cite{Bal05,Bal06}.

Recently, optical quantum information experiments have made use of
the spatial mode structure of light~\cite{Mai01,Vaz03,Lan04}; use of
this degree of freedom requires a shared reference frame for both
position and orientation.  Using spatial modes, it is possible to
restrict attention to states of a single photon with a fixed orbital
angular momentum (the standard basis for which is the
Laguerre-Gauss-Vortex modes~\cite{Sie86}). Encodings into a subspace
of fixed orbital angular momentum will be invariant under rotations
about the direction of propagation, and thus will not require a
shared reference frame for orientation about this direction. These
encodings do require a shared reference frame for the direction of
propagation, and also a precise determination of the separation
between parties in order to compensate for the relative phase (Gouy
shift) acquired between different states of fixed orbital angular
momentum during propagation~\cite{Spe04}.

\subsubsection{Communication without a shared Cartesian frame}
\label{subsec:NoSRFCommSpins}

We now turn our attention to the problem of how Alice and Bob can
perform both classical and quantum communication through the
exchange of spin-1/2 systems (qubits) when they lack a shared
Cartesian frame~\cite{BRS03}. This problem has a much richer
structure than the phase-reference case investigated above, due to
the existence of decoherence-free \emph{subsystems} (rather than
subspaces).  For simplicity, we consider a noiseless channel that
transmits these spin-1/2 systems from Alice to Bob; these results
can be extended to noisy channels or higher-dimensional (spin
$>1/2$) systems~\cite{Enk06,Byr06}.

The group of transformations of orientation relative to a Cartesian
frame is SO(3), which we will extend to SU(2) to allow for spinor
representations.  We will denote an element of SU(2) by $\Omega,$
which might represent, for instance, a set of three Euler angles. In
the case of a single spin-1/2 system, the Wigner rotation operators
$R(\Omega)$ provide an irreducible representation of SU(2). If Alice
sends $N$ spin-1/2 systems to Bob, and she describes these, relative
to her Cartesian frame, by $\rho$ then Bob describes these same
spins relative to his Cartesian frame by
\begin{equation}
    \mathcal{E}_{N}[\rho]
    =\int\mathrm{d}\Omega\,R(\Omega)^{\otimes N}\rho R^{\dag}(\Omega)^{\otimes N}\,.
    \label{eq:NQubitDecoheringChannel}
\end{equation}
That is, he averages over all passive rotations that might relate
his frame to Alice's, and every rotation acts on each of the $N$
spins identically as $R(\Omega)^{\otimes N}$ because each spin
experiences the same rotation by virtue of the fact that each is
prepared relative to the same Cartesian frame.  We refer to this
representation of SO(3) as \emph{collective}.  Thus, Bob's lack of
Alice's Cartesian frame has the same effect as collective noise on
the channel. It is still possible for Alice and Bob to communicate
by encoding in the \emph{relational} degrees of freedom of the
qubits, as we shall see.

The problem of determining the communication capacities in the
presence of this restriction is quite simple if we decompose the
Hilbert space in the manner dictated by Eqs.~\eqref{eq:Hdecomp1} and
\eqref{eq:Hdecomp2}. We begin with some simple examples,
illustrating the basic techniques and some few-qubit schemes for
classical and quantum communication, before presenting the general
results.

\paragraph{One transmitted qubit.}

Given that $R(\Omega)$ is an irreducible representation on
$\mathcal{H}_{1/2},$ by Schur's lemma, the SU(2)-twirling on one
qubit is equivalent to the completely depolarizing operation,
\begin{equation}
    \label{eq:twirling1qubit}
    \mathcal{E}_{1}=\mathcal{D}_{\mathcal{H}_{1/2}}\,.
\end{equation}
Thus, if Alice prepares a single qubit in the state $\rho$ and
transmits it to Bob, he represents the state of this received qubit
as the completely mixed state
\begin{equation}
    \label{eq:twirling1qubit_states}
    \mathcal{E}_{1}[\rho]=\tfrac{1}{2}I\,.
\end{equation}
Consequently, Bob can infer nothing about $\rho$ from the outcome of
any measurement.  So, without a shared RF, Alice cannot communicate
any information to Bob using only a single qubit.

\paragraph{Two transmitted qubits: a classical channel.}

The unitary representation $R(\Omega)^{\otimes 2}$ of SU(2) is
reducible.  To decompose it into irreducible representations, we
briefly review the representation theory of SU(2).

The inequivalent representations of SU(2) are labeled by the total
angular momentum $J^{2}$ quantum number $j$.  The carrier spaces of
these representations are the charge sectors $\mathcal{H}_{j}$. The
carrier spaces of the irreducible representations, the gauge spaces,
are denoted $\mathcal{M}_{j}$.  Such spaces have dimensionality
$2j+1,$ and may be decomposed into a basis $|j,m\rangle$ of
eigenstates of $J_z$ with eigenvalues $\hbar m$ where
$m\in\{-j,-j+1,\dots ,j\}.$ The multiplicity spaces
$\mathcal{N}_{j}$ arise when there are different ways of coupling
multiple systems to a given total angular momentum. A pair of spins
with angular momentum numbers $j_{1}$ and $j_{2}$ couple to any
total angular momenta $j$ satisfying $|j_{1}-j_{2}|\leq j\leq
j_{1}+j_{2}$.  We summarize this as $j_{1}\otimes j_{2}= |j_{1}
-j_{2}|\oplus\cdots\oplus(j_{1}+j_{2})$.

It follows that for a pair of spin 1/2 systems, we have $(\frac{1}
{2})^{\otimes2}=\frac{1}{2}\otimes\frac{1}{2}=0\oplus1.$  The
possible total angular momenta are $j=0$ and $j=1$ and each has
multiplicity $1$.  The joint eigenstates of total angular momentum
operators $J^{2}$ and $J_z$, denoted $|j,m\rangle$, form a basis of
the Hilbert space (the coupled representation).  We can relate this
coupled basis to the joint eigenstates of $J_{1}^{2}$, $J_{1z}$,
$J_{2}^{2}$, $J_{2z}$, denoted by $\left\vert
j_{1},m_{1}\right\rangle \otimes\left\vert j_{2},m_{2}\right\rangle
$ (the uncoupled representation) by
\begin{align}
\left\vert 1,1\right\rangle  & =\left\vert 00\right\rangle \\
\left\vert 1,0\right\rangle  & =\left(  \left\vert
01\right\rangle +\left\vert 10\right\rangle \right)/\sqrt{2}  \\
\left\vert 1,-1\right\rangle  & =\left\vert 11\right\rangle \\
\left\vert 0,0\right\rangle  & =\left(  \left\vert 01\right\rangle
-\left\vert 10\right\rangle \right)/\sqrt{2} \equiv\left\vert \psi
^{-}\right\rangle
\end{align}
where $\left\vert 0\right\rangle$ ($\left\vert 1\right\rangle$) is
the quantum information-theoretic shorthand for $\left\vert
1/2,\pm1/2\right\rangle$, and $\left\vert 01\right\rangle
\equiv\left\vert 0\right\rangle \otimes\left\vert 1\right\rangle$,
etcetera. These are the $j=1$ (symmetric) triplet states and the
$j=0$ (antisymmetric) singlet state.

Suppressing multiplicity spaces when they are 1-dimensional (because
$\mathcal{H}_{j}=\mathcal{M}_{j}\otimes\mathbb{C}
=\mathcal{M}_{j}$), we have
\begin{equation}
    \label{eq:Hdecomp2qubits}
    \underset{\mathbf{4}}{(\mathcal{H}_{1/2})^{\otimes2}}
    =\underset{\mathbf{3}}{\mathcal{H}_{j=1}}\oplus
    \underset{\mathbf{1}}{\mathcal{H}_{j=0}} \,,
\end{equation}
where the dimensionality of each space is expressed in bold
underneath each subspace. Writing
$R(\Omega)^{\otimes2}=R_{j=1}(\Omega)\oplus R_{j=0}(\Omega)$, and
applying Schur's lemma, we infer that
\begin{equation}
    \label{eq:twirling2qubits}
    \mathcal{E}_{2}=(\mathcal{D}_{\mathcal{M}_{j=1}}
    \circ\mathcal{P}_{j=1})+\mathcal{P}_{j=0}\,,
\end{equation}
where $\mathcal{P}_{j}[\rho]=\Pi_{j}\rho\Pi_{j}$ and $\Pi_{j}$ is
the projector onto the subspace $\mathcal{H}_{j}.$ This equation
asserts that the coherence between the singlet and triplet spaces is
eliminated and the triplet space is depolarized.

Thus, if Alice transmits two qubits and she assigns the state $\rho$
to the pair, Bob describes the pair by
\begin{equation}
    \label{eq:twirling2qubits_states}
    \mathcal{E}_{2}[\rho]=p_{j=1}(\tfrac{1}{3}\Pi_{j=1})+p_{j=0}
    |\psi^{-}\rangle\langle\psi^{-}|\,,
\end{equation}
where $p_{j}=\mathrm{Tr}(\rho\Pi_{j})$.  Note that Bob can
distinguish perfectly between the antisymmetric state
$|\psi^{-}\rangle\langle\psi^{-}|$ and a state $\rho_{S}$ which lies
in the symmetric subspace because
$\mathcal{E}_{2}[|\psi^{-}\rangle\langle\psi^{-}|]
=|\psi^{-}\rangle\langle\psi^{-}|$ and
$\mathcal{E}_{2}[\rho_{S}]=\tfrac{1}{3}\Pi_{j=1}$, and these two
images are orthogonal.

Thus, Alice can communicate one classical bit to Bob with every two
transmitted qubits by implementing the following protocol: Alice
sends Bob the antisymmetric state $|\psi^{-}\rangle$ to communicate
$b=0$ and any state in the symmetric subspace (for example, the
state $|00\rangle$) for $b=1$. Bob then performs a projective
measurement onto the antisymmetric and symmetric subspaces and
recovers $b$ with certainty.

\paragraph{Three transmitted qubits: a quantum channel.}

We must determine how $R(\Omega)^{\otimes3}$ is decomposed into
irreducible representations. To see how three spin-1/2 systems
couple to total spin, imagine coupling the first pair to a spin
$j_{1}$ and then coupling this to the third:
$(\frac{1}{2})^{\otimes3}=(0\oplus1)\otimes\frac{1}{2}=\frac
{1}{2}\oplus\frac{1}{2}\oplus\frac{3}{2}.$ Note that because the
third spin 1/2 can couple to either $j_{1}=0$ or $j_{1}=1$ to yield
$j=1/2,$ the latter representation has multiplicity $2.$ We let
$\left\vert 1/2,\pm1/2,\lambda \right\rangle $ denote a basis of
$\mathcal{H}_{j=1/2}$ in the coupled representation, where $\lambda$
is a degeneracy index which by convention we take to be $0$ if the
coupling was to $j_{1}=0$ and $1$ if the coupling was to $j_{1}=1.$
These states can be given explicitly in terms of the three spin-1/2
systems as
\begin{align}
|\tfrac{1}{2},\tfrac{1}{2},0\rangle &
=\frac{1}{\sqrt{2}}(|011\rangle
-|101\rangle)\,,\\
|\tfrac{1}{2},-\tfrac{1}{2},0\rangle &
=\frac{1}{\sqrt{2}}(|010\rangle
-|100\rangle)\,,\\
|\tfrac{1}{2},\tfrac{1}{2},1\rangle &
=\frac{1}{\sqrt{6}}(2|110\rangle
-|101\rangle-|011\rangle)\,,\\
|\tfrac{1}{2},-\tfrac{1}{2},1\rangle &
=\frac{1}{\sqrt{6}}(-2|001\rangle +|010\rangle+|100\rangle)\,.
\end{align}
We can then define an isomorphism $\mathcal{H}_{j=1/2}=\mathcal{M}
_{j=1/2}\otimes\mathcal{N}_{j=1/2}$ through $\left\vert
m\right\rangle \otimes\left\vert \lambda\right\rangle
\equiv|\frac{1}{2},m,\lambda\rangle$ with $\left\vert m\right\rangle
$ a basis of $\mathcal{M}_{j=1/2}$ and $\left\vert
\lambda\right\rangle $ a basis of the multiplicity space
$\mathcal{N}_{j=1/2}.$  Thus the total Hilbert space decomposes as
\begin{equation}
    \label{eq:Hdecomp3qubits}
    \underset{\mathbf{8}}{(\mathcal{H}_{1/2})^{\otimes3}}
    =\underset{\mathbf{4}}{\mathcal{H}_{j=3/2}}
    \oplus\big(\underset{\mathbf{2}}{\mathcal{M}_{j=1/2}}
    \otimes\underset{\mathbf{2}}{\mathcal{N}_{j=1/2}}\big)\,,
\end{equation}
where again we have included the dimensions of each subsystem.

An application of Schur's lemma along the lines presented in
Sec.~\ref{subsec:GeneralSSR} implies that
\begin{equation}
    \label{eq:twirling3qubits}
    \mathcal{E}_{3}=\mathcal{D}_{\mathcal{M}_{j=3/2}}
    \circ\mathcal{P}_{j=3/2}+(\mathcal{D}_{\mathcal{M}_{j=1/2}}
    \otimes\mathcal{I}_{\mathcal{N}_{j=1/2}})\circ\mathcal{P}_{j=1/2}\,.
\end{equation}
where $\mathcal{I}$ is the identity map.  Note that the operation
$\mathcal{D}_{\mathcal{M}_{j}}\otimes\mathcal{I}_{\mathcal{N}_{j}}$
is only defined on the space of operators acting on
$\mathcal{M}_{j}\otimes \mathcal{N}_{j}=\mathcal{H}_{j}$, but it is
always preceded by $\mathcal{P} _{j}$, which projects into this
space. If Alice prepares three qubits in the state $\rho$, then Bob
assigns to them the state
\begin{equation}
    \label{eq:twirling3qubits_states}
    \mathcal{E}_{3}[\rho]=p_{3/2}\left( \frac{1}{4}\Pi_{j=3/2}\right)
    +p_{1/2}\left(\frac{1}{2}I_{\mathcal{M}_{j=1/2}}
    \otimes\rho_{\mathcal{N}_{j=1/2}}\right)\,,
\end{equation}
where $p_{j}=\mathrm{Tr}\left(\rho\Pi_{j}\right)$ and
\begin{equation}
    \rho_{\mathcal{N}_{j=1/2}}
    =p_{1/2}^{-1}\mathrm{Tr}_{\mathcal{M}_{j=1/2}}\left(
    \Pi_{j=1/2} \rho\Pi_{j=1/2}\right)\,.
\end{equation}

We note that the subsystem $\mathcal{N}_{j=1/2}$ is unaffected by
the decohering superoperator $\mathcal{E}_{3}$; i.e., it is a
\emph{decoherence-free subsystem}. Thus, Alice can encode a logical
qubit into this subsystem~\cite{Kem01}. That is, she can prepare
states of the form $\sigma\otimes \rho$ on
$\mathcal{M}_{j=1/2}\otimes\mathcal{N}_{j=1/2}$, where $\rho$ is the
logical qubit state she wishes to transmit to Bob, and Bob can
access this decoherence-free subsystem and retrieve the quantum
information without a shared RF. Thus, one logical qubit can be
transmitted using three physical qubits without a shared RF.

\paragraph{Asymptotic behaviour.}

The above two schemes demonstrate that classical and quantum
communication are possible without a shared RF. The efficiency of
the above schemes can be increased through the use of more qubits,
because the sizes of the decoherence-free subsystems can grow
exponentially with increasing number of qubits.

For simplicity, we consider only the case where $N$ is even. The
collective (tensor) representation of SU(2) on $N$ spin-$1/2$
systems, $R(\Omega)^{\otimes N}$, can again be decomposed into a
direct sum of SU(2) irreps, each with angular momentum quantum
number $j$ ranging from $0$ to $N/2$. That is, we can decompose the
Hilbert space as
\begin{equation}
    \underset{\mathbf{2^N}}{(\mathcal{H}_{1/2})^{\otimes N}}
    =\bigoplus_{j=0}^{N/2}\underset{\mathbf{2j+1}}{\mathcal{M}_{j}}
    \otimes\underset{\mathbf{c_{j}^{(N)}}}{\mathcal{N}_{j}}\,,
    \label{eq:SU(2)Decomposition}
\end{equation}
where we have indicated the dimensions of the various spaces.  The
multiplicity of the irrep $j$, which is the dimension of
$\mathcal{N}_{j}$, is found from representation theory to be
\begin{equation}
    c_{j}^{(N)}=\binom{N}{N/2-j}
    \frac{2j+1}{N/2+j+1}\,.
    \label{MultiplicitySU(2)}
\end{equation}

Relative to this decomposition, the SU(2)-twirling operation
$\mathcal{E}_{N}$ has the form
\begin{equation}
    \label{eq:twirlingNqubits}
    \mathcal{E}_{N}=\sum_{j}(\mathcal{D}_{\mathcal{M}_{j}}\otimes
    \mathcal{I}_{\mathcal{N}_{j}})\circ\mathcal{P}_{j}\,,
\end{equation}
as can be inferred from the result for arbitrary groups in
Sec.~\ref{subsec:GeneralSSR}. The carrier spaces for the irreducible
representations of SU(2), the $\mathcal{M}_{j},$ are the
decoherence-full subsystems for $\mathcal{E}_{N}$, while the
multiplicity spaces $\mathcal{N}_{j}$, which carry the trivial
representation of SU(2), are the decoherence-free subsystems for
$\mathcal{E}_{N}$.

Alice can choose to transmit classical messages by preparing
orthogonal states as follows: for each irrep $j$, she can choose one
arbitrary state from each multiplicity. Thus it is possible to
transmit, without a shared RF, a number of classical messages equal
to the number $C^{(N)}$ of SU(2) irreps in the direct sum
decomposition of the tensor representation of SU(2) on $N$ qubits,
which is given by
\begin{equation}
    C^{(N)}=\sum_{j=0}^{N/2}c_{j}^{(N)}=\binom{N}{N/2}\,.
    \label{eq:TotalMult}
\end{equation}
In fact, this is the \emph{maximum} number of classical messages
that can be sent; for a proof, see \textcite{BRS03}. Thus, the
number of classical bits that can be transmitted per qubit using the
above scheme is $N^{-1}\log _{2}C^{(N)}$, which tends asymptotically
to $1-(2N)^{-1}\log_{2}N$; in the large $N$ limit, one classical bit
can be transmitted for every qubit sent. Remarkably, this rate is
equivalent to what can be accomplished if Alice and Bob \emph{do}
possess a shared RF.

To determine the optimal scheme for transmitting quantum (rather
than classical) information, again using $N$ qubits and under the
restriction of no shared RF, we identify the largest
decoherence-free subsystem for $\mathcal{E}_{N}$. This is the
subsystem $\mathcal{N}_{j}$ with the greatest multiplicity
$c_{j}^{(N)}$. Asymptotically, this is found to occur at
$j_{\mathrm{max}}=\sqrt{N}/2$, and the number $N^{-1}\log_{2}
c_{j_{\mathrm{max}}}^{(N)}$ of logical qubits encoded per physical
qubit in $N$ physical qubits behaves as $1-N^{-1}\log_{2}(N)$,
approaching unity for large $N$. Full details can be found in
\textcite{Kem01}. This remarkable result proves that quantum
communication without a shared RF is asymptotically as efficient as
quantum communication with a shared RF, and is the communication
analog of ``asymptotic universality''~\cite{Kni00}.  In addition, we
note that the algorithm for encoding/decoding quantum information
into the decoherence-free subsystems can be done
\emph{efficiently}~\cite{Bac06a,Bac06b}.

\paragraph{Relativistic considerations.}

The ability to perform quantum information processing in a
relativistic setting has also been of recent interest (see
\textcite{Per04} for a review), and in this context it is natural to
consider whether parties who do not share an inertial frame (i.e., a
reference frame for the Poincar\'{e} group) can still perform
quantum communication, etc., and at what efficiency. It has been
shown that classical and quantum communication can be performed at
the same rate as demonstrated above using indistinguishable massive
spin-1/2 particles, or using photons, if appropriately localized
wavepackets for these particles are used~\cite{Bar05}. In addition,
continuous-variable quantum information can be shared using related
methods~\cite{Kok05}.

\subsubsection{Consequences for quantum information processing}
\label{subsec:ConsequencesQIP}

The communication schemes presented above imply that Alice and Bob
can share entangled states in the absence of any particular shared
RF. Consider the case of lacking a shared Cartesian frame as an
example. Denoting the logical qubit that can be encoded using three
physical qubits in Alice's (Bob's) possession by
$\{|0_L\rangle_{A(B)},|1_L\rangle_{A(B)}\}$, a triple of physical
qubits in Alice's possession can be maximally entangled with a
triple in Bob's possession using the state
$\frac{1}{\sqrt{2}}(|0_L\rangle_A|0_L\rangle_B+|1_L\rangle_A|1_L\rangle_B)$.
Because Alice and Bob can perform any measurement in their
respective logical qubit Hilbert spaces, they can demonstrate
quantum nonlocality (Bell's theorem) despite having no shared
Cartesian RF~\cite{BRS03,Cab03}. It also follows that such entangled
states can be used for quantum teleportation of logical qubits,
which implies that the latter does not rely upon the existence of a
shared Cartesian RF either, contrary to some
expectations~\cite{Enk01b}. In fact, for \emph{any} quantum
information task that assumes some shared RF, it is possible to make
use of logical encodings to perform the task \emph{without} this
shared RF. (Any task, that is, which deals with speakable rather
than unspeakable quantum information; the alignment of RFs, for
instance, obviously cannot be achieved in this way.)  It should be
noted, however, that although one can achieve quantum information
tasks without any \emph{particular} kind of shared RF, \emph{some}
form of shared RF is always required. For instance, in the example
just described, Alice and Bob must agree on the ordering of the
three physical qubits, and this agreement constitutes a kind of
shared RF~\cite{BRS04a}.

Several recent experiments have demonstrated the key techniques
required for quantum information processing without a shared
Cartesian frame. These experiments make use of single-photon
polarization qubits. Lacking a shared RF for polarization means that
Bob's polarizing elements (such as calcite crystals) are
uncorrelated with Alice's. The relevant group is also SU(2), and
thus the analysis presented above applies to this scenario as well.
\textcite{Ban04} have demonstrated that two orthogonal entangled
states of two single-photon polarization qubits remain perfectly
distinguishable between two parties who do not share a reference
frame for polarization, thereby demonstrating the classical
communication protocol in Sec.~\ref{subsec:NoSRFCommSpins}b. In
addition, \textcite{Bou04} have demonstrated non-orthogonal
entangled states -- states of a logical qubit encoded in four
single-photon polarization qubits -- that are identical in any
reference frame; see also \textcite{Zou06}. These states demonstrate
the basic principles of a decoherence-free subsystem that are needed
for quantum communication without a shared RF.

\subsection{QKD without a shared reference frame}
\label{subsec:QKD}

The possibility of performing secure communication through the use
of quantum key distribution (QKD) is one of the most celebrated
applications of quantum information science~\cite{Gis02}. Because
of its advanced state of development, it is also one of the first
quantum protocols to require explicit consideration of shared
reference frames, or the lack thereof, between communicating
parties.  All practical QKD protocols are based on the exchange of
quantum states of light, and as discussed in
Sec.~\ref{subsubsec:NoSRFPhase}, essentially any identification of
a mode structure (either spatial, time-bin, or polarization)
requires a reference frame of some sort.  For example, in all
single-photon implementations of QKD, a shared clock is necessary
in order to agree upon a short time window for communication;
otherwise, dark counts from the photodetectors can greatly reduce
security and efficiency~\cite{Bra00}.

QKD schemes that obviate the need for certain shared reference
frames (and that are robust against other forms of noise) have
recently been developed, and make use of the techniques of
decoherence-free subspaces and subsystems~\cite{Wal03}.  Consider
the following proposal of~\textcite{Boi04}. Alice (the sender) and
Bob (the receiver) wish to perform QKD using the polarization
states of single photons.  This choice avoids the stabilization
problems inherent in phase-based schemes, but presents a problem
of its own: if an optical fibre is used as the quantum channel,
the polarization of a transmitted photon is rotated by a random
amount due to optical birefringence. Although this random rotation
fluctuates with time, it can be considered constant on a short
time scale so that all photons in a pulse are subject to the same
rotation. Thus, the problem becomes equivalent to one in which
Alice communicates to Bob using a noiseless channel, but in which
they do not share a reference frame for polarization. The
communication scenario, then, becomes equivalent to that analyzed
in Sec.~\ref{subsec:NoSRFCommSpins}.

Alice can perform quantum communication with Bob without a shared RF
for polarization through the use of decoherence-free subspaces or
subsystems.  We now briefly outline two straightforward and
experimentally-accessible QKD protocols using these techniques; the
first protocol makes use of a four-photon decoherence-free subspace,
and the second makes use of a three-photon decoherence-free
subsystems.

The smallest non-trivial decoherence-free subspace for the
superoperator $\mathcal{E}_N$ of
Eq.~\eqref{eq:NQubitDecoheringChannel} occurs for $N=4$.  It is the
two-dimensional $j=0$ (singlet) subspace.  A simple QKD scheme using
this subspace is as follows. Define the state
$|\psi^-\rangle_{\mu\nu} = (|0\rangle_\mu|1\rangle_\nu -
|1\rangle_\mu|0\rangle_\nu)/\sqrt{2}$ to be the two-photon singlet
state of photons $\mu$ and $\nu$ ($\mu,\nu\in\{1,2,3,4\}$). Define
three four-photon states as products of singlet states of differing
photons, i.e.,
\begin{align}
    |\Psi_1\rangle &= |\psi^-\rangle_{12} |\psi^-\rangle_{34} \,, \nonumber \\
    \label{eq:4PhotonQKD}
    |\Psi_2\rangle &= |\psi^-\rangle_{13} |\psi^-\rangle_{24} \,, \\
    |\Psi_3\rangle &= |\psi^-\rangle_{14} |\psi^-\rangle_{23} \,.
    \nonumber
\end{align}
Clearly, all three states are $j=0$ (singlet) states in the $N=4$
decoherence-free subsystem.  Thus, each of the states
$|\Psi_a\rangle$ prepared by Alice is represented the same way by
Bob, even though they do not share a reference frame for
polarization.

Note that these states are also non-orthogonal, satisfying
$|\langle\Psi_a|\Psi_b\rangle| = 1/2$ for $a\neq b$.  Thus, if
Alice restricts her transmitted states to a pair of these, then
they can implement a B92-type QKD protocol~\cite{Ben92}.  In
addition, this protocol can be defined in such a way that Bob need
only perform single-photon measurements in some fixed polarization
basis (i.e., without the need for entangling measurements);
see~\textcite{Boi04} for details.

As noted in Sec.~\ref{subsec:NoSRFCommSpins}, there exists a
two-dimensional decoherence-free subsystem with $N=3$.  There is a
simple modification of the above QKD protocol which makes use of
this subsystem.  Define the following three \emph{mixed} states,
obtained from the three pure states of Eq.~\eqref{eq:4PhotonQKD} by
discarding the last photon, i.e.,
\begin{equation}\label{3PhotonQKD}
    \rho_a = {\rm Tr}_4[|\Psi_a\rangle\langle\Psi_a|] \,.
\end{equation}
In terms of the decomposition of the three-qubit Hilbert space of
Sec.~\ref{subsec:NoSRFCommSpins}c, all three of these states lie on
the $\mathcal{H}_{j=1/2}$ subspace, and in terms of the tensor
product structure $\mathcal{H}_{j=1/2} = \mathcal{M}_{j=1/2} \otimes
\mathcal{N}_{j=1/2}$, these states are products of the completely
mixed state $\frac{1}{2}I$ on $\mathcal{M}_{j=1/2}$ and one of three
pure non-orthogonal states on $\mathcal{N}_{j=1/2}$.  Again, if
Alice restricts her transmitted states to a pair of these, then they
can implement a B92-type QKD protocol without the need for a shared
RF for polarization.

The unconditional security of the QKD schemes of \textcite{Boi04},
which are based on the use of the above states, has been
proven~\cite{Boi05}.  In addition, a BB84-version of this QKD
scheme, which does not require a shared reference frame for
polarization, has been demonstrated experimentally~\cite{Che06}. We
note that the essential concept of this scheme -- to use the
techniques of decoherence-free subspaces or subsystems to obviate
the necessity for a shared reference frame in QKD -- can be applied
to any system and RF.  In particular, it has been proposed to use
the spatial encodings of optical modes discussed at the end of
Sec.~\ref{subsubsec:NoSRFPhase} to perform QKD~\cite{Spe04}.

\subsection{Entanglement without a shared reference frame}
\label{subsec:Entanglement}

Entanglement is often considered the key resource in quantum
information processing, and so it is valuable to consider the role
of shared reference frames in both qualitative and quantitative
properties of bipartite entanglement.  As we will demonstrate in
this section, the very meaning of entanglement between parties who
do not share a reference frame must be reassessed, with some
surprising results.

\subsubsection{Entanglement without a shared phase reference}
\label{sec:optics}

As an example, we again consider a number of optical modes shared
between two parties, Alice and Bob, who do not share a common phase
reference.  We will consider all states and operations to be
described relative to the phase reference of a third party, Charlie,
which is assumed to be uncorrelated with both Alice's and Bob's
local phase references.  (For many of the issues we wish to
consider, we could dispense with Charlie and describe everything
relative to either Alice or Bob, but this introduces an artificial
asymmetry into the formalism which easily leads to confusion. We
therefore opt to describe all states relative to Charlie, whether he
participates in the protocol or not.)  As such, Alice redescribes
states prepared relative to Charlie's phase reference by mixing over
all possible phase shifts. Bob does the same, and because Alice and
Bob's phase references are uncorrelated, the phases over which they
mix are independent. Recalling the results of
Sec.~\ref{subsec:PNSSR}, the mixing over phases yields a
photon-number superselection rule, and the independence implies that
Alice and Bob are subject to \emph{local} photon-number
superselection rules. In this case, all of Alice's operations
commute with the local map $\mathcal{U}_A$, defined as in
Eq.~\eqref{eq:LPNSSR2} as
\begin{equation}\label{eq:LocalAU(1)}
    \mathcal{U}_A[\rho_A] \equiv \sum_n \Pi^A_n \rho_A \Pi^A_n \,,
\end{equation}
where $\Pi^A_n$ in the projector onto the eigenspace of total photon
number $n$ on Alice's local modes.  All of Bob's operations commute
with the local map $\mathcal{U}_B$, defined similarly.

In such situations, there has been considerable debate over the
entanglement properties of certain types of states, such as the
two-mode single-photon state~\cite{Tan91,Har94,Gre95,Har95,Enk05b},
\begin{equation}
  \label{V-EPR}
  (|0\rangle_A |1\rangle_B + |1\rangle_A |0\rangle_B)/\sqrt{2}\,.
\end{equation}
There is a temptation to say that this state is entangled simply
because of its non-product form.  However, it is far more useful to
consider whether or not this state satisfies certain
\emph{operational} notions of entanglement.  One such notion is
whether a state can be used to violate a Bell inequality.  Another
is whether it is useful as a resource for quantum information
processing, for instance, to teleport qubits or implement a dense
coding protocol.  In the context of a local photon-number
superselection rule, this two-mode single-photon state fails to
satisfy either of these notions of entanglement, because all such
tasks would require Alice and Bob to violate the local photon-number
superselection rule.  A different but equally common notion of
entanglement is that a state is entangled if it cannot be prepared
by LOCC.  The two-mode single-photon state certainly \emph{does} fit
\emph{this} notion because the pure non-product states cannot be
prepared by LOCC. Thus we see that operational notions of
entanglement that coincided for pure states under unrestricted LOCC,
namely being not locally preparable and being useful as a resource
for tasks such as teleportation or violating a Bell inequality, do
not coincide under a local photon-number superselection rule, and
the state in question is judged entangled by one notion and not the
other.\footnote{Of course, if there is no local photon-number
superselection rule, this state would satisfy all of these notions
of entanglement, as emphasized by \textcite{Enk05b}.  In particular,
no such superselection rule would apply if all parties share a
common phase reference.}

Another class of states whose entanglement properties have been
discussed in the quantum optics literature are those that are
separable but not locally preparable under a local photon-number
superselection rule~\cite{Rud01a,Ver03}. Examples of such states are
\begin{equation}
\label{refbits}
  |{+}\rangle_A|{+}\rangle_B\,, \qquad |{-}\rangle_A|{-}\rangle_B\,,
\end{equation}
where $|\pm\rangle = (|0\rangle\pm|1\rangle)/\sqrt{2}$.
(\textcite{Rud01a} and \textcite{Ver03} considered states such as
the equal mixture of $|{+}\rangle_A|{+}\rangle_B$ and
$|{-}\rangle_A|{-}\rangle_B$.  For simplicity, we restrict our
attention to pure states.)  Because of the superselection rule,
these states cannot be prepared locally. However, because they are
product states, they clearly cannot be used for tasks such as
teleportation or violating a Bell inequality. We will return our
attention to states such as these in Sec.~\ref{sec:QRFs}.

In contrast, consider a state of the form
\begin{equation}
  \label{E-EPR} (|01\rangle_A |10\rangle_B +
  |10\rangle_A |01\rangle_B)/\sqrt{2} \,.
\end{equation}
This state is certainly not locally preparable.  In addition, it
\emph{can} be used to violate a Bell inequality, implement dense
coding, and so on, despite the superselection rule.  This is because
Alice and Bob can still implement any measurements they please in
the 2-dimensional subspaces spanned by $|01\rangle$ and
$|10\rangle$.  Thus, this state is unambiguously entangled by any
reasonable notion.

We see, then, that the remarkable and often confusing entanglement
properties of states when parties do not share a reference frame can
be understood by recognizing that different operational notions of
entanglement do not coincide in this case.  Specifically, for pure
quantum-optical states in a situation where Alice and Bob to not
share a phase reference, there exists a proper gap between states
that are locally-preparable under LOCC, and states that are useful
for performing quantum information tasks such as teleportation and
violating a Bell inequality.  The existence of this proper gap is
reminiscent of a similar situation for \emph{mixed} quantum states:
that of bound entanglement~\cite{Hor98}. This analogy can be
extended further; in the following section, we demonstrate that some
of the strange phenomena from mixed-state entanglement --
activation, and multi-copy entanglement distillation -- are present
as well in pure-state quantum optics with a local photon-number SSR.
This analogy is pursued in detail in \textcite{BDSW06}.

\subsubsection{Activation and entanglement distillation}
\label{subsec:Activation}

In this section, we demonstrate that there exist analogous processes
of activation~\cite{Hor99} and multi-copy entanglement
distillation~\cite{Wat04} using pure bipartite quantum-optical
states when Alice and Bob do not share a phase reference.  An
understanding of these processes and their relation to the
above-mentioned gap between two commonly-used notions of
entanglement is key to resolving several recent controversies
regarding the entanglement of quantum-optical
states~\cite{Enk05,BDSW06}.

We now demonstrate that, to achieve a Bell inequality violation with
the state
\begin{equation}
  \label{eq:SinglePhotonEnt}
  (|0\rangle_A |1\rangle_B + |1\rangle_A |0\rangle_B)/\sqrt{2}\,,
\end{equation}
it is necessary to use a process that is analogous to activation.
Understanding the necessity of an additional resource for this
process resolves the controversy over the use of the state to
demonstrate quantum nonlocality~\cite{Tan91,Har94,Gre95,Har95}.

As we have shown, this state cannot be used for tasks such as
violating a Bell inequality when Alice and Bob do not share a phase
reference, i.e., when a local photon-number superselection rule
applies. However, combining $(|0\rangle_{A} |1\rangle_{B} +
|1\rangle_{A} |0\rangle_{B})/\sqrt{2}$ with $|{+}\rangle_{A}
|{+}\rangle_{B}$, one obtains a state that \emph{is} useful for such
tasks. The state $|{+}\rangle_{A} |{+}\rangle_{B}$ is said to
\emph{activate} the entanglement of $(|0\rangle_{A} |1\rangle_{B} +
|1\rangle_{A} |0\rangle_{B})/\sqrt{2}$. This is seen as follows. Let
Alice and Bob both perform a quantum non-demolition measurement of
local photon number on \emph{both} of their local modes, and
post-select the case where they both find a local photon number of
one.  The resulting state is
\begin{multline}
    \Pi^A_1 \otimes \Pi^B_1 [ \tfrac{1}{\sqrt{2}}(|0\rangle_{A} |1\rangle_{B}
    + |1\rangle_{A} |0\rangle_{B}) |{+}\rangle_{A}
    |{+}\rangle_{B} ] \\
    \propto \tfrac{1}{\sqrt{2}}( |01\rangle_{A} |10\rangle_B +
    |10\rangle_A |01\rangle_B) \,.
\end{multline}

Violations of a Bell inequality have recently been demonstrated
experimentally using the state~\eqref{eq:SinglePhotonEnt} by
\textcite{Hes04} and \textcite{Bab04}. One can take two different
perspectives on such an experiment.  It is illustrative to
consider them both.

In \textcite{Hes04}, in addition to the
state~\eqref{eq:SinglePhotonEnt}, a correlated pair of coherent
states $|\alpha\rangle_A |\alpha\rangle_B$, where $|\alpha\rangle
\equiv \sum_n ( e^{-|\alpha|^2/2} \alpha^n / \sqrt{n!}) |n\rangle$,
are assumed to be shared between Alice and Bob. These modes are used
as the local oscillators in the homodyne detections at each site.
Noting that neither $(|0\rangle_{A} |1\rangle_{B} + |1\rangle_{A}
|0\rangle_{B})/\sqrt{2}$ nor $|\alpha\rangle_A |\alpha\rangle_B$ can
be used individually for violating a Bell inequality, it is unclear
how it is possible to do so using such resources. The resolution of
the puzzle is that a pair of correlated coherent states
$|\alpha\rangle_A |\alpha\rangle_B$, much like the state
$|{+}\rangle_A |{+}\rangle_B$ discussed above, \emph{activates} the
entanglement of the two-mode single photon state.

An experimental demonstration of nonlocality using the two-mode
single photon state can also be described as in \textcite{Bab04}.
Rather than treating the local oscillators as coherent states, they
are treated as correlated classical phase references. In this case,
they constitute an additional resource that ``lifts'' the
restriction of the local photon-number superselection rule, and the
state $(|0\rangle_{A} |1\rangle_{B} + |1\rangle_{A}
|0\rangle_{B})/\sqrt{2}$ becomes unambiguously entangled. These two
alternative descriptions are equally valid; see~\textcite{BRS06}.

The existence of such activation processes also resolves a
controversy concerning the source of entanglement in the
experimental realization of \textcite{Fur98} of continuous-variable
quantum teleportation.  Again, it is illustrative to consider two
different perspectives of this experiment.

The first perspective is a variant of the one presented by
\textcite{Rud01a}.  In our language, it can be synopsized as
follows. Alice and Bob are presumed to be restricted in the
operations they can perform by a local photon-number superselection
rule.  They share a two-mode squeezed state $|\gamma\rangle
=\sqrt{1-\gamma^{2}}\sum_{n=0}^{\infty}\gamma^{n}|n,n\rangle$ where
$0\leq\gamma\leq 1$.  In addition, they share two other modes
prepared in a product of correlated coherent states $|\alpha\rangle
|\alpha\rangle$.\footnote{The state assigned by \textcite{Rud01a} is
simply a mixed version (mixed over the phase of the pump beam) of
$|\gamma\rangle|\alpha\rangle|\alpha\rangle$.} The former is the
purported entanglement resource in the teleportation protocol, while
the latter is a quantum version of a shared phase reference. These
states are analogous to $(|0\rangle_{A} |1\rangle_{B} +
|1\rangle_{A} |0\rangle_{B})/\sqrt{2}$ and $|{+}\rangle_{A}
|{+}\rangle_{B}$ respectively -- neither can be used as a resource
for teleportation when considered on its own. So the question arises
as to how teleportation could possibly have been achieved. The
answer is that the product of coherent states \emph{activates} the
entanglement in the two-mode squeezed
state.\footnote{\textcite{vEF02a} suggest a similar protocol to the
one we describe here, for the mixed states discussed in the previous
footnote.  Homodyne measurement is performed on the pump beam with
respect to an external phase reference, and the measurement result
will yield a two-mode squeezed state that is unambiguously entangled
with respect to this external RF.}

The second perspective is one wherein the shared phase reference is
treated classically; this perspective was taken in~\textcite{Fur98}.
As described above, this classical shared phase reference acts as a
resource that lifts the superselection rule, and causes the two-mode
squeezed state to become unambiguously entangled.

An analogue of multi-copy entanglement distillation can also be
demonstrated in our quantum optical example.  Two copies of the
state $(|0\rangle_{A} |1\rangle_{B} + |1\rangle_{A}
|0\rangle_{B})/\sqrt{2}$ can be used to obtain free entanglement
(i.e., not bound) in the presence of the SSR, whereas only one copy
cannot. The protocol, introduced in \textcite{Wis03} and discussed
in greater detail in \textcite{Vac03}, is as follows. As in the
activation example above, Alice and Bob both perform a quantum
non-demolition measurement of local photon number (on both local
modes) and post-select the case where they both find a local photon
number of one.  The resulting state is
\begin{multline}
    \Pi^A_1 \otimes \Pi^B_1 [\tfrac{1}{\sqrt{2}}(|0\rangle_{A}
    |1\rangle_{B}
    + |1\rangle_{A} |0\rangle_{B})]^{\otimes 2} \\
    \propto \tfrac{1}{\sqrt{2}}( |01\rangle_{A} |10\rangle_B +
    |10\rangle_A |01\rangle_B) \,,
\end{multline}
where $|\psi\rangle^{\otimes 2} = |\psi\rangle |\psi\rangle$. A
process very similar to this 2-copy entanglement distillation has
been demonstrated in quantum optics experiments
(c.f.~\textcite{Shi88,Ou88}), where correlated but unentangled
photon pairs from parametric downconversion were made incident on
the two input modes of a beamsplitter, so each photon transforms to
a state of the form $(|0\rangle_{A} |1\rangle_{B} + |1\rangle_{A}
|0\rangle_{B})/\sqrt{2}$.  Subsequently, measurements on the two
output modes are postselected for one photon detection at each
output mode. The fact that their postselected results are consistent
with a description of an entangled state demonstrates that the
entanglement of the state $(|0\rangle_{A} |1\rangle_{B} +
|1\rangle_{A} |0\rangle_{B})/\sqrt{2}$ has been distilled by making
use of two copies.

Finally, we consider the analogue of multi-copy entanglement
distillation from two copies of the two-mode squeezed state
$|\gamma\rangle
=\sqrt{1-\gamma^{2}}\sum_{n=0}^{\infty}\gamma^{n}|n,n\rangle$.
Homodyne measurements by Alice and Bob (relative to their
uncorrelated local oscillators) can be performed on one copy of this
state to establish a shared phase reference, which then lifts the
superselection rule and causes the second copy to become
unambiguously entangled.

\subsubsection{Quantifying bi-partite entanglement without a shared reference frame}
\label{subsubsec:QuantEnt}

As we have seen above, operational notions of entanglement for a
bipartite pure state no longer coincide when parties do not share
a reference frame.  How, then, does one \emph{quantify} the amount
of entanglement of a bipartite state in such a situation?
Entanglement measures can be defined in the presence of such a
restriction again by being operational.  In the following, we
discuss one such operational measure which quantifies the
distillable entanglement under a local Abelian superselection
rule. (This measure is directly related to the \emph{entanglement
of particles}~\cite{Wis03b}.) We note that these results apply
directly to a general (possibly non-Abelian) SSR, with local
operations restricted as in
Eq.~\eqref{eq:GInvariantOperations}~\cite{Bar03}; however, for
simplicity, we focus here on the Abelian case.

We continue with the scenario of the previous section. Consider a
bipartite state $\rho_{AB}$ shared by Alice and Bob and defined
relative to Charlie's phase reference. We assume that in addition to
this bipartite system, Alice and Bob each possess a number of
\emph{quantum registers}, not subject to any SSR, with total Hilbert
space dimension equal to or greater than that of their respective
systems.  (For example, these registers could be standard qubits
over which Alice and Bob have complete control.) These registers are
initiated in a pure product state $\varrho_{AB}$.

The \emph{entanglement in the presence of an SSR} of the state
$\rho_{AB}$ is quantified through a measure $E_{\textrm{SSR}}$,
which is defined by the maximum amount of entanglement that Alice
and Bob can produce between their quantum registers using local
U(1)-invariant operations and classical communication (U(1)-LOCC).
The latter can be quantified by an appropriate standard measure $E$;
it seems most appropriate to use the \emph{entanglement of
distillation}.

We now prove that the entanglement in the presence of an SSR,
$E_{\textrm{SSR}}(\rho_{AB})$, is given by the entanglement
$E(\mathcal{U_{\text{loc}}}[\rho_{AB}])$ that they can produce from
the state $\mathcal{U_{\text{loc}}}[\rho_{AB}]$ by
\emph{unconstrained} LOCC, where $\mathcal{U_{\text{loc}}} \equiv
\mathcal{U}_A \otimes \mathcal{U}_B$.

The proof is illustrative, so we present it here.  Let $O =
\{\mathcal{O}\}$ be the set of all LOCC operations by Alice and Bob
that commute with $\mathcal{U_{\text{loc}}}$. Note that, for any
quantum operation $\mathcal{E}$, the composite operation
$\bar{\mathcal{E}} \equiv \mathcal{U_{\text{loc}}}\circ
\mathcal{E}\circ \mathcal{U_{\text{loc}}}$ is in the set $O$. Let
$\mathcal{O} \in O$ be some operation on the initial state
$\rho_{AB} \otimes \varrho_{AB}$. The final state of the registers
is given by $\varrho_{AB}' = {\rm Tr}_{\rm sys}\bigl( \mathcal{O}
[\rho_{AB} \otimes \varrho_{AB}] \bigr)$, where the trace is over
the shared system. The maximum entanglement produced between the
registers is given by maximizing $E(\varrho_{AB}')$ over all
operations in $O$. Thus,
\begin{align}
  \label{MaxEntTrans}
  E_{\textrm{SSR}}&(\rho_{AB}) \nonumber \\
  &= \max_{\mathcal{O}}\ E\bigl({\rm
    Tr}_{\rm sys}\bigl(
  \mathcal{O}[\rho_{AB} \otimes \varrho_{AB}] \bigr) \bigr) \nonumber \\
  &= \max_{\mathcal{O}}\ E\bigl({\rm Tr}_{\rm sys} \bigl(
  (\mathcal{U_{\text{loc}}}\circ \mathcal{O}\circ\mathcal{U_{\text{loc}}}) [\rho_{AB} \otimes
  \varrho_{AB}]\bigr) \bigr) \nonumber \\
  &= \max_{\mathcal{E}}\ E\bigl({\rm Tr}_{\rm sys} \bigl(
  (\mathcal{U_{\text{loc}}}\circ \mathcal{E}\circ\mathcal{U_{\text{loc}}}) [\rho_{AB} \otimes
  \varrho_{AB}]
  \bigr) \bigr) \nonumber \\
  &= \max_{\mathcal{E}}\ E\bigl({\rm Tr}_{\rm sys} \bigl( \mathcal{E}
  \bigl[\mathcal{U_{\text{loc}}} [\rho_{AB}] \otimes \varrho_{AB}\bigr]
    \bigr)\bigr) \, ,
\end{align}
where the second line follows from the properties of the trace and
by applying the definition~\eqref{eq:LPNSSR1} to $\mathcal{U}_A$ and
$\mathcal{U}_B$, and the last line follows from the properties of
the trace.  The latter maximization is over \emph{all} LOCC (not
just operations that commute with $\mathcal{U_{\text{loc}}}$), and
gives the entanglement $E(\mathcal{U_{\text{loc}}}[\rho_{AB}])$ that
Alice and Bob can produce between their registers from the state
$\mathcal{U_{\text{loc}}}[\rho_{AB}]$ by unconstrained LOCC.

\subsubsection{Extensions and application to other systems}

The general perspective discussed above for investigating
entanglement without a shared reference frame can be applied to
other situations, although for the most part, this issue has not
been explored.  Condensed matter systems is one area where these
results can be directly applied, because these systems possess a
number of practical restrictions on operations.  Local
particle-number superselection rules often apply in practice; for
example, as noted by several authors, the single-electron two-mode
Fock state $(|0\rangle_A|1\rangle_B + |1\rangle_A
|0\rangle_B)/\sqrt{2}$ has ambiguous entanglement properties under
this restriction~\cite{Wis03b,Sam05,Bee05,Dow06}. For this reason,
most proposals for creating bi-partite entangled states make use of
spin or orbital angular momentum degrees of freedom of multiple
particles~\cite{Sam03,Bee03,Sam04}.  We note, however, that the
two-mode single-electron Fock state is an entanglement resource akin
to the two-mode single-photon state, which we have shown to be
useful through activation or multi-copy entanglement distillation;
also, a suitable shared U(1) reference frame could ``lift'' the
restriction of the superselection rule, and the two-mode
single-electron Fock state would be unambiguously entangled with
such a resource.  (Determining a suitable quantum state of such a
shared U(1) RF consisting of fermions is an outstanding problem in
general.  The 2-copy entanglement distillation protocol described
above applies equally well to fermionic states of the form
$(|0\rangle_A|1\rangle_B + |1\rangle_A |0\rangle_B)/\sqrt{2}$. The
activation protocols described above, however, do not appear to have
precise fermionic analogues; specifically, there are several
challenges in defining an analogue of the optical coherent state for
fermions. See~\textcite{Dow06b}.) Moreover, entangled states between
angular momentum degrees of freedom of different particles will
yield no real advantage over the two-mode single-electron Fock state
in situations wherein there is a local SU(2) superselection rule.
Such a superselection rule will be in force, for instance, if the
parties fail to share a Cartesian frame for spatial orientations. As
with quantum optical systems, such considerations emphasize the need
to be operational when classifying or quantifying entanglement.

The theory of entanglement for indistinguishable particles is
another situation where considerations of entanglement without a
shared reference frame are relevant.  States of indistinguishable
particles can appear entangled due to the necessary symmetrization
or anti-symmetrization of the wavefunction.  For example, in the
position representation of two indistinguishable particles, a
wavefunction of the two particles is expressed as
\begin{equation}\label{eq:IndistinguishableWavefunction}
    \psi_{12}(\mathbf{x}_1,\mathbf{x}_2) =
    \frac{1}{\sqrt{2}}\big(\psi_1(\mathbf{x}_1) \psi_2(\mathbf{x}_2) \pm
    \psi_1(\mathbf{x}_2) \psi_2(\mathbf{x}_1)\big) \,,
\end{equation}
where the $\pm$ cases correspond to bosons and fermions.  The
entanglement properties of such a state are the subject of some
debate~\cite{Pas01,Sch01,Wis03b,Dow06}.  From the perspective of
this review, one can view the indistinguishability of particles as a
lack of a \emph{reference ordering}, i.e., lack of a reference frame
to uniquely label the
particles~\cite{Eis00,Bar03,Kor04,Jon05,Jon06}. For example, if the
particles described in the above state were distinguishable through
another degree of freedom, such as their spin, then the entanglement
of the above state would be unambiguous. Thus, in many condensed
matter systems, it may be worthwhile to consider the possibility of
``lifting'' the restriction of indistinguishability, viewed as a
lack of a reference ordering, through an appropriate reference
frame.  (Such a reference frame would necessarily make use of some
physical degrees of freedom to uniquely label the particles.)

\subsection{Private shared reference frames as cryptographic key}
\label{subsec:PrivateSRFs}

Two parties, Alice and Bob, are said to possess a private shared RF
for some degree of freedom if their reference frames are perfectly
correlated with each other, and are completely \emph{uncorrelated}
with any other party.  Such private shared RFs can be used as a
novel kind of key for cryptography.  To illustrate the general idea,
consider the case where Alice and Bob share a private Cartesian
frame.\footnote{Although it is difficult to imagine how a Cartesian
frame defined by the fixed stars might be made private, it is clear
that if the Cartesian frame is defined by a set of gyroscopes,
privacy amounts to no other party having gyroscopes that are
correlated with those of Alice and Bob.}  They can achieve some
private classical communication as follows: Alice transmits to Bob
an orientable physical system (e.g., a pencil or a gyroscope) after
encoding her message into the relative orientation between this
system and her local reference frame (for instance, by turning her
bit string into a set of Euler angles). Bob can decrypt the message
by measuring the relative orientation between this system and his
local reference frame. Because an eavesdropper (Eve) does not have a
reference frame correlated with theirs, she cannot infer any
information about the message from the transmission.

We shall consider the quantum version of this example, where Alice
sends spin-1/2 particles to Bob via a noiseless channel, as in the
communication problem of Sec.~\ref{subsec:NoSRFCommSpins}.  Note
that whereas in that problem Bob lacked the RF with respect to which
the spins were prepared, here Bob shares the RF and it is \emph{Eve}
who lacks it. Thus, the superoperator $\mathcal{E}_N$ of
Eq.~(\ref{eq:NQubitDecoheringChannel}) now describes the restriction
that \emph{Eve} faces by virtue of lacking the private shared RF. In
Sec.~\ref{subsec:NoSRFCommSpins} we sought to determine how Alice
could encode information in such a way that it remained accessible
to someone who lacked her RF, whereas here we are interested in the
opposite problem: how to encode information in such a way that it is
\emph{inaccessible} to someone who lacks her RF (but accessible to
someone who has it).  We follow~\textcite{BRS04a}, to which the
reader is directed for a more complete analysis.

A few points are worth noting before presenting the results. First,
private communication using a private SRF is similar in some ways to
private-key cryptography, specifically, the Vernam cipher (one-time
pad).  For example, the secret key in the Vernam cipher can be used
only once to ensure perfect security. Similarly, for our
communication schemes, only a single plain-text (classical or
quantum) can be encoded using a single private SRF. If the same
private SRF is used to encode two plain-texts, then the
\emph{relation} that holds between the two cipher-texts carries
information about the plain-texts, and because it is possible to
learn about this relation without making use of the SRF, Eve can
obtain this information. This is akin to the fact that in our
example of the classical pencil or gyroscope, Eve can measure the
angular separation of the two pencils.

This analogy prompts us to raise and dismiss the possibility that a
private SRF is \emph{equivalent}, as a resource, to some amount of
secret key or entanglement. It is true that a private SRF may,
through public communication, yield secret key. Conversely, as will
be seen in Sec.~\ref{subsec:PrivateUnspeakable}, a secret key may,
through public communication, yield a private SRF. Moreover, if,
contrary to what has been assumed here and in
Sec.~\ref{subsec:PrivateUnspeakable}, the parties possess a
\emph{public} SRF, then a private SRF is equivalent to an unbounded
amount of secret key (in practice, the size of the key is limited by
the bounded size of the physical systems that define the SRF or the
bounded degree of correlation in the SRF).  For instance, the
parties can measure the Euler angles relating the private SRF to the
public SRF and then express these in binary to obtain secret bits.
Nonetheless ---and this is the critical point--- in the absence of
either a public SRF or public communication of unspeakable
information, there is no procedure for interconverting secret key
and private SRF.  Thus the two resources are not equivalent.
Similarly, one can show that the resource of a private SRF is
distinct from that of entanglement.

\paragraph{One qubit.}

Consider the transmission of a single qubit from Alice to Bob. As
they share an RF, Bob represents states of this single qubit in the
same way as Alice. On the other hand, Eve, who does not share
Alice's RF, describes the state $\rho$ as $\mathcal{E}_1(\rho) =
\tfrac{1}{2}I$, as in Eq.~(\ref{eq:twirling1qubit_states}).  She
consequently cannot correlate the outcomes of her measurements with
Alice's preparations. It follows that using this single qubit and
their private shared RF, Alice and Bob can privately communicate one
logical qubit, and thus also one logical classical bit.

\paragraph{Two qubits: Decoherence-full subspaces.}

If multiple qubits are transmitted, it is possible for Eve to
acquire some information about the preparation even without access
to the private shared RF by performing \emph{relative} measurements
on the qubits.  For two transmitted qubits in the state $\rho$,
Eve's description is $\mathcal{E}_2(\rho) =  p_{j=1}
(\tfrac{1}{3}\Pi_{j=1}) + p_{j=0} \Pi_{j=0}$ as in
  Eq.~(\ref{eq:twirling2qubits_states}). Despite not sharing the RF, Eve
can still discriminate the singlet and triplet subspaces and thus
acquire information about the preparation. Nonetheless, Alice can
achieve some private quantum communication by encoding the state of
a qutrit (a 3-dimensional generalization of the qubit) into
$\mathcal{H}_{j=1}$, the triplet subspace. Bob, sharing the private
RF, can recover this qutrit with perfect fidelity.  However, Eve
identifies all such qutrit states with $\tfrac{1}{3}\Pi_{j=1}$, and
therefore cannot infer anything about Alice's preparation.

The property of $\mathcal{H}_{j=1}$ that is key for this scheme is
that the two-qubit superoperator $\mathcal{E}_2$ is completely
depolarizing on it, as seen explicitly from
Eq.~(\ref{eq:twirling2qubits}), which we repeat:
$\mathcal{E}_{2}=(\mathcal{D}_{\mathcal{M}_{j=1}}\circ\mathcal{P}_{j=1})
+\mathcal{P}_{j=0}$. We define subspaces with this property to be
\emph{decoherence-full subspaces}, consistent with the terminology
presented in Sec.~\ref{subsec:GeneralSSR}.

Now consider how many \emph{classical} bits of information Alice can
transmit privately to Bob. An obvious scheme is for her to encode a
classical trit as three orthogonal states within the triplet
subspace.  However, this is not the most efficient scheme. Suppose
instead that Alice encodes two classical bits as the four orthogonal
states
\begin{equation}
  \label{eq:fourprivatestates}
  \left| i\right\rangle =\frac{1}{2}\left| \psi ^{-}\right\rangle
  +\frac{\sqrt{3}}{2}\left| {\bf n}_{i}\right\rangle \left| {\bf
  n}_{i}\right\rangle \, ,\quad i=1,\ldots,4 \, ,
\end{equation}
where $\left| \psi ^{-}\right\rangle $ is the singlet state and the
$\left| {\bf n}_{i}\right\rangle \left| {\bf n}_{i}\right\rangle $
are four states in the triplet subspace with both spins pointed in
the same direction, with the four directions forming a tetrahedron
on the Bloch sphere; specifically,
\begin{align}\label{eq:Tetrahedron}
    |\mathbf{n}_1\rangle &= |0\rangle \,, \\
    |\mathbf{n}_2\rangle &=
    \frac{i}{\sqrt{3}}\Big(|0\rangle+\sqrt{2}|1\rangle\Big) \,, \\
    |\mathbf{n}_3\rangle &=
    \frac{-i}{\sqrt{3}}\Big(|0\rangle+e^{2\pi i/3}\sqrt{2}|1\rangle\Big) \,, \\
    \label{eq:Tetrahedron4}
    |\mathbf{n}_4\rangle &=
    \frac{i}{\sqrt{3}}\Big(|0\rangle+e^{-2\pi i/3}\sqrt{2}|1\rangle\Big) \,,
\end{align}
as in~\textcite{Mas95}.  It is straightforward to verify that
\begin{equation}
  \mathcal{E}_{2}(|i\rangle\langle i|)=\tfrac{1}{4}I\,, \quad
  \forall\ i \,,
\end{equation}
i.e., all four states are represented by Eve as the completely mixed
state. As these four states are orthogonal, they are completely
distinguishable by Bob and so provide an optimal private classical
communication scheme.

\paragraph{Three qubits:  Decoherence-full subsystems.}

Consider the case where Alice transmits three qubits to Bob. The
Hilbert space $\mathcal{H}_{1/2}^{\otimes 3}$ and the superoperator
$\mathcal{E}_3$ decompose into irreps as
\begin{equation}
  \underset{\mathbf{8}}{(\mathcal{H}_{1/2})^{\otimes3}}
  =\underset{\mathbf{4}}{\mathcal{H}_{j=3/2}}
  \oplus\big(\underset{\mathbf{2}}{\mathcal{M}_{j=1/2}}
  \otimes\underset{\mathbf{2}}{\mathcal{N}_{j=1/2}}\big)\, ,
\end{equation}
and
\begin{multline}
  \mathcal{E}_{3}=\mathcal{D}_{\mathcal{M}_{j=3/2}}\circ\mathcal{P}_{j=3/2} \\
  +(\mathcal{D}_{\mathcal{M}_{j=1/2}}\otimes
  \mathcal{I}_{\mathcal{N}_{j=1/2}})\circ\mathcal{P}_{j=1/2}\,,
\end{multline}
as in Eqs.~(\ref{eq:Hdecomp3qubits}) and (\ref{eq:twirling3qubits}).
Clearly, the four-dimensional subspace $\mathcal{H}_{j=3/2}$ is a
decoherence-full subspace.  We also see that any state on
$\mathcal{H}_{j=1/2}$ that is of the product form $\rho \otimes
\sigma$ with respect the factorization
$\mathcal{H}_{j=1/2}=\mathcal{M}_{j=1/2} \otimes
\mathcal{N}_{j=1/2}$ is mapped by $\mathcal{E}_3$ to the state
$\tfrac{1}{2}I_{\mathcal{M}_{j=1/2}} \otimes \sigma$ (see also
Eq.~\eqref{eq:twirling3qubits_states}). Thus, every state of the
virtual subsystem $\mathcal{M}_{j=1/2}$ is mapped to the completely
mixed state on that subsystem. Such a subsystem is an example of a
\emph{decoherence-full subsystem}.

Alice can therefore achieve private communication of two qubits
using the decoherence-full subspace $\mathcal{H}_{j=3/2}$ or a
single qubit using the decoherence-full subsystem
$\mathcal{M}_{j=1/2}$.  Note, however, that for greater numbers of
transmitted qubits, the decoherence-full subsystems typically have
greater dimensionality than the decoherence-full subspaces, and
schemes that encode within them are necessary to achieve optimal
efficiency, as discussed below.

For private \emph{classical} communication, the question of optimal
efficiency is much more complex.  One scheme would be for Alice to
encode two (classical) bits into four orthogonal states within the
$j=3/2$ decoherence-full subspace. Alice can also encode two bits
into four orthogonal \emph{maximally entangled} states on the
virtual tensor product $\mathcal{M}_{j=1/2} \otimes
\mathcal{N}_{j=1/2}$, because the depolarization on
$\mathcal{M}_{j=1/2}$ is sufficient to map all of these to
$\tfrac{1}{2}I_{\mathcal{M}_{j=1/2}} \otimes
\tfrac{1}{2}I_{\mathcal{N}_{j=1/2}}$, making them indistinguishable
to Eve.

It turns out that the optimally efficient scheme for private
classical communication uses \emph{both} the $j=3/2$ and $j=1/2$
subspaces.  Let $|j{=}3/2,\mu\rangle$, $\mu=1,\ldots,4$ be four
orthogonal states on the $j=3/2$ subspace, and let
$|j{=}1/2,\mu\rangle$, $\mu=1,\ldots,4$ be four maximally entangled
states (as described above) on the $j=1/2$ subspace.  Define the
eight orthogonal states
\begin{equation}
  \label{eq:EightOnThreeQubits}
  |b,\mu\rangle = \frac{1}{\sqrt{2}}\bigl(|j{=}3/2,\mu\rangle + (-1)^b
   |j{=}1/2,\mu\rangle\bigr) \, ,
\end{equation}
where $b=1,2$ and $\mu=1,\ldots,4$.  Alice can encode 3 bits into
these eight states, which are completely distinguishable by Bob. It
is easily shown that the decohering superoperator $\mathcal{E}_3$
maps all of these states to the completely mixed state on the total
Hilbert space; thus, these states are completely indistinguishable
from Eve's perspective.

\paragraph{General results.}

In general, an optimally efficient private quantum communication
scheme for $N$ spin-1/2 systems is given by encoding into the
largest decoherence-full subsystem for $\mathcal{E}_N$ of
Eq.~\eqref{eq:NQubitDecoheringChannel}. The largest is
$\mathcal{M}_{j=N/2}$ and has dimension $N+1$. Thus, given a private
Cartesian frame and the transmission of $N$ qubits, Alice and Bob
can privately communicate $\log_2(N+1)$ qubits.

The general results for private classical communication are much
more complex, and beyond the scope of this review.  (Observe the
complexity of even the three-qubit example above.)  Here, we simply
state the result, which is that the number of private classical bits
that can be communicated using a private shared Cartesian frame and
$N$ qubits is $3\log_2 N$~\cite{BRS04a}.

These results show that, asymptotically, the private classical
capacity ($3\log_2 N$) is three times the private quantum capacity
($\log_2 N$).  By relaxing the requirement of \emph{perfect}
privacy, it is possible to use the properties of random subspaces to
nearly triple the private quantum capacity, almost closing the gap
between the private classical and quantum capacities~\cite{BHS05}.
Finally, we note in passing that bipartite entangled states of $2N$
spins, completely mixed on the total-$J{=}0$ subspace, have been
identified as a resource for private quantum and classical
communication; it is illustrative to view such states as quantum
private shared RFs~\cite{Liv06}.

\section{Quantum treatment of reference frames}
\label{sec:QRFs}

As we have seen in the previous two sections, the lack of a
reference frame has the effect of inducing a superselection rule.
We have explored examples of how the lack of a phase reference in
quantum optics experiments leads to an Abelian SSR and how the lack
of a Cartesian frame leads to a non-Abelian SSR.

However, some SSRs are typically viewed as being \emph{axiomatic}; a
canonical example is a SSR for electric charge, which forbids
superpositions of eigenstates of different
charge~\cite{Wic52,WWW70,Str74}.  In a classic paper,
\textcite{Aha67} challenged the necessity of this SSR, and outlined
a gedanken experiment for exhibiting a coherent superposition of
charge eigenstates as an example of how this SSR can be obviated in
practice.  This gedanken experiment highlights the requirement of an
appropriate reference frame in order to exhibit superpositions
between eigenstates of superselected quantities, and as a result it
can be argued that an SSR is simply a \emph{practical} limitation
due to the lack of such a reference frame.  This point has been
repeated by several authors~\cite{Mir69,Mir70,Lub70,Giu00a,Giu00b}.

In this section, we demonstrate that this result is general:  any
SSR associated with a unitary representation of a compact group can
be viewed as the lack of an appropriate reference frame, and can be
overcome by using an appropriate quantum system to serve as a
reference frame.

\subsection{Relational descriptions of phase}
\label{subsec:Quantizationphaseref}

\subsubsection{Quantization of a phase reference}
\label{subsec:QuantizationphaserefA}

Suppose we have a system that transforms under U(1) and where the
associated eigenstates are denoted by $|n\rangle$. For concreteness,
we shall imagine these to be eigenstates of photon number, or of
number of bosonic atoms, in some mode.  If there is no SSR for U(1),
then we can prepare states such as
\begin{align}
\label{eq:psi0psi1}
    |\psi_{0}\rangle &=\big(|0\rangle+|1\rangle \big)/\sqrt{2} \,,  \\
    |\psi_{\pi}\rangle &=\big(|0\rangle-|1\rangle \big)/\sqrt{2} \,,
\end{align}
which differ only in their phases.  We distinguish such a state,
which has coherence between $|0\rangle$ and $|1\rangle$, from the
incoherent mixture $I/2=\frac{1}{2}|0\rangle\langle 0| +\frac{1}{2}
|1\rangle \langle 1|$ by measuring an ensemble of such systems in
the basis $\{|\psi_{0}\rangle,|\psi_{\pi}\rangle\}$ and observing
whether the outcome is random or not.

Now suppose instead that there is a SSR for U(1) in force for this
system.  For optical systems, this corresponds as discussed above to
the situation where one lacks the phase reference with which these
states were prepared. If the states $|n\rangle$ are eigenstates of
bosonic atom number (such as are used in describing Bose-Einstein
condensates), then such a SSR is often assumed to be an axiomatic
restriction (cf.~\textcite{Wic52,Cir96,Leg01,Wis03b}). In either
case, it becomes impossible to prepare a coherent superposition of
eigenstates of total number. Nonetheless, it is still possible to
prepare a \emph{pair} of systems in such a way that they have a well
defined \emph{relative phase}~\cite{Nem03}. We consider the pair
consisting of our original system, which we denote by $S$, and a new
system, which we denote by $R$. Defining the states
\begin{equation}
    |\chi_{0(\pi)}\rangle_R = \big(|n-1\rangle_R \pm |n\rangle_R \big)/\sqrt{2}\,,
\end{equation}
on $R$ (with $n\geq1$), we may then define states on the pair with
relative phases $0$ and $\pi$ respectively,
\begin{align}
    |\Psi_{0}\rangle_{RS} = \big( |\chi_{0}\rangle_R |\psi_0\rangle_S -
    |\chi_{\pi}\rangle_R |\psi_{\pi}\rangle_S \big)/\sqrt{2}\,, \\
    |\Psi_{\pi}\rangle_{RS} = \big( |\chi_{0}\rangle_R |\psi_\pi\rangle_S-
    |\chi_{\pi}\rangle_R |\psi_{0}\rangle_S \big)/\sqrt{2}\,.
\end{align}
Noting that these states can also be expressed as
\begin{align}
    \label{eq:OpticalStateRelPhase0}
    |\Psi_{0}\rangle_{RS} &= \big( |n\rangle_R |0\rangle_S +
    |n-1\rangle_R |1\rangle_S \big)/\sqrt{2}\,, \\
    |\Psi_{\pi}\rangle_{RS} &= \big( |n\rangle_R |0\rangle_S -
    |n-1\rangle_R |1\rangle_S \big)/\sqrt{2}\,,
\end{align}
it is clear that both of these states are eigenstates of total
number with eigenvalue $n$ and are therefore valid preparations
under the SSR.

Moreover, within the eigenvalue $n$ eigenspace, one can measure the
basis $\{\left\vert \Psi_{0}\right\rangle ,\left\vert
\Psi_{\pi}\right\rangle \}$ in order to statistically distinguish
states with a well-defined relative phase from those, like
$\frac{1}{2}\left\vert n+1\right\rangle \left\langle n+1\right\vert
\otimes\left\vert 0\right\rangle \left\langle 0\right\vert
+\frac{1}{2}\left\vert n\right\rangle \left\langle n\right\vert
\otimes \left\vert 1\right\rangle \left\langle 1\right\vert$, which
do not have a well-defined relative phase.  Clearly, this
measurement is also valid within the constraints of the SSR.

In fact, for every preparation, operation and measurement of the
system that is not U(1)-invariant, one can find an equivalent
preparation, operation and measurement for the relation between the
pair of systems that \emph{is} U(1)-invariant.  To do so, we simply
use the map
\begin{align}
\label{eq:quantizationmapforphase}
    \left\vert 0\right\rangle  & \rightarrow\left\vert
    n\right\rangle_{R}\left\vert 0\right\rangle_S \nonumber \\
    \left\vert 1\right\rangle  & \rightarrow\left\vert n-1\right\rangle
    _{R}\left\vert 1\right\rangle_S\, ,
\end{align}
so that in particular, we have
\begin{equation}
    \label{eq:StoRS}
    a|0\rangle + b|1\rangle
    \rightarrow a|n\rangle_{R}|0\rangle_S
    + b |n-1\rangle_{R}|1\rangle_S \,,
\end{equation}
for $|a|^2+|b|^2 =1$.

It is straightforward to generalize the quantization map of
Eq.~\eqref{eq:quantizationmapforphase} to the case of a system which
may have more than one photon. If it has at most $m_{\max }$
photons, we simply use the map
\begin{equation}
  \left\vert m\right\rangle
  \rightarrow \left\vert n-m\right\rangle_{R}
  \left\vert m\right\rangle _{S}
\end{equation}
where we require that $n\geq m_{\max }$. In this case,
\begin{equation}
\label{eq:StoRS2}
    \sum_{m=0}^{m_{\max }}c_{m}| m\rangle
    \rightarrow \sum_{m=0}^{m_{\max }}c_{m}|n-m\rangle_{R}|m\rangle_{S}\,.
\end{equation}

This extension of the Hilbert space corresponds physically to
\emph{incorporating the phase reference into the quantum formalism}.
In other words, it describes the \emph{internalization} or
\emph{quantization} of the reference frame. To see this, consider
the following analogy with classical mechanics. Suppose a ball is
bounced off of a wall. If we do not treat the wall as a dynamical
entity, but rather as an external potential that appears in the
equations of motion of the ball, then the solutions to the equations
of motion are not translationally-invariant. Specifically, if we
take a given bouncing trajectory for the ball and translate it in
such a way that the bounce no longer coincides with the location of
the wall, we do not obtain another solution -- the external
potential breaks the translation-invariance. However, if we
\emph{internalize} the wall, that is, treat its position as a
dynamical degree of freedom, then we find that the equations of
motion, and the solutions, will be invariant under translations of
the entire system (consisting of the ball and the wall).

Similarly, when one writes down a state such as
$|\psi_{0(\pi)}\rangle = (|0\rangle\pm|1\rangle)/\sqrt{2}$, the
phase of this state is only defined relative to an external phase
reference. We can view this external phase reference as a type of
external potential, which provides the means for preparing states
and performing operations (i.e., giving solutions to the
quantum-mechanical equations of motion) that are not invariant under
phase shifts. However, if we incorporate the phase reference as an
internal system (and we do not compare our internal systems to any
other external phase reference), then the only empirically
meaningful states and operations are invariant under phase shifts of
the entire system (including the internalized phase reference).

Whether one treats the wall in our classical example as an external
potential or an internal dynamical system is a choice of the
physicist.  Similarly, one can treat a reference frame internally or
externally; with either choice, one can obtain an empirically
adequate description of the experiment~\cite{BRS06}.

\subsubsection{Dequantization of a phase reference}
\label{dequantphaseref}

It is useful to consider the opposite problem to the one considered
above, namely, given a description of an experiment wherein the
phase reference is being treated internally, how does one obtain a
description wherein it is treated externally?  In our classical
example of a ball bouncing off a wall, this involves finding the
equations of motion for the relative position of the ball to the
wall.

We would like to determine the quantum analogue of this process. In
the context of a quantum reference frame for spatial location, it is
relatively straightforward. To externalize the reference frame, one
defines a novel tensor product structure of the Hilbert space in
terms of the commuting pair of observables $q_{R}-q_{S}$ and
$p_{R}+p_{S},$ where $q_{R},p_{R}$ and $q_{S},p_{S}$ are the
position and momentum operators for the reference frame and system
respectively.

The procedure is a bit more subtle in our U(1) example, but also
involves identifying a novel tensor product structure of the Hilbert
space.  The original tensor product structure, corresponding to the
reference frame and system division, will be denoted
$\mathcal{H=H}_{R}\otimes\mathcal{H}_{S}$.  The product states with
respect to this structure, $\left\vert n\right\rangle _{R}\left\vert
m\right\rangle_{S}$, are simultaneous eigenstates of $\hat{N}_{R}$,
the number operator for $R$, and $\hat{N}_{S}$, the number operator
for $S$ with eigenvalues $n$ and $m$ respectively.  The operators
$\hat{N}_{R}$ and $\hat{N}_{S}$ form a complete set of commuting
operators for the Hilbert space $\mathcal{H}$.

By choosing a different complete set of commuting operators, we can
define an alternate tensor product structure for the Hilbert space.
Specifically, we choose $\hat{N}_{S}$, the number operator for $S$,
and $\hat{N}_{\mathrm{tot}}=\hat{N}_{S}+\hat{N}_{R}$, the total
number operator. The state $|m\rangle_{R}|n\rangle_{S}$ is also a
joint eigenstate of this pair, with eigenvalues $m$ and $m+n$
respectively.  Given that $\hat{N}_{S}$ and $\hat{N}_{\mathrm{tot}}$
form a complete set of commuting observables, we may label an
element of the basis $\{|n\rangle_{R}|m\rangle_{S}\}$ instead by the
eigenvalues of $\hat{N}_{S}$ and $\hat{N}_{\mathrm{tot}}$, that is,
$|n\rangle_{R}|m\rangle_{S}=|N_{\mathrm{tot}}=m+n,N_{S}=m\rangle$.

Now, if it were the case that any pair of values, one drawn from the
spectra of $N_{S}$ and the other drawn from the spectra of
$N_{\mathrm{tot}}$, could be simultaneous eigenvalues of $N_{S}$ and
$N_{\mathrm{tot}}$, then we could define a new tensor product
structure by $|N_{\mathrm{tot}} = l,N_{S}=m\rangle = |l\rangle
\otimes|m\rangle $. However, any pair $(l,m)$ with $m>l$ cannot be
simultaneous eigenvalues.  This problem can be resolved by
restricting our attention to states $|n\rangle_{R}|m\rangle_{S}$
where the minimum value of $n$ is larger than the maximum value of
$m$. Recalling the physical significance of these eigenvalues, we
see that this corresponds to assuming that the RF has more
excitations than the system.

Assuming a system with at most $m_{\max}$ excitations, and a
reference with a number of excitations that is at least $m_{\max}$,
we may focus upon the subspace
$\mathcal{H}'=\mathrm{span}\{|n\rangle_{R}|m\rangle_{S}\,,
m=0,\ldots,m_{\max},n\ge m_{\max}\}$. It is then straightforward to
introduce a tensor product structure on $\mathcal{H}'$ as follows.
We define an $m_{\max}$-dimensional Hilbert space
$\mathcal{H}_{\text{rel}}$ with an orthonormal basis
$|m\rangle_{\text{rel}}$ labeled by the eigenvalue $m$ of
$\hat{N}_{S}$. We call this the \emph{relational} Hilbert space. We
also define a Hilbert space $\mathcal{H}_{\text{gl}}$ with an
orthonormal basis $|l\rangle_{\text{gl}}$ labeled by the eigenvalue
of $\hat{N}_{\text{\textrm{tot}}}$.  We call this the \emph{global}
Hilbert space.  We then have a vector space isomorphism
\begin{equation}
    \mathcal{H}'\cong\mathcal{H}_{\text{gl}}
    \otimes\mathcal{H}_{\text{rel}}\,,
\end{equation}
which is made by identifying
\begin{equation}
    |l\rangle_{\text{gl}}|m\rangle_{\text{rel}}
    \equiv|N_{\mathrm{tot}}{=}l,N_{S}{=}m\rangle\,,
\end{equation}
for all $m\leq m_{\max}$ and $l\geq m_{\max}$.

We can therefore define a linear map from the subspace
$\mathcal{H}^{\prime}$ of $\mathcal{H}_{R}\otimes\mathcal{H}_{S}$ to
$\mathcal{H}_{\mathrm{gl}}\otimes\mathcal{H}_{\mathrm{rel}}$ in
terms of their respective basis states as
\begin{equation}
    |n\rangle_{R}\left\vert m\right\rangle_{S}
    \mapsto|m+n\rangle_{\mathrm{gl}}|n\rangle_{\mathrm{rel}}\,.
\end{equation}
Under this map, we have
\begin{equation}
    a\left\vert n+1\right\rangle _{R}|0\rangle_{S}
    +b|n\rangle_{R}\left\vert 1\right\rangle _{S}
    \mapsto\left\vert n\right\rangle_{\mathrm{gl} } \otimes (a|0\rangle_{\mathrm{rel}}
    +b|1\rangle_{\mathrm{rel}})
    \,.
    \label{eq:MapEntanglementToCoherence}
\end{equation}
Any U(1)-invariant state on $\mathcal{H}_R \otimes \mathcal{H}_S$
will lead to a state on
$\mathcal{H}_{\mathrm{rel}}\otimes\mathcal{H}_{\mathrm{gl}}$ that
commutes with $\hat{N}_{\textrm{tot}}$, i.e., the state will be
diagonal in the number basis of $\mathcal{H}_{\mathrm{gl}}$. By
discarding the global degrees of freedom and considering only the
reduced density matrix on $\mathcal{H}_{\textrm{rel}}$, we are
essentially moving to a paradigm of description wherein the RF is
not treated within the quantum formalism. We call this procedure
\emph{externalizing} or \emph{dequantizing} the reference frame. For
instance, if we follow the map of
Eq.~(\ref{eq:MapEntanglementToCoherence}) by a trace over
$\mathcal{H}_{\mathrm{gl}}$, we obtain the map
\begin{equation}
    a\left\vert n+1\right\rangle _{R}|0\rangle_{S}
    +b|n\rangle_{R}\left\vert 1\right\rangle _{S}
    \mapsto a|0\rangle_{\mathrm{rel}}
    +b|1\rangle_{\mathrm{rel}}
    \,,
    \label{eq:MapEntanglementToCoherence2}
\end{equation}
which is the inverse of Eq.~\eqref{eq:StoRS}, the map describing the
internalization or quantization of the phase reference.

\subsubsection{The optical coherence controversy}

This simple analysis of the quantum treatment of reference frames is
useful for resolving a controversy concerning whether quantum
coherences between photon number eigenstates are fact or
fiction~\cite{Mol97,Gea98,Mol98,Rud01a,vEF02a,vEF02b,Rud01b,Nem02,Nem03,Nem04,San03,Wis03,Wis04,Fuj03,Smo04,SS03,BRS06}.
It is standard practice in quantum optics to model the state of the
electromagnetic field generated by a laser to be a coherent state,
which is a coherent superposition of photon number eigenstates. One
justification that may be given for such an approach is that if one
imagines the source of the radiation to be a classical oscillating
dipole (which seems a reasonable assumption) then a simple
calculation shows that the field is left in a coherent state. On the
other hand, if one quantizes the dipole moments in the gain medium
and assumes that these are initially in a thermal state (which must
have zero expectation value of the dipole moment operator), and that
the coupling between the gain medium and the radiation field
conserves photon number (which again seem like reasonable
assumptions), then the reduced density operator of the field is
found to be in an \emph{incoherent mixture} of photon number
eigenstates~\cite{Mol97}.  The fact that distinct states are
obtained by the two analyses has led many researchers to conclude
that the two descriptions are inconsistent and that one must be
wrong.

To gain insight into this controversy, it is useful to consider the
gain medium as a phase reference for the radiation field.  Rather
than considering this case in detail, we return to the example of
the previous section, which provides a simplified version of the
controversial phenomena. Recall that we also considered two distinct
paradigms of description for a system $S$ and a phase reference $R$.
In the first description -- the \emph{external-$R$} paradigm -- only
$S$ was treated quantum mechanically, so that the total Hilbert
space was $\mathcal{H}_S$. In the second description -- the
\emph{internal-$R$} paradigm -- both $S$ and $R$ were treated
quantum mechanically, so that the total Hilbert space was
$\mathcal{H}_S \otimes \mathcal{H}_R$. Moreover, the state on
$\mathcal{H}_{\text{S}}$ is different in the two cases. For
instance, if the state of $S$ in the external-$R$ paradigm is
$|\psi_{0}\rangle=(|0\rangle+|1\rangle)/\sqrt{2}$ of
Eq.~\eqref{eq:psi0psi1}, after internalizing $R$, the joint state is
$|\Psi_{0}\rangle$ of Eq.~\eqref{eq:OpticalStateRelPhase0}, and the
state on $\mathcal{H}_{S}$ is $\frac{1}{2} |0\rangle\langle 0|
+\frac{1}{2} |1\rangle\langle 1|$.

Thus, just as the classical and quantum treatments of the gain
medium in the generation of laser light led to distinct state
ascriptions for the radiation field, our classical and quantum
treatments of the phase reference $R$ lead to distinct state
ascriptions for the system $S$.  It is a mistake however to conclude
that the two descriptions are inconsistent. As we saw in the
previous subsection, both descriptions are valid. To resolve the
confusion explicitly, we elaborate on the physical interpretation of
the states in these Hilbert space.

In the external-$R$ paradigm, we saw that the phase of the quantum
state of $S$ (that is, the phase of the ratio of amplitudes of
$|0\rangle$ and $|1\rangle$) can only be given meaning relative to
the external phase reference $R$. So it is clear that the state on
$\mathcal{H}_S$ describes not just the intrinsic properties of $S$,
but some of its extrinsic properties as well, specifically, its
relation to $R$.

In the internal-$R$ paradigm, any phase of the quantum state of $S$
can also only be given meaning relative to an external phase
reference, but $R$ is no longer an external RF, and any phase
reference that is still treated externally, say $R'$, has been
assumed not to be correlated with $S$. Thus, we expect $S$ to not
have a well-defined phase in this case.  The point is that in the
internal-$R$ paradigm, $\mathcal{H}_S$ also describes extrinsic
properties of $S$, but in this case it is the relation of $S$ to
$R'$, rather than $R$.

Thus, the fact that the quantum states on $\mathcal{H}_S$ are
distinct in the two paradigms is not an inconsistency because
despite the common notation, they describe different degrees of
freedom: one describes the relation of $S$ to $R$ and the other the
relation of $S$ to $R'$.

Moreover, if one wishes to recover the quantum state describing the
relation of $S$ to $R$ in the internal-$R$ paradigm, it is clear
that one should not look at the quantum state on $\mathcal{H}_S$
because this amount to tracing over $\mathcal{H}_R$ which
corresponds to ignoring $R$, and one clearly cannot ignore a system
when one seeks to find the relation between it and another.  But
where then is information about the relation between $S$ and $R$
found in $\mathcal{H}_S\otimes\mathcal{H}_R$?  The answer is that it
is found in a \emph{virtual subsystem}, specifically in the Hilbert
space $\mathcal{H}_{\text{rel}}$.  The resolution of the optical
coherence controversy is achieved in an analogous manner.

The key insight for resolving these sorts of confusions is that
quantum states of systems in an external RF paradigm do not simply
describe its intrinsic properties, but also the \emph{relation} of
the system to the external RF; further discussion on this issue can
be found in~\cite{Wis04,BRS06}.

\subsubsection{Generalization to composite systems}

The generalization of the quantization procedure in
Sec.~\ref{subsec:QuantizationphaserefA} to the case of multiple
systems (i.e., modes) is not so straightforward. The problem is that
if we wish to describe a pair of systems $S_{1}$ and $S_{2}$
relative to an RF $R$, then the reduced density operators on
$RS_{1}$ and on $RS_{2}$ cannot both be pure entangled
states~\cite{Cof00}. This fact is known as the monogamy of pure
entanglement.  As a result, the quantum description of RF and system
that was presented above is only adequate if the system in question
is the only one that will ever be compared to the RF. However, the
most general notion of an RF is something with respect to which the
orientation of \emph{many} systems can be defined. We consider such
a generalization presently.

First, note that if we demand that there be no limit on the number
of systems that can be correlated with the RF, and that the degree
of correlation with the RF be equal for all the systems, then the
reduced density operator on $RS$ for an arbitrary system $S$ must be
unentangled, that is, a \emph{separable} state.  At first glance,
this might seem problematic, because it might seem that the
entanglement in Eqs.~\eqref{eq:StoRS} and \eqref{eq:StoRS2} is
critical for the RF quantization procedure to work.  It is true that
if we restrict ourselves to a subspace of $\mathcal{H}_{R}$ of the
same dimension as the system of interest, as we did in
Eqs.~\eqref{eq:StoRS} and \eqref{eq:StoRS2}, then we cannot obtain a
faithful representation. However, by allowing ourselves to make use
of a larger subspace, we can obtain a good representation, and in
the limit of arbitrarily large dimension, we obtain a perfect
representation, as before. Defining the unnormalized states
\begin{equation}
    |\chi_\phi\rangle_{R}
    =\sum_{n=0}^{\infty} e^{i\phi n}|n\rangle_{R}\,,
\end{equation}
which have well-defined phase, the quantization map takes the form
\begin{equation}
    \rho \rightarrow \int {\rm d}\phi |\chi_\phi \rangle_{R}
    \langle \chi_\phi | \otimes U(\phi ) \rho U^{\dag}(\phi)\,.
\end{equation}
We must demonstrate that this is a faithful representation of the
system that satisfies the U(1)-SSR. Rather than doing so for the
phase reference case individually, we proceed directly to present
the generalization of this quantization map to an arbitrary group,
and prove that the latter has the properties we desire.

\subsection{Quantization of a general reference frame}
\label{subsec:QuantizationGeneral}


Consider the quantum description of a system with Hilbert space
$\mathcal{H}_{S}$, in the case where one possesses an external
reference frame for a degree of freedom associated with the group
$G$. The Born rule predicts that, for a preparation associated with
density operator $\rho$ followed by a transformation associated with
operation $\mathcal{E}$ and finally a measurement associated with
the POVM $\{E_{k}\}_{k}$, the probability of the measurement outcome
$k$ is $\mathrm{Tr}[\mathcal{E}(\rho)E_{k}]$.

Now consider a quantum description for the same system, but where a
SSR for $G$ is in force.  This SSR implies a restriction on the
states, transformations, and measurements.  However, as we now
demonstrate, it is possible to append this system with another
quantum system $R$ which serves as a quantum reference frame in such
a way that, although a SSR for $G$ applies to the entire composite
system $RS$, the RF allows us to \emph{effectively} describe the
system as if the SSR did not exist~\cite{Kit04}.

Our aim is to map the elements of the old representation (of
preparations, operations, measurements) to elements in a new,
$G$-invariant representation on $\mathcal{H}_{R}
\otimes\mathcal{H}_{S}$.  Thus, we seek a map
\begin{align}
    \rho &\rightarrow\rho^{\mathrm{inv}}\,,\\
    \{E_{k}\}_{k} &\rightarrow\{E_{k}^{\mathrm{inv}}\}_{k}\,,\\
    \mathcal{E} &\rightarrow\mathcal{E}^{\mathrm{inv}}\,.
\end{align}
such that $\rho^{\mathrm{inv}}=\mathcal{G}(\rho^{\mathrm{inv}})$ and
$E_{k} ^{\mathrm{inv}}=\mathcal{G}(E_{k}^{\mathrm{inv}})$ are both
$G$-invariant operators on $\mathcal{H}_{R} \otimes\mathcal{H}_{S}$,
and $\mathcal{E}^{\mathrm{inv}}$ is a $G$-invariant superoperator
 on $\mathcal{B}(\mathcal{H}_{R}) \otimes \mathcal{B}(\mathcal{H}_{S})$
(this implies that when acting on a $G$-invariant operator
$\tilde{A}$, $\mathcal{E}^{\mathrm{inv}}$ satisfies $(\mathcal{G}
\circ \mathcal{E}^{\mathrm{inv}} \circ \mathcal{G})[\tilde{A}] =
\mathcal{E}^{\mathrm{inv}}[\tilde{A}])$.

In addition, we would like this map to preserve the statistical
predictions of the old representation; if all the statistics of the
Born rule can be reproduced in this new representation, then it is
equivalent to the old one.  Specifically, we want this map to be
such that
\begin{equation}
  \mathrm{Tr}_{RS}[\mathcal{E}^{\mathrm{inv}}(\rho^{\mathrm{inv}})
  E_{k}^{\mathrm{inv}}]=\mathrm{Tr}_{S}[\mathcal{E}(\rho)E_{k}]\,,
\label{eq:Bornrule}
\end{equation}
for all states $\rho$, operations $\mathcal{E}$ and measurements
$\{E_k\}_k$. Such a map does exist (assuming we can allow $d_{R}$,
the dimension of $\mathcal{H}_R$, to be arbitrarily large), as we
now demonstrate.

First, the quantum system $R$ that will constitute the RF for $G$
must clearly transform under $G$ in some nontrivial manner.  Thus,
$\mathcal{H}_{R}$ must carry a representation of $G$, denoted $U_R$,
which in general will be reducible.  In order for $R$ to serve as a
\emph{complete} quantum RF for $G$, the state $|g\rangle$ on
$\mathcal{H}_R$ corresponding to the configuration $g\in G$ must not
possess a non-trivial invariant subgroup, i.e., if $U_R(g')|g\rangle
\propto|g\rangle$ then $g'$ must be the identity. It follows that
the states of $R$ transform as,
\begin{equation}
  U_R(g')|g\rangle =|g'g\rangle\, ,\quad \forall\ g,g'\in G\,.
\end{equation}
For this quantum system to function as a \emph{perfect} reference
frame for $G$, the different configurations $|g\rangle$ must all be
distinguishable.  Thus, we require that states for different
configurations are orthogonal
\begin{equation}
  \label{eq:DeltaFunction}
  \langle g|g'\rangle = \delta(g^{-1}g')\,,
\end{equation}
where $\delta(g)$ is the delta-function on $G$ defined by $\int
\mathrm{d}g\,\delta(g)f(g)=f(e)$ for any continuous function $f$ of
$G$, where $e$ is the identity element in $G$.  The above
requirements are the defining properties of the \emph{left regular
representation} of $G$.  In the case of a Lie group, the
dimensionality of $\mathcal{H}_{R}$ must be infinite for such states
to exist.  We refer to such an infinite-dimensional quantum RF as
\emph{unbounded}.\footnote{For finite groups, one need only assume
that $\langle g|g'\rangle = \delta_{g,g'}$ where $\delta_{g,g'}$ is
the Kronecker-delta.}

We now present the map from operators on $\mathcal{H}_{S}$ to
$G$-invariant operators on $\mathcal{H}_{R}\otimes\mathcal{H}_{S}$:
\begin{equation}
  \label{eq:MoneyMap}
  \$:A\mapsto\int_{G}\mathrm{d}g\, |g\rangle\langle g|
  \otimes U_S(g)AU_S^{\dag}(g)\,,
\end{equation}
where $U_S$ is the representation of $G$ on the system.

Using this map $\$$, we define the invariant versions of density
operators, elements of POVMs and Kraus operators respectively as
\begin{align}
    \label{eq:InvDensity}
    \rho^{\mathrm{inv}}  &  =\frac{1}{d_R}\$(\rho)\,, \\
    E^{\mathrm{inv}}  &  =\$(E)\,,\\
    K^{\mathrm{inv}}  &  =\$(K)\,,
\end{align}
where $d_R$ is the dimensionality of the Hilbert space
$\mathcal{H}'_R\equiv \mathrm{span}\{\left\vert g\right\rangle,g\in
G\}$ spanned by the orbit of the RF states, which may be a subspace
of $\mathcal{H}_R$. (One can easily check that
$\mathrm{Tr}_{RS}[\rho^{\mathrm{inv}}] = 1$ if
$\mathrm{Tr}_S[\rho]=1$.)

The following are properties of the $\$$ map:
\begin{enumerate}
  \item $\$(A)$ is $G$-invariant;
  \item $\$(A+B)=\$(A)+\$(B)$ and $\$(AB)=\$(A)\$(B),$ so the algebra
of operators is reproduced.
\end{enumerate}
The $G$-invariance of $\$(A)$ follows from
\begin{align}
    (U_R(g^{\prime})&\otimes
    U_S(g^{\prime}))\$(A)(U_R^{\dag}(g^{\prime})\otimes
    U_S^{\dag}(g^{\prime}))\nonumber\\
    &=\int_{G}\mathrm{d}g |g'g\rangle\langle g'g |
    \otimes U_S(g'g)AU_S^{\dag}(g'g) \nonumber \\
    &=\$(A)\,,
\end{align}
where the final equality follows from the invariance of the Haar
measure $\mathrm{d}g$.  To prove property 2, we note that $\$$ is
linear by definition, and that
\begin{align}
    \$(A)\$(B)  &  =\int \mathrm{d}g\,\mathrm{d}g'\, |g\rangle
    \langle g|g'\rangle \langle g'| \nonumber \\
    &\qquad \otimes U_S(g)AU_S^{\dag}(g)U_S(g^{\prime})BU_S^{\dag}(g^{\prime})
    \nonumber \\
    &  =\int \mathrm{d}g\, |g\rangle
    \langle g| \otimes U_S(g)ABU_S^{\dag}(g)
    \nonumber \\&=\$(AB)\,,
\end{align}
where we have used Eq.~\eqref{eq:DeltaFunction}.

From these properties, one can show that if $\rho$ is a density
operator, then so is $\rho^{\text{\textrm{inv}}}$, if $\{E_{k}\}$ is
a POVM, then so is $\{E_{k}^{\mathrm{inv}}\}$, and that if
$\mathcal{E}$ is a superoperator with Kraus operators $\{K_{\mu}\}$
satisfying $\sum_{\mu}K_{\mu}^{\dag}K_{\mu}=E$, then the
superoperator $\mathcal{E}^{\mathrm{inv}}$ having Kraus operators
$\{K_{\mu}^{\mathrm{inv}}\}$ satisfies
$\sum_{\mu}(K_{\mu}^{\mathrm{inv}})^{\dag}
K_{\mu}^{\mathrm{inv}}=E^{\mathrm{inv}}$.  Most importantly, one can
prove that the new representation satisfies Eq.~(\ref{eq:Bornrule})
and therefore reproduces the quantum statistics:
\begin{align}
  \mathrm{Tr}_{RS}[&\rho^{\mathrm{inv}}E_k^{\mathrm{inv}}] \notag \\
  &= d_{R}^{-1}\mathrm{Tr}_{RS}[\$(\rho)\$(E_k)]  \notag \\
  &= d_{R}^{-1}\mathrm{Tr}_{RS}[\$(\rho E_k)]  \notag \\
  &= d_{R}^{-1}\mathrm{Tr}_{RS}\Bigl[\int_{G}\mathrm{d}g\, |g\rangle\langle g|
  \otimes U_S(g)\rho E_k U_S^{\dag }(g) \Bigr]  \notag \\
  &= d_{R}^{-1}\mathrm{Tr}_{R}\Bigl[\int_{G}\mathrm{d}g\,|g\rangle\langle g| \Bigr]
  \mathrm{Tr}_S[\rho E_k]  \notag \\
  &= \mathrm{Tr}_S[\rho E_k]\, .  \label{eq:InfiniteBorn}
\end{align}
The case where there is a nontrivial operation $\mathcal{E}$ can be
dealt with similarly.

This is a remarkable result.  It proves that superselection rules
cannot provide any fundamental restrictions on quantum theory.  This
has particular implications for quantum cryptography as we discuss
below.  It also proves that \emph{all} superselection rules
associated with unitary representations of compact groups result
from a lack of an appropriate reference frame, because, as we have
shown, including an unbounded quantum reference frame reproduces a
quantum theory that is equivalent to one in which the superselection
rule does not apply.

\subsection{Are certain superselection rules fundamental?}

We now return to the question, introduced at the beginning of this
section, of whether certain superselection rules are more
fundamental than others.  That is, are certain SSRs
\emph{axiomatic}, as opposed to those which arise in practice when
there is not an appropriate RF?  This issue bears on several
controversies that are the counterparts of the optical coherence
controversy in other contexts. It has arisen in the context of
coherence between charge eigenstates in
superconductivity~\cite{And86,Haa62,Ker74} and of coherence between
atom number eigenstates in Bose-Einstein
condensation~\cite{Jav96,Hos96,Cas97,Yoo97,Leg01}. Here, however,
the intuition for the coherences being a fiction is based on the
notion that the superselection rule for charge and for baryon number
are axiomatic, so that any quantum state that violates this SSR does
not represent reality.

To make the discussion definite, let us compare on the one hand
quantities such as charge and baryon number, for which axiomatic
SSRs are conventionally assumed to apply, and on the other,
quantities such as linear momentum, angular momentum and photon
number, for which SSRs are generally not assumed to apply.  We wish
to consider whether our conclusion, that it is possible to
effectively lift a superselection rule, should apply to both of
these equally.

Certainly, the example of the phase reference provided in
Sec.~\ref{subsec:Quantizationphaseref} applies equally well in the
case of atom number (and thus baryon number) as it does to the case
of photon number. In both cases, one can certainly create
well-defined relative phases between a pair of systems.  Moreover,
the reasons for interpreting the larger of the two systems as a
reference frame for the other are just as valid in the case of atom
number as they are in the case of photon number.  Finally, in both
cases one can recover a description of the relational degree of
freedom, wherein one effectively has lifted the SSR.

Given the generalization to non-Abelian groups, provided in
Sec.~\ref{subsec:QuantizationGeneral}, it would appear that all such
superselection rules may be lifted in practice.  Of course, the
technical challenge in doing so is to build a reference frame for
the degree of freedom in question. Admittedly, it may be more
difficult to construct good reference frames for some degrees of
freedom, but there is nothing in principle preventing their
construction. For instance, to lift the superselection rule
associated with charge, one must simply have a large reference
system with respect to which one can coherently exchange charge, as
argued by \textcite{Aha67}. As another example, the experimental
realization of Bose-Einstein condensation in alkali atoms provided a
reference frame for the phase that is conjugate to atom
number~\cite{Dow06b}. We see no obstacle in principle to lifting
more general sorts of superselection rules as well.

What sets the two categories apart in practice seems to be the fact
that some reference frames, such as those for spatial location or
angular position, are ubiquitous, whereas others, such as a frame
for the quantity conjugate to charge, tend not to arise through
natural causes and are difficult to prepare and maintain.  But this
may be only a practical and not a fundamental difference.

Another motivation might be given for treating the two categories
differently, specifically, that a superselection rule for linear
momentum would seem to imply that objects could not be localized in
space, and \emph{this}, one might think, would be contrary to what
is observed. However, all that is ever observed empirically is the
localization of systems relative to other systems, and this is
consistent with a superselection rule for total linear momentum. If
one seeks to describe the entire universe quantum mechanically, as
is typically done in quantum cosmology and some approaches to
quantum gravity, then it is natural to assume SSRs for all global
transformations, so that there is no distinction between charge and
linear momentum.  One can reach this conclusion by noting that all
physical systems that could serve as RFs have been quantized.
Alternatively, one can appeal to one of the central lessons of
general relativity: that all observable quantities ought to be
relational.

\subsection{Superselection rules and quantum cryptography}
\label{subsec:SSRandCrypto}

Information-theoretic security is a form of security that does not
rely on assumptions about the computational capabilities of one's
adversary. The appeal of quantum cryptography is that it offers
protocols achieving this sort of security where classical protocols
fail. Quantum key distribution is the primary example of a task for
which this is the case. On the other hand, there exist cryptographic
tasks, such as bit commitment, for which it has been shown that even
quantum protocols cannot achieve information-theoretic security. The
possibility of quantum key distribution arises ultimately from a
\emph{restriction} imposed by the laws of quantum mechanics on
would-be eavesdroppers -- namely, that quantum information cannot be
cloned. By definition, SSRs also impose restrictions on the
accessible quantum states and operations. For instance, a SSR for
charge forbids the creation of superpositions of eigenstates of
differing total charge. It is conceivable therefore, as first
suggested by \textcite{Pop02}, that SSRs could place restrictions on
would-be cheaters and thereby achieve greater security for some
tasks (for instance, unconditional security for bit commitment)
\cite{May02,Ver03,Kit04,DiV04}.

To motivate the intuition that SSRs might improve the security of
quantum protocols, we consider the case of a partially binding and
partially concealing bit commitment protocol \cite{SR01} in the
presence of a superselection rule for SO(3). Alice prepares two
qubits in either the singlet state $\left\vert \psi
^{-}\right\rangle $, which has total spin 0, or the triplet state
$|11\rangle $, which has total spin 1, according to whether she
wants to commit a bit $b=0$ or $1$ respectively. She sends Bob one
of the two qubits as a token of her commitment. Bob cannot
distinguish the reduced states $I/2$ and $\left\vert 1\right\rangle
\left\langle 1\right\vert $ with certainty and so the protocol is
partially concealing. At a later stage, she sends him the second
qubit, at which point Bob checks her honesty by performing a
projective measurement to discriminate $\left\vert
\psi^{-}\right\rangle $ from $|11\rangle$. There is no cheating
strategy that allows Alice to unveil an arbitrary bit value, so the
protocol is partially binding. Clearly each step in the honest
protocol respects the SSR. However it is quite plausible, at first
sight, that an optimal cheating strategy for Alice will not respect
the SSR -- either because she must prepare a state which is a
superposition of two different angular momenta, such as $(\left\vert
\psi ^{-}\right\rangle +\left\vert 11\right\rangle )/\sqrt{2}$, or
because prior to sending the second qubit to Bob she must apply to
it some local operation that violates the SSR. If all of Alice's
optimal cheating strategies required SSR violation, then the degree
of bindingness against Alice and thus the security of the protocol
would be greater by virtue of the SSR.

Despite the plausibility of this notion, it turns out that SSRs
\emph{do not}, in general, offer the possibility of cryptographic
protocols with greater security~\cite{Kit04}.  This result can be
proven using the general framework of
Sec.~\ref{subsec:QuantizationGeneral}. We begin by demonstrating
this for the case of arbitrary two-party cryptographic protocols.
Such protocols can be formulated as follows. Alice and Bob each hold
a local system in their laboratories, called $A$ and $B$
respectively, and exchange a message system $M$ back and forth. At
the outset, they share a product state $\rho
_{A}\otimes\rho_{M}\otimes\rho_{B}$ and in each round of the
protocol, one of the parties applies a joint operation on their
local systems and the message system and then sends the message
system to the other party. At the end, both parties perform a
measurement on their local system.

Security in this context is a restriction on the degree to which a
cheating Alice can influence the probability distribution over the
outcomes of the final measurement of an honest Bob, and a similar
restriction with the roles of Alice and Bob reversed. (no
restrictions are guaranteed for the case where both parties cheat.)
We consider the case of a cheating Alice here.

Because we may include any ancillas used by Alice and Bob in the
local systems $A$ and $B,$ we can assume that all operations are
unitary. \ In the honest protocol, the first operation implemented
by Alice is $V_{A_{1}},$ the first implemented by Bob is
$V_{B_{1}},$ the second by Alice is $V_{A_{2}},$ and so forth. We
denote the POVM associated with Bob's final measurement by
$\{E_{B,k}\}$ where $k$ labels the possible outcomes. The
probability of outcome $k$ is
\begin{equation}
    p_{B}(k)=\mathrm{Tr}\left(
    E_{B,k}V(\rho_{A}\otimes\rho_{M}\otimes\rho _{B})V^{\dag}\right)\,,
\end{equation}
where
\begin{equation}
    V=V_{B_{n}}V_{A_{n}}\cdots V_{B_{2}}V_{A_{2}}V_{B_{1}}V_{A_{1}}\,.
\end{equation}

Suppose that the honest protocol respects the SSR, and that the SSR
is associated with a group $G.$  In this case, all the states,
unitaries and POVM elements described above are $G$-invariant
operators.  We now show that a cheating strategy that violates the
SSR can always be simulated by a cheating strategy that respects the
SSR, and consequently a cheater that faces a SSR does not suffer any
disadvantage in cheating ability compared to one who does not.

Suppose that Alice's optimal SSR-violating cheating strategy is one
wherein she replaces each $G$-invariant operation $V_{A_{j}}$ with
an operation $V_{A_{j}}^{\prime}$ that need not be $G$-invariant.
She thereby can cause the probability of outcome $k$ to be
$p_{B}^{\prime}(k)=\mathrm{Tr}\left(
E_{B,k}V^{\prime}(\rho_{A}\otimes\rho_{M}\otimes\rho_{B})V^{\prime\dag
}\right)  $ where $V^{\prime}=V_{B_{n}}V_{A_{n}}^{\prime}\cdots
V_{B_{2} }V_{A_{2}}^{\prime}V_{B_{1}}V_{A_{1}}^{\prime}.$ \ We now
demonstrate that there is an SSR-\emph{respecting} cheating strategy
that also leads to $p_{B}^{\prime}(k).$ \ The trick is to use the
construction of Sec.~\ref{subsec:QuantizationGeneral}. \ Alice
simply extends her local system $A$ to $RA,$ where $R$ is a system
that will play the role of a local
reference frame. She replaces the $G$-noninvariant operation $V_{A_{j}%
}^{\prime},$ which acts nontrivally on $AM,$ with
$\$(V_{A_{j}}^{\prime}),$ where $\$$ is the map defined in
Eq.~(\ref{eq:MoneyMap}).  $\$(V_{A_{j}}^{\prime})$ is a
$G$-invariant unitary operator that acts nontrivially on $RAM.$ \
Bob's operations must be trivial on $R,$ so that we can write these
as $I_{R}\otimes V_{B_{j}}.$ \ It is useful to note however that
because the $V_{B_{j}}$ are $G$-invariant, it follows that
$I_{R}\otimes V_{B_{j}}=\$(V_{B_{j}})$. \ \ Moreover, given that
$\$$ preserves the algebra of operators (property (2) of the $\$$
map), we have
$\$(V_{B_{n}})\$(V_{A_{n}}^{\prime})\cdots\$(V_{B_{1}})\$(V_{A_{1}}^{\prime
})=\$(V_{B_{n}}V_{A_{n}}^{\prime}\cdots V_{B_{1}}V_{A_{1}}^{\prime
})=\$(V^{\prime}).$\ The initial state and Bob's final measurement
are also trivial on $R$ and $G$-invariant. \ It follows that
$d_R^{-1} I_{R}\otimes\rho
_{A}\otimes\rho_{M}\otimes\rho_{B}=d_{R}^{-1}\$(\rho_{A}\otimes\rho
_{M}\otimes\rho_{B})$ and $I_{R}\otimes E_{B,k}=\$(E_{B,k}).$


Thus, the probability of outcome $k$ in Alice's SSR-respecting
cheating strategy is
\begin{align}
    d_{R}^{-1} &\mathrm{Tr}\left(  \$(E_{B,k})\$(V^{\prime})\$(\rho_{A}
    \otimes\rho_{M}\otimes\rho_{B})\$(V^{\prime\dag})\right) \nonumber \\
    & =\mathrm{Tr}\left(
    E_{B,k}V^{\prime}\rho_{A}\otimes\rho_{M}\otimes\rho
    _{B}V^{\prime\dag}\right)  =p_{B}^{\prime}(k)\,,
\end{align}
where we have used Eq.~(\ref{eq:InfiniteBorn}). Thus, any
probability distribution achieved by a cheating strategy that
violates the SSR can also be achieved by one that respects it.

It is straightforward to generalize this result to the case of an
$n$-party protocol with $k$ cheating parties.  We begin with the
case of a pair of cheating parties (Alice and David).  If their
SSR-violating cheating strategies consist of unitaries
$V_{A_{j}}^{\prime}$ and $V_{D_{j}}^{\prime},$ then, by the same
reasoning as applied above, they can achieve an equivalent degree of
success using SSR-respecting cheating strategies consisting of
unitaries $\$(V_{A_{j}}^{\prime})$ and $\$(V_{D_{j}}^{\prime})$
which are nontrivial on $RA$ and $RD$ respectively.  Because only
one of these parties is ever implementing an operation at a given
time, they can achieve this strategy by passing the RF $R$ back and
forth between them.  Moreover, even if Alice and David are prevented
from implementing such transmissions during the protocol, there is a
resource they may share prior to the protocol, namely, a shared RF,
which allows them to do just as well.  Suppose their shared RF is
constituted of a pair of systems, $R_{1}$ and $R_{2}$, in the state
\begin{equation}
  \rho_{R_{1}R_{2}}=\int \text{d}g\, |g\rangle_{R_{1}}\langle g|
  \otimes|g\rangle_{R_{2}}\langle g|\,.
\end{equation}
We show that by using $R_{2}$ alone, David can achieve the same
operations on $MD$ as could be achieved if Alice had passed him
$R_{1}$.  We define $\$_{1}$ and $\$_{2}$ as the generalizations of
$\$$ for $R=R_{1}$ and $R=R_{2}$ respectively.  If Alice had passed
David a copy of $R_{1}$, he could replace $V_{D_{j}}^{\prime}$ by
the operation $\$_{1}(V_{D_{j}}^{\prime})$.  But given that
\begin{align}
    \$_{1}(V_{D_{j}}^{\prime})
    &\rho_{R_{1}R_{2}}\otimes\rho_{MD}\$_{1}(V_{D_{j}}^{\prime\dag}) \nonumber \\
    & =\int \text{d}g\, |g\rangle_{R_{1}}\langle g|
    \otimes|g\rangle_{R_{2}}\langle g| \nonumber \\
    &\qquad\otimes(U(g)V_{D_{j}}U^{\dag}(g))\rho_{MD}(U(g)V_{D_{j}}^{\dag}U^{\dag}(g)) \nonumber \\
    &=\$_{2}(V_{D_{j}}^{\prime})\rho_{R_{1}R_{2}}\otimes\rho_{MD}\$_{2}
    (V_{D_{j}}^{\prime\dag})\,,
\end{align}
it follows that David achieves the same effect by replacing
$V_{D_{j}}^{\prime}$ by the operation $\$_{2}(V_{D_{j}}^{\prime}),$
which is something that he can achieve locally. The generalization
of this argument to an arbitrary number of cheating parties is
straightforward.

It should be noted that for Lie groups, the states we have been
considering are, strictly speaking, not normalizable.  However, one
can introduce a sequence of normalizable approximations to these
states, parametrized by an integer $N$, corresponding to RFs of
bounded size, such that the results described here are reproduced in
the limit $N\rightarrow\infty,$ that is, the limit of an unbounded
RF.  See \textcite{Kit04} for details.

Superselection rules that are not associated with a compact symmetry
group have also been considered, for instance, the superselection
rule for univalence which denies the possibility of a coherent
superposition of boson and fermion \cite{Dop90} and superselection
rules that can arise in two-dimensional systems that admit
non-Abelian anyons. Using different methods, it has been shown that
such SSRs also fail to yield any advantages for two-party
cryptography~\cite{Kit04}, but the question remains open for
multi-party protocols.

\section{Aligning reference frames}
\label{sec:Aligning}

Separated parties often require the use of a shared reference frame.
For instance, they may require their clocks to be synchronized or
their Cartesian frames to be aligned.  Furthermore, although it was
shown in Sec.~\ref{sec:NoSRF} that lacking a reference frame does
not prevent one from achieving information-theoretic tasks such as
communication, cryptography and computation, this restriction can
decrease the (non-asymptotic) efficiency with which they can be
achieved and often requires more sophisticated encodings.  Thus,
separated parties might opt to initially devote their communication
resources to setting up a shared reference frame and thereafter use
a standard encoding, rather than perpetually circumventing the lack
of such an RF with a relational encoding.

We refer to the process by which observers correlate their local
reference frames, that is, by which they refine their knowledge of
the relation between them, as \emph{reference frame alignment}. In
order to do so, the parties must exchange systems with the relevant
degrees of freedom, which serve as finite samples of the sender's
local RF and can be compared to the receiver's local RF to obtain
some information about the relative orientation of the two frames.
For example, through the exchange of spin-1/2 particles, Alice and
Bob can align their local Cartesian frames. Exchanging quantum
states of an optical mode allows them to align their phase
references.

We discussed in Sec.~\ref{sec:QRFs} how a quantum system of
unbounded size can play the same role as a classical reference
frame.  By the transmission of such a system, one can achieve
perfect alignment of separated classical reference frames. However,
one is often restricted to sending systems of bounded size, either
to economize on communication resources or because of the
impracticality of encodings that require a joint preparation of too
many systems. It is therefore of great interest to determine the
fundamental quantum limits on the alignment precision that can be
achieved for given communication resources.  This is the question we
shall address in this section.

It should be noted that if the communication resources are bounded,
the end result of an alignment scheme is \emph{partial} correlation
between the local reference frames. The operational consequence of
not having complete correlation is that the parties must contend
with the decoherence that arises from the \emph{weighted}
$G$-twirling operation discussed in Sec.~\ref{subsec:GeneralSSR}. As
the imprecision in this alignment approaches zero, the weighted
$G$-twirling operation approaches the identity map.  We shall be
concerned here with schemes that minimize this imprecision.

Consider an alignment scheme for some form of reference frame, which
makes use of a number $N$ of transmitted quantum systems.  The
expected error in alignment, measured by the variance, can be
theoretically determined as a function of $N$.  The problem has
close connections with the field of quantum parameter estimation
\cite{Hol82} and quantum metrology (see \textcite{Gio04a}), and a
general feature of this behaviour is closely related to well-studied
results in phase estimation.  Specifically, if the $N$ quantum
systems are used \emph{independently} (i.e., entangled signal states
are not used, and measurements on individual systems are
independent) one can only achieve an error, quantified by the
variance, that scales as $1/N$. This result, which is a consequence
of the central limit theorem, is commonly known as the
\emph{standard quantum limit}.  In contrast, strategies which make
use of entanglement between the $N$ systems, as well as joint
measurements, can achieve an error (variance) that scales as
$1/N^2$. This result, commonly known as the \emph{Heisenberg limit},
represents the fundamental limit to the scaling of accuracy as
allowed by the laws of quantum physics~\cite{Gio06}.

We begin with a discussion of a simple example to provide some
intuition about what sorts of states are optimal for the alignment
problem. Heuristically, they are states that, when mixed over the
action of the group, have significant support on the largest
possible dimensionality of Hilbert space, thereby making them as
distinguishable as possible. This intuition can be made rigorous in
the context of a simple figure of merit: the maximum likelihood of a
correct guess. After introducing a more useful figure of merit, the
fidelity, we describe in detail strategies for the alignment of
phase references, spatial directions, and Cartesian frames, and
demonstrate how the Heisenberg limit can be achieved. We also
overview results on the alignment problem for a few other sorts of
reference frames. For alternate overviews of techniques for aligning
directions and Cartesian frames, see \textcite{Per02b} and
\textcite{Bag06}.

\subsection{Example: sending a direction with two spins}
\label{subsec:GisinPopescuexample}

Suppose Alice and Bob have uncorrelated Cartesian frames and they
wish to align their $\mathbf{z}$ axes by Alice transmitting a pair
of spin-1/2 particles to Bob.  What state of these two spins should
Alice prepare, and what measurement should Bob perform, in order to
optimize their expected success in this task?

A seemingly reasonable strategy would be for Alice to send parallel
spins aligned with her $\mathbf{z}$ axis.  Assuming Alice's
$\mathbf{z}$ axis points in the $\mathbf{n}$ direction relative to
Bob's frame, this strategy corresponds to sending Bob the state
$|\mathbf{n}\rangle|\mathbf{n}\rangle$, where
$(\hat{\mathbf{S}}\cdot\mathbf{n})|\mathbf{n}\rangle =
\frac{\hbar}{2}|\mathbf{n}\rangle$, relative to his local frame.
Bob's task is now one of state estimation -- to optimally estimate
the pure state $|\mathbf{n}\rangle$ given two copies~\cite{Mas95}.
First, we note that the set of states from which Bob must measure
are all on the three-dimensional symmetric $j=1$ subspace
$\mathcal{H}_{j=1}$ of two spins; he can thus restrict his
measurement to a POVM on this Hilbert space.  We consider the case
where Bob performs a \emph{covariant} measurement, i.e., a
continuously-parametrized POVM of the form
\begin{equation}\label{eq:TwoSpinCovPOVM}
    \big\{ E_\Omega = R(\Omega)^{\otimes 2} E_0 R^\dag(\Omega)^{\otimes
    2}\,, \ \Omega \in {\rm SU(2)} \big\} \,,
\end{equation}
where $E_0 = |e\rangle\langle e|$ is a positive rank-1 operator.
(Any POVM of higher rank can be simulated by a rank-1 POVM followed
by classical post-processing of the result.) To form a POVM, the
vector $|e\rangle$ must satisfy the normalization condition
\begin{equation}\label{eq:TwoSpinCovPOVMNormalize}
    \int {\rm d}\Omega \, E_\Omega = I_{j=1} \,,
\end{equation}
where $I_{j=1}$ is the identity operator on $\mathcal{H}_{j=1}$.  It
is straightforward to show (for example, by using
Eq.~\eqref{eq:twirling2qubits}) that this condition completely
constraints the form of the POVM, i.e., it requires that $|e\rangle
= \sqrt{3}|00\rangle$ up to an arbitrary choice of single-spin basis
$\{ |0\rangle, |1\rangle \}$. Let Bob choose $|0\rangle$ to be
aligned with his ${+}\mathbf{z}$ direction.

If Alice sends two spins in the state
$|\mathbf{n}\rangle|\mathbf{n}\rangle$, and Bob performs the
measurement~\eqref{eq:TwoSpinCovPOVM}, then the probability of Bob
obtaining the measurement outcome $\Omega$ is given by the Born
rule,
\begin{align}\label{eq:TwoSpinBornRule}
    p(\Omega|\mathbf{n}) &= {\rm
    Tr}[E_{\Omega}|\mathbf{n}\rangle|\mathbf{n}\rangle
    \langle\mathbf{n}|\langle\mathbf{n}|] \,, \\
    &= 3\cos^4(\beta/2) \,,
\end{align}
where $\beta = \cos^{-1} (\mathbf{n}\cdot\Omega\mathbf{z})$ is the
angle between $\mathbf{n}$ and $\Omega\mathbf{z}$.  If Bob obtains
the measurement outcome $\Omega$, then his best guess as to the
direction $\mathbf{n}$ is $\mathbf{n}_g = \Omega\mathbf{z}$. A
natural way with which to quantify the quality of Bob's guess is to
use the fidelity $(1+\mathbf{n}\cdot \mathbf{n}_g)/2 =
\cos^2(\beta/2)$, which gives a value of $1$ if he guesses correctly
($\mathbf{n}_g = \mathbf{n}$), a value of $0$ if he guesses the
opposite direction ($\mathbf{n}_g = - \mathbf{n}$), and which
decreases monotonically between these two limits.  A random guess
would give an average fidelity of $1/2$.

The average fidelity of Bob's guess, then, is given by averaging
over the distribution of transmitted states by Alice (chosen to be
uniformly sampled from the sphere) and all possible measurement
outcomes by Bob, weighted by the fidelity,
\begin{equation}\label{eq:TwoSpinAvgFid}
    F = \int {\rm d}\mathbf{n} \int {\rm d}\Omega \,
    p(\Omega|\mathbf{n}) \frac{1+\mathbf{n}\cdot \Omega\mathbf{z}}{2}
    \,.
\end{equation}
As the probability $p(\Omega|\mathbf{n})$ and the fidelity depend
only on the angle $\beta = \cos^{-1}
(\mathbf{n}\cdot\Omega\mathbf{z})$, this expression simplifies to
\begin{equation}
    F = \frac{1}{2}\int_0^\pi {\rm d}\beta\, \sin\beta\,
    \big(3\cos^4(\beta/2) \bigr)
    \frac{1+\cos\beta}{2} = \frac{3}{4} \,.
\end{equation}
We note that the same average fidelity can be achieved with a finite
(4-element) PVM in the basis given by
Eq.~\eqref{eq:fourprivatestates}~\cite{Mas95}.

Remarkably, this method where Alice sends two \emph{parallel} spins
is not optimal; a higher average fidelity can be achieved if Alice
instead sends two \emph{anti-parallel} spins~\cite{Gis99}, as we now
demonstrate. Let $|\mathbf{n}\rangle|{-}\mathbf{n}\rangle$ be the
two-qubit state transmitted by Alice; again, Bob must perform a type
of state-estimation to determine $\mathbf{n}$.  Note that the set of
possible states is no longer contained within the $j=1$ symmetric
subspace, and thus Bob must now perform a measurement on the entire
two-spin Hilbert space.  Again, choosing a covariant POVM of the
form~\eqref{eq:TwoSpinCovPOVM}, the new normalization condition now
becomes
\begin{equation}\label{eq:TwoSpinCovPOVMNormalize2}
    \int {\rm d}\Omega \, E_\Omega = I \,,
\end{equation}
where $I$ is now the identity on the full two-spin Hilbert space.
Choosing $E_0 = |e\rangle\langle e|$ to be rank-1, this
normalization again completely constrains the POVM (up to an
arbitrary choice of single-spin basis by Bob) to be
\begin{equation}\label{eq:TwoSpinSeed}
    |e\rangle = \sqrt{3}|\psi^+\rangle + |\psi^-\rangle \,,
\end{equation}
where $|\psi^\pm\rangle =
\frac{1}{\sqrt{2}}(|01\rangle\pm|10\rangle)$.

Again, the probability that Bob obtains the measurement outcome
$\Omega$ given that Alice prepared
$|\mathbf{n}\rangle|{-}\mathbf{n}\rangle$ is a function only of the
angle $\beta$ between $\mathbf{n}$ and $\Omega\mathbf{z}$, given in
this case by
\begin{equation}\label{eq:TwoAntiSpinBornRule}
    p(\Omega|\mathbf{n}) = \frac{(1+\sqrt{3}\cos\beta)^2}{2} \,.
\end{equation}
This leads to a fidelity of $F = (1+\sqrt{3})/(2\sqrt{3}) \simeq
0.789$, which is greater that that achieved for the parallel spin
case.  (We note that this fidelity can also be achieved with a
finite (4-outcome) PVM of the form
\begin{equation}\label{eq:AntiParallelPVM}
    |i\rangle = \frac{\sqrt{3}}{2}
    \frac{|\mathbf{n}_i\rangle|{-}\mathbf{n}_i\rangle +
    |{-}\mathbf{n}_i\rangle|\mathbf{n}_i\rangle}{\sqrt{2}} +
    \frac{1}{2}|\psi^-\rangle \,,
\end{equation}
where $|\mathbf{n}_i\rangle$ are along the four direction of the
tetrahedron, given in
Eqs.~(\ref{eq:Tetrahedron})-(\ref{eq:Tetrahedron4}).  An experiment
demonstrating this protocol has been performed by~\textcite{Jef06},
wherein it was referred to as ``quantum orienteering''.)

A heuristic explanation of why the anti-parallel spins are superior
to the parallel spins for this task is obtained by investigating the
\emph{orbits} of the transmitted states under the relevant group, in
this case, the group of rotations SU(2).  Consider the orbit under
the group of a state of parallel spins
$|\mathbf{n}\rangle|\mathbf{n}\rangle$,
\begin{equation}\label{eq:OrbitParallel}
    M_{\rm par} = \big\{ \big(R(\Omega)\otimes R(\Omega)\big)
    |\mathbf{n}\rangle|\mathbf{n}\rangle\,, \ \Omega \in \text{SU(2)}
    \big\}\,.
\end{equation}
This orbit has support entirely on the three-dimensional symmetric
$j=1$ subspace of the two-spin Hilbert space.  In contrast, the
orbit under the group of a state of anti-parallel spins
$|\mathbf{n}\rangle|{-}\mathbf{n}\rangle$,
\begin{equation}\label{eq:OrbitAntiParallel}
    M_{\rm anti} = \big\{ \big(R(\Omega)\otimes R(\Omega)\big)
    |\mathbf{n}\rangle|{-}\mathbf{n}\rangle\,, \ \Omega \in \text{SU(2)}
    \big\}\,,
\end{equation}
has support on the full four-dimensional two-spin Hilbert space.
The latter orbit spans a larger space, therefore its elements are,
loosely speaking, more orthogonal and consequently easier to
discriminate. We will see in the following that this heuristic
idea can be formalized to provide optimal methods for the
alignment of any type of reference frame.

\subsection{General approach to aligning reference frames}
\label{subsec:GeneralAlignment}

Consider the general problem of aligning an RF associated with the
group $G$ using the one-way transmission of a quantum system
(composed, say, of a number of elementary systems) with Hilbert
space $\mathcal{H}$. For instance, one might be trying to
communicate information about a Cartesian frame, associated with the
group SU(2), using $N$ spin-1/2 particles and corresponding Hilbert
space $\mathcal{H}= (\mathcal{H}_{1/2})^{\otimes N}$.  The problem
is to devise an optimal protocol for this task, given the allowed
communication resources, for some given figure of merit.

The most general statements we can make about optimal RF
distribution schemes concern the form of an optimal POVM for a given
covariant set of signal states, provided the figure of merit
satisfies some very general and natural properties.  Optimal states
for a general task of this sort can be taken to be pure, given that
any mixed state scheme is a convex sum of pure state schemes and
therefore can do no better than the best pure state scheme.  Thus,
the signal states form an orbit of pure states
\begin{equation}
  |\psi(g)\rangle =U(g)|\psi\rangle\, ,
\end{equation}
where $U(g)$ is the representation of $G$ on $\mathcal{H}$.  We call
$|\psi\rangle$ the \emph{fiducial state}.

We now consider a general form for the fiducial state in terms of
the decomposition of $\mathcal{H}$ into irreps of $G$, given by
Eqs.~\eqref{eq:Hdecomp1} and~\eqref{eq:Hdecomp2}.  In terms of the
charge sectors $\mathcal{H}_q$, we express the fiducial state as
\begin{equation}\label{eq:FiducialStateOnChargeSectors}
    |\psi\rangle = \sum_q \beta_q |\psi_q\rangle\,,
\end{equation}
with $\beta_q$ satisfying $\sum_q |\beta_q|^2 =1$, and where
$|\psi_q\rangle \propto \Pi_q|\psi\rangle$ is the normalized
component of the fiducial state on each charge sector
$\mathcal{H}_q$ (where we recall that $\Pi_q$ is the projector onto
the $q$th charge sector). Each state $|\psi_q\rangle$ can be viewed
as a (generally entangled) state on the tensor product decomposition
$\mathcal{H}_q = \mathcal{M}_q \otimes \mathcal{N}_q$ of
Eq.~\eqref{eq:Hdecomp2}. Let
\begin{equation}\label{eq:Schmidt}
    |\psi_q\rangle = \sum_{m=1}^{d_q} \lambda^{(q)}_m |\phi^{(q)}_m\rangle
    \otimes |r_m^{(q)}\rangle \,,
\end{equation}
be a Schmidt decomposition of this state on $\mathcal{M}_q \otimes
\mathcal{N}_q$, where
\begin{equation}
    d_q \equiv \min\{\dim\mathcal{M}_{q},\dim\mathcal{N}_{q}\}\,.
\end{equation}
We note that, if $|\psi_q\rangle$ does not have maximal Schmidt rank
(meaning some of the $\lambda^{(q)}_m$ are zero), then the Schmidt
vectors are not unique.  Let $\tilde{\mathcal{N}}_q \subseteq
\mathcal{N}_q$ be the $d_q$-dimensional space spanned by the Schmidt
vectors $\{ |r_m^{(q)}\rangle \}$.  Let $\{ |\phi^{(q)}_m\rangle \}$
be a basis for $\mathcal{M}_q$, obtained using the Schmidt vectors
from~\eqref{eq:Schmidt} and, if necessary, completing this set
arbitrarily to a basis.

A general expression for the fiducial state is thus
\begin{equation}\label{eq:GeneralFiducial}
    |\psi\rangle = \sum_q \beta_q \sum_{m=1}^{d_q} \lambda^{(q)}_m
    |\phi^{(q)}_m\rangle \otimes |r_m^{(q)}\rangle \,,
\end{equation}
which lies on the subspace $\tilde{\mathcal{H}} \subset \mathcal{H}$
given by
\begin{equation}\label{eq:GeneralSupport}
    \tilde{\mathcal{H}}
    \equiv \sum_q \mathcal{M}_{q}\otimes\tilde{\mathcal{N}}_{q}\,.
\end{equation}

In addition, the support of the orbit of the fiducial state, i.e.,
the space
\begin{align}
  \mathcal{H}^{\psi}
  &= \mathrm{span}\big\{U(g)|\psi\rangle\,, \ g\in G\big\} \nonumber \\
  &= \text{\textrm{supp}}\big[\mathcal{G}(|\psi\rangle\langle\psi |
  )\big] \,,
\end{align}
will also lie within $\tilde{\mathcal{H}}$,
\begin{equation}\label{eq:HpsisubseteqHtilde}
  \mathcal{H}^{\psi} \subseteq \tilde{\mathcal{H}} \,.
\end{equation}

Thus, for any choice of a fiducial state $|\psi\rangle$, the
measurement may be described by a POVM that is restricted to
$\tilde{\mathcal{H}}$. In addition, if the figure of merit we are
attempting to optimize satisfies some general conditions (which we
discuss below), the optimal POVM can be chosen to be
\emph{$G$-covariant}.\footnote{We note that choosing a covariant
POVM is sufficient to obtain an optimal protocol, but not necessary.
For practical schemes, it may be valuable to identify finite-element
POVMs that also obtain the optimum, as we saw in
Sec.~\ref{subsec:GisinPopescuexample}.} Moreover, its elements can
be taken to be rank-1.\footnote{For a proof of this,
see~\textcite{Chi05}.  In general, any non-rank-1 POVM can be
simulated by a rank-1 POVM followed by classical post-processing of
the result; however, this simulation need not be covariant.} Thus,
the optimal POVM must have the form $\{E(g)\}$, given by
\begin{equation}
  E(g)=U(g)|e\rangle\langle e |U(g)^{\dag}\,.
\end{equation}
We call $| e\rangle\langle e |$ the \emph{fiducial POVM element}.
Given that these elements form a resolution of identity on
$\tilde{\mathcal{H}}$, we have $\int {\rm d}g\, E(g) =
I_{\tilde{\mathcal{H}}}$, or equivalently,
\begin{equation}
  \mathcal{G}[| e\rangle\langle e |]
  =I_{\tilde{\mathcal{H}}} \,.
  \label{eq:twirledfiducialeffect}
\end{equation}
Thus an RF alignment scheme of this sort is specified by a fiducial
state $|\psi\rangle$ and a fiducial POVM element $| e\rangle\langle
e |$.  In order to determine an optimal scheme, we first determine
the optimal $| e\rangle\langle e |$ for a given $|\psi\rangle$.

The constraint Eq.~(\ref{eq:twirledfiducialeffect}) completely fixes
the form of $| e\rangle$ to be
\begin{equation}
  | e\rangle =\sum_q \sqrt{\dim
  (\mathcal{M}_{q})}\sum_{m=1}^{d_{q}} |\phi_{m}^{(q)}\rangle
  \otimes |r_{m}^{(q)}\rangle \,.
  \label{eq:optfiducialeffect}
\end{equation}
This vector $| e\rangle$ of Eq.~\eqref{eq:optfiducialeffect} can be
described as follows: it is a coherent superposition across the
charge sectors where $|\psi\rangle$ has support, with the amplitude
squared in each such charge sector given by the dimensionality of
$\mathcal{M}_{q}$, and the projections in each such charge sector
given by maximally entangled states across
$\mathcal{M}_{q}\otimes\tilde{\mathcal{N}}_{q}$.

We now demonstrate why $| e\rangle$ must take this form.
Eq.~(\ref{eq:twirledfiducialeffect}) can be expressed as
\begin{equation}
  \sum_q \mathcal{D}_q \otimes \mathcal{I}_q \big[ \Pi_q |e\rangle
  \langle e| \Pi_q \bigr]
  = \sum_q I_{\mathcal{M}_q} \otimes
  I_{\tilde{\mathcal{N}}_q}\,.
  \label{eq:fff}
\end{equation}
In terms of the charge sectors, we write
\begin{equation}
  | e\rangle = \sum_q
  c_{q}| e_{q}\rangle\,,\label{eq:eps}
\end{equation}
where the $c_{q}$ are nonzero and $| e_{q}\rangle \equiv \Pi_q|
e\rangle$ is a normalized state in the $j$th sector. Projecting
Eq.~(\ref{eq:fff}) onto a single charge sector and tracing over
$\mathcal{M}_{q}$, we find
\begin{equation}
  \mathrm{Tr}_{\mathcal{M}_{q}}\big[ | e_{q}\rangle
  \langle  e_{q} |\big]
  =I_{\tilde{\mathcal{N}}_{q}}/d_{q}\,,
\end{equation}
which tells us that the reduced density operator on
$\tilde{\mathcal{N}}_{q}$ of $| e_{q}\rangle$ is the completely
mixed state, and consequently that $| e_{q}\rangle$ is a maximally
entangled states across
$\mathcal{M}_{q}\otimes\tilde{\mathcal{N}}_{q}$.  This may be
written as
\begin{equation}
  | e_{q}\rangle
  = \frac{1}{\sqrt{d_{q}}} \sum_{m=1}^{d_{q}}
  |\phi_{m}^{(q)}\rangle \otimes |r_{m}^{(q)}\rangle \,.
  \label{eq:epsj}
\end{equation}
Projecting Eq.~(\ref{eq:fff}) onto a single charge sector and
tracing over both $\tilde{\mathcal{N}}_{q}$ and $\mathcal{M}_{q}$,
we conclude that
\begin{equation}
  |c_{q}|^{2}=\mathrm{Tr}[I_{\mathcal{M}_{q}}]
  \mathrm{Tr}[I_{\tilde{\mathcal{N}}_{q}}]
  =\dim(\mathcal{M}_{q})d_{q}\,.
  \label{eq:c_j}
\end{equation}
We may define $|\phi_{m}^{(q)}\rangle$ in such a way that the
coefficients $c_{q}$ can be taken to be real and positive. Combining
Eqs.~(\ref{eq:eps}), (\ref{eq:epsj}) and (\ref{eq:c_j}), we recover
Eq.~(\ref{eq:optfiducialeffect}).

The covariant POVM, then, is fixed by the problem, and it remains
only to determine the optimal fiducial state $|\psi\rangle$. To do
so, one needs to specify a figure of merit.

\subsection{Maximum likelihood estimation}
\label{subsec:MaxLikelihood}

We now consider a particular choice for a figure of merit: the
maximum likelihood of a correct guess~\cite{Chi04c}.  Because our
standard example is a continuous group, the likelihood of a correct
guess is infinitesimal, and so we must look at the \emph{maximum
likelihood density} -- the probability density $\mu$ of obtaining
the POVM outcome $E(g)$ given that the signal state is
$|\psi(g)\rangle$, averaged over the prior distribution over signal
states, which we take to be uniform.  (A more simple analysis is
possible for finite groups.) This density takes the simple form
\begin{align}
  \mu &  =\int {\rm d}g\, \mathrm{Tr}\big[ E(g)
  |\psi(g)\rangle\langle \psi(g)|\big] \nonumber \\
  &=|\langle  e|\psi\rangle |^{2}\,.
  \label{eq:MuMax}
\end{align}
As the fiducial POVM element $|e\rangle$ is fixed, optimization is
achieved by taking $|\psi\rangle$ to be parallel to $| e\rangle$,
\begin{equation}
  |\psi\rangle = \frac{| e\rangle}{\| e\|}\,,
\end{equation}
where $\| e\| =\sqrt{\langle e| e\rangle}$. It follows from
Eq.~(\ref{eq:optfiducialeffect}) that
\begin{equation}
  \| e\|
  = \sqrt{\sum_{q}\dim (\mathcal{M}_{q})d_{q}}
  = \sqrt{ \dim\tilde{\mathcal{H}} }\,.
\end{equation}
The optimal fiducial state, which has the form of
Eq.~\eqref{eq:GeneralFiducial}, must have all Schmidt coefficients
$\lambda^{(q)}_m$ equal and non-zero, and the coefficients $\beta_q$
are completely fixed by the optimal $|e\rangle$; i.e., the optimal
fiducial state is
\begin{equation}
  |\psi\rangle
  = \sum_j \sqrt{\frac{\dim(\mathcal{M}_{q})}{\dim\tilde{\mathcal{H}}}}
  \sum_{m=1}^{d_{q}}
  |\phi_{m}^{(q)}\rangle \otimes |r_{m}^{(q)}\rangle \,.
\end{equation}
Note that this state satisfies
\begin{equation}
  \label{eq:Gtwirledpsi}
  \mathcal{G}[|\psi\rangle\langle\psi|]
  =I_{\tilde{\mathcal{H}}}/\dim\tilde{\mathcal{H}}\,.
\end{equation}
We can conclude that the maximum likelihood density of a correct
guess takes the general form
\begin{align}
  \mu_{\max}&=\| e\|^2 = \dim\tilde{\mathcal{H}}
  = \text{rank}(\mathcal{G}[|\psi\rangle\langle\psi|]) \nonumber \\
  &=\sum_{q}\dim\mathcal{M}_{q}\times\min\{\dim\mathcal{M}_{q}
  ,\dim\mathcal{N}_{q}\}\,.
  \label{eq:pmaxdimHpsi}
\end{align}

Given Eq.~(\ref{eq:Gtwirledpsi}) and Eq.~(\ref{eq:pmaxdimHpsi}), we
can interpret this result as follows: we maximize the likelihood of
a correct guess by choosing the fiducial signal state $|\psi\rangle$
to be such that, under $G$-averaging, the weights of the state
$\mathcal{G}[|\psi\rangle\langle\psi|]$ are spread uniformly over
the largest possible space.  Thus, at least for the case of maximum
likelihood estimation, the intuition behind why antiparallel spins
do better than parallel spins is found to have a rigorous
counterpart, and indeed this intuition is found to generalize to the
alignment of any RF whose configurations correspond to the elements
of a group.  By choosing a fiducial state in this way, the signal
states are made as distinguishable as possible.

\subsubsection{Maximum likelihood estimation of a phase reference}
\label{subsec:AlignPhaseRef1}

With the general results above, the optimal performance of any
particular alignment protocol quantified by maximizing the
likelihood can be directly and simply calculated.  Suppose, for
example, one seeks to align phase references by transmitting at most
$n_{\max}$ photons in a single mode.  The relevant group in this
case is U(1). The charge sectors correspond to total photon number,
so we use $n$ rather than $q$ to denote them. Because the irreps of
U(1) are one-dimensional, we have $\dim\mathcal{M}_{n}=1$.  In this
case, the optimal fiducial POVM element and the optimal fiducial
signal state are
\begin{equation}
  | e\rangle =\sum_{n=0}^{n_{\max}}|n\rangle\,,\quad
  |\psi\rangle = \frac{1}{\sqrt{n_{\max}+1}} \sum_{n=0}^{n_{\max}}
  |n\rangle\,.
\end{equation}
Clearly,
\begin{equation}
  \mathcal{G}\big[|\psi\rangle\langle\psi|\big]
  =\sum_{n=0}^{n_{\max}}\frac{1}{n_{\max}+1}|n\rangle
  \langle n|\,,
\end{equation}
so that the maximum likelihood density of a correct guess is
\begin{equation}
  \mu_{\max}=\mathrm{rank}\big(\mathcal{G}\big[|\psi\rangle
  \langle\psi| \big]\big)= n_{\max}+1 \,.
\end{equation}
Note that for multiple modes, the subspaces $\mathcal{H}_{n}$ may be
multi-dimensional, and in this case the basis states can be chosen
to be any set of eigenstates of $\hat{N}_{\rm tot}$, i.e., any
multi-mode states that are eigenstates of total photon number.

For comparison, it is useful to consider the maximum likelihood that
could be achieved using a coherent state $|\alpha\rangle$ with mean
photon number $n_{\max}/2$ (we cut off the amplitude for
$n>n_{\max}$ which is negligible for sufficiently large values of
$n_{\max}$),
\begin{equation}
  \mu_{CS} =|\langle\alpha|e\rangle|^{2}
  =\Big|\sum_{n=0}^{n_{\rm max}}
  \frac{e^{-|\alpha|^2/2}\alpha^{n}}{\sqrt{n!}}\Big|^{2}\,,
\end{equation}
which behaves as $|\alpha|=\sqrt{n_{\max}/2}$ for large values of
$n_{\max}$.

Thus, the phase eigenstate offers a quadratic improvement in
$n_{\max}$ over the coherent state.  Heuristically, this is due to
the fact that for the Poissonian number distribution of the coherent
state most of the support lies within $\pm\sqrt{\bar{n}}$ of the
mean photon number $\bar{n}=n_{\max}/2$. Thus, the majority of the
support of the U(1)-orbit of a coherent state is carried by a
subspace of the Hilbert space with dimension that scales as
$\sqrt{\bar{n}}$. For the optimal state $|\psi\rangle=
\frac{1}{\sqrt{2\bar{n}+1}} \sum_{n=0}^{2\bar{n}} |n\rangle$, in
contrast, the dimensionality of this subspace scales as $\bar{n}$.

The quadratic improvement achievable in such cases is typically
explained by noting that there is an uncertainty relation between
phase and photon number, and to achieve the smallest possible
variance in phase, one requires the largest possible variance in
photon number. This is certainly a useful tool for understanding the
successes of different schemes.  Nonetheless, as we shall show
presently, whereas the optimal strategies for aligning RFs
associated with non-Abelian groups can also be understood in terms
of the support of the group orbit of the signal state, it is at
present unclear whether an argument in terms of an uncertainty
principle can be provided in such cases.

\subsubsection{Maximum likelihood estimation of a Cartesian frame}
\label{subsec:MaxLCartesian}

We now consider the task of optimally aligning a full spatial
(Cartesian) frame through the exchange of spin-1/2 particles, based
on maximizing the likelihood of a correct estimation.  For this
example, we will be required to use the multiplicity of irreducible
representations of SO(3) that occur in $(\mathcal{H}_{1/2})^{\otimes
N}$ to obtain the optimal scheme.

We restrict ourselves to the case of $N$ an even number for
simplicity. $N$ spin-1/2 particles carry a tensor representation
$R^{\otimes N}$ of SO(3); this representation is reducible, and
the irreducible representations (labeled by $j$) appear with
nontrivial multiplicities.  As analyzed in
Sec.~\ref{subsec:NoSRFCommSpins}, the total Hilbert space of $N$
spin-1/2 particles can be decomposed as
\begin{equation}\label{NqubitHilbertSpaceSU(2)Decomp}
    (\mathcal{H}_{1/2})^{\otimes N} = \bigoplus_{j=0}^{N/2}
    \mathcal{M}_j \otimes \mathcal{N}_j \,,
\end{equation}
where $\mathcal{M}_j$ carry irreducible representations $R_j$ of
SO(3) and have dimensionality $2j+1$, and $\mathcal{N}_j$ carry the
trivial representation of SO(3) and have dimensionality given by
Eq.~(\ref{MultiplicitySU(2)}). For this example, the dimension of
each decoherence-free subsystem $\mathcal{N}_j$ is greater than or
equal to the dimension of the corresponding decoherence-full
subsystem $\mathcal{M}_j$ for all $j$ \emph{except} $j=N/2$ (where
$\dim\mathcal{N}_j=1$). Thus, the maximum dimension of
$\mathcal{H}^\psi$ is, from Eqs.~\eqref{eq:HpsisubseteqHtilde} and
\eqref{eq:pmaxdimHpsi},
\begin{align}
  \label{eq:HpsiBoundSO(3)}
  \dim\mathcal{H}^{\psi}
  &= (N+1) + \sum_{j=0}^{N/2-1} (2j+1)^2 \nonumber \\
  &= \tfrac{1}{6}N^3 + \tfrac{5}{6}N + 1 \,.
\end{align}
For each $j<N/2$ we choose a $(2j+1)$-dimensional subspace
$\mathcal{N}^\psi_j \subset \mathcal{N}_j$ with basis
$|j,\alpha(m)\rangle$.  The optimal signal state thus has the form
\begin{multline}
  |\psi^{(N)}\rangle = \frac{1}{\dim\mathcal{H}^{\psi}}
  \Big(\sqrt{N+1} |N/2,N/2\rangle \\
  + \sum_{j=0}^{N/2-1}\sqrt{2j+1}\sum_{m=-j}^{j}
  |j,m\rangle \otimes |j,\alpha(m)\rangle \Big) \,,
\end{multline}
and the maximum likelihood density of a correct guess is
\begin{equation}
  \mu_{\max}=\tfrac{1}{6}N^3 + \tfrac{5}{6}N + 1 \,,
\end{equation}
which scales as $N^3/6$ for large $N$.

\subsection{General figures of merit}

As discussed above, maximizing the likelihood of the correct guess
led us directly to a general principle for choosing the fiducial
signal state.  However, as a figure of merit, the maximum likelihood
density is not a very practical choice -- it rewards only a
perfectly correct guess.  In many situations, one would desire a
figure of merit that would quantify the performance of a scheme by
the amount of Shannon information gained by the recipient about the
sender's reference frame.  However, such figures of merit usually
lead to intractable optimization problems.

A more common and tractable approach is to introduce a \emph{payoff
function} $f(g',g)$ which specifies the payoff for guessing group
element $g'$ when the actual group element is $g$~\cite{Chi05}.
Assuming a uniform prior for the signal states, the figure of merit
for the alignment scheme can then be the \emph{average payoff}
\begin{equation}
  \label{eq:GeneralAveragePayoff}
  \bar{f}=\int\mathrm{d}g\, \mathrm{d}g' \, p(g'|g)f(g',g)\,,
\end{equation}
where $p(g'|g)$ is the probability of guessing $g'$ when the signal
state is $g$ for the scheme in question.  In particular, the
commonly-used \emph{fidelity}, which quantifies the variance of the
average guess and which leads to direct comparisons with the
standard quantum limit, is one choice of payoff function; we will
determine this fidelity and explore protocols that optimize its
average in the examples that follow.

The task of reference frame alignment imposes some natural
constraints on the form of payoff functions. First, we note that the
group elements $g$ and $g'$ denote an orientation relative to some
background RF, i.e., the identity group element corresponds to
``aligned with this background RF''. However, it is desirable to
construct protocols that are independent of any background RF; for
example, if the background RF was transformed by a group element $h
\in G$, and $g$ and $g'$ were now defined with respect to this
transformed background RF, the payoff function should be the same.
Such protocols are associated with a payoff function $f(g',g)$ that
is \emph{right-invariant}, i.e., which satisfies
\begin{equation}
  f(g'h^{-1},gh^{-1}) =f(g',g)\,, \quad \forall\
  h \in G \,.
\end{equation}
In addition, the payoff function should be a function only of the
\emph{relative} transformation relating the transmitted state
(determined by $g$) and the measurement outcome (determined by
$g'$).  This requirement of a protocol demands that the payoff
function be \emph{left-invariant},
\begin{equation}
  f(hg',hg) =f(g',g)\,, \quad \forall\ h \in G \,.
\end{equation}
Payoff functions that are left-invariant are also referred to as
\emph{covariant}.  It is always possible to find a covariant POVM
that is optimal for any estimation problem with a covariant
(left-invariant) payoff function~\cite{Hol82}, and it is for this
reason that we focussed our attention on covariant POVMs early in
this section.

Any function $f(g',g)$ that is left-invariant can be written as a
function $f(g',g) = \tilde{f}(g^{\prime -1} g)$; if this function is
also right-invariant, then it satisfies $\tilde{f}(hg^{\prime
-1}gh^{-1}) = \tilde{f}(g^{\prime -1} g)$, and thus is a \emph{class
function}, i.e., a function on the conjugacy classes of $G$. (Recall
two group elements, $g_{1}$ and $g_{2}$, are in the same conjugacy
class if there exists another group element $h$ such that
$g_{1}=hg_{2}h^{-1}$.)  Any class function $\tilde{f}$ can be
expanded as a sum of the characters\footnote{The characters
$\chi_q(g)$ of a group $G$ form a basis of class functions; they are
given by the trace of the irreducible representations $T_q$ of $G$,
i.e., $\chi_q(g) = {\rm Tr}[T_q(g)]$.} $\chi_q(g)$ of $G$ as
\begin{equation}
  \label{eq:ExpandCharacters}
  \tilde{f}(g^{\prime-1}g) = \sum_{q}a_{q}\chi_{q}(g^{\prime-1}g)\,,
\end{equation}
where the $a_q$ are arbitrary coefficients.  We restrict our
attention to real, positive-valued payoff functions, which will
allow us to perform a simple maximization.

We note that the maximum likelihood estimation task described above
corresponds to choosing a payoff function $f(g',g) =
\delta(g^{\prime-1}g)$, a delta function.  This payoff function is
both left- and right-invariant, and its expansion in terms of
characters as in Eq.~\eqref{eq:ExpandCharacters} corresponds to
choosing all $a_q$ positive and equal to the dimension of the irrep.

As a consequence of the covariance of both the set of signal states
and the POVM, the probability $p(g'|g)$ is also a function of
$g^{\prime-1}g$, i.e.,
\begin{align}
  p(g^{\prime}|g)&=|\left\langle  e\right\vert U(g^{\prime-1}
  g)\left\vert \psi\right\rangle |^{2}\nonumber \\
  &\equiv \tilde{p}(g^{\prime-1}g) \,.
  \label{eq:GeneralBornRuleAlignment}
\end{align}
The average payoff of Eq.~\eqref{eq:GeneralAveragePayoff} then
simplifies as
\begin{align}
  \bar{f} & =\int \mathrm{d}g\, \mathrm{d}g'\, \tilde{p}(g^{\prime-1}g)
  \tilde{f}(g^{\prime-1}g)\nonumber \\
  &=\int\mathrm{d}g\, \tilde{p}(g)\tilde{f}(g)\,,
  \label{eq:GeneralAveragePayoff2}
\end{align}
which follows from the invariance of the measure $\mathrm{d}g$.
Using the explicit form of Eq.~\eqref{eq:GeneralBornRuleAlignment},
we have
\begin{equation}
  \label{eq:fbar}
\bar{f} = \int\mathrm{d}g\, \langle \psi | U^{\dag}(g) | e \rangle
\langle e | U(g) | \psi \rangle \tilde{f}(g)\,.
\end{equation}
Defining
\begin{equation}
  \label{eq:MatrixM}
  M \equiv \int \mathrm{d}g\, U^{\dag}(g)| e\rangle\langle e|
  U(g) \tilde{f}(g)\,,
\end{equation}
we may rewrite $\bar{f}$ as
\begin{equation}
  \label{eq:AvgFMaxEigen}
  \bar{f}= \langle \psi|M |\psi\rangle \,.
\end{equation}
This expression is the generalization of Eq.~\eqref{eq:MuMax} to an
arbitrary covariant payoff function.  As the fiducial POVM element
is completely constrained to be of the form of
Eq.~\eqref{eq:optfiducialeffect}, the operator $M$ is therefore
determined by the figure of merit. In order to maximize the average
payoff $\bar{f}$, then, one must find a fiducial state
$|\psi\rangle$ of the form~\eqref{eq:GeneralFiducial} that lies in
the eigenspace of $M$ with the largest eigenvalue.  Specifically, we
solve the eigenvalue equation
\begin{equation}
    M|\psi\rangle = \lambda^{\rm max}|\psi\rangle \,,
\end{equation}
and the use of this state yields a maximal average payoff of
\begin{equation}
\bar{f}^{\rm max} = \lambda^{\rm max} \, .
\end{equation}

For the problem of optimally aligning reference frames using a left-
and right-invariant payoff function, we will use the following
result of \textcite{Chi05} without proof\footnote{The proof relies
on determining an upper bound on the average payoff, and then
demonstrating that states of the form~\eqref{eq:ChiribellaFiducial}
saturate this bound.}: that the optimal fiducial signal state can be
chosen to have the form
\begin{equation}\label{eq:ChiribellaFiducial}
    |\psi\rangle = \sum_q \beta_q \sum_{m=1}^{d_q}
    |\phi^{(q)}_m\rangle \otimes |r_m^{(q)}\rangle \,,
\end{equation}
for coefficients $\beta_q$ satisfying $\sum_q |\beta_q|^2 =1$. These
coefficients are determined by the specific choice of payoff
function.  We note, however, that this result greatly simplifies the
optimization problem:  the number of coefficients is now given by
the number of irreps appearing in the decomposition of $U$, rather
than by the dimension of the Hilbert space.

\subsubsection{Fidelity of aligning a phase reference}
\label{subsec:SharedPhaseRef}

We now reconsider the problem of aligning a phase reference, as in
Sec.~\ref{subsec:AlignPhaseRef1}, but with an alternate (and
commonly used) payoff function:  the function $f(\theta',\theta) =
\cos^{2}[(\theta'-\theta)/2]$, which takes the value $1$ for the
correct guess ($\theta' = \theta$) and $0$ for $\theta' = \theta +
\pi$.  Note that this payoff function is left- and right-invariant,
and can be written as $\tilde{f}(\theta) = \cos^2(\theta/2)$, where
$\theta$ now denotes the relative angle between the signal and
guess. This figure of merit is commonly referred to as the
\emph{fidelity}.

The Hilbert space for this task will be restricted as follows:  we
allow arbitrarily few or many modes, but the maximum total photon
number is restricted to $n_{\rm max}$.  (An alternate approach would
be to bound the mean photon number; however, this adds considerable
complexity to the problem.)

As mentioned above, for any alignment scheme based on
\emph{independent} uses of $N$ modes with at most a single photon
in each, that is, $N$ single-rail qubits, the average fidelity
will approach $\bar{f} = 1$ as $1/N$, from the central limit
theorem. This scaling is referred to as the standard quantum
limit; the optimal scheme outperforms this scaling, as we will now
demonstrate. We now optimize over choices of signal state
$|\psi^{(N)}\rangle$ in order to maximize the expected payoff,
quantified by the average fidelity $\bar{f}$ of
Eq.~(\ref{eq:AvgFMaxEigen}).

Let $\{|n\rangle\,,\, n=0,1,\ldots,n_{\rm max}\}$ be an arbitrary
set of eigenstates of the total number operator $\hat{N}^{\rm tot}$;
the details of these states, including their mode structure, is
irrelevant to the task. The fiducial POVM element is of the form $|
e\rangle = \sum_{n=0}^{n_{\rm max}} |n\rangle$.

The operator $M$ of Eq.~\eqref{eq:MatrixM} is given in this instance
by the matrix
\begin{align}
  M_{nn^{\prime }} &= \langle n|M|n'\rangle \nonumber \\
  &=\int_0^{2\pi} \frac{\mathrm{d}\theta }{2\pi }
  e^{i(n-n')\theta}\cos ^{2}(\theta /2) \nonumber \\
  &=\tfrac{1}{2}\delta _{n,n^{\prime }}
  +\tfrac{1}{4}\delta _{n,n^{\prime}+1}
  + \tfrac{1}{4}\delta _{n+1,n^{\prime }}\,.
\end{align}
Note that
\begin{equation}
  M=\tfrac{1}{4}\tilde{M}+\tfrac{1}{2}I\,,
\end{equation}
where $\tilde{M}_{nn'} = \delta_{n,n'+1}+\delta_{n+1,n'}$. As any
eigenvector of $\tilde{M}$ is an eigenvector of $M$, it suffices to
find the eigenvalues and eigenvectors of $\tilde{M}$.  The maximum
average fidelity is then
\begin{equation}
  \bar{f}^{\max} =\tfrac{1}{2} + \tfrac{1}{4}\lambda^{\max }(\tilde{M})\,,
\end{equation}
and is achieved when $|\psi\rangle$ is the eigenvector of
$\tilde{M}$ associated with the maximum eigenvalue
$\lambda^{\max}(\tilde{M})$.

The characteristic equation we must solve is
\begin{equation}
  \det (\tilde{M}-\lambda I)=0\,,
\end{equation}
where $I$ is the identity. Defining $G_{k}\equiv \tilde{M}-\lambda
I$, where $k$ is the dimension of the vector space on which
$\tilde{M}$ acts, one finds that
\begin{equation}
  \det G_{k}=-\lambda \det G_{k-1}-\det G_{k-2}\,,
\end{equation}
for which the solution is
\begin{equation}
  \det G_{k}=U_{k}(-\lambda /2)\,,
\end{equation}
where the $U_{k}$ are the Chebyshev polynomials of the second kind,
given by
\begin{equation}
    U_{k}(\cos \theta)=\frac{\sin[(k+1)\theta ]}{\sin\theta}\,.
\end{equation}
Given that $U_{k}(x)=\pm U_{k}(-x)$, it follows that the
characteristic equation is $U_{N+1}(\lambda /2)=0$, and thus the
largest eigenvalue is $\lambda^{\max} = 2\cos(\pi/(N+2))$. The
maximum average fidelity for the distribution of a phase reference
is thus
\begin{equation}
  \bar{f}^{\max}
  = \tfrac{1}{2}\bigl(1+\cos(\pi/(N+2))\bigr) \,.
\end{equation}

To find the eigenvector $|\psi\rangle$ associated with the largest
eigenvalue, we must solve
\begin{equation}
  \tilde{M}|\psi\rangle =\lambda^{\max} |\psi\rangle\,.
\end{equation}
Let $\beta_n$ be the coefficients of $|\psi\rangle =
\sum_{n=0}^{n_{\rm max}} \beta_n |n\rangle$.  The definition of
$\tilde{M}$ leads to
\begin{equation}
  \beta_{n+1}+\beta_{n-1}=\lambda^{\max}\beta_{n}\,,
\end{equation}
for $1\leq j\leq N-1$.  At $n=0$, we have
$\beta_{1}=\lambda^{\max}\beta_{0}$, and at $n=N$, we have
$\beta_{N-1}=\lambda^{\max }\beta_{N}$.  The solution is $\beta_{n}
= U_{n}(\lambda^{\max}/2)$, and the coefficients $\beta_{n}$ fall to
zero at $N+1$ as required.  The optimal state thus has the form
\begin{equation}
  \label{eq:U(1)OptimalState}
  | \psi^{(N)} \rangle =
  \mathcal{N} \sum_{n=0}^{N}\sin
  \left[ \frac{(n+1)\pi }{N+2}\right] |n\rangle\,,
\end{equation}
where the normalization $\mathcal{N}$ is approximately $\mathcal{N}
\simeq (N/2+1)^{-1/2}$ in the large-$N$ limit~\cite{Ber00}.

In the limit of large $N$, the average fidelity behaves as
\begin{equation}
  \bar{f}^{\max}
  \simeq 1-\frac{\pi ^{2}}{4N^2}\,, \quad \text{for}\ N\gg 1 \,.
\end{equation}
Thus, this optimal protocol for the alignment of a phase reference
has an error (variance) which decreases as $1/N^2$, i.e., at the
Heisenberg limit.

\subsubsection{Fidelity of aligning a Cartesian frame}
\label{subsec:AligningCartesian}

We now consider the task of aligning a Cartesian frame through the
exchange of spin-1/2 particles, using the fidelity as the figure of
merit~\cite{Chi04,Bag04b,Chi05}.

We first develop the payoff function, which we require to be both
left- and right-invariant.  One such possibility is to use the mean
deviation between Alice's coordinate axes and Bob's, i.e.,
\begin{equation}\label{SU(2)PayoffFunction}
    f(\Omega'|\Omega) = 1 - \frac{1}{8} \sum_{i=x,y,z} \bigl|\Omega\mathbf{n}_{iA} -
    \Omega'\mathbf{n}_{iB}\bigr|^2\,,
\end{equation}
where $\Omega\mathbf{n}$ denotes the vector obtained by rotating the
vector $\mathbf{n}$ by $\Omega \in$ SO(3). This function can be
expressed in terms of the characters $\chi_j(\Omega)$ for SO(3);
because these characters will be useful in the following, we briefly
review them here.  The characters of SO(3) are given by the trace of
the irreps $R_j$ as
\begin{equation}\label{eq:SU(2)Character}
    \chi_j(\Omega) = {\rm Tr}[R_j(\Omega)] \,.
\end{equation}
Recall that any rotation $\Omega$ in SO(3) can be expressed as a
rotation by $\omega$, in the range $0\leq \omega < 2\pi$, about
\emph{some} axis. Conjugation by another rotation in SO(3) simply
changes the axis, not the value of $\omega$.  Thus, conjugacy
classes are labeled by an angle $\omega$ (a rotation by $\omega$
about some axis), and characters being class functions are functions
only of $\omega$.  Explicitly, they are given by
\begin{equation}\label{eq:SU(2)CharacterExplicit}
    \chi_j(\Omega) = \chi_j(\omega) =
    \frac{\sin[(2j+1)\omega/2]}{\sin(\omega/2)} \,.
\end{equation}
The payoff function can be expressed in terms of the character
$\chi_{j=1}$ of $R_{j=1}$ (the representation of SO(3) that acts on
spatial vectors), as
\begin{equation}\label{eq:SU(2)PayoffCharacter}
  f(\Omega',\Omega) = \frac{1}{4} + \frac{1}{4} \chi_1(\Omega'^{-1}\Omega) \,.
\end{equation}
As it is a covariant function only on the conjugacy class, we can
express it as
\begin{equation}
    \tilde{f}(\omega) = \frac{1}{4} + \frac{1}{4} \chi_1(\omega)\,.
\end{equation}

The fiducial POVM element can be written as
\begin{equation}
  | e^{(N)}\rangle = \sqrt{N+1} |\tfrac{N}{2},\tfrac{N}{2}\rangle
  + \sum_{j=0}^{N/2-1}(2j+1) |e_j\rangle \,,
\end{equation}
where
\begin{equation}
    |e_j\rangle \equiv \frac{1}{\sqrt{2j+1}}\sum_{m=-j}^{j}
  |j,m\rangle \otimes |j,\alpha(m)\rangle \,,
\end{equation}
are maximally entangled.

From Eq.~\eqref{eq:ChiribellaFiducial}, the optimal fiducial signal
state has the form
\begin{equation}
  \label{eq:ChiribellaFiducialCartesian}
  |\psi^{(N)}\rangle = \beta_{N/2} |\tfrac{N}{2},\tfrac{N}{2}\rangle
  + \sum_{j=0}^{N/2-1}\beta_j |e_j\rangle \,,
\end{equation}
where the coefficients $\beta_j$ are to be determined.  For
simplicity and brevity, we will only solve this eigenvalue problem
in the limit of large $N$.  In this limit, the $\beta_{N/2}$ term
(the only exceptional term) can be ignored.

As with the phase distribution problem, the goal is to find the
state $|\psi\rangle$ that maximizes
\begin{align}
  \bar{f} &= \langle\psi |M |\psi\rangle\nonumber \\
  &= \sum_{j,j'} \beta_j^* \beta_{j'} M_{jj'}\,,
  \label{eq:SU(2)FidelityBetaM}
\end{align}
where $M_{jj'}$ is the matrix
\begin{equation}
    M_{jj'} = \int {\rm d}\Omega\, \langle e_j| R_j(\Omega)|e_j\rangle \langle
    e_{j'}|R_{j'}(\Omega)^\dag |e_{j'}\rangle \tilde{f}(\Omega) \,.
\end{equation}
We note that
\begin{align}
  \langle e_j &|R_j(\Omega)|e_j\rangle \nonumber \\
  &= \frac{1}{2j+1}\sum_{m,m'=-j}^j \langle j,m|R_j(\Omega)|j,m'\rangle \langle
  \alpha(m)|\alpha(m')\rangle \nonumber \\
  &= \frac{1}{2j+1}\sum_{m=-j}^j \langle j,m|R_j(\Omega)|j,m\rangle
  \nonumber \\
  &= \frac{1}{2j+1}\chi_j(\Omega) \,.
\end{align}
Thus,
\begin{equation}
  M_{jj'} \propto \int \text{d}\omega\,
  \chi_j(\omega)\chi_{j'}^*(\omega)(\frac{1}{4} + \frac{1}{4} \chi_1(\omega))\,.
\end{equation}
To evaluate this integral, one can make use of the orthogonality
properties of group characters; see~\textcite{Chi04} for details. We
find that
\begin{equation}
  M_{jj'} \propto \frac{1}{4} \delta_{j,j'} + \frac{1}{4} \big(\delta_{j,j'-1} +\delta_{j-1,j'}\big)\,.
\end{equation}
The eigenvalue problem, then, is essentially identical to that
solved for the distribution of a phase reference in the previous
section. In this limit, then, this maximum average fidelity scales
as
\begin{equation}
    \label{MaxAvgFidSU(2)Asymptotic}
    \bar{f}^{\max} \simeq 1 - \frac{\pi^2}{N^2} \,,  \quad \text{for}\ N\gg 1
    \,.
\end{equation}
Thus, this scheme also scales at the Heisenberg limit.

We note that this particular task has given rise to some controversy
and errors in the literature.  In particular, a mistaken claim of
optimality for this task in~\textcite{Bag01b}, which resulted from a
failure to include the multiplicity of irreducible representations,
led to some confusion over the use of covariant measurements in this
task~\cite{Per02a}.

\subsection{Reference frames associated with coset spaces}
\label{subsec:Coset}

A directional RF, for the $\mathbf{z}$-axis say, can be obtained
from a full Cartesian RF by throwing away the information about the
azimuthal angle.  To specify a direction, therefore, it is
sufficient to specify an equivalence class of Cartesian frames,
those related by an SO(2) transformation about this axis.  Hence, a
directional RF is associated with an element of the coset space
SO(3)/SO(2). This coset space is equivalent to $S_{2}$, the space of
points on a three-dimensional sphere, which corresponds to the
possible directions in space.

Thus, certain reference frames have distinct configurations which do
not correspond to the elements of a group, but rather those of a
coset space of a group.  If we consider a reference frame for a
group $G$ but are unconcerned about the difference between those
related by a subgroup $G_0$ of transformations, then we can speak of
a reference frame for the coset space $G/G_0$.  We may incorporate
such cases into the framework specified above by choosing our figure
of merit to reflect the unimportance of the subgroup in the
estimation task; i.e., choose a payoff function
$\tilde{f}(g^{\prime-1}g)$ that satisfies
\begin{equation}\label{eq:InvariantPayoff}
    \tilde{f}(g^{\prime-1}gg_0) = \tilde{f}(g^{\prime-1}g) \,,
    \quad \forall\ g_0 \in G_0\,.
\end{equation}

In other words, we imagine choosing signal states and POVMs that are
covariant for a group $G$ that is a covering group for the coset
space in question.  Let $z$ be a set of coset representatives, i.e.,
$z \in G/G_0$, and let $\mathrm{d}z$ be a left-invariant measure on
$G/G_0$.  Then, using Eq.~\eqref{eq:GeneralAveragePayoff2},
\begin{align}
  \bar{f}  &=\int\mathrm{d}g\, \tilde{p}(g)\tilde{f}(g)\nonumber \\
  &= \int_{G/G_0} \mathrm{d}z\, \Big(
  \int_{G_0}\mathrm{d}g_0\,\tilde{p}(zg_0) \Big)\tilde{f}(z) \nonumber \\
  &= \int_{G/G_0} \mathrm{d}z \, \tilde{p}_{\rm inv}(z)\tilde{f}(z)
  \,.
\end{align}
Here, we have defined
\begin{align}
  \tilde{p}_{\rm inv}(z) &\equiv
  \int_{G_0}\mathrm{d}g_0\,\tilde{p}(zg_0) \nonumber \\
  & =\mathrm{Tr}\Big[  \big(
  \int_{G_0} \mathrm{d}g_0\, U(zg_0)\left\vert e\right\rangle
  \left\langle e\right\vert U^{\dag}(zg_0)\big)  \left\vert
  \psi\right\rangle \left\langle \psi\right\vert \Big] \nonumber
  \\
  &= \mathrm{Tr}\big[ U(z) E_{\rm inv} U^\dag(z) |\psi\rangle\langle
  \psi| \big] \,,
\end{align}
where
\begin{equation}
    E_{\rm inv} = \int_{G_0} \mathrm{d}g_0\, U(g_0)\left\vert e\right\rangle
  \left\langle e\right\vert U^{\dag}(g_0) \,,
\end{equation}
is $G_0$-invariant.  Thus, for any covariant measurement with
fiducial POVM element $|e\rangle$ that achieves the optimum figure
of merit, there exists a $G_0$-invariant covariant measurement with
fiducial POVM element $E_{\rm inv}$ that achieves the \emph{same}
optimum. For this reason, we may as well restrict the fiducial
signal state and fiducial POVM element to be $G_0$-invariant.

We note that, if the group $G_0$ is non-Abelian, it may not be
possible to find a pure state that is invariant under the subgroup.
In such a situation, if one wishes to work with $G_0$-invariant
states and measurements, then one will have to use \emph{mixed}
fiducial states and POVM elements~\cite{Chi04b}.  We now consider an
example with an Abelian group $G_0$, for which these complications
do not arise.

\subsubsection{Aligning a direction}

Consider the task of optimally aligning a direction in space through
the exchange of spin-1/2 particles.  This was first considered for
just two particles by \textcite{Gis99,Mas00}, as discussed in
Sec.~\ref{subsec:GisinPopescuexample}. The problem was subsequently
considered for an arbitrary number of particles
by~\textcite{Bag00,Per01a,Bag01a}. (For a related investigation,
wherein it is addressed how to perform this task using product
states, see~\textcite{Bag01b}.)

Let $|\psi^{(N)}\rangle$ be the fiducial signal state.  Again, we
restrict our attention to the case of $N$ even.  Because we are
concerned only with aligning a direction and not a full Cartesian
frame, we can choose $|\psi^{(N)}\rangle$ to be invariant under
rotations about the $z$-axis without loss of generality.  Any such
pure invariant state is an eigenstate of $J_{z}$; thus, choose
$|\psi_{m}^{(N)}\rangle$ to be an eigenstate of $J_{z}$ with
eigenvalue $\hbar m$. Clearly, $m$ must be in the range $-N/2,\ldots,N/2$.

First, some notation.  It is standard to express a rotation in SO(3)
in terms of its Euler angles $(\alpha,\beta,\gamma)$. Specifically,
a unitary rotation operator can be expressed as
\begin{equation}\label{EulerAngles}
    R(\alpha,\beta,\gamma) = R_z(\alpha)R_y(\beta)R_z(\gamma)\,,
\end{equation}
where $R_y$ and $R_z$ are SO(2) rotations about the $y$- and
$z$-axes, respectively.  For any element in SO(3), a set of Euler
angles can be found in the range $0\leq \alpha,\gamma < 2\pi$ and $0
\leq \beta < \pi$.  The invariant subgroup is $G_0 = $SO(2) in this
problem, rotations about the $z$-axis; thus, the parameters
$(\alpha,\beta)$ provide coordinates for the coset space
SO(3)/SO(2).

\paragraph{Maximum likelihood.}

We now maximize the likelihood of a correct guess.  Restricting the
fiducial POVM element to be SO(2)-invariant, it takes the form
\begin{equation}
    \label{eq:DirectionFiducialPOVMstate}
    | e_m^{(N)}\rangle =\sum_{j=m}^{N/2}\sqrt{2j+1} |j,m\rangle \,.
\end{equation}
As we wish to include all possible irreps $j$, following the general
construction of Sec.~\ref{subsec:GeneralAlignment}, we should choose
$m=0$, i.e., a fiducial POVM element
\begin{equation}
    \label{eq:DirectionFiducialPOVMstateM=0}
    | e^{(N)}\rangle =\sum_{j=0}^{N/2}\sqrt{2j+1} |j,0\rangle \,.
\end{equation}
The signal state should then be parallel to this vector, of the form
\begin{equation}
    \label{eq:DirectionFiducialSignalMaxLikely}
    |\psi^{(N)}\rangle =\frac{1}{(N/2+1)^2}\sum_{j=0}^{N/2}
    \sqrt{2j+1} |j,0\rangle \,,
\end{equation}
and the maximum likelihood density of a correct guess is
\begin{equation}
  \mu_{\max}= (N/2+1)^2 \,.
\end{equation}

\paragraph{Fidelity.}

A natural payoff function for this problem is the inner product
between Bob's guess direction $\mathbf{n}_g$ and Alice's
transmitted $\mathbf{n}$, given by $\tilde{f}(\theta) =
(1+\mathbf{n}_g\cdot\mathbf{n}) = \cos^{2}(\theta/2)$, where
$\theta$ is the angle between their directions.  This payoff
function is also known as the \emph{fidelity}.  We provide the
details for this optimization as well. Note that this is the
generalization of the example provided in
Sec.~\ref{subsec:GisinPopescuexample} from two to an arbitrary
number of spin-1/2 systems.

Again, the fiducial POVM element is essentially\footnote{The choice
of $m=0$, necessary to optimize the maximum likelihood problem, is
not \emph{a priori} optimal for maximizing the average fidelity.
However, $m$ can be left free and then optimized at the end, with
the result that $m=0$ is indeed optimal for this
task~\cite{Per01a}.} constrained to be that of
Eq.~\eqref{eq:DirectionFiducialPOVMstateM=0}, and now the signal
state takes the general form
\begin{equation}
    \label{eq:DirectionFiducialSignalFidelity}
    |\psi^{(N)}\rangle = \sum_{j=0}^{N/2} b_j |j,0\rangle \,,
\end{equation}
where the coefficients $b_j$ are to be determined. The Born rule
yields
\begin{equation}
  \tilde{p}(\alpha,\beta,0)
  =| \langle \psi^{(N)} |
  R^{\otimes N}(\alpha,\beta,0)| e^{(N)}\rangle|^{2}\,.
\end{equation}
This quantity is independent of $\alpha$, and thus the relevant
conditional probability is
\begin{equation}
  \tilde{p}(\beta)
  =|\langle \psi^{(N)}| R^{\otimes N}(0,\beta,0 )
  | e^{(N)}\rangle|^{2}\,.
  \label{eq:CovariantBornRuleSU(2)}
\end{equation}
Note that $R^{\otimes N}(0,\beta,0) = R_y^{\otimes N}(\beta)$ (a
rotation about the $y$-axis) and the reduced Wigner matrix
$d_{00}^j(\beta)$ is given by
\begin{align}\label{Wignerd}
    d_{00}^j(\beta) &\equiv \langle j,0|R_y(\beta)|j,0\rangle \nonumber \\
    &= P_j(\cos\beta) \,,
\end{align}
where $P_j(x)$ is a Legendre polynomial.

The operator $M$ of Eq.~\eqref{eq:MatrixM} is given in this instance
by the matrix
\begin{align}
  M_{jj^{\prime }} &= \langle j,0|M|j',0\rangle \nonumber \\
  &=\tfrac{1}{2}\int_0^\pi \sin\beta\, \mathrm{d}\beta\,
  P_j(\cos\beta)P_{j'}(\cos\beta) \cos^{2}(\beta/2) \nonumber \\
  &= \tfrac{1}{4}\int_{-1}^1 \mathrm{d}x\,
  P_j(x)P_{j'}(x)(P_0(x) + P_1(x)) \nonumber \\
  &= \tfrac{1}{4}\Big( \tfrac{2}{2j+1}\delta_{j,j'}
  + \tfrac{2j}{(2j+1)(2j'+1)}\delta_{j,j'+1} \nonumber \\
  &\qquad\qquad +
  \tfrac{2j'}{(2j+1)(2j'+1)}\delta_{j,j'-1} \Big) \,,
\end{align}
where we have expanded the payoff function in terms of Legendre
polynomials.

This eigenvalue problem is essentially the same as those solved in
the previous section.  The maximum average fidelity is given by
\begin{equation}
    \label{MaxAvgFidS2Analytic}
    \bar{f}^{\max} = \frac{1+x_{N/2+1}}{2} \,,
\end{equation}
where $x_{N/2+1}$ is the largest zero of the Legendre polynomial
$P_{N/2+1}(x)$. In the limit of large $N$, this maximum average
fidelity scales as
\begin{equation}
    \label{MaxAvgFidS2Asymptotic}
    \bar{f}^{\max} \simeq 1 - \frac{\zeta^2}{N^2} \,,
    \quad \text{for}\ N\gg 1 \,,
\end{equation}
where $\zeta \simeq 2.4$.  Thus, this optimal scheme also scales at
the Heisenberg limit~\cite{Per01a,Bag01a}.

\subsection{Relation to phase/parameter estimation}

We note that the task of aligning a phase reference is essentially
equivalent to the task of estimating an unknown phase.
Specifically, instead of viewing the problem of \emph{noiseless}
transmission of a quantum system between parties who do not share
a phase reference, the problem could instead be viewed as one of
transmission of the same quantum system between parties who
\emph{do} share a phase reference, but where the transmitting
channel induces an unknown phase shift on the system.

In this light, we note that the protocol presented in
Sec.~\ref{subsec:SharedPhaseRef} is equivalent to the optimal
solution for phase estimation using the same figure of
merit~\cite{Ber00}. Techniques for quantum-limited phase estimation
have been well studied, and there exist a wide variety of alternate
methods that could each be applied, in some form, to the task of
aligning a phase reference. For an overview of phase estimation
techniques from different viewpoints, we refer the reader to the
review article on quantum metrology by~\textcite{Gio04a}, or the
text of~\textcite{Nie00} which discusses phase estimation techniques
from a quantum algorithm perspective.  Also, see~\textcite{Gio06}
for a unified framework of these techniques.

Similarly, the task of aligning a reference frame for $G$ through
the transmission of a quantum system is essentially equivalent to
estimating an unknown element $g \in G$ given a quantum channel that
acts on the same quantum system with the unitary $U(g)$.  This
latter task is generally referred to as \emph{parameter estimation}.

We note, then, that the scheme for aligning a Cartesian
frame presented in Sec.~\ref{subsec:AligningCartesian} is closely
related to a method for estimating an unknown SU(2) (or more
generally SU($d$)) transformation~\cite{Aci01}.  We briefly review
this latter scheme, because of its close relation to the topic at
hand.  Let $R(\Omega)$ be the unitary representation of an unknown
rotation $\Omega \in$ SU(2), which acts on states of a Hilbert space
$\mathcal{H}$; the task is to estimate $\Omega$ through one
application of $R(\Omega)$ to some quantum state.  For this problem,
we allow the use of an \emph{ancillary} system, with Hilbert space
$\mathcal{K}$ of arbitrary dimension; this ancilla is assumed to
transform trivially under SU(2).  (That is, SU(2) acts as
$R(\Omega)\otimes I$ on $\mathcal{H}\otimes\mathcal{K}$.)  Without
loss of generality, one can choose $\mathrm{dim}\,\mathcal{K} =
\mathrm{dim}\,\mathcal{H}$, and choose a basis for $\mathcal{K}$
with the same labels as $\mathcal{H}$.  We choose the standard SU(2)
angular momentum basis $|j,m,\alpha\rangle$, where $\alpha$ labels
the multiplicity.

An optimal state $|\psi\rangle$ on $\mathcal{H}\otimes\mathcal{K}$
for maximizing the likelihood of a correct guess, up to a
normalization constant, is
\begin{equation}
    |\psi\rangle \propto \sum_{j=0}^{N/2-1} \frac{1}{\sqrt{(2j+1)c_j}}
    \sum_{m=-j}^j \sum_{\alpha=1}^{c_j}|j,m,\alpha\rangle_{\mathcal{H}}
    |j,m,\alpha\rangle_{\mathcal{K}} \,,
\end{equation}
where $c_j$ is the multiplicity of the $j$th representation.  The
fiducial POVM element will be parallel to this vector.  We note that
this state is a superposition over irreps $j$ of a
maximally-entangled state between an irrep $j$ on $\mathcal{H}$ and
an equally-sized space on $\mathcal{K}$.
The optimality of this state for alignment follows from the general
arguments presented in Sec.~\ref{subsec:MaxLikelihood}: the optimal
fiducial state is the one that maximizes the dimension of the group
orbit. Without the help of an ancilla, this is achieved within a
given irrep $j$ by entangling the gauge space $\mathcal{M}_j$ (on
which the group acts nontrivially) with the multiplicity space
$\mathcal{N}_j$ (on which it acts trivially). In the present
context, it is achieved by entangling the system with the ancilla.

As such methods for parameter estimation have importance for quantum
computing in terms of the characterization of quantum gates, it is
interesting to consider how the methods of reference frame alignment
may be applied to such characterization problems as well.

Finally, work on magnetometry -- the use of magnets as direction
indicators to determine the strength and direction of a magnetic
field -- is also a problem of parameter estimation, closely related
to the problem of reference frame distribution.  Practical proposals
for quantum-limited magnetometry make use of spin-squeezed clouds of
cold atoms to measure the three components of an unknown magnetic
field through a form of phase estimation~\cite{Pet05}.  It would be
interesting to investigate whether spin-squeezed states of
indistinguishable particles (e.g., atoms) can be used for the
distribution of a direction or frame as efficiently as the optimal
protocols derived above, which make use of (distinguishable) qubits
in highly-entangled states and corresponding entangling
measurements.

\subsection{Communication complexity of alignment}
\label{subsec:CommComplexity}

We have thus far only considered protocols for RF alignment wherein
there is a single round of communication from Alice to Bob.  We now
consider multi-round protocols~\cite{Rud03,deB05,Gio06}. Whereas the
single-round protocols generally require entanglement between the
transmitted systems to achieve the Heisenberg limit, multi-round
protocols have the advantage that they can achieve this limit
despite using no entanglement.

With multi-round protocols, it is natural to frame the problem as
one of \emph{communication complexity}, wherein one investigates the
resources of \emph{rounds of communication} along with the standard
resources of number of transmitted quantum or classical bits.  To
conform with standard notions of communication complexity, it is
useful to consider the alignment problem with two departures from
the approach adopted in analyzing the previous alignment protocols.
First, we consider the \emph{worst case} scenarios (rather than the
\emph{average} case considered above); second, we avoid the use of
payoff functions, such as the fidelity, with a view to obtaining a
more precise estimate of how well any given instance of the protocol
has performed.  As such, we consider strategies for aligning spatial
reference frames that allow Bob to directly determine the angle
which relates his and Alice's RFs to some specified accuracy with a
bounded probability of error in the worst case scenario. More
precisely, if $\theta$ is an angle relating Alice and Bobs' RFs, and
$\theta'$ is the estimation of $\theta$ inferred by Bob, then we
will be interested in the amount (and type) of communication
required for protocols that achieve $P_{\rm
error}=\text{Pr}[|\theta-\theta'|\ge \delta]\le \epsilon$, for some
fixed $\epsilon,\delta>0$. By setting $\delta=1/2^{k+1}$ we say that
with probability $(1-\epsilon)$ Bob has a $k$-bit approximation to
$\theta$.

We now describe such a protocol for the case of sharing a phase
reference through the exchange of qubits, i.e., the same task as
investigated in Sec.~\ref{subsec:SharedPhaseRef}.  This protocol can
also be applied to the task of aligning a Cartesian
frame~\cite{Rud03}.  The effects of decoherence on these protocols
has been shown to be equivalent to that of decoherence on the
``standard'' protocol of
Sec.~\ref{subsec:SharedPhaseRef}~\cite{Boi06}.


Let $\theta_{BA}$ be the unknown angle (misalignment) that relates
Bob's phase reference to Alice's.  In this protocol, Alice and Bob
use an algorithm that estimates each bit of the phase angle
$\theta_{BA}$ independently. We define the phase angle $\theta_{BA}
= \pi T$, where $T$ has the binary expansion $T =
0.t_1t_2t_3\cdots$.  Alice and Bob will attempt to determine $T$ to
$k$ bits of precision, and accept a total error probability $P_{\rm
error} \leq \epsilon$.  If the total error probability is to be
bounded by $\epsilon$, then each $t_i$, $i=1,\ldots,k$, must be
estimated with an error probability of $\epsilon /k$. (An error in
any one bit causes the protocol to fail, so the total error
probability in estimating all $k$ bits is $P_{\rm error} =
1-(1-\epsilon /k)^k \le \epsilon$.)

To estimate the first bit $t_1$, Alice prepares a single qubit in
the state $(|0\rangle_A + |1\rangle_A)/\sqrt{2}$ (relative to her
phase reference) and sends the qubit to Bob.  Bob then performs his
operation $X_B$ and sends the qubit back to Alice, where $X_B$ is
the Pauli bit-flip operator according to Bob. She then performs her
operation $X_A$.  The resulting combined transformation $X_A X_B$ is
described in Alice's frame as
\begin{align}
  X_A X_B &= X_A (e^{-i\theta_{BA} Z/2} X_A e^{+i\theta_{BA}Z/2})
  \nonumber \\
  &= e^{+i \theta_{BA} Z} \,,
\end{align}
where $Z$ is the Pauli phase-flip operator.  Finally, Alice performs
a Hadamard transformation $H_A$ (in her frame) and measures the
observable $O_A = -Z$.  The expected value of this observable is
$\langle O_A \rangle = \cos(2\theta_{BA})$, with an uncertainty of
\begin{equation}
  \Delta O_A = \sqrt{\langle O_A^2 \rangle -
  \langle O_A \rangle ^2} = \sin{(2\theta_{BA})}\,.
\end{equation}

Expressing $\langle O_A\rangle$ in terms of $T$, we have
\begin{equation}
  \langle O_A \rangle = \cos{(2 \theta_{BA})}
  = \cos{(2\pi 0.t_1 t_2\cdots)} \,.
\end{equation}
By repeating this procedure $n_1$ times, i.e., sending $n_1$
independent qubits and averaging the results, Alice obtains
$\overline{\langle O_A \rangle}$, the estimate of $\langle O_A
\rangle$. If $n_1$ is chosen such that $|\overline{\langle {O}_A
\rangle} - \langle O_A \rangle| \le 1/2$ with some error
probability, then $|\overline{T} - T  | \le 1/4$, thus determining
the first bit $t_1$ with this same probability. The required number
of iterations $n_1$ to achieve the desired error is given by the
Chernoff bound, with $\delta = 1/4$.  That is, the probability that
the first bit of Alice's estimate $\overline{\langle O_A \rangle}$
differs from the first bit of the actual value $\langle O_A \rangle$
decreases exponentially in the number of repetitions $n_1$, and is
bounded explicitly by
\begin{equation}
    \label{FirstBitError}
    {\rm Pr}\Bigl[|\overline{\langle O_A \rangle} - \langle O_A \rangle|
    \ge 1/2\Bigr] \le \epsilon/k \le 2e^{-n_1/32} \,.
\end{equation}
Thus, allowing a probability of error $\epsilon/k$ in this bit, we
require $n_1 \ge 32\ln(2k/\epsilon)$ iterations.

Now we define a similar procedure for estimating an arbitrary bit,
$t_{j+1}$. Alice prepares the energy eigenstate $|0\rangle$, and
performs her $H_A$ operation. Alice and Bob then pass the qubit back
and forth to each other $2^j$ times, each time Bob performs his
$X_B$ operation and Alice performs her $X_A$ operation. That is,
they jointly implement the operation $(X_A X_B)^{2^j}$. Finally
Alice performs her $H_A$ operation. Expressing these operations in
Alice's frame, the protocol to estimate $t_{j+1}$ produces the state
\begin{align}
 |\psi_j\rangle_A &= H_A (X_A X_B)^{2^j} H_A |0\rangle \nonumber \\
 &= H_A (e^{+i\omega t_{AB} Z})^{2^j} H_A |0\rangle \nonumber \\
 &= H_A e^{+i2^j \omega t_{BA} Z} H_A |0\rangle \nonumber \\
   &= \bigl[ i\sin (2^j \omega t_{BA})|0\rangle +
  \cos (2^j \omega t_{BA} )|1\rangle\bigr]_A \,.
\end{align}
Alice then measures the observable $O_A = -Z$. The expected value of
this observable is:
\begin{align}
     \langle O_A \rangle
     &= \cos{(2^{j+1} \omega t_{BA})} \nonumber \\
     &= \cos(2^j[2\pi 0.t_1t_2\cdots] ) \nonumber \\
     &= \cos(2\pi t_1t_2\cdots t_j.t_{j+1}t_{j+2}\cdots ) \nonumber \\
     &= \cos(2\pi 0.t_{j+1}t_{j+2}\cdots ) \,.
     \label{Eqn:ExpectOImproved}
\end{align}
This expression has the same form as one iteration of the scheme to
estimate the first bit $t_1$; Alice and Bob simply require more
exchanges to implement $(X_A X_B)^{2^j}$.  To get a probability
estimate for each bit $t_{j+1}$, this more complicated procedure is
repeated $n_{j+1}$ times.  Because we require equal probabilities
for correctly estimating each bit, we can set all $n_{j+1}$ equal to
the same value, $n \sim 32 \ln(2k/\epsilon)$. Thus the total amount
of qubit communication $N_c$ required to obtain bits $t_1$ through
$t_k$ by this procedure is
\begin{equation}
    N_c= n\times \sum_{j=1}^{k}2^{j-1}=n(2^{k}-1)=O(2^{k}\ln
    (2k/\epsilon ))\,.
\end{equation}


To facilitate comparison with the previous sections, wherein the
focus was on maximizing the average fidelity, we imagine that Alice
and Bob use the above protocol to obtain, with probability
$(1-\epsilon)$, an angle $\theta'$ which is a ``$k$-bit'' estimator
of the true angle $\theta$, i.e., they obtain $|\theta-\theta'|\le
2\pi/2^{k+1}$ with probability $(1-\epsilon)$.  The fidelity of this
estimate is $f=\cos^2((\theta-\theta')/2)$. Since $\cos x\ge 1-x^2$
we have that $\bar{f}\ge 1- (\frac{2\pi}{2^k})^2$.  To compare with
the average case fidelity computed previously, we assume that when
the protocol fails (which happens with probability $\epsilon$) the
fidelity obtained is 0, i.e., worse than a random guess.  We have
then that the expected fidelity from this protocol satisfies
\begin{equation}
    \bar{f} \ge (1-\epsilon)\Big[
    1-\Big(\frac{2\pi}{2^k}\Big)^2\Big]\,.
\end{equation}

If we take $\epsilon =1/2^{2k},$ then the total number of qubit
communications scales as $k2^{k}$ for large $k$, while the expected
fidelity $\bar{f}$ scales as $2^{-2k} \simeq 1- (\log N_c)/N_c^{2}$.
Thus, remarkably, this protocol beats the standard quantum limit of
$1/N_c$, yet does not require entangled states or collective
measurements.  This is achieved at the cost of an increased
complexity in coherent qubit communications.

\subsection{Clock synchronization}

If Alice and Bob share a common frequency standard, then they can
use techniques for phase reference alignment that were outlined
above to perform clock synchronization, i.e., aligning a temporal
reference~\cite{Joz00}. To see the relation between phase alignment
and clock synchronization, it is easiest to work in a rotating frame
(interaction picture) in which states are described as
$|\psi\rangle_I = e^{i H_0 t / \hbar} |\psi\rangle$, and observables
and transformations as $A_I = e^{iH_0 t / \hbar} A e^{-iH_0 t /
\hbar}$, where $H_0$ is the free Hamiltonian.  In this rotating
frame, states are stationary under free evolution, and the problem
of clock synchronization is reduced to one of aligning a phase
reference.

In one class of clock synchronization protocols based on phase
estimation, the systems exchanged are \emph{ticking qubits:}
nondegenerate two-level quantum systems that undergo time
evolution~\cite{Chu00}.  Much like phase estimation, the use of
entangling operations and/or measurements can lead to different
scalings in the synchronization accuracy.  For example, a protocol
that uses only separable (unentangled) ticking qubits and
single-qubit measurements requires $O(2^{2k})$ ticking qubit
communications (a coherent transfer of a single qubit from Alice to
Bob) to achieve an accuracy in the time offset of Alice's and Bob's
clocks of $k$ bits. That is, the synchronization accuracy scales as
the standard quantum limit. In comparison, a protocol
by~\textcite{Chu00} makes use of the Quantum Fourier Transform and
an exponentially large range of qubit ticking frequencies.  This
protocol requires only $O(k)$ quantum messages to achieve $k$ bits
of precision, an exponential advantage over the standard quantum
limit.  Although this protocol gives insight into the ways that
quantum resources may allow an advantage in clock synchronization,
it is unsatisfactory for two reasons: (1) its use of exponentially
demanding physical resources is arguably the origin of the enhanced
efficiency~\cite{Chu00,Gio01}; and (2) Alice and Bob need to \emph{a
priori} share a synchronized clock in order to implement the
required operations as defined in \textcite{Chu00}. These problems
are not present in subsequent protocols based on ticking qubits,
which used the techniques for phase estimation presented in
Sec.~\ref{subsec:CommComplexity} to design a clock synchronization
protocol that operates near the Heisenberg limit~\cite{deB05}.  The
Heisenberg limit for clock synchronization can be achieved by making
use of the phase estimation protocol of
Sec.~\ref{subsec:SharedPhaseRef}.

Distinct from the approaches based on phase estimation, there has
been considerable interest in another class of clock synchronization
protocols which make use of entanglement~\cite{Joz00}; see also
\textcite{Bur01,Joz01,Pre00,Yur02}. In a variant of this approach, a
third party Charlie distributes a large number of boxes to Alice and
Bob, where each box contains one spin of a spin singlet.  Each box
also contains a classical magnetic field aligned in the
$z$-direction, such that the free Hamiltonian for each spin is $H_0
= \chi \sigma_z$ for some constant energy $\chi$. (We note that this
establishes a shared RF between Alice and Bob for this particular
direction.) Clock synchronization can be achieved by Alice
performing measurements on her spins in her $x$-direction at time
$t=0$.  By informing Bob (via any classical channel) as to the
sub-ensemble of the singlets for which she obtained the $+x$
outcome, Bob can identify the subset of his particles which are all
precessing (via the free Hamiltonian) around the common
$z$-direction in phase with Alice's clock. However because Bob does
not necessarily share a common $x$-direction with Alice, he cannot
actually read out this phase information.  This complication can be
neatly circumvented with a slight modification -- on half of the
particles, Alice and Bob use a different magnetic field strength in
the $z$-direction to establish two different precession frequencies.
Bob can now choose \emph{any} $x$-direction to measure each
subensemble, because both ensembles of precessing spins are offset
by the same unknown phase shift with respect to Alice's spins, and
he can achieve synchronization by merely observing the beats between
the oscillations.

In the language of the final section of this review, we can
understand this protocol as one in which standard \emph{refbits}
(see Sec.~\ref{subsec:QuantifyingBoundedRFs} for more details) are
being distilled from the initial singlets and put to use as a
bounded shared RF.  Note that if such synchronization was Charlie's
intention all along, then such a protocol would not be a
particularly efficient use of resources -- he could just as simply
have distributed to Alice and Bob a shared RF state (such as those
given in Eq.~\eqref{refbits}, for example) which require no
entanglement whatsoever~\cite{Pre00}.\footnote{An analogue of this
scheme for Cartesian frame alignment has been proposed~\cite{Rud99},
and similar reservations apply; however, at least this latter
approach was motivated by a funny story \cite{Rud99b}.}  This fact,
together with the result that this entanglement cannot be
\emph{purified} (an issue we return to in
Sec.~\ref{subsec:Purification}), suggest that shared entanglement
between two parties does \emph{not} provide an advantage for clock
synchronization (or other forms of RF alignment).

A third class of protocols for clock synchronization make use of
precise timing of light signals exchanged between parties, and for
which the quantum limits have recently been investigated. Instead of
classical coherent state light pulses for the signals, one can use
highly entangled states of many photons and beat the standard
quantum limit~\cite{Gio01}. Essentially, the advantage is due to
entanglement-induced bunching in arrival time of individual photons,
enabling more accurate timing measurements. The key disadvantage of
this technique is that the loss of a single photon destroys the
entanglement and renders the measurement
useless~\cite{Gio04a,Gio01}, although techniques have been developed
to ``trade off'' the quantum advantage in return for robustness
against loss~\cite{Gio02}. Furthermore, the effect of dispersion is
known to be an important issue with such protocols, with the use of
entanglement possibly offering an advantage here as
well~\cite{Fit02,Gio04b}.  We note that such protocols differ from
those based on phase estimation in that they make use of
relativistic principles (specifically, the constancy of the speed of
light).

\subsection{Other instances of alignment}

Above, we considered the alignment of Cartesian frames using $N$
spin-1/2 systems.  A different approach to this alignment problem is
to use a single Hydrogen atom~\cite{Per01b}. The analysis of this
task is similar to that presented above, with the notable difference
that multiplicities of representations of SO(3) are \emph{not}
available with the hydrogen atom.  Also, the use of elliptic Rydberg
states of the hydrogen atom have been considered for this problem,
with resulting fidelity comparable with that of the optimal scheme
for a hydrogen atom~\cite{Lin03}.

Although a phase reference (clock) and spatial direction (or
Cartesian frame) may be the most ubiquitous types of reference
systems, it is possible to distribute more general and exotic types
of reference frames through the exchange of appropriate quantum
systems.

For example, consider the distribution of a \emph{reference
ordering}, i.e., a labeling of $N$ objects~\cite{Kor04}. Through the
use of techniques similar to those described in
Sec.~\ref{subsec:AligningCartesian} for the distribution of a
Cartesian frame, one can construct an optimal protocol that
distributes an ordering of $N$ particles using $N$ systems with
dimensionality $N/e$.  (In contrast, the classical problem requires
$N$ systems each with $N$ distinguishable states.)

Another problem is the distribution of a reference frame for
chirality~\cite{Dio00,Gis04,Col05}.  Such quantum systems have been
given the moniker ``quantum gloves''. Clearly, a full Cartesian
frame includes a reference for chirality; however, distributing a
full Cartesian frame purely for the purposes of distributing a
reference chirality is not very economical, and more efficient
methods are possible.  Also, a chiral reference can be distributed
\emph{perfectly}, i.e., with no error, with only finite quantum
resources.  Methods for the distribution of a chiral reference using
only two kinds of particle (i.e., a proton and an electron) and only
four spinless particles, along with other interesting combinations,
can be found in~\textcite{Col05}.

One can also consider the problem of \emph{secret sharing} of
unspeakable information~\cite{Bag06c}.  In such a protocol, a
quantum state is shared between several parties with the aim that a
reference frame can only be determined if all of the parties come
together (to perform collective operations and measurements).
Parties working alone, or together using only LOCC, cannot determine
the reference frame with the same precision.

\subsection{Private communication of unspeakable information}
\label{subsec:PrivateUnspeakable}

With the development of optimal schemes for the distribution of a
reference frame, it is natural to consider how two parties, Alice
and Bob, can perform such a distribution \emph{privately} by using
some number of shared secret bits -- a classical key -- to randomize
the signal state~\cite{Chi06}.  In other words, we consider the
problem of how well two parties can convert a private classical key
into a private shared RF (of bounded size) using a public channel
and given that they do not previously possess a shared RF (private
or public). For concreteness, we investigate the private
communication of a Cartesian frame.

Consider the optimal scheme for the distribution of a Cartesian
frame using fidelity as the figure of merit (which achieves the
Heisenberg limit). The fiducial signal state for this scheme is
given in Eq.~\eqref{eq:ChiribellaFiducialCartesian}.
(The scheme will be essentially identical for any figure of merit.)
As with any private quantum communication, Alice and Bob can choose
unitary operators from a set of unitaries, based on their classical
key, to \emph{randomize} any quantum state as viewed by an
eavesdropper Eve who does not share this key, using the techniques
of~\textcite{Amb00}.  In general, to completely randomize a state on
a Hilbert space of dimension $d = 2^N$, Alice and Bob require a key
consisting of $2N$ secret bits.  However, we note that Alice and Bob
do not share a reference frame to begin with and thus they can only
perform correlated operations on the multiplicity spaces
$\mathcal{N}_j$.
Despite this restriction, a complete randomization can be achieved
for the fiducial state of
Eq.~\eqref{eq:ChiribellaFiducialCartesian}. To see this, recall that
this state can be expressed as a coherent superposition, over all
representations (charge sectors) of SO(3), of
\emph{maximally-entangled} states across the representation space
$\mathcal{M}_j$ and an equally-sized subspace
$\tilde{\mathcal{N}}_j$ of the multiplicity space. Because of this
particular structure of the fiducial state, a complete randomization
over just the subsystems $\tilde{\mathcal{N}}_j$ will take every
signal state to the same mixed state (and thus achieves a complete
randomization on the Hilbert space that is the span of the supports
of the signal states). The dimension of each subsystem
$\tilde{\mathcal{N}}_j$ is equal to that of $\mathcal{M}_j$, namely,
$d_j=2j+1$, thus requiring $2\log_2 (2j+1)$ secret bits to
completely randomize. The total number of secret bits required to
completely randomize all the signal states is
\begin{equation}
    \log_2 \Bigl[ \sum_{j=0}^{N/2} (2j+1)^2 \Bigr] \simeq 3\log_2 N \,.
\end{equation}

Thus, through the transmission of $N$ qubits on a public channel and
using $3 \log_2 N$ classical bits of private key, one can achieve
the distribution of a private Cartesian frame at the Heisenberg
limit, which is to say with an error that scales as $1/N^2$.  We
note that this number $3 \log_2 N$ is identical to the number of
classical bits that can be transmitted privately given a private
shared Cartesian frame as key, as discussed in
Sec.~\ref{subsec:PrivateSRFs}.

\subsection{Dense coding of unspeakable information}
\label{subsec:DenseCoding}

Consider the following problem. Alice wants to send Bob classical
information, but at the time that Alice learns which message she
would like to send Bob, the cost of using the quantum channel is
very high, whereas earlier, before Alice learns the message, the
cost of using the channel is low. \emph{Dense coding} allows Alice
to make use of the channel at the early time, prior to learning the
message she wishes to send, in order to increase the amount of
information she succeeds at transmitting to Bob at the later time.

Suppose that Alice wishes to send to Bob a direction in space
rather than a classical message.  Suppose moreover that at the
time where Alice learns the direction she would like to send Bob,
the cost of using the quantum channel is very high, whereas
earlier, before Alice learns the direction, the cost of using the
channel is low.  One would have a natural analogue of dense coding
to unspeakable information (directional information in this case)
if use of the channel at the early time allowed Bob to estimate
Alice's direction with greater accuracy at the later time.

The following is such an analogue.  Alice prepares a pair of
spin-1/2 systems in a singlet state, and in the first use of the
channel, sends one of these to Bob. Later, when she has a sample
of the classical direction $\hat{\mathbf{n}}$ that she would like
to send to Bob, she implements a unitary rotation of $\pi$ degrees
about $\hat{\mathbf{n}}$ on her spin-1/2 system and sends it to
Bob. Through Alice's operation, the singlet is transformed into
$(|{+}\hat{\mathbf{n}}\rangle|{-}\hat{\mathbf{n}}\rangle +
|{-}\hat{\mathbf{n}}\rangle|{+}\hat{\mathbf{n}}\rangle)/\sqrt{2}$,
which is a two-spin state that can be used to indicate the
direction $\hat{\mathbf{n}}$; in fact, given that the image of
such a state under SU(2)-averaging covers the entire symmetric
subspace, this state is as good a direction indicator as the
parallel spin state of Sec.~\ref{subsec:GisinPopescuexample}.  In
her second use of the quantum channel, she sends her spin to Bob,
and Bob estimates the direction. The optimal average fidelity that
can be achieved for such a state and measurement was shown in
Sec.~\ref{subsec:GisinPopescuexample} to be $3/4$, which is
greater than the fidelity of $2/3$ that could be achieved using a
single spin-1/2 system.\footnote{The optimal average fidelity that
can be achieved for a symmetric product state consisting of $N$
spin-1/2 systems (a parallel state) is $(N+1)/(N+2)$, and this
result generalizes to any symmetric pure state~\cite{Mas95}.}
Thus, this scheme provides an analogue of dense coding for
unspeakable quantum information. The optimization of this sort of
dense coding scheme has not been investigated to date.

A slightly different analogue of dense coding of unspeakable
information was considered in~\textcite{Bag04a}, building on the
results of~\textcite{Aci01}. This protocol involves Alice initially
sending Bob half of an entangled state over multiple spin systems.
It is assumed that subsequently the entire lab of the sender is
subject to the same SU(2) transformation that her half of the
entangled pairs are subject to. Under this assumption the three
parameters describing the relation of her spin-1/2 system to that of
the receiver also describe the relation of her local Cartesian frame
to that of the receiver.  In this scenario, the sender is
essentially passive: both the spin and the local Cartesian frame
must be acted upon by some external agency. Unfortunately, it is not
clear whether this is of practical significance in the most common
case where the SU(2) transformation acting on the local Cartesian
frame is a rotation in space.  For instance, if a rotation of the
entire laboratory is realized by an external torque, it is not clear
that the state of a spin-1/2 system stored in this laboratory (i.e.,
in some trapping potential) will necessarily undergo the same
rotation.\footnote{That is, unless the spin degrees of freedom are
coupled to other fields in the lab.  This coupling itself would
negate the protocol however, as it implies some ongoing active
transformations on the stored spins.} Nonetheless, the optimal
solution for this sort of scheme has been provided for an arbitrary
number of spin-1/2 systems~\cite{Bag04a}. The optimal state bears a
strong similarity to the optimal state for aligning Cartesian RFs,
presented in Sec.~\ref{subsec:AligningCartesian}.

\subsection{Error correction of unspeakable information?}

We end this section with a cautionary note on the potential use of
quantum methods for aligning reference frames, first made
by~\textcite{Pre00} for the specific task of clock synchronization:
that the standard techniques of quantum error correction
\emph{cannot} be directly applied to unspeakable information.

Consider a situation wherein Alice and Bob wish to align their
respective frames by exchanging quantum systems via some
\emph{noisy} quantum channel.  Let $\mathcal{F}$ be the decohering
superoperator describing the channel.  The form of this noise is
critical to their ability to complete this task; here, we consider
only two extreme cases.  If the noise is of the form $\mathcal{F} =
\oplus_q I_{\mathcal{M}_q} \otimes \mathcal{D}_{\mathcal{N}_q}$ in
terms of the decomposition of Eq.~\eqref{eq:HdecompFull}, where
$\mathcal{D}_\mathcal{N}$ is the completely depolarizing
superoperator on $\mathcal{N}$.  This noise affects only the
multiplicity subsystems; in other words, it acts only on the
relational degrees of freedom of the transmitted systems.  In such a
case, RF alignment is still possible (although possibly at a
decreased efficiency, as the optimal protocols took advantage of
these multiplicity subsystems). Alice and Bob can choose to transmit
states that are encoded entirely within the gauge subsystems
$\mathcal{M}_q$, as these subsystems are decoherence-free in terms
of the noise.

On the other hand, if the noise is of the form $\mathcal{F} =
\oplus_q \mathcal{D}_{\mathcal{M}_q} \otimes I_{\mathcal{N}_q}$ in
terms of the decomposition of Eq.~\eqref{eq:HdecompFull}, then the
gauge subsystems $\mathcal{M}_q$ will experience complete
decoherence. However, Alice and Bob cannot choose to execute their
alignment protocol entirely within the decoherence-free multiplicity
subsystems $\mathcal{N}_q$, because these subsystems cannot carry
unspeakable information (at least, not of this type).  Whereas
speakable information can be encoded into any desired subsystem,
unspeakable information must be encoded into subsystems carrying the
appropriate degree of freedom.

This latter case can be worded as a simple physical example.
Consider the alignment of a phase reference, using a noisy channel
that simply adds a constant but unknown phase shift. If Alice and
Bob use one of the techniques of this section to attempt to align
their phase references using this channel, Bob will acquire an
estimate of the phase difference between his and Alice's RFs.
However, because Bob knows this estimate may differ from the actual
difference by some unknown shift, caused by the channel, he in fact
has learnt nothing about the relation between his phase reference
and Alice's. There is no protocol that they can perform that will
distinguish the unknown phase shift relating their RFs and the
unknown phase shift applied by the channel, and thus alignment
cannot be performed using this channel~\cite{Pre00,Yur02}.

\section{Quantum information with bounded reference frames}
\label{sec:BoundedRFs}

In the reference frame alignment schemes of Sec.~\ref{sec:Aligning},
we determined which quantum states of a given bounded size were
optimal in serving as a sample of the sender's classical reference
frame.  The systems were ultimately measured relative to the
receiver's classical reference frame, so that the unspeakable
information that they contained was essentially amplified to the
macroscopic scale with some associated uncertainty.  However, there
will be situations for which this amplification process is not
ideal, and instead one should make direct use of the quantum RF
itself.\footnote{For example, \textcite{Jan03} consider the
constraints on amplifying and copying quantum RFs for phase.}
Whatever purpose the recipient had in mind in trying to align his
classical RF with that of the sender's, one can ask to what extent
he could achieve this same purpose by storing his quantum sample of
the sender's reference frame in his lab and thereafter using it in
place of his classical RF.

Furthermore, many quantum experiments involve mesoscopic or even
microscopic systems that can be understood as playing the role of a
reference frame.  For instance, a Bose-Einstein condensate may act
as a reference frame for the phase conjugate to atom number, even
though it may contain a relatively small number of atoms.  We are
therefore led to consider the question of how well a bounded-size
quantum system may stand in for a classical reference frame.

In Sec.~\ref{sec:QRFs} we have already considered the problem of
treating reference frames within the quantum formalism, but the
system instantiating the reference frame was assumed to be of
unbounded size. Here we shall be interested in bounded-size quantum
reference frames. We shall focus in particular on the implications
of such RFs on one's ability to perform quantum-information
processing tasks, specifically: the fundamental primitive of quantum
state estimation, operations and measurements in quantum
computation, and the quantum cryptographic protocols of data hiding
and bit commitment. Furthermore, we demonstrate that for bounded
shared RFs, like entanglement, it is possible to develop a general
theory of the manner in which this resource is distributed,
transformed from one form to another, distilled, degraded with use,
quantified, etcetera.

\subsection{Measurements and state estimation with bounded reference frames}

State estimation is a fundamental primitive of quantum information
processing.  In this section, we discuss the role of reference
frames in performing measurements required for state estimation, and
the effect of bounding the size of this RF.

\subsubsection{A directional example}
\label{subsec:RelDirMeas}

Consider the task of estimating whether the state of a spin-1/2
system is aligned or anti-aligned with some perfect (unbounded)
directional RF, given the promise that it is one of the two. If
one is able to compare the system with this RF, then this task can
be easily achieved, as it corresponds simply to discriminating a
pair of orthogonal states, $|{+}\mathbf{z}\rangle$ and
$|{-}\mathbf{z}\rangle$, where we take $\mathbf{z}$ to be the axis
defined by the directional RF. Specifically, a measurement of
$\mathbf{S}\cdot\mathbf{z}$, the spin along $\mathbf{z}$,
determines the answer with certainty.  In contrast, if one is not
able to make use of this RF, then a superselection rule is in
force, the measurement of $\mathbf{S}\cdot\mathbf{z}$ becomes
impossible, and the states $|{+}\mathbf{z}\rangle$ and
$|{-}\mathbf{z}\rangle$ become completely indistinguishable.

There is an intermediate scenario between these two extreme cases,
however, wherein one only has access to a sample of the RF -- one
that is of bounded size.  In this case, $|{+}\mathbf{z}\rangle$ and
$|{-}\mathbf{z}\rangle$ become partially distinguishable, as we now
demonstrate with an example.  We consider the case wherein the
directional RF is a spin-$j$ system, for some arbitrary but finite
$j$, prepared in an SU(2) coherent state $|j\mathbf{z}\rangle$ (the
eigenstate of $\mathbf{J}\cdot\mathbf{z}$ associated with the
maximum eigenvalue).

Because the task is to estimate the \emph{relations} between the
bounded RF and the system, it is possible to restrict the
measurement to one that is invariant under collective rotations
(i.e., rotations of both the bounded RF and the spin-1/2 system by
the same amount).  In other words, one can consider a global
superselection rule associated with the group SU(2) to apply,
because the system serving as an RF for direction is treated
internally.  As a result, the form of the measurement is highly
constrained. Note that the joint Hilbert space $\mathcal{H}_j
\otimes \mathcal{H}_{1/2}$ of the bounded RF and system decomposes
into a sum $\mathcal{H}_{j+1/2} \oplus \mathcal{H}_{j-1/2}$ of a
$J=j+1/2$ and a $J=j-1/2$ irreducible representation of SU(2), the
group of collective rotations.  By Schur's lemmas (see the Proof in
Sec.~\ref{sec:Formal}) a positive operator on this space that is
SU(2)-invariant must have the form $p_{+}\Pi_{j+1/2} +
p_{-}\Pi_{j-1/2}$, where $\Pi_{j\pm1/2}$ is the projector onto
$\mathcal{H}_{j\pm 1/2}$. Thus, a rotationally-invariant measurement
is represented by a POVM with elements of this form. However, any
such POVM may be obtained by classical post-processing of the
outcome of the two-element projective measurement
$\{\Pi_{j+1/2},\Pi_{j-1/2}\}$, so that the latter is the most
informative rotationally-invariant POVM.  The POVM elements
$\Pi_{j+1/2}$ and $\Pi_{j-1/2}$ are associated with the measurement
outcomes ``aligned'' and ``anti-aligned,'' respectively.

Denote the probability that the state $|{\pm}\mathbf{z}\rangle$ is
found to be aligned with the bounded RF by $p(+|\pm)$ and the
probability that it is found to be anti-aligned by $p(-|\pm)$. The
Born rule
\begin{equation}
  p(\pm|\pm)= \langle j\mathbf{z}|\langle\pm\mathbf{z}|
  \Pi_{j\pm1/2}|j\mathbf{z}\rangle|{\pm}\mathbf{z}\rangle\,,
\end{equation}
yields
\begin{align}
  p({+}|{+}) &=\frac{2j+1}{2j+2}\,,\quad &
  p({-}|{+}) &=\frac{1}{2j+2} \,,\nonumber\\
  p({+}|{-}) &=0\,, &
  p({-}|{-}) &=1\,.
\end{align}
Assuming equal prior probabilities for $|{+}\mathbf{z}\rangle$ and
$|{-}\mathbf{z}\rangle$, the average probability of successful
discrimination is
\begin{equation}
  p_{\mathrm{success}}
  =\frac{1}{2}p({+}|{+})+\frac{1}{2}p({-}|{-})
  =1 - \frac{1}{4(j+1)}\,.
\end{equation}
The smallest possible RF corresponds to taking $j=1/2$, in which
case $p_{\mathrm{success}}=5/6$; see~\textcite{Pry05} for an
experiment based on this example. For large $j$, we have a
probability of success that approaches $1$ linearly in $1/j$, and we
recover perfect distinguishability as $j\to\infty$, corresponding to
the case of an unbounded RF.

This example can be extended to the problem of estimating the
relative angle between a spin $j_{1}$ and a spin $j_{2}$; the
optimal measurement is the projective measurement $\{\Pi_{J},\,
J=|j_1-j_2|,\ldots,j_1+j_2\}$ onto the subspaces of total angular
momentum $J$, for the same reasons as above~\cite{BRS04b}.  The
optimization problem becomes nontrivial when we allow for states of
the bounded RF and/or the system that span multiple irreps, i.e.,
states that are not eigenstates of $\hat{J}^2$.  Measuring a
spin-$j$ system relative to a bounded directional RF consisting of a
pair of spins is considered in \textcite{Lin06,Bag06b}.

These sorts of results serve to illustrate how, for Lie groups at
least, a measurement relative to a bounded RF cannot perfectly
simulate one relative to an unbounded RF.  Note that we have only
considered the inferential but not the transformative aspect of the
measurement, that is, we have not considered how the quantum state
of the system is updated as a result of the measurement.  The work
of~\textcite{Wig52} and of~\textcite{AY60} demonstrates, however,
that for rank-1 projective measurements one cannot perfectly
simulate a von Neumann update rule when an unbounded RF is replaced
by a bounded one.

\subsubsection{Measuring relational degrees of freedom}

We note that measurements relative to a bounded quantum RF are
example of measurements of \emph{relational} degrees of freedom.
While a complete discussion of relational formulations of quantum
theory is beyond the scope of this review, we briefly make some
connections between problems involving reference frames and
relational ones.

The estimation of relative parameters for various degrees of freedom
encompasses such natural tasks as estimating the distance between
two massive particles, the phase between two modes of an
electromagnetic field (the essential aim of a homodyne measurement),
or the angle between a pair of spins as described above, all of
which are clearly related to issues of bounded reference frames.
Such measurements have been discussed recently in connection with
their ability to induce a relation between quantum systems that had
no relation prior to the measurement, e.g., inducing a relative
phase between two Fock states~\cite{Jav96,Mol97,San03} or a relative
position between two momentum eigenstates~\cite{Rau03,Cab05}.

Also, measurements of relative parameters are critical for achieving
\emph{programmable} quantum measurements~\cite{Dus02,Fiu02}.  Such
measurements use the state of a quantum system (the reference
system) to determine the form of a measurement performed on another
system.  For example, a reference system in an SU(2) coherent state
as above can be used to ``program'' the choice of measurement basis
of a spin-1/2 system.  The problem of optimizing programmable
quantum measurements, including determining the form of the joint
measurement and the optimal state of the reference system, is
directly related to the problem of optimizing measurements of
relative parameters between a system and a bounded reference frame.

Note that problems of relative parameter estimation are
complementary to those of determining the optimal measurement
schemes for estimating \emph{collective} parameters for a rotational
degree of freedom, the subject of Sec.~\ref{sec:Aligning}.


\subsection{Quantum computation with bounded reference frames}

\subsubsection{Precision of quantum gates}
\label{subsec:Precision}

In the majority of architectures proposed for quantum computation,
external classical fields are utilized to implement single qubit
logical operations.  As an example, we will focus on the use of
coherent states of the electromagnetic field, which are particularly
ubiquitous for quantum computing architectures -- in the form of
either lasers or radio frequency fields generated by an oscillating
current.  From the perspective of this review, we interpret such
fields as defining a reference frame -- in this case, a clock --
with respect to which coherent superpositions of the computational
basis (energy eigenstate) states are necessarily defined.  Note that
the fact that the RF is interacting directly with the qubits does
not weaken such a viewpoint -- at \emph{some} stage in the quantum
information processing a clock must physically interact with the
quantum computer (perhaps via intermediary systems); if it did not,
then there would be no operational difference if we enforced a
superselection rule for energy.\footnote{Previously, we have
concerned ourselves primarily with SSRs for photon number when
dealing with optical examples.  However, in this section, as we are
discussing the coupling of atoms to the optical fields under
(additively) energy conserving Hamiltonians, we thus extend the type
of SSR under consideration to one for total energy.}

If the reference frame is bounded -- quantified in this example as
a finite mean photon number of the laser field -- the operations
performed with respect to this bounded RF may have imperfect
precision, and generally the system and the field become
entangled.  A simple and standard model of a single two-level atom
resonantly interacting with a single mode (cavity) field via a
Jaynes-Cummings interaction serves to illustrate the basic
idea~\cite{Enk01}. The interaction Hamiltonian takes the form
\begin{equation}
    \label{Eq:JaynesCummings}
    H=i\hbar g\big( \hat{S}^{+} \hat a - \hat a^{\dagger}\hat S^{-}\big)\,,
\end{equation}
where $\hat a$ is the field mode annihilation operator, and $\hat
S^{\pm}$ are the atomic raising and lowering operators.  If the
field mode is initially prepared as a coherent state $|\alpha
\rangle=e^{-\frac{1}{2}|\alpha|^{2}}\sum_{n} (\alpha^{n}/\sqrt{n!})
|n\rangle$ with very large amplitude $|\alpha|^2 \to \infty$, it is
common to replace the field operators $\hat a,\ \hat a^{\dagger}$
with classical c-numbers $\alpha$ and $\alpha^*$.  In the language
of this review, this is the process of externalizing the reference
frame. For an atom initially in the excited state, evolution under
this classical field then yields the well-known Rabi oscillations
between the ground $|0\rangle$ and excited state $|1\rangle$.
Specifically, the state at time $t$ is
\begin{equation}
  \label{eq:ClassicalJCSolution}
  |\psi_{C}(t)\rangle=\sin(g|\alpha|t)
  |0\rangle+\cos(g|\alpha|t)|1\rangle \,.
\end{equation}

Consider now what occurs if we choose not to externalize the
driving field, and in particular describe it via a finite
amplitude coherent state. The evolution of the atom and field
under the Hamiltonian of Eq.~(\ref{Eq:JaynesCummings}) can be
solved exactly, yielding
\begin{equation}
  |\psi_{Q}(t)\rangle=\sum_{n=1}^{\infty}A_{n}(t)|0\rangle|n\rangle
  +B_{n-1}(t)|1\rangle|n-1\rangle\,,
\end{equation}
where
\begin{align}
  A_{n}(t) & =\frac{e^{-\frac{1}{2}|\alpha|^{2}}\alpha^{n-1}}{\sqrt{(n-1)!}}
  \sin(g\sqrt{n}t) \,, \\
  B_{n-1}(t) & =\frac{e^{-\frac{1}{2}|\alpha|^{2}}\alpha^{n-1}}{\sqrt{(n-1)!}}
  \cos(g\sqrt{n}t)\,.
\end{align}

To compare this evolution under a bounded reference frame to the
ideal (unbounded) case we should dequantize the reference
frame\footnote{This dequantization was not carried out in
\textcite{Enk01}, although for this example it does not make a
quantitative difference to the overall conclusions about gate
fidelity.} using the techniques of Sec.~\ref{dequantphaseref}.
Following the procedure outlined therein, we now move into the
tensor product structure induced by energy difference (relational)
versus total energy (global). In terms of this tensor product
structure we have
\begin{equation}
  |\psi_{Q}(t)\rangle
  =\sum_{n=0}^{\infty}|\phi_{n}(t)\rangle_{\mathrm{rel}}|n-1\rangle_{\mathrm{gl}}\,,
\end{equation}
where $|\phi_{n}(t)\rangle_{\mathrm{rel}}=(A_{n}(t)|0\rangle_{\mathrm{rel}}
  +B_{n-1}(t)|1\rangle_{\mathrm{rel}})$ is unnormalized.

The reduced density matrix of the relational system (i.e., we trace
out the global degree of freedom) is therefore a mixed state
\begin{equation}
  \rho_Q(t)=\sum_{n=0}^{\infty}|\phi_{n}(t)\rangle\langle\phi_{n}(t)|\,,
\end{equation}
and it is this mixed state we wish to compare with the pure state
(\ref{eq:ClassicalJCSolution}) expected in the unbounded,
externalized description.

We perform the comparison by computing the fidelity $F(t) =
\langle\psi_{C}(t)|\rho_Q(t)|\psi_{C}(t)\rangle$ between the mixed
state obtained through the full quantum treatment and the pure state
of Eq.~\eqref{eq:ClassicalJCSolution} obtained via the above
approximation.  If we consider the specific choice of evolution time
$t=\pi/(2|\alpha|g)$, which corresponds to performing a $\sigma_{x}$
gate, then this fidelity is very well approximated (even for small
values of $|\alpha|$) by~\cite{Har84}
\begin{align}
  F\big( t=\tfrac{\pi}{2|\alpha|g}\big)  & \cong \frac{1}{2}\Big(
  1-\cos\big( \tfrac{\pi\sqrt{1+|\alpha|^{2}}}{|\alpha|}\big)
  e^{-\frac{\pi^{2}}{8(|\alpha|^{2}+1)}}\Big) \nonumber \\
  & =1-\tfrac{\pi^{2}}{16|\alpha|^{2}}
  +O\big(\tfrac{1}{|\alpha|^{4}}\big)\,.
\end{align}
We see that the gate operation results in a state that is in error
(as quantified by the fidelity) by an amount that is inversely
proportional to the mean number of photons in the driving field.

The extent to which such a model captures the essential features of
currently-proposed quantum computing architectures has been the
subject of considerable debate,
cf.~\textcite{Gea02,Enk01,Sil03,Ita03,Gea02b,Nha05,Gea05}. What is
clear is that such effects are generally about two orders of
magnitude smaller than the typical spontaneous emission rates in
these systems.  However, in situations wherein the reference frame
is small (for example, if quantum computers together with the
control fields were to be built on chips in an integrated manner) or
in systems which have negligible spontaneous emission, then it is
not unreasonable that such considerations will have to be
incorporated into analyses of fault tolerance.

\subsubsection{Degradation of a quantum reference frame}

As we have demonstrated, a bounded reference frame can result in
non-trivial limitations on one's ability to perform operations and
measurements on quantum systems, and thus limitations on quantum
information processing tasks such as quantum computing.  However,
this imprecision is not the only limitation enforced by quantum
mechanics.  In addition, any measurement that acquires information
about the relations between the system and RF must necessarily
disturb them uncontrollably. The resulting disturbance to the RF
can be understood as a measurement \emph{back action}. The effect
of this back-action has been studied for reference frames for
spatial position~\cite{Aha84}, for directional reference
frames~\cite{Aha98} and for clocks~\cite{Cas00}. Here, we
investigate how measurement back-action on a bounded RF can lead
to its \emph{degradation}, i.e., a reduction of its suitability to
perform future measurements.

The conventional approach wherein reference frames suffer no back
action may yield a poor approximation to the full quantum
treatment, as suggested above.  This issue may be particularly
important for quantum computation, where a large number of
high-precision measurements must be performed.  In some
implementations, such measurements are performed relative to a
reference frame that is usually described by a finite quantum
system; for example, the proposed single-spin measurement
technique using magnetic resonance force microscopy~\cite{Rug04},
or the single-electron transistors used for measurement of
superconducting qubits~\cite{Mak01}.  We now demonstrate that the
number of measurements for which a quantum reference frame can be
used scales quadratically rather than linearly in the size of the
reference frame, which is a promising result for the prospect of
using microscopic or mesoscopic reference frames in performing
repeated high-precision measurements.

In the following example, we investigate the degradation of a
quantum reference direction as it is used for repeated
measurements. We use a spin-$j$ system for our quantum reference
direction (the RF), with Hilbert space $\mathcal{H}_{j}$.  We
choose the initial quantum state of the spin-$j$ system to be
$\rho^{(0)} = |j,j\rangle\langle j,j|$; this choice of initial
state simplifies the analysis, and in addition it has been
determined to be the initial state that maximizes the initial
success probability~\cite{BRST06}. This quantum RF is aligned in
the $+z$ direction relative to a background frame.

The systems to be measured will be spin-1/2 systems, each with a
Hilbert space $\mathcal{H}_{1/2}$. We choose the initial state of
each such system to be the completely mixed state $I/2$, and our
quantum RF will be used to measure many such independent spin-1/2
systems sequentially.  We shall assume trivial dynamics between
measurements, and thus our time index will simply be an integer
specifying the number of measurements that have taken place. The
state of the RF following the $n$th measurement is denoted
$\rho^{(n)}$, with $\rho^{(0)}$ denoting the initial state of the RF
prior to any measurement. We consider the state of the RF from the
perspective of someone who has \emph{not} kept a record of the
outcomes of previous measurements.  Thus, at every measurement, we
average over the possible outcomes with their respective weights to
obtain the final density operator.

The measurement which optimally determines whether a spin-1/2
particle is aligned or anti-aligned to a spin-$j$ system was
determined in Sec.~\ref{subsec:RelDirMeas} to be the two-outcome
projective measurement $\{\Pi _{+}\equiv \Pi _{j+1/2},\Pi _{-}\equiv
\Pi _{j-1/2}\}$ on $\mathcal{H}_{j}\otimes \mathcal{H}_{1/2}$.  We
use this measurement here.  It can be shown that of the many ways of
implementing this measurement, the update rule that degrades the
reference frame the least is the standard L\"{u}ders update
rule~\cite{BRST06}. Thus, the resulting evolution of the quantum RF
as a result of the $n$th measurement is
\begin{equation}
  \rho^{(n+1)}=\mathcal{E}(\rho^{(n)})\,,
  \label{eq:DecohMap}
\end{equation}
where the superoperator $\mathcal{E}$ is given by
\begin{equation}
  \mathcal{E}(\rho)=\text{Tr}_{S}\Bigl(\sum_{c\in \{+,-\}}\Pi _{c}(\rho
  \otimes I/2)\Pi_{c}\Bigr)\,,  \label{eq:DecohMap2}
\end{equation}
with $\text{Tr}_{S}$ the partial trace over $\mathcal{H}_{1/2}$.

The map $\mathcal{E}$ can be written using the operator-sum
representation as
\begin{equation}
  \mathcal{E}(\rho )=\frac{1}{2}\sum_{c\in \{+,-\}}\sum_{a,b\in
  \{0,1\}}E_{ab}^{c}\rho E_{ab}^{c\dag }\,,  \label{eq:RFdecoherence}
\end{equation}
where $E_{ab}^{c}\equiv \langle a|\Pi _{c}|b\rangle$ is a Kraus
operator on $\mathcal{H}_{j}$ and $\{|0\rangle,|1\rangle \}$ is a
basis for $\mathcal{H}_{1/2}$. These operators can be
straightforwardly determined in terms of Clebsch-Gordon
coefficients.

We quantify the quality of a quantum RF as the average probability
of a successful estimation of the orientation of a fictional
``test'' spin-1/2 system which is, with equal probability, either
aligned or anti-aligned with the background $+z$-axis.  Denote the
pure state of the test spin-1/2 system that is aligned
(anti-aligned) with the initial RF by $|0\rangle$ ($|1\rangle$). For
a spin-1/2 system prepared in the state $|0\rangle$ or $|1\rangle$
with equal probability, the average probability of success using a
quantum RF state $\rho$ is
\begin{equation}
  \overline{P}_{\rm s}
  =\frac{1}{2}\mathrm{Tr}_{R}(\rho(E_{00}^{+}+E_{11}^{-}))\,.
\end{equation}
The solution for $\rho^{(n)}$, given the initial state $\rho^{(0)} =
|j,j\rangle\langle j,j|$, yields an average probability of success
$\overline{P}_{\text{s}}(n)$ that decreases with $n$ as
\begin{equation}
  \overline{P}_{\text{s}}(n)=\frac{1}{2}+\frac{j}{2j+1}\Bigl(1-\frac{2}{(2j+1)^{2}}
  \Bigr)^{n}\,.
  \label{eq:PsuccessSU(2)CS}
\end{equation}
The initial slope $R$ of this function bounds the rate of
degradation. It is
\begin{equation}
  R \equiv \overline{P}_{\text{s}}(1)-\overline{P}_{\text{s}}(0) =-2j/(2j+1)^{3}\,.
\end{equation}
Thus, in the large $j$ limit, we have the rate of degradation with
$n$ satisfying $R\geq -1/(4j^{2})$.

Let $\epsilon <1$ be a fixed allowed error probability for the
spin-1/2 direction estimation problem. After $n$ measurements, the
probability of successful estimation is lower bounded by $1+nR$, so
the number of measurements required to ensure that this bound be
greater than $1-\epsilon$ is $-\epsilon/R$. Consequently, the number
of measurements that can be implemented relative to the spin-$j$ RF
with probability of error less than $\epsilon$ is
\begin{equation}
  n_{\mathrm{max}}\simeq \epsilon j^{2}\,.
\end{equation}
This result implies that the number of measurements for which an RF
is useful, that is, the longevity of an RF, increases
\emph{quadratically} rather than linearly with the size of the RF.
Thus, in order to maximize the number of measurements that can be
achieved with a given error threshold, one should combine all of
one's RF resources into a single large RF and perform all
measurements relative to it, rather than use a number of smaller RFs
individually.  We note that this degradation, as quantified by the
decreasing average probability of success
$\overline{P}_{\text{s}}(n)$, can be modeled precisely as the
distribution of a \emph{classical} reference direction undergoing a
random walk~\cite{BRST06b}.

Using similar methods, it has also been demonstrated that a bounded
quantum phase reference, realized as a single-mode quantum state of
the electromagnetic field with bounded photon number, also leads to
a longevity that scales quadratically in this size (mean photon
number)~\cite{BRST06}.  It is an open problem to determine if this
quadratic scaling is a general result.

We have discussed the degradation associated with a loss of purity
of the reference frame state. Another mechanism of degradation is
for the reference frame to become misaligned with the background
reference frame of which it is a token. \textcite{Pou06} have shown
how a reference frame can suffer this sort of misalignment when the
systems being measured have a non-zero polarization (that is, are
not described by the completely mixed state), and that in the
presence of such drift, the longevity scales linearly, rather than
quadratically, in the size of the reference system.

\subsection{Quantum cryptography with bounded reference frames}

In Sec.~\ref{subsec:SSRandCrypto}, we demonstrated that SSRs cannot
provide any fundamental limitations on quantum cryptographic
protocols, essentially because quantum reference systems which obey
the SSR can enable it to be effectively lifted. However, this result
does not mean that SSRs are uninteresting for cryptography. In
quantum cryptography, it is typical to focus on unconditional
security -- security not premised upon assumptions about the
resources of one's adversaries, but only upon the validity of the
laws of quantum mechanics.  In classical cryptography, in contrast,
security is typical \emph{conditional} -- it is generally premised
upon assumptions about the computational capabilities of one's
opponents.  Other types of conditional security can be premised upon
assumptions about other non-computational capabilities or resources
available to the adversarial parties. In this section, we consider
the specific case where the physical resource about which
assumptions are being made is some kind of RF, the lack of which in
turn induces an effective SSR. This is effectively an assumption of
bounded resources, because given unlimited resources the SSR can be
lifted as in Sec.~\ref{subsec:QuantizationGeneral}. We use the
specific examples of data hiding and bit commitment to illustrate
protocols that achieve this sort of security.\footnote{We note in
passing that a different type of assumption, namely that Alice and
Bob share partially misaligned reference frames, can be used as a
kind of guaranteed noisy channel, and, as in classical cryptography,
such channels can be used for secure two party
protocols~\cite{Har06}.}

\subsubsection{Data hiding with a superselection rule}

In a quantum data hiding protocol, one party (Charlie) wants to
share a single bit of data by distributing systems amongst two other
parties (Alice and Bob) in such a way that the bit can only be
recovered if the parties have some mechanism for performing joint
measurements on the distributed systems. Such measurements could be
performed by the parties coming together, or by using a quantum
channel, or by performing teleportation (using prior entanglement)
with a classical channel.  These possibilities are generally
considered equivalent.  Thus, Charlie must assume that the two
parties have no access to the specific physical resource of a
quantum channel.  It has been proven that \emph{perfect} quantum
data hiding is not possible even with this assumption~\cite{Ter01}.

If, in addition, Charlie can assume that the two parties do not
share a phase reference (that is, they are subject to a local
Abelian SSR) then perfect data hiding can be
achieved~\cite{Ver03}.  For example, without a shared phase
reference for their optical modes as in Sec.~\ref{sec:optics},
Alice and Bob cannot distinguish the pair of orthogonal pure
states $|\psi^\pm\rangle = (|01\rangle \pm|10\rangle)/\sqrt{2}$
using LOCC because $\mathcal{U}_A \otimes \mathcal{U}_B
(|\psi^+\rangle) = \mathcal{U}_A \otimes \mathcal{U}_B
(|\psi^-\rangle)$, and so this pair of states could be used to
encode the classical bit.

As per the discussion in Sec.~\ref{subsec:QuantizationGeneral}, it
is also clear that shared reference systems could be used by Alice
and Bob to break such a data hiding protocol.  Such reference
systems need not be entangled, which shows that breaking of the data
hiding in this case is quite different to the case of using
entanglement to implement a quantum channel.  However, if Charlie
has reason to believe that the reference systems shared by Alice and
Bob are bounded in size, then it is still possible for him to
achieve data hiding. He does this by using such a large number of
systems to encode the bit that any bounded shared reference does not
suffice to extract all the required data.

\subsubsection{Ancilla-free bit commitment}

A particularly simple class of bit commitment protocols \cite{SR01}
involve Alice preparing one of two orthogonal states
$|\chi_{0,1}\rangle,$ according to whether she wishes to commit a
bit $b=0,1.$ Here $|\chi _{0,1}\rangle$ are (generally entangled)
states over a ``proof'' system and a ``token'' system. She sends the
token system to Bob as her commitment. To unveil the bit, she sends
Bob the proof system, and he verifies her commitment by projecting
onto $|\chi_{0,1}\rangle$.  A simple version of such a protocol was
discussed in Sec.~\ref{subsec:SSRandCrypto} above.

Consider the situation wherein Alice and Bob are constrained such
that they cannot make use of a reference frame, either shared or
local.  This constraint enforces the protocol to be \emph{ancilla
free} -- for instance we do not allow either party to prepare
ancillary systems which could then act as an effective RF in the
manner described in Sec.~\ref{subsec:QuantizationGeneral}. It
turns out that under such a constraint, arbitrarily secure bit
commitment is possible~\cite{DiV04}.

It is illustrative to first consider a case where such a restriction
does \emph{not} help.  Consider ancilla-free bit commitment in the
case that Alice and Bob lack a phase reference.  As Alice must
prepare the initial states $|\chi_{0,1}\rangle$, they must each lie
completely in a single superselection sector, i.e., eigenstates of
total photon number, and take the form
\begin{equation}
  |\chi_{b}\rangle=\sum_{n}c_{n}^{b}|N-n\rangle_{P}|n\rangle_{T}\,.
\end{equation}
Because the reduced density matrices of the token system
$\rho_{b}=\sum_{n}|c_{n}^{b}|^2|n\rangle_T\langle n|$ are diagonal
in the number basis, the fact that Bob can only perform measurements
diagonal in this basis actually is no restriction on him at all --
he can cheat (by trying to distinguish these states) just as well as
he could in an unconstrained protocol.

Consider now if Alice is cheating, i.e., she attempts to commit
her bit only after the commitment stage.  An optimal cheating
strategy for Alice is to prepare a state
$|\tilde{\chi}\rangle\propto|\chi_{0}\rangle+U_{P}\otimes
I_T|\chi_{1}\rangle,$ where the unitary matrix $U_{P}$ on the
proof system is one which maximizes the overlap
$\langle\chi_{0}|U_{P}\otimes I_T|\chi_{1}\rangle$~\cite{SR01}.
If, after the commitment stage, she decides to commit $b=0$, then
she simply sends the proof system as is.  If instead she decides
to commit $b=1$ then she applies $U_{P}^{\dagger}$ to the proof
system before sending it to Bob.  The question is whether Alice
can perform this optimal cheating strategy despite the SSR.
Consider first the unitary $U_{P}$ which maximizes
$\langle\chi_{0}|U_{P}\otimes I_T|\chi_{1}\rangle$.  If
$U_{P}|N-n\rangle\equiv|A_{N-n}\rangle$, where $|A_{i}\rangle$ is
a state not necessarily respecting the SSR, then
$\langle\chi_{0}|U_{P}\otimes I_T|\chi_{1}
\rangle=\sum_{n}|c_{n}^{b}|^2 \langle A_{N-n}|N-n\rangle$. Clearly
the maximization of this expression will be achieved by choosing
$|A_{N-n}\rangle =e^{i\phi_{N-n}}|N-n\rangle$, i.e. for a unitary
$U_{P}$ which is diagonal in the number state basis. Furthermore
the state $|\tilde{\chi}\rangle$ then takes the generic form
$\sum_{n}c_{n}|N-n\rangle_{P}|n\rangle_{T}$ which respects the
SSR. Under the assumptions of an ancilla-free SSR protocol, any
state prepared by Alice or any unitary operator she applies is
constrained to be diagonal in the number basis -- as we have seen,
in this case she can still achieve the optimal cheating strategy
despite such a constraint.

The Abelian SSR induced by lack of a phase reference therefore does
not help devise a more secure ancilla-free bit commitment. It can be
shown, however, that a different type of Abelian SSR \emph{does}
lead to arbitrarily secure ancilla-free bit commitment.  We
follow~\textcite{DiV04}.  Consider a number of spin systems, with a
local Abelian SSR given as follows:  all local operations must
commute with the total local angular momentum operator $J^2$. Thus
all states and operations must be diagonal in total spin quantum
number $j$. However, they can have coherence between the different
$m$ eigenvalues of $J_{z}$. This superselection rule is distinct
from any that we have considered in this review, and does not appear
to be related to the lack of an appropriate reference frame.  The
key property of this Abelian SSR that will be useful for
ancilla-free bit commitment, and is distinct from the other Abelian
SSRs we consider, is that the quantum number $j$ labeling the local
superselection sectors is \emph{non-additive}.

Using the standard $|j,m\rangle$ notation for the uncoupled basis,
consider the bit commitment protocol which is defined by the
following two states $|\chi_{b}\rangle$ of total spin $j=1:$
\begin{multline}
  |\chi_{b}\rangle=\tfrac{1}{\sqrt{2}}|1,1\rangle_{P}|\phi_{0}^{b}\rangle
  _{T} \\ +\tfrac{1}{\sqrt{3}}|1,0\rangle_{P}|\phi_{1}^{b}\rangle_{T}+\tfrac{1}
  {\sqrt{6}}|1,-1\rangle_{P}|\phi_{2}^{b}\rangle_T\,,
\end{multline}
where
\begin{align}\label{eq:BitCommStates}
    |\phi_{0}^{b}\rangle_T &= \tfrac{2}{3}|0,0\rangle_T+(-1)^{b}\tfrac{1}
    {2}|1,0\rangle_T+\tfrac{\sqrt{2}}{6}|2,0\rangle_T \,, \\
    |\phi_{1}^{b}\rangle_T &= (-1)^{b}\tfrac{\sqrt{3}}{2}|1,1\rangle_T
    -\tfrac{1}{2}|2,1\rangle_T \,, \\
    |\phi_{2}^{b}\rangle_T &=|2,2\rangle_{T} \,,
\end{align}
We note that, although the proof system is also an eigenstate of
$J_T^2$ with eigenvalue $j=1$, the token system is \emph{not}; this
is a result of the non-additive nature of this SSR.  If we look at
the reduced density matrices $\rho_{0,1}$ on the token system in the
uncoupled basis, we see that they are block-diagonal (incoherent
mixtures) in the eigenspaces of $J_z$, with eigenvalues $m=0,1,2$.
Within each block, the diagonal elements of
$|\phi_{m}^{b}\rangle\langle\phi_{m}^{b}|$ are the same -- that is,
they are indistinguishable by their total spin.  Under the SSR, Bob
is restricted to performing measurements which are diagonal in total
spin, and so these two states are completely indistinguishable.

In a general bit commitment scenario, indistinguishability of the
token systems by Bob would imply that Alice has complete control --
that she should be able to perfectly change her commitment after the
commitment stage.  However, this is not the case for this example --
the two states $\rho_{0,1}$ have a non-unit fidelity $F\left(
\rho_{0},\rho_{1}\right)  <1$. Because the fidelity sets a bound
(for these type of protocols) on how well Alice can control the
outcome (regardless of any restrictions on her), we see that some
security against Alice will be possible.

Generalizations of the above pair of states $|\chi_{b}\rangle$ can
be defined for which, as $j$ becomes large, $F\left(
\rho_{0},\rho_{1}\right) \rightarrow 0$, implying perfect security
against Alice~\cite{DiV04}.  We believe that a fruitful avenue for
future research would be to determine if such constraints on
ancilla-free bit commitment can be achieved using a non-Abelian
superselection rule of the form discussed in this review.

\subsection{Quantifying bounded shared reference frames}
\label{subsec:QuantifyingBoundedRFs}

Much of quantum information theory is concerned with tradeoffs in
the utilization of various types of fundamental resources. The
canonical example is quantum teleportation, which demonstrates that
one ebit (Bell pair) of shared entanglement plus two communicated
classical bits are equivalent to the communication of a single
qubit.  In this section, we demonstrate that a shared reference
frame is also a quantifiable resource, akin to entanglement, which
allows parties to perform tasks that they were unable to perform
without it, or to perform tasks more efficiently.

Consider the ``activation'' example of Sec.~\ref{subsec:Activation},
which involved two parties (Alice and Bob) who do not share the
phase reference of a third party (Charlie) or each other. In this
context, the two-mode single-photon state $(|1\rangle_A|0\rangle_B +
|0\rangle_A|1\rangle_B)/\sqrt{2}$ could not be used to perform
quantum teleportation or to violate a Bell inequality.  However, if
Alice and Bob were also provided with the bipartite state
$|+\rangle_A|+\rangle_B$, where $|+\rangle =
(|0\rangle+|1\rangle)/\sqrt{2}$ described relative to Charlie's
phase reference, they could activate the entanglement in the former
state through LOCC. Although Alice and Bob do not share Charlie's
phase reference, clearly the state $|+\rangle_A|+\rangle_B$ provides
a bounded version of it. This bounded shared phase reference can be
used to activate the entanglement of the two-mode single-photon
state, as can an unbounded classical shared phase reference.
However, unlike the latter, the bounded shared phase reference
$|+\rangle_A|+\rangle_B$ can only activate the entanglement with
probability $1/2$, and in addition, is \emph{consumed} in the
process; it is a shared reference frame that can be depleted, in
this case, through a single use.  The state $|+\rangle_A|+\rangle_B$
can be considered an elementary unit of Charlie's phase reference,
much like an ebit (a Bell pair) is considered an elementary unit of
entanglement. As a result of this analogy, the state
$|+\rangle_A|+\rangle_B$ has been denoted a \emph{refbit}.

Continuing this example, we note that the two-mode single-photon
state $(|1\rangle_A|0\rangle_B + |0\rangle_A|1\rangle_B)/\sqrt{2}$
can also be viewed as a resource for ``activating'' another copy of
this same state.  (This process can alternatively be viewed as
2-copy entanglement distillation, as in
Sec.~\ref{subsec:Activation}.) This state is invariant under global
phase changes, and thus is completely uncorrelated with Charlie's
(or any other party's) phase reference, but it nevertheless provides
a bounded version of a \emph{shared} phase reference for Alice and
Bob.  It is useful to view this state as the elementary unit of a
shared phase reference between Alice and Bob, uncorrelated with any
other.

Because of the dual purpose of this state -- either as as an
elementary unit of a shared phase reference, or as a state from
which entanglement can be activated with the use of a shared phase
reference -- it has been named and categorized in many different
ways depending on its intended use.  So which way should it be
viewed? The answer is that this state can serve as a resource for
\emph{both} entanglement and a shared reference frame, and that one
must trade off its usefulness for one purpose against the other. In
fact, a wide variety of tradeoffs between refbits, ebits, cbits, and
other resources can be derived~\cite{Enk05,Enk06}, which emphasizes
the utility of thinking of reference frames as yet another form of
resource.

Now that we have identified ``standard'' elementary unit(s) of a
shared reference frame, we can quantify how well a given quantum
state serves as a shared reference frame by the state's asymptotic
interconvertibility to this standard form, using local operations
and classical communication.\footnote{For an alternate measure of
how well a quantum state can serve as a shared reference frame,
based on entropic properties, see \textcite{Vac05}.} A remarkable
property of entanglement of pure bipartite states is that, by
observing the properties of asymptotically reversible
transformations using LOCC, entanglement can be quantified by a
single additive measure:  the reversible conversion efficiency to a
standard form of entanglement, the ebit. For an Abelian
superselection rule, the resource of a quantum shared reference
frame can be quantified by a single additive measure in a similar
fashion.  Thus, the nonlocal properties of pure bipartite quantum
states in the presence of an Abelian superselection rule are
completely characterized by two additive measures:  the entanglement
(for which we can use an operational measure such as $E_{\rm SSR}$,
the entanglement in the presence of a SSR, discussed in
Sec.~\ref{subsubsec:QuantEnt}) and another measure quantifying the
state's ability to serve as a shared RF. In the following, we
investigate one such measure for the latter, the
\emph{superselection induced variance} (SIV)~\cite{Sch04a,Sch04b}.
We note that measures for quantifying quantum shared RFs in the
presence of \emph{non-Abelian} superselection rules have not been
explored to date.

Let Alice and Bob each have in their possession a number of optical
modes, and consider a situation as in Sec.~\ref{sec:optics} where
they do not share a phase reference.  Thus, Alice and Bob are
restricted by an Abelian local superselection rule for photon
number.  Let $\mathcal{H}_A$ ($\mathcal{H}_B$) be the local Hilbert
space for Alice's (Bob's) modes, and $\hat{N}_A$ ($\hat{N}_B$) be
the total local photon number operator for these modes.  Consider a
bipartite quantum state $|\phi\rangle$ on $\mathcal{H}_A \otimes
\mathcal{H}_B$ that is an eigenstate of total photon number
$\hat{N}_A + \hat{N}_B$.  (This condition ensures that the state is
not correlated with another party's phase reference.)  The
superselection induced variance $V(\phi)$ of this state is defined
to be the variance in the local photon number
\begin{equation}\label{eq:SIV}
    V(\phi) \equiv 4\big(\langle \phi|\hat{N}_A^2
    \otimes I_B|\phi\rangle
    - \langle\phi|\hat{N}_A \otimes I_B|\phi\rangle^2\big)
    \,.
\end{equation}
This SIV satisfies the following properties: (1) it is additive,
meaning $V(\phi\otimes\phi') = V(\phi) + V(\phi')$ for any
$|\phi\rangle,|\phi'\rangle$; (2) it is symmetric under exchange of
$A$ and $B$; and (3) it is a bipartite monotone, in that it is
non-increasing under LOCC operations that can be performed by Alice
and Bob without a shared phase reference.  The state
$(|1\rangle_A|0\rangle_B + |0\rangle_A|1\rangle_B)/\sqrt{2}$, which
we identified above as an elementary unit of shared phase reference,
has an SIV of $1$.

Two measures -- the entanglement and the SIV -- completely quantify
the nonlocal resources of a bipartite state.  To prove this result,
the general idea is to show that Alice and Bob can, through LOCC
restricted by the superselection rule, reversibly convert an
asymptotic number of copies of the bipartite state into a number of
states with only the first type of resource (entanglement) and none
of the second (SIV), and a number of states with only the second and
none of the first.  Let Alice and Bob share $N$ copies of a
bipartite state $|\phi\rangle$, which has entanglement $E_{\rm
SSR}(\phi)$ and SIV $V(\phi)$.  In addition, let Alice and Bob each
have in their possession an arbitrary number of quantum registers --
quantum systems that are not restricted by any superselection rule,
such as were discussed in Sec.~\ref{subsubsec:QuantEnt}; these
registers are initiated in an arbitrary unentangled state
$|\tilde{0}\rangle|\tilde{0}\rangle$.  Then the transformation
\begin{multline}\label{AsymptoticSIREnt}
    |\phi\rangle^{\otimes N} \otimes
    \big(|\tilde{0}\rangle|\tilde{0}\rangle\big)^{\otimes
    E_{\rm SSR}(\phi)N} \\ \rightarrow \Big( \sum_n c_n |n\rangle
|N-n\rangle \Big)
    \otimes |\tilde\psi^-\rangle^{\otimes E_{\rm SSR}(\phi)N} \,,
\end{multline}
is asymptotically reversible, and Alice and Bob can perform this
transformation with LOCC restricted by the SSR, where the
coefficients $c_n$ are Gaussian-distributed with variance
$NV(\phi)/4$, and $|\tilde\psi^-\rangle$ is a maximally-entangled
Bell state of a pair of qubits of Alice's and Bob's quantum
registers.

Let's analyze the two states on the right side of
Eq.~\eqref{AsymptoticSIREnt}.  The first state, $\sum_n c_n
|n\rangle |N-n\rangle$, has SIV of $N V(\phi)$.  Such a state serves
as a good ``standard'' shared RF for large $N$~\cite{Vac03}.  Also,
although it is a non-separable pure state, the entanglement in the
presence of a SSR, $E_{\rm SSR}$, of this state is zero.  In
contrast, the state of the unrestricted registers
$|\tilde\psi^-\rangle^{\otimes E_{\rm SSR}(\phi)N}$ clearly contains
an amount of entanglement equal to $E_{\rm SSR}(\phi)N$ standard
ebits. As this system is a quantum register, and not a system with a
phase degree of freedom, it clearly has no function as a shared
phase reference; thus, the SIV of this state is zero.  The two
states on the right side of Eq.~\eqref{AsymptoticSIREnt}, then,
represent standard forms for each type of nonlocal resource --
superselection induced variance, and entanglement in the presence of
a SSR -- and contain none of the other type.

A proof that the transformation~\eqref{AsymptoticSIREnt} is
asymptotically reversible with LOCC restricted by the SSR can be
found in~\textcite{Sch04a}. In their proof, they used the entropy of
entanglement $E$ rather than $E_{\rm SSR}$; we note that in the
asymptotic limit for an Abelian SSR, $E_{\rm SSR}(\phi^{\otimes N})
\rightarrow E(\phi^{\otimes N})$ for any pure state
$|\phi\rangle$~\cite{Wis03b}.  Thus, their proof applies directly to
the above statement.   This result can be extended to apply to mixed
states~\cite{Sch04b}.  Finally, we note that explicit protocols for
activation -- creating states with $E_{\rm SSR} \neq 0$ using states
with $E_{\rm SSR} = 0$ using a quantum shared reference frame state
-- have been developed~\cite{Vac03,BDSW06}.

\subsection{Purification of bounded shared reference frames?}
\label{subsec:Purification}

As noted in the previous section, the state $(|1\rangle_A|0\rangle_B
+ |0\rangle_A|1\rangle_B)/\sqrt{2}$ can be viewed as the elementary
unit of a shared phase reference between Alice and Bob, uncorrelated
with any other.  This state has the appearance of a
maximally-entangled Bell state (see Sec.~\ref{subsec:Entanglement}),
and so a natural question is to ask whether a number of imperfect
(noisy) states can be \emph{purified} to a smaller number of
superior states of this form.  Of course, for such a process to be
of any use, it would need to be implementable without the use of
some other, unbounded shared RF.  An affirmative answer would mean
that shared RFs are a resource that can be purified, just like
entanglement. Unfortunately, however, such a task does not appear to
be possible, as we now demonstrate, following~\textcite{Pre00}.

Consider a noisy shared RF state that is a mixture of the state
$(|1\rangle_A|0\rangle_B + |0\rangle_A|1\rangle_B)/\sqrt{2}$ with
probability $p>1/2$ and the state $(|1\rangle_A|0\rangle_B -
|0\rangle_A|1\rangle_B)/\sqrt{2}$ with probability $1-p$.  Let Alice
and Bob share two copies of this mixed state.  With these states,
they attempt to perform the following simple entanglement
purification protocol~\cite{Ben96}:  they each apply a CNOT on the
two qubits in their possession, and then perform an $X$ measurement
on the target qubit.  Each party obtains a measurement outcome $\pm
1$, which they communicate with each other classically, and compare
whether the results are the same or different.  Effectively, though
this process, they have measured the joint non-local operator
\begin{equation}\label{eq:EntDistillationOperator}
    (X_A X_B)_1 \cdot (X_A X_B)_2 \,.
\end{equation}
In the standard entanglement purification protocol, if Alice and Bob
keep only those states where they obtain the same measurement
results, the resulting states will have higher fidelity with the
state $(|1\rangle_A|0\rangle_B + |0\rangle_A|1\rangle_B)/\sqrt{2}$.

Note, however, that this protocol requires operations which are not
U(1)-invariant.  For example, a measurement of $X$ must be performed
relative to that party's local phase reference.  Let Alice and Bob
make use of unbounded local phase references in this protocol.  Note
that the operator $X_A$ is defined with respect to Alice's phase
reference, and $X_B$ with respect to Bob's. If their phase
references differ by a phase shift $\theta_{BA}$, then these two
operations are related by $X_B = e^{-i \theta_{BA}Z/2} X_A e^{+i
\theta_{BA}Z/2}$; see Sec.~\ref{subsec:CommComplexity}.  Thus, the
state to which they are purifying in this instance is
\begin{equation}
    (|1\rangle_A|0\rangle_B + e^{-i
    \theta_{BA}}|0\rangle_A|1\rangle_B)/\sqrt{2} \,.
\end{equation}
If Alice's and Bob's local phase references are \emph{uncorrelated},
as we assumed, then $\theta_{BA}$ is completely unknown, and the
protocol does \emph{not} yield a state with higher fidelity with
$(|1\rangle_A|0\rangle_B + |0\rangle_A|1\rangle_B)/\sqrt{2}$.

\subsection{Treating bounded reference frames as decoherence}
\label{subsec:Decoherence}

We conclude this section with a discussion of a promising approach
to describing the effect of using bounded RFs.  As demonstrated
above, bounded RFs limit one's ability to prepare states and to
perform quantum operations and measurements on a system, and the
nature of these limitations is similar in many ways to that of
decoherence.  One is led, then, to ask whether it is possible to
treat bounded RFs externally rather than internally (in the sense of
Sec.~\ref{dequantphaseref}) by positing an unavoidable decoherence.
In other words, if such a description existed, then the bounded size
of the RF could be said to effectively reduce the purity and/or
coherence of systems described with respect to it.

While no completely general description of treating bounded RFs in
this manner has yet been developed, specific examples of such
decoherence mechanisms and their consequences have been discussed in
various relational approaches to quantum
theory~\cite{Gam04,Gam04b,Pou05,Mil06,Pag83}.  These discussions
have primarily focussed on the tricky issue of internalizing time in
quantum theory.  Unsurprisingly, given the interpretation of certain
types of phase references as clocks, these relational formulations
generally follow along the lines of the procedures we have already
reviewed.  One begins by treating all systems which can serve as a
clock as internal, constructs (pure or mixed) states that are
invariant under global time shifts, identifies relational spaces in
the decoherence-free subsystems, re-factorizes the Hilbert space in
terms of the induced tensor product, and finally interprets the new
formulation as the `true' dynamical description.  As expected, a
form of decoherence in this new description is found whenever the
size of the internalized reference system(s) is bounded.  It is an
interesting open problem to identify the appropriate decoherence
maps (if they exist) that describe the dynamics of a system relative
to a bounded (particularly non-Abelian) RF. While one can debate the
appropriateness of these approaches from a foundational perspective,
such an approach would certainly be useful for addressing questions
in the field of quantum information.

\section{Outlook}
\label{sec:Outlook}

The study of reference frames and superselection rules in the
context of quantum information theory is an unfinished task. In this
section, we provide an overview of the topics we have discussed
together with some open problems and research directions, while
outlining the practical and foundational significance of this sort
of investigation.

It is useful to divide the practical applications into two broad
categories corresponding to whether their purpose is the
manipulation of speakable information or of unspeakable information,
that is, corresponding to the nature of their inputs and outputs.

The first category contains the standard problems of interest in
quantum information theory, both those that use quantum systems to
manipulate classical information, and those whose inputs and outputs
are themselves quantum information.  Even though these ultimately
process speakable information (whether classical or quantum),
protocols for such tasks must always encode this information using
\emph{some} degree of freedom, which requires some form of RF. Thus
we are led to ask how much the absence of a particular RF or of a
shared RF among separated parties decreases the efficiency of
various information-processing tasks, or increases the practical
difficulty of implementing them.  What is the answer to such
questions in the case where one has an imprecise RF or two parties
share RFs that are only partially correlated?  Such questions have
been considered here for a variety of tasks, such as quantum and
classical communication (Sec.~\ref{subsec:NoSRFComm}), quantum key
distribution (Sec.~\ref{subsec:QKD}), and implementing quantum gates
(Sec.~\ref{subsec:Precision}). There are many more tasks that could
be considered. Also, most of the communication and cryptographic
problems considered to date have determined the efficiency only in
the case where one demands perfect fidelity encoding and decoding
and perfect security. Furthermore, these sorts of questions have
been scarcely addressed for the case of shared RF that are partially
correlated. Finally, although there have been a few experiments
demonstrating the viability of some of these schemes, such as
relational encodings (Sec.~\ref{subsec:ConsequencesQIP}), the
development of realistic physical implementations remains as much a
source of experimental challenges as any other quantum technology.

The second category of applications consists of tasks that
explicitly involve the manipulation of unspeakable information, such
as clock synchronization or the alignment of Cartesian frames.
Quantum considerations become important to achieve the optimal
precision and it is the tools of quantum information theory that are
best suited to a treatment of the problem. We may describe the
alignment of remote reference frames as the communication of
unspeakable information, and as soon as one starts describing and
thinking about such tasks in the language of information theory,
many new tasks suggest themselves. Examples mentioned in this review
are: dense coding of unspeakable information
(Sec.~\ref{subsec:DenseCoding}), using private shared RFs as a
cryptographic key (Sec.~\ref{subsec:PrivateSRFs}), the private
communication of unspeakable information
(Sec.~\ref{subsec:PrivateUnspeakable}), and secret sharing of
unspeakable information. Many more analogies of this sort could be
considered. Indeed, for almost any information-theoretic task of
interest today, it is interesting to muse about possible analogues
of it for unspeakable information. (A particularly intriguing
question to consider is whether there is such an analogue for
computation.) On the experimental side, the implementation of
quantum protocols for even the best-studied of these sorts of tasks,
the alignment of reference frames, has, with the exception of phase
estimation, only just begun to be investigated.

As emphasized in the introduction, imposing a restriction on
operations generically leads to the identification of a novel
resource to overcome this restriction, and we are then compelled to
develop a theory for how that resource may be manipulated.  For
instance, under the restriction of LOCC, entanglement becomes a
resource, and the theory of how this resource can be manipulated
--- the theory of entanglement --- has been the subject of a significant
amount of work in recent years.  Others have considered the theory
of communication under natural restrictions such as local operations
and \emph{public} communication~\cite{Col02} or restrictions to only
Gaussian quantum-optical states and operations~\cite{Eis03}. A
superselection rule (either local and global) is another sort of
natural restriction, and under this restriction, any quantum state
that acts as an RF becomes a resource. The theory of such resources
might aptly be called the \emph{theory of quantum reference frames}
or the \emph{theory of unspeakable quantum information}. It
endeavors to answer questions such as how this resource is depleted
with use, transformed from one form to another, shared among several
parties, etcetera. Such a theory has only begun to be developed. The
limited results on bounded quantum reference frames, described in
Sec.~\ref{sec:BoundedRFs} (see also Sec.~\ref{subsec:Entanglement})
are evidence of this. Moreover, in a sense there is a family of
theories to be developed, because we obtain different results
depending on the group $G$ with which the superselection rule is
associated. Many investigations to date have applied only to the
cases of the group U(1) and/or the group SU(2). Ultimately, one
would like to have a generic theory of unspeakable information that
applies to any group; in particular, non-compact groups (such as the
Lorentz or Poincar\'e groups) may require more general mathematical
tools than those discussed here.

It is worth noting that this research is not necessarily driven by
applications. In this sense, it is similar to the study of
entanglement in quantum information theory, which although initially
motivated by its practical applications, has increasingly become an
interesting subject in its own right. Of course, just as the
development of the theory of entanglement has led to many unforeseen
practical dividends, we may expect this of a general theory of
unspeakable information as well.

Finally, applying the tools of quantum information theory to the
study of RFs and SSRs can shed light on foundational issues in
quantum theory. Examples from this review include whether there
exist axiomatic superselection rules and the controversy over the
nonlocality of a single photon.  Another issue which is likely to be
clarified by an analysis in terms of quantum reference frames is
that of spontaneous symmetry breaking, which is significant both in
condensed matter physics and quantum field theory and the
foundational status of which is notoriously murky. Foundational
issues related to particle statistics may also benefit from such an
analysis. For instance, an interesting open question is whether the
univalence superselection rule, which forbids coherent
superpositions of bosons and fermions~\cite{Giu96}, may be lifted by
an appropriate reference frame~\cite{Dow06b}. Because this SSR is
not associated with a compact group, answering this question
requires a formalism more general than the one presented here.

\begin{acknowledgments}
We thank Steven van Enk, Gilad Gour, and Pieter Kok for helpful
comments on the manuscript. SDB acknowledges support from the
Australian Research Council. TR acknowledges support from the
Engineering and Physical Sciences Research Council of the United
Kingdom.  RWS acknowledges support from the Royal Society.
\end{acknowledgments}


\begin{thebibliography}{159}
\expandafter\ifx\csname
natexlab\endcsname\relax\def\natexlab#1{#1}\fi
\expandafter\ifx\csname bibnamefont\endcsname\relax
  \def\bibnamefont#1{#1}\fi
\expandafter\ifx\csname bibfnamefont\endcsname\relax
  \def\bibfnamefont#1{#1}\fi
\expandafter\ifx\csname citenamefont\endcsname\relax
  \def\citenamefont#1{#1}\fi
\expandafter\ifx\csname url\endcsname\relax
  \def\url#1{\texttt{#1}}\fi
\expandafter\ifx\csname urlprefix\endcsname\relax\def\urlprefix{URL
}\fi \providecommand{\bibinfo}[2]{#2}
\providecommand{\eprint}[2][]{\url{#2}}

\bibitem[{\citenamefont{Acin} \emph{et~al.}(2001)\citenamefont{Acin, Jan{\'e},
  and Vidal}}]{Aci01}
\bibinfo{author}{\bibnamefont{Acin}, \bibfnamefont{A.}},
  \bibinfo{author}{\bibfnamefont{E.}~\bibnamefont{Jan{\'e}}}, and
  \bibinfo{author}{\bibfnamefont{G.}~\bibnamefont{Vidal}},
  \bibinfo{year}{2001}, \bibinfo{journal}{Phys. Rev. A}
  \textbf{\bibinfo{volume}{64}}, \bibinfo{eid}{050302}.

\bibitem[{\citenamefont{Aharonov and Kaufherr}(1984)}]{Aha84}
\bibinfo{author}{\bibnamefont{Aharonov}, \bibfnamefont{Y.}}, and
  \bibinfo{author}{\bibfnamefont{T.}~\bibnamefont{Kaufherr}},
  \bibinfo{year}{1984}, \bibinfo{journal}{Phys. Rev. D}
  \textbf{\bibinfo{volume}{30}}, \bibinfo{pages}{368}.

\bibitem[{\citenamefont{Aharonov} \emph{et~al.}(1998)\citenamefont{Aharonov,
  Kaufherr, Popescu, and Reznik}}]{Aha98}
\bibinfo{author}{\bibnamefont{Aharonov}, \bibfnamefont{Y.}},
  \bibinfo{author}{\bibfnamefont{T.}~\bibnamefont{Kaufherr}},
  \bibinfo{author}{\bibfnamefont{S.}~\bibnamefont{Popescu}}, and
  \bibinfo{author}{\bibfnamefont{B.}~\bibnamefont{Reznik}},
  \bibinfo{year}{1998}, \bibinfo{journal}{Phys. Rev. Lett.}
  \textbf{\bibinfo{volume}{80}}, \bibinfo{pages}{2023}.

\bibitem[{\citenamefont{Aharonov and Susskind}(1967)}]{Aha67}
\bibinfo{author}{\bibnamefont{Aharonov}, \bibfnamefont{Y.}}, and
  \bibinfo{author}{\bibfnamefont{L.}~\bibnamefont{Susskind}},
  \bibinfo{year}{1967}, \bibinfo{journal}{Phys. Rev.}
  \textbf{\bibinfo{volume}{155}}, \bibinfo{pages}{1428}.

\bibitem[{\citenamefont{Ambainis} \emph{et~al.}(2000)\citenamefont{Ambainis,
  Mosca, Tapp, and de Wolf}}]{Amb00}
\bibinfo{author}{\bibnamefont{Ambainis}, \bibfnamefont{A.}},
  \bibinfo{author}{\bibfnamefont{M.}~\bibnamefont{Mosca}},
  \bibinfo{author}{\bibfnamefont{A.}~\bibnamefont{Tapp}}, and
  \bibinfo{author}{\bibfnamefont{R.}~\bibnamefont{de Wolf}}, in
  \emph{Proc. 41$^{st}$ Annual Symposium on Foundations of Computer
  Science}, (IEEE, Los Alamitos, 2000), p. 547.  Also eprint quant-ph/0003101.

\bibitem[{\citenamefont{Anderson}(1986)}]{And86}
\bibinfo{author}{\bibnamefont{Anderson}, \bibfnamefont{P.~W.}},
  \bibinfo{year}{1986}, \emph{\bibinfo{title}{The Lesson of Quantum Theory}}
  (\bibinfo{publisher}{Elsevier}).

\bibitem[{\citenamefont{Araki and Yanase}(1960)}]{AY60}
\bibinfo{author}{\bibnamefont{Araki}, \bibfnamefont{H.}}, and
  \bibinfo{author}{\bibfnamefont{M.~M.}~\bibnamefont{Yanase}},
  \bibinfo{year}{1960}, \bibinfo{journal}{Phys. Rev.}
  \textbf{\bibinfo{volume}{120}}, \bibinfo{eid}{622}.

\bibitem[{\citenamefont{Babichev} \emph{et~al.}(2004)\citenamefont{Babichev,
  Appel, and Lvovsky}}]{Bab04}
\bibinfo{author}{\bibnamefont{Babichev}, \bibfnamefont{S.~A.}},
  \bibinfo{author}{\bibfnamefont{J.}~\bibnamefont{Appel}}, and
  \bibinfo{author}{\bibfnamefont{A.~I.} \bibnamefont{Lvovsky}},
  \bibinfo{year}{2004}, \bibinfo{journal}{Phys. Rev. Lett.}
  \textbf{\bibinfo{volume}{92}}, \bibinfo{eid}{193601}.

\bibitem[{\citenamefont{Bacon} \emph{et~al.}(2006a)\citenamefont{Bacon,
  Chuang, and Harrow}}]{Bac06a}
\bibinfo{author}{\bibnamefont{Bacon}, \bibfnamefont{D.}},
  \bibinfo{author}{\bibfnamefont{I.~L.}~\bibnamefont{Chuang}}, and
  \bibinfo{author}{\bibfnamefont{A.~W.}~\bibnamefont{Harrow}},
  \bibinfo{year}{2006}, \bibinfo{journal}{Phys. Rev. Lett.}
  \textbf{\bibinfo{volume}{97}}, \bibinfo{eid}{170502}.

\bibitem[{\citenamefont{Bacon} \emph{et~al.}(2006b)\citenamefont{Bacon,
  Chuang, and Harrow}}]{Bac06b}
\bibinfo{author}{\bibnamefont{Bacon}, \bibfnamefont{D.}},
  \bibinfo{author}{\bibfnamefont{I.~L.}~\bibnamefont{Chuang}}, and
  \bibinfo{author}{\bibfnamefont{A.~W.}~\bibnamefont{Harrow}},
  \bibinfo{year}{2006}, e-print quant-ph/0601001.

\bibitem[{\citenamefont{Bagan}
  \emph{et~al.}(2000)\citenamefont{Bagan, Baig, Brey,
  Munoz-Tapia, and Tarrach}}]{Bag00}
\bibinfo{author}{\bibnamefont{Bagan}, \bibfnamefont{E.}},
  \bibinfo{author}{\bibfnamefont{M.}~\bibnamefont{Baig}},
  \bibinfo{author}{\bibfnamefont{A.}~\bibnamefont{Brey}},
  \bibinfo{author}{\bibfnamefont{R.}~\bibnamefont{Munoz-Tapia}}, and
  \bibinfo{author}{\bibfnamefont{R.}~\bibnamefont{Tarrach}},
  \bibinfo{year}{2000}, \bibinfo{journal}{Phys. Rev. Lett.}
  \textbf{\bibinfo{volume}{85}}, \bibinfo{eid}{5230}.


\bibitem[{\citenamefont{Bagan}
  \emph{et~al.}(2001{\natexlab{a}})\citenamefont{Bagan, Baig, Brey,
  Munoz-Tapia, and Tarrach}}]{Bag01a}
\bibinfo{author}{\bibnamefont{Bagan}, \bibfnamefont{E.}},
  \bibinfo{author}{\bibfnamefont{M.}~\bibnamefont{Baig}},
  \bibinfo{author}{\bibfnamefont{A.}~\bibnamefont{Brey}},
  \bibinfo{author}{\bibfnamefont{R.}~\bibnamefont{Munoz-Tapia}}, and
  \bibinfo{author}{\bibfnamefont{R.}~\bibnamefont{Tarrach}},
  \bibinfo{year}{2001}{\natexlab{a}}, \bibinfo{journal}{Phys. Rev. A}
  \textbf{\bibinfo{volume}{63}}, \bibinfo{eid}{052309}.

\bibitem[{\citenamefont{Bagan}
  \emph{et~al.}(2001{\natexlab{b}})\citenamefont{Bagan, Baig, and
  Munoz-Tapia}}]{Bag01b}
\bibinfo{author}{\bibnamefont{Bagan}, \bibfnamefont{E.}},
  \bibinfo{author}{\bibfnamefont{M.}~\bibnamefont{Baig}}, and
  \bibinfo{author}{\bibfnamefont{R.}~\bibnamefont{Munoz-Tapia}},
  \bibinfo{year}{2001}{\natexlab{b}}, \bibinfo{journal}{Phys. Rev. A}
  \textbf{\bibinfo{volume}{64}}, \bibinfo{pages}{022305}.

\bibitem[{\citenamefont{Bagan}
  \emph{et~al.}(2004{\natexlab{a}})\citenamefont{Bagan, Baig, and
  Munoz-Tapia}}]{Bag04a}
\bibinfo{author}{\bibnamefont{Bagan}, \bibfnamefont{E.}},
  \bibinfo{author}{\bibfnamefont{M.}~\bibnamefont{Baig}}, and
  \bibinfo{author}{\bibfnamefont{R.}~\bibnamefont{Munoz-Tapia}},
  \bibinfo{year}{2004}{\natexlab{a}}, \bibinfo{journal}{Phys. Rev. A}
  \textbf{\bibinfo{volume}{69}}, \bibinfo{eid}{050303}.

\bibitem[{\citenamefont{Bagan}
  \emph{et~al.}(2004{\natexlab{b}})\citenamefont{Bagan, Baig, and
  Munoz-Tapia}}]{Bag04b}
\bibinfo{author}{\bibnamefont{Bagan}, \bibfnamefont{E.}},
  \bibinfo{author}{\bibfnamefont{M.}~\bibnamefont{Baig}}, and
  \bibinfo{author}{\bibfnamefont{R.}~\bibnamefont{Munoz-Tapia}},
  \bibinfo{year}{2004}{\natexlab{b}}, \bibinfo{journal}{Phys. Rev. A}
  \textbf{\bibinfo{volume}{70}}, \bibinfo{eid}{030301}.

\bibitem[{\citenamefont{Bagan} \emph{et~al.}(2006{\natexlab{a}})\citenamefont{Bagan,
  Iblisdir, and Munoz-Tapia}}]{Bag06b}
\bibinfo{author}{\bibnamefont{Bagan}, \bibfnamefont{E.}},
  \bibinfo{author}{\bibfnamefont{S.}~\bibnamefont{Iblisdir}}, and
  \bibinfo{author}{\bibfnamefont{R.}~\bibnamefont{Munoz-Tapia}},
  \bibinfo{year}{2006}{\natexlab{a}}, \bibinfo{journal}{Phys. Rev. A}
  \textbf{\bibinfo{volume}{73}}, \bibinfo{eid}{022341}.

\bibitem[{\citenamefont{Bagan} \emph{et~al.}(2006{\natexlab{b}})\citenamefont{Bagan,
  Calsamiglia, Demkowicz-Dobrza{\'n}ski, and Munoz-Tapia}}]{Bag06c}
\bibinfo{author}{\bibnamefont{Bagan}, \bibfnamefont{E.}},
  \bibinfo{author}{\bibfnamefont{J.}~\bibnamefont{Calsamiglia}},
  \bibinfo{author}{\bibfnamefont{R.}~\bibnamefont{Demkowicz-Dobrza{\'n}ski}}, and
  \bibinfo{author}{\bibfnamefont{R.}~\bibnamefont{Munoz-Tapia}},
  \bibinfo{year}{2006}{\natexlab{b}},
  e-print quant-ph/0606165.

\bibitem[{\citenamefont{Bagan and Munoz-Tapia}(2006)}]{Bag06}
\bibinfo{author}{\bibnamefont{Bagan}, \bibfnamefont{E.}}, and
  \bibinfo{author}{\bibfnamefont{R.}~\bibnamefont{Munoz-Tapia}},
  \bibinfo{year}{2006}, \bibinfo{journal}{Int. J. of Quantum Information}
  \textbf{\bibinfo{volume}{4}}, \bibinfo{pages}{5}.

\bibitem[{\citenamefont{Ball and Banaszek}(2005)}]{Bal05}
\bibinfo{author}{\bibnamefont{Ball}, \bibfnamefont{J.~L.}}, and
  \bibinfo{author}{\bibfnamefont{K.}~\bibnamefont{Banaszek}},
  \bibinfo{year}{2005}, \bibinfo{journal}{Open Syst. Inf. Dyn.}
  \textbf{\bibinfo{volume}{12}}, \bibinfo{pages}{121}.

\bibitem[{\citenamefont{Ball and Banaszek}(2006)}]{Bal06}
\bibinfo{author}{\bibnamefont{Ball}, \bibfnamefont{J.~L.}}, and
  \bibinfo{author}{\bibfnamefont{K.}~\bibnamefont{Banaszek}},
  \bibinfo{year}{2006}, \bibinfo{journal}{J. Phys. A: Math. Gen.}
  \textbf{\bibinfo{volume}{39}}, \bibinfo{pages}{L1}.

\bibitem[{\citenamefont{Banaszek} \emph{et~al.}(2004)\citenamefont{Banaszek,
  Dragan, Wasilewski, and Radzewicz}}]{Ban04}
\bibinfo{author}{\bibnamefont{Banaszek}, \bibfnamefont{K.}},
  \bibinfo{author}{\bibfnamefont{A.}~\bibnamefont{Dragan}},
  \bibinfo{author}{\bibfnamefont{W.}~\bibnamefont{Wasilewski}}, and
  \bibinfo{author}{\bibfnamefont{C.}~\bibnamefont{Radzewicz}},
  \bibinfo{year}{2004}, \bibinfo{journal}{Phys. Rev. Lett.}
  \textbf{\bibinfo{volume}{92}}, \bibinfo{eid}{257901}.

\bibitem[{\citenamefont{Bartlett}
  \emph{et~al.}(2006{\natexlab{a}})\citenamefont{Bartlett, Doherty, Spekkens,
  and Wiseman}}]{BDSW06}
\bibinfo{author}{\bibnamefont{Bartlett}, \bibfnamefont{S.~D.}},
  \bibinfo{author}{\bibfnamefont{A.~C.} \bibnamefont{Doherty}},
  \bibinfo{author}{\bibfnamefont{R.~W.} \bibnamefont{Spekkens}}, and
  \bibinfo{author}{\bibfnamefont{H.~M.} \bibnamefont{Wiseman}},
  \bibinfo{year}{2006}{\natexlab{a}}, \bibinfo{journal}{Phys. Rev. A}
  \textbf{\bibinfo{volume}{73}}, \bibinfo{eid}{022311}.

\bibitem[{\citenamefont{Bartlett} \emph{et~al.}(2005)\citenamefont{Bartlett,
  Hayden, and Spekkens}}]{BHS05}
\bibinfo{author}{\bibnamefont{Bartlett}, \bibfnamefont{S.~D.}},
  \bibinfo{author}{\bibfnamefont{P.}~\bibnamefont{Hayden}}, and
  \bibinfo{author}{\bibfnamefont{R.~W.} \bibnamefont{Spekkens}},
  \bibinfo{year}{2005}, \bibinfo{journal}{Phys. Rev. A}
  \textbf{\bibinfo{volume}{72}}, \bibinfo{eid}{052329}.

\bibitem[{\citenamefont{Bartlett} \emph{et~al.}(2003)\citenamefont{Bartlett,
  Rudolph, and Spekkens}}]{BRS03}
\bibinfo{author}{\bibnamefont{Bartlett}, \bibfnamefont{S.~D.}},
  \bibinfo{author}{\bibfnamefont{T.}~\bibnamefont{Rudolph}}, and
  \bibinfo{author}{\bibfnamefont{R.~W.} \bibnamefont{Spekkens}},
  \bibinfo{year}{2003}, \bibinfo{journal}{Phys. Rev. Lett.}
  \textbf{\bibinfo{volume}{91}}, \bibinfo{eid}{027901}.

\bibitem[{\citenamefont{Bartlett}
  \emph{et~al.}(2004{\natexlab{a}})\citenamefont{Bartlett, Rudolph, and
  Spekkens}}]{BRS04a}
\bibinfo{author}{\bibnamefont{Bartlett}, \bibfnamefont{S.~D.}},
  \bibinfo{author}{\bibfnamefont{T.}~\bibnamefont{Rudolph}}, and
  \bibinfo{author}{\bibfnamefont{R.~W.} \bibnamefont{Spekkens}},
  \bibinfo{year}{2004}{\natexlab{a}}, \bibinfo{journal}{Phys. Rev. A}
  \textbf{\bibinfo{volume}{70}}, \bibinfo{eid}{032307}.

\bibitem[{\citenamefont{Bartlett}
  \emph{et~al.}(2004{\natexlab{b}})\citenamefont{Bartlett, Rudolph, and
  Spekkens}}]{BRS04b}
\bibinfo{author}{\bibnamefont{Bartlett}, \bibfnamefont{S.~D.}},
  \bibinfo{author}{\bibfnamefont{T.}~\bibnamefont{Rudolph}}, and
  \bibinfo{author}{\bibfnamefont{R.~W.} \bibnamefont{Spekkens}},
  \bibinfo{year}{2004}{\natexlab{b}}, \bibinfo{journal}{Phys. Rev. A}
  \textbf{\bibinfo{volume}{70}}, \bibinfo{eid}{032321}.

\bibitem[{\citenamefont{Bartlett}
  \emph{et~al.}(2006{\natexlab{b}})\citenamefont{Bartlett, Rudolph, and
  Spekkens}}]{BRS06}
\bibinfo{author}{\bibnamefont{Bartlett}, \bibfnamefont{S.~D.}},
  \bibinfo{author}{\bibfnamefont{T.}~\bibnamefont{Rudolph}}, and
  \bibinfo{author}{\bibfnamefont{R.~W.} \bibnamefont{Spekkens}},
  \bibinfo{year}{2006}{\natexlab{b}}, \bibinfo{journal}{Int. J. of Quantum
  Information} \textbf{\bibinfo{volume}{4}},
  \bibinfo{pages}{17}.

\bibitem[{\citenamefont{Bartlett}
  \emph{et~al.}(2006{\natexlab{c}})\citenamefont{Bartlett, Rudolph, Spekkens,
  and Turner}}]{BRST06}
\bibinfo{author}{\bibnamefont{Bartlett}, \bibfnamefont{S.~D.}},
  \bibinfo{author}{\bibfnamefont{T.}~\bibnamefont{Rudolph}},
  \bibinfo{author}{\bibfnamefont{R.~W.} \bibnamefont{Spekkens}}, and
  \bibinfo{author}{\bibfnamefont{P.~S.} \bibnamefont{Turner}},
  \bibinfo{year}{2006}{\natexlab{c}}, \bibinfo{journal}{New J. Phys.}
  \textbf{\bibinfo{volume}{8}}, \bibinfo{pages}{58}.

\bibitem[{\citenamefont{Bartlett}
  \emph{et~al.}(2006{\natexlab{d}})\citenamefont{Bartlett, Rudolph, Sanders,
  and Turner}}]{BRST06b}
\bibinfo{author}{\bibnamefont{Bartlett}, \bibfnamefont{S.~D.}},
  \bibinfo{author}{\bibfnamefont{T.}~\bibnamefont{Rudolph}},
  \bibinfo{author}{\bibfnamefont{B.~C.} \bibnamefont{Sanders}}, and
  \bibinfo{author}{\bibfnamefont{P.~S.} \bibnamefont{Turner}},
  \bibinfo{year}{2006}{\natexlab{d}}, eprint quant-ph/0607107.

\bibitem[{\citenamefont{Bartlett and Terno}(2005)}]{Bar05}
\bibinfo{author}{\bibnamefont{Bartlett}, \bibfnamefont{S.~D.}}, and
  \bibinfo{author}{\bibfnamefont{D.~R.} \bibnamefont{Terno}},
  \bibinfo{year}{2005}, \bibinfo{journal}{Phys. Rev. A}
  \textbf{\bibinfo{volume}{71}}, \bibinfo{eid}{012302}.

\bibitem[{\citenamefont{Bartlett and Wiseman}(2003)}]{Bar03}
\bibinfo{author}{\bibnamefont{Bartlett}, \bibfnamefont{S.~D.}}, and
  \bibinfo{author}{\bibfnamefont{H.~M.} \bibnamefont{Wiseman}},
  \bibinfo{year}{2003}, \bibinfo{journal}{Phys. Rev. Lett.}
  \textbf{\bibinfo{volume}{91}}, \bibinfo{eid}{097903}.

\bibitem[{\citenamefont{Beenakker}(2005)}]{Bee05}
\bibinfo{author}{\bibnamefont{Beenakker}, \bibfnamefont{C.~W.~J.}}, in
  \emph{Proceedings of the International School of Physics
  ``Enrico Fermi''}, Course CLXII (IOS, Amsterdam).  Also e-print cond-mat/0508488.

\bibitem[{\citenamefont{Beenakker} \emph{et~al.}(2003)\citenamefont{Beenakker,
  Emary, Kindermann, and van Velsen}}]{Bee03}
\bibinfo{author}{\bibnamefont{Beenakker}, \bibfnamefont{C.~W.~J.}},
  \bibinfo{author}{\bibfnamefont{C.}~\bibnamefont{Emary}},
  \bibinfo{author}{\bibfnamefont{M.}~\bibnamefont{Kindermann}}, and
  \bibinfo{author}{\bibfnamefont{J.~L.} \bibnamefont{van Velsen}},
  \bibinfo{year}{2003}, \bibinfo{journal}{Phys. Rev. Lett.}
  \textbf{\bibinfo{volume}{91}}, \bibinfo{eid}{147901}.

\bibitem[{\citenamefont{Bennett}(1992)}]{Ben92}
\bibinfo{author}{\bibnamefont{Bennett}, \bibfnamefont{C.~H.}},
  \bibinfo{year}{1992}, \bibinfo{journal}{Phys. Rev. Lett.}
  \textbf{\bibinfo{volume}{68}}, \bibinfo{pages}{3121}.

\bibitem[{\citenamefont{Bennett}
  \emph{et~al.}(1996)\citenamefont{Bennett, DiVincenzo, Smolin,
  and Wootters}}]{Ben96}
\bibinfo{author}{\bibnamefont{Bennett}, \bibfnamefont{C.~H.}},
  \bibinfo{author}{\bibfnamefont{D.~P.} \bibnamefont{DiVincenzo}},
  \bibinfo{author}{\bibfnamefont{J.~A.} \bibnamefont{Smolin}}, and
  \bibinfo{author}{\bibfnamefont{W.~K.} \bibnamefont{Wootters}},
  \bibinfo{year}{1996}, \bibinfo{journal}{Phys. Rev. A}
  \textbf{\bibinfo{volume}{54}}, \bibinfo{pages}{3824}.

\bibitem[{\citenamefont{Berry and Wiseman}(2000)}]{Ber00}
\bibinfo{author}{\bibnamefont{Berry}, \bibfnamefont{D.~W.}}, and
  \bibinfo{author}{\bibfnamefont{H.~M.} \bibnamefont{Wiseman}},
  \bibinfo{year}{2000}, \bibinfo{journal}{Phys. Rev. Lett.}
  \textbf{\bibinfo{volume}{85}}, \bibinfo{pages}{5098}.

\bibitem[{\citenamefont{Boileau} \emph{et~al.}(2004)\citenamefont{Boileau,
  Gottesman, Laflamme, Poulin, and Spekkens}}]{Boi04}
\bibinfo{author}{\bibnamefont{Boileau}, \bibfnamefont{J.-C.}},
  \bibinfo{author}{\bibfnamefont{D.}~\bibnamefont{Gottesman}},
  \bibinfo{author}{\bibfnamefont{R.}~\bibnamefont{Laflamme}},
  \bibinfo{author}{\bibfnamefont{D.}~\bibnamefont{Poulin}}, and
  \bibinfo{author}{\bibfnamefont{R.~W.} \bibnamefont{Spekkens}},
  \bibinfo{year}{2004}, \bibinfo{journal}{Phys. Rev. Lett.}
  \textbf{\bibinfo{volume}{92}}, \bibinfo{eid}{017901}.

\bibitem[{\citenamefont{Boileau} \emph{et~al.}(2005)\citenamefont{Boileau,
  Tamaki, Batuwantudawe, Laflamme, and Renes}}]{Boi05}
\bibinfo{author}{\bibnamefont{Boileau}, \bibfnamefont{J.-C.}},
  \bibinfo{author}{\bibfnamefont{K.}~\bibnamefont{Tamaki}},
  \bibinfo{author}{\bibfnamefont{J.}~\bibnamefont{Batuwantudawe}},
  \bibinfo{author}{\bibfnamefont{R.}~\bibnamefont{Laflamme}}, and
  \bibinfo{author}{\bibfnamefont{J.~M.} \bibnamefont{Renes}},
  \bibinfo{year}{2005}, \bibinfo{journal}{Phys. Rev. Lett.}
  \textbf{\bibinfo{volume}{94}}, \bibinfo{eid}{040503}.

\bibitem[{\citenamefont{Boixo} \emph{et~al.}(2006)\citenamefont{Boixo,
  Caves, Datta, and Shaji}}]{Boi06}
\bibinfo{author}{\bibnamefont{Boixo}, \bibfnamefont{S.}},
  \bibinfo{author}{\bibfnamefont{C.~M.}~\bibnamefont{Caves}},
  \bibinfo{author}{\bibfnamefont{A.}~\bibnamefont{Datta}}, and
  \bibinfo{author}{\bibfnamefont{A.} \bibnamefont{Shaji}},
  \bibinfo{year}{2006}, \bibinfo{journal}{Laser Physics}
  \textbf{\bibinfo{volume}{16}}, \bibinfo{pages}{1525}.

\bibitem[{\citenamefont{Bourennane}
  \emph{et~al.}(2004)\citenamefont{Bourennane, Eibl, Gaertner, Kurtsiefer,
  Cabello, and Weinfurter}}]{Bou04}
\bibinfo{author}{\bibnamefont{Bourennane}, \bibfnamefont{M.}},
  \bibinfo{author}{\bibfnamefont{M.}~\bibnamefont{Eibl}},
  \bibinfo{author}{\bibfnamefont{S.}~\bibnamefont{Gaertner}},
  \bibinfo{author}{\bibfnamefont{C.}~\bibnamefont{Kurtsiefer}},
  \bibinfo{author}{\bibfnamefont{A.}~\bibnamefont{Cabello}}, and
  \bibinfo{author}{\bibfnamefont{H.}~\bibnamefont{Weinfurter}},
  \bibinfo{year}{2004}, \bibinfo{journal}{Phys. Rev. Lett.}
  \textbf{\bibinfo{volume}{92}}, \bibinfo{eid}{107901}.

\bibitem[{\citenamefont{Brassard} \emph{et~al.}(2000)\citenamefont{Brassard,
  L\"utkenhaus, Mor, and Sanders}}]{Bra00}
\bibinfo{author}{\bibnamefont{Brassard}, \bibfnamefont{G.}},
  \bibinfo{author}{\bibfnamefont{N.}~\bibnamefont{L\"utkenhaus}},
  \bibinfo{author}{\bibfnamefont{T.}~\bibnamefont{Mor}}, and
  \bibinfo{author}{\bibfnamefont{B.~C.} \bibnamefont{Sanders}},
  \bibinfo{year}{2000}, \bibinfo{journal}{Phys. Rev. Lett.}
  \textbf{\bibinfo{volume}{85}}, \bibinfo{pages}{1330}.

\bibitem[{\citenamefont{de~Burgh and Bartlett}(2005)}]{deB05}
\bibinfo{author}{\bibnamefont{de~Burgh}, \bibfnamefont{M.}}, and
  \bibinfo{author}{\bibfnamefont{S.~D.} \bibnamefont{Bartlett}},
  \bibinfo{year}{2005}, \bibinfo{journal}{Phys. Rev. A}
  \textbf{\bibinfo{volume}{72}}, \bibinfo{eid}{042301}.

\bibitem[{\citenamefont{Burt} \emph{et~al.}(2001)\citenamefont{Burt, Ekstrom,
  and Swanson}}]{Bur01}
\bibinfo{author}{\bibnamefont{Burt}, \bibfnamefont{E.~A.}},
  \bibinfo{author}{\bibfnamefont{C.~R.} \bibnamefont{Ekstrom}}, and
  \bibinfo{author}{\bibfnamefont{T.~B.} \bibnamefont{Swanson}},
  \bibinfo{year}{2001}, \bibinfo{journal}{Phys. Rev. Lett.}
  \textbf{\bibinfo{volume}{87}}, \bibinfo{eid}{129801}.

\bibitem[{\citenamefont{Byrd}(2006)}]{Byr06}
\bibinfo{author}{\bibnamefont{Byrd}, \bibfnamefont{M.~S.}},
  \bibinfo{year}{2006}, \bibinfo{journal}{Phys. Rev. A}
  \textbf{\bibinfo{volume}{73}}, \bibinfo{eid}{032330}.

\bibitem[{\citenamefont{Cabello}(2003)}]{Cab03}
\bibinfo{author}{\bibnamefont{Cabello}, \bibfnamefont{A.}},
  \bibinfo{year}{2003}, \bibinfo{journal}{Phys. Rev. A}
  \textbf{\bibinfo{volume}{68}}, \bibinfo{eid}{042104}.

\bibitem[{\citenamefont{Cable} \emph{et~al.}(2005)\citenamefont{Cable, Knight,
  and Rudolph}}]{Cab05}
\bibinfo{author}{\bibnamefont{Cable}, \bibfnamefont{H.}},
  \bibinfo{author}{\bibfnamefont{P.~L.} \bibnamefont{Knight}}, and
  \bibinfo{author}{\bibfnamefont{T.}~\bibnamefont{Rudolph}},
  \bibinfo{year}{2005}, \bibinfo{journal}{Phys. Rev. A}
  \textbf{\bibinfo{volume}{71}}, \bibinfo{eid}{042107}.

\bibitem[{\citenamefont{Casher and Reznik}(2000)}]{Cas00}
\bibinfo{author}{\bibnamefont{Casher}, \bibfnamefont{A.}}, and
  \bibinfo{author}{\bibfnamefont{B.}~\bibnamefont{Reznik}},
  \bibinfo{year}{2000}, \bibinfo{journal}{Phys. Rev. A}
  \textbf{\bibinfo{volume}{62}}, \bibinfo{pages}{042104}.

\bibitem[{\citenamefont{Castin and Dalibard}(1997)}]{Cas97}
\bibinfo{author}{\bibnamefont{Castin}, \bibfnamefont{Y.}}, and
  \bibinfo{author}{\bibfnamefont{J.}~\bibnamefont{Dalibard}},
  \bibinfo{year}{1997}, \bibinfo{journal}{Phys. Rev. A}
  \textbf{\bibinfo{volume}{55}}, \bibinfo{pages}{4330}.

\bibitem[{\citenamefont{Chen} \emph{et~al.}(2006)\citenamefont{Chen, Zhang,
  Boileau, Jin, Yang, Zhang, Yang, Laflamme, and Pan}}]{Che06}
\bibinfo{author}{\bibnamefont{Chen}, \bibfnamefont{T.-Y.}},
  \bibinfo{author}{\bibfnamefont{J.}~\bibnamefont{Zhang}},
  \bibinfo{author}{\bibfnamefont{J.-C.} \bibnamefont{Boileau}},
  \bibinfo{author}{\bibfnamefont{X.-M.} \bibnamefont{Jin}},
  \bibinfo{author}{\bibfnamefont{B.}~\bibnamefont{Yang}},
  \bibinfo{author}{\bibfnamefont{Q.}~\bibnamefont{Zhang}},
  \bibinfo{author}{\bibfnamefont{T.}~\bibnamefont{Yang}},
  \bibinfo{author}{\bibfnamefont{R.}~\bibnamefont{Laflamme}}, and
  \bibinfo{author}{\bibfnamefont{J.-W.} \bibnamefont{Pan}},
  \bibinfo{year}{2006}, \bibinfo{journal}{Phys. Rev. Lett.}
  \textbf{\bibinfo{volume}{96}}, \bibinfo{eid}{150504}.

\bibitem[{\citenamefont{Chiribella and D'Ariano}(2004{\natexlab{a}})}]{Chi04b}
\bibinfo{author}{\bibnamefont{Chiribella}, \bibfnamefont{G.}}, and
  \bibinfo{author}{\bibfnamefont{G.~M.} \bibnamefont{D'Ariano}},
  \bibinfo{year}{2004{\natexlab{a}}}, \bibinfo{journal}{J. Math. Phys.}
  \textbf{\bibinfo{volume}{45}}, \bibinfo{pages}{4435}.

\bibitem[{\citenamefont{Chiribella}
  \emph{et~al.}(2004{\natexlab{b}})\citenamefont{Chiribella, D'Ariano, Perinotti, and
  Sacchi}}]{Chi04}
\bibinfo{author}{\bibnamefont{Chiribella}, \bibfnamefont{G.}},
  \bibinfo{author}{\bibfnamefont{G.~M.} \bibnamefont{D'Ariano}},
  \bibinfo{author}{\bibfnamefont{P.}~\bibnamefont{Perinotti}}, and
  \bibinfo{author}{\bibfnamefont{M.~F.} \bibnamefont{Sacchi}},
  \bibinfo{year}{2004{\natexlab{b}}}, \bibinfo{journal}{Phys. Rev. Lett.}
  \textbf{\bibinfo{volume}{93}}, \bibinfo{eid}{180503}.

\bibitem[{\citenamefont{Chiribella}
  \emph{et~al.}(2004{\natexlab{c}})\citenamefont{Chiribella, D'Ariano, Perinotti, and
  Sacchi}}]{Chi04c}
\bibinfo{author}{\bibnamefont{Chiribella}, \bibfnamefont{G.}},
  \bibinfo{author}{\bibfnamefont{G.~M.} \bibnamefont{D'Ariano}},
  \bibinfo{author}{\bibfnamefont{P.}~\bibnamefont{Perinotti}}, and
  \bibinfo{author}{\bibfnamefont{M.~F.} \bibnamefont{Sacchi}},
  \bibinfo{year}{2004{\natexlab{c}}}, \bibinfo{journal}{Phys. Rev. A}
  \textbf{\bibinfo{volume}{70}}, \bibinfo{eid}{062105}.

\bibitem[{\citenamefont{Chiribella}
  \emph{et~al.}(2005)\citenamefont{Chiribella, D'Ariano, and
  Sacchi}}]{Chi05}
\bibinfo{author}{\bibnamefont{Chiribella}, \bibfnamefont{G.}},
  \bibinfo{author}{\bibfnamefont{G.~M.} \bibnamefont{D'Ariano}}, and
  \bibinfo{author}{\bibfnamefont{M.~F.} \bibnamefont{Sacchi}},
  \bibinfo{year}{2005}, \bibinfo{journal}{Phys. Rev. A}
  \textbf{\bibinfo{volume}{72}}, \bibinfo{eid}{042338}.

\bibitem[{\citenamefont{Chiribella}
  \emph{et~al.}(2006)\citenamefont{Chiribella, Maccone, and
  Perinotti}}]{Chi06}
\bibinfo{author}{\bibnamefont{Chiribella}, \bibfnamefont{G.}},
  \bibinfo{author}{\bibfnamefont{L.} \bibnamefont{Maccone}}, and
  \bibinfo{author}{\bibfnamefont{P.} \bibnamefont{Perinotti}},
  \bibinfo{year}{2006}, e-print quant-ph/0608042.

\bibitem[{\citenamefont{Chuang}(2000)}]{Chu00}
\bibinfo{author}{\bibnamefont{Chuang}, \bibfnamefont{I.~L.}},
  \bibinfo{year}{2000}, \bibinfo{journal}{Phys. Rev. Lett.}
  \textbf{\bibinfo{volume}{85}}, \bibinfo{pages}{2006}.

\bibitem[{\citenamefont{Cirac} \emph{et~al.}(1996)\citenamefont{Cirac,
  Gardiner, Naraschewski, and Zoller}}]{Cir96}
\bibinfo{author}{\bibnamefont{Cirac}, \bibfnamefont{J.~I.}},
  \bibinfo{author}{\bibfnamefont{C.~W.}~\bibnamefont{Gardiner}},
  \bibinfo{author}{\bibfnamefont{M.}~\bibnamefont{Naraschewski}}, and
  \bibinfo{author}{\bibfnamefont{P.}~\bibnamefont{Zoller}},
  \bibinfo{year}{2005}, \bibinfo{journal}{Phys. Rev. A}
  \textbf{\bibinfo{volume}{54}}, \bibinfo{eid}{R3714}.

\bibitem[{\citenamefont{Coffman}
  \emph{et~al.}(2000)\citenamefont{Coffman, Kundu, and
  Wootters}}]{Cof00}
\bibinfo{author}{\bibnamefont{Coffman}, \bibfnamefont{V.}},
  \bibinfo{author}{\bibfnamefont{J.} \bibnamefont{Kundu}}, and
  \bibinfo{author}{\bibfnamefont{W.~K.} \bibnamefont{Wootters}},
  \bibinfo{year}{2000}, \bibinfo{journal}{Phys. Rev. A}
  \textbf{\bibinfo{volume}{61}}, \bibinfo{eid}{052306}.

\bibitem[{\citenamefont{Collins and Popescu}(2002)}]{Col02}
\bibinfo{author}{\bibnamefont{Collins}, \bibfnamefont{D.}},and
  \bibinfo{author}{\bibfnamefont{S.}~\bibnamefont{Popescu}},
  \bibinfo{year}{2002}, \bibinfo{journal}{Phys. Rev. A}
  \textbf{\bibinfo{volume}{65}}, \bibinfo{eid}{032321}.

\bibitem[{\citenamefont{Collins} \emph{et~al.}(2005)\citenamefont{Collins,
  Diosi, Gisin, Massar, and Popescu}}]{Col05}
\bibinfo{author}{\bibnamefont{Collins}, \bibfnamefont{D.}},
  \bibinfo{author}{\bibfnamefont{L.}~\bibnamefont{Diosi}},
  \bibinfo{author}{\bibfnamefont{N.}~\bibnamefont{Gisin}},
  \bibinfo{author}{\bibfnamefont{S.}~\bibnamefont{Massar}}, and
  \bibinfo{author}{\bibfnamefont{S.}~\bibnamefont{Popescu}},
  \bibinfo{year}{2005}, \bibinfo{journal}{Phys. Rev. A}
  \textbf{\bibinfo{volume}{72}}, \bibinfo{eid}{022304}.

\bibitem[{\citenamefont{Diosi}(2000)}]{Dio00}
\bibinfo{author}{\bibnamefont{Diosi}, \bibfnamefont{L.}}, \bibinfo{year}{2000},
  e-print quant-ph/0007046.

\bibitem[{\citenamefont{DiVincenzo}
  \emph{et~al.}(2004)\citenamefont{DiVincenzo, Smolin, and Terhal}}]{DiV04}
\bibinfo{author}{\bibnamefont{DiVincenzo}, \bibfnamefont{D.~P.}},
  \bibinfo{author}{\bibfnamefont{J.~A.} \bibnamefont{Smolin}}, and
  \bibinfo{author}{\bibfnamefont{B.~M.} \bibnamefont{Terhal}},
  \bibinfo{year}{2004}, \bibinfo{journal}{New J. Phys.}
  \textbf{\bibinfo{volume}{6}}, \bibinfo{pages}{80}.

\bibitem[{\citenamefont{Doplicher and Roberts}(1990)}]{Dop90}
\bibinfo{author}{\bibnamefont{Doplicher}, \bibfnamefont{S.}}, and
  \bibinfo{author}{\bibfnamefont{J.~E.} \bibnamefont{Roberts}},
  \bibinfo{year}{1990}, \bibinfo{journal}{Comm. Math. Phys.}
  \textbf{\bibinfo{volume}{131}}, \bibinfo{pages}{51}.

\bibitem[{\citenamefont{Dowling} \emph{et~al.}(2006)\citenamefont{Dowling,
  Doherty, and Wiseman}}]{Dow06}
\bibinfo{author}{\bibnamefont{Dowling}, \bibfnamefont{M.~R.}},
  \bibinfo{author}{\bibfnamefont{A.~C.} \bibnamefont{Doherty}}, and
  \bibinfo{author}{\bibfnamefont{H.~M.} \bibnamefont{Wiseman}},
  \bibinfo{year}{2006}, \bibinfo{journal}{Phys. Rev. A}
  \textbf{\bibinfo{volume}{73}}, \bibinfo{eid}{052323}.

\bibitem[{\citenamefont{Dowling} \emph{et~al.}(2006)\citenamefont{Dowling,
  Bartlett, Rudolph, and Spekkens}}]{Dow06b}
\bibinfo{author}{\bibnamefont{Dowling}, \bibfnamefont{M.~R.}},
  \bibinfo{author}{\bibfnamefont{S.~D.}~\bibnamefont{Bartlett}},
  \bibinfo{author}{\bibfnamefont{T.}~\bibnamefont{Rudolph}}, and
  \bibinfo{author}{\bibfnamefont{R.~W.} \bibnamefont{Spekkens}},
  \bibinfo{year}{2006}, \bibinfo{journal}{Phys. Rev. A}
  \textbf{\bibinfo{volume}{74}}, \bibinfo{pages}{052113}.

\bibitem[{\citenamefont{Du\v{s}ek and Bu\v{z}ek}(2002)}]{Dus02}
\bibinfo{author}{\bibnamefont{Du\v{s}ek}, \bibfnamefont{M.}}, and
  \bibinfo{author}{\bibfnamefont{V.}~\bibnamefont{Bu\v{z}ek}},
  \bibinfo{year}{2002}, \bibinfo{journal}{Phys. Rev. A}
  \textbf{\bibinfo{volume}{66}}, \bibinfo{pages}{022112}.

\bibitem[{\citenamefont{Eisert} \emph{et~al.}(2000)\citenamefont{Eisert,
  Felbinger, Papadopoulos, Plenio, and Wilkens}}]{Eis00}
\bibinfo{author}{\bibnamefont{Eisert}, \bibfnamefont{J.}},
  \bibinfo{author}{\bibfnamefont{T.} \bibnamefont{Felbinger}},
  \bibinfo{author}{\bibfnamefont{P.} \bibnamefont{Papadopoulos}},
  \bibinfo{author}{\bibfnamefont{M. B.} \bibnamefont{Plenio}}, and
  \bibinfo{author}{\bibfnamefont{M.} \bibnamefont{Wilkens}},
  \bibinfo{year}{2000}, \bibinfo{journal}{Phys. Rev. Lett.}
  \textbf{\bibinfo{volume}{84}}, \bibinfo{pages}{1611}.

\bibitem[{\citenamefont{Eisert and Plenio}(2003)}]{Eis03}
\bibinfo{author}{\bibnamefont{Eisert}, \bibfnamefont{J.}}, and
  \bibinfo{author}{\bibfnamefont{M. B.} \bibnamefont{Plenio}},
  \bibinfo{year}{2003}, \bibinfo{journal}{Int. J. of Quantum Information}
  \textbf{\bibinfo{volume}{1}}, \bibinfo{pages}{479}.

\bibitem[{\citenamefont{van Enk}(2001)}]{Enk01b}
\bibinfo{author}{\bibnamefont{van Enk}, \bibfnamefont{S.~J.}},
  \bibinfo{year}{2001}, \bibinfo{journal}{J. Mod. Opt.}
  \textbf{\bibinfo{volume}{48}}, \bibinfo{pages}{2049}.

\bibitem[{\citenamefont{van Enk}(2005{\natexlab{a}})}]{Enk05}
\bibinfo{author}{\bibnamefont{van Enk}, \bibfnamefont{S.~J.}},
  \bibinfo{year}{2005}{\natexlab{a}}, \bibinfo{journal}{Phys. Rev. A}
  \textbf{\bibinfo{volume}{71}}, \bibinfo{eid}{032339}.

\bibitem[{\citenamefont{van Enk}(2005{\natexlab{b}})}]{Enk05b}
\bibinfo{author}{\bibnamefont{van Enk}, \bibfnamefont{S.~J.}},
  \bibinfo{year}{2005}{\natexlab{b}}, \bibinfo{journal}{Phys. Rev. A}
  \textbf{\bibinfo{volume}{72}}, \bibinfo{eid}{064306}.

\bibitem[{\citenamefont{van Enk}(2006)}]{Enk06}
\bibinfo{author}{\bibnamefont{van Enk}, \bibfnamefont{S.~J.}},
  \bibinfo{year}{2006}, \bibinfo{journal}{Phys. Rev. A}
  \textbf{\bibinfo{volume}{73}}, \bibinfo{eid}{042306}.

\bibitem[{\citenamefont{van Enk and Fuchs}(2002{\natexlab{a}})}]{vEF02a}
\bibinfo{author}{\bibnamefont{van Enk}, \bibfnamefont{S.~J.}}, and
  \bibinfo{author}{\bibfnamefont{C.~A.} \bibnamefont{Fuchs}},
  \bibinfo{year}{2002}{\natexlab{a}}, \bibinfo{journal}{Phys. Rev. Lett.}
  \textbf{\bibinfo{volume}{88}}, \bibinfo{eid}{027902}.

\bibitem[{\citenamefont{van Enk and Fuchs}(2002{\natexlab{b}})}]{vEF02b}
\bibinfo{author}{\bibnamefont{van Enk}, \bibfnamefont{S.~J.}}, and
  \bibinfo{author}{\bibfnamefont{C.~A.} \bibnamefont{Fuchs}},
  \bibinfo{year}{2002}{\natexlab{b}}, \bibinfo{journal}{Quantum Info. Comp.}
  \textbf{\bibinfo{volume}{2}}, \bibinfo{pages}{151}.

\bibitem[{\citenamefont{van Enk and Kimble}(2001)}]{Enk01}
\bibinfo{author}{\bibnamefont{van Enk}, \bibfnamefont{S.~J.}}, and
  \bibinfo{author}{\bibfnamefont{H.~J.} \bibnamefont{Kimble}},
  \bibinfo{year}{2001}, \bibinfo{journal}{Quantum Info. Comp.}
  \textbf{\bibinfo{volume}{2}}, \bibinfo{pages}{1}.

\bibitem[{\citenamefont{Fitch and Franson}(2002)}]{Fit02}
\bibinfo{author}{\bibnamefont{Fitch}, \bibfnamefont{M.~J.}}, and
  \bibinfo{author}{\bibfnamefont{J.~D.} \bibnamefont{Franson}},
  \bibinfo{year}{2002}, \bibinfo{journal}{Phys. Rev. A}
  \textbf{\bibinfo{volume}{65}}, \bibinfo{eid}{053809}.

\bibitem[{\citenamefont{Fiur\'a\v{s}ek}
  \emph{et~al.}(2002)\citenamefont{Fiur\'a\v{s}ek, Du\v{s}ek, and
  Filip}}]{Fiu02}
\bibinfo{author}{\bibnamefont{Fiur\'a\v{s}ek}, \bibfnamefont{J.}},
  \bibinfo{author}{\bibfnamefont{M.}~\bibnamefont{Du\v{s}ek}}, and
  \bibinfo{author}{\bibfnamefont{R.}~\bibnamefont{Filip}},
  \bibinfo{year}{2002}, \bibinfo{journal}{Phys. Rev. Lett.}
  \textbf{\bibinfo{volume}{89}}, \bibinfo{pages}{190401}.

\bibitem[{\citenamefont{Fujii}(2003)}]{Fuj03}
\bibinfo{author}{\bibnamefont{Fujii}, \bibfnamefont{M.}}, \bibinfo{year}{2003},
  \bibinfo{journal}{Phys. Rev. A}
  \textbf{\bibinfo{volume}{68}}, \bibinfo{eid}{050302}.

\bibitem[{\citenamefont{Fulton and Harris}(1991)}]{FultonHarris}
\bibinfo{author}{\bibnamefont{Fulton}, \bibfnamefont{W.}}, and
  \bibinfo{author}{\bibfnamefont{J.}~\bibnamefont{Harris}},
  \bibinfo{year}{1991}, \emph{\bibinfo{title}{Representation Theory: A First
  Course}}, number \bibinfo{number}{129} in \bibinfo{series}{Graduate Texts in
  Mathematics} (\bibinfo{publisher}{Springer}).

\bibitem[{\citenamefont{Furusawa} \emph{et~al.}(1998)\citenamefont{Furusawa,
  S{\o}rensen, Braunstein, Fuchs, Kimble, and Polzik}}]{Fur98}
\bibinfo{author}{\bibnamefont{Furusawa}, \bibfnamefont{A.}},
  \bibinfo{author}{\bibfnamefont{J.~L.} \bibnamefont{S{\o}rensen}},
  \bibinfo{author}{\bibfnamefont{S.~L.} \bibnamefont{Braunstein}},
  \bibinfo{author}{\bibfnamefont{C.~A.} \bibnamefont{Fuchs}},
  \bibinfo{author}{\bibfnamefont{H.~J.} \bibnamefont{Kimble}}, and
  \bibinfo{author}{\bibfnamefont{E.~S.} \bibnamefont{Polzik}},
  \bibinfo{year}{1998}, \bibinfo{journal}{Science}
  \textbf{\bibinfo{volume}{282}}, \bibinfo{pages}{706}.

\bibitem[{\citenamefont{Gambini}
  \emph{et~al.}(2004{\natexlab{a}})\citenamefont{Gambini, Porto, and
  Pullin}}]{Gam04}
\bibinfo{author}{\bibnamefont{Gambini}, \bibfnamefont{R.}},
  \bibinfo{author}{\bibfnamefont{R.~A.} \bibnamefont{Porto}}, and
  \bibinfo{author}{\bibfnamefont{J.}~\bibnamefont{Pullin}},
  \bibinfo{year}{2004}{\natexlab{a}}, \bibinfo{journal}{Phys. Rev. Lett.}
  \textbf{\bibinfo{volume}{93}}, \bibinfo{eid}{240401}.

\bibitem[{\citenamefont{Gambini}
  \emph{et~al.}(2004{\natexlab{b}})\citenamefont{Gambini, Porto, and
  Pullin}}]{Gam04b}
\bibinfo{author}{\bibnamefont{Gambini}, \bibfnamefont{R.}},
  \bibinfo{author}{\bibfnamefont{R.~A.} \bibnamefont{Porto}}, and
  \bibinfo{author}{\bibfnamefont{J.}~\bibnamefont{Pullin}},
  \bibinfo{year}{2004}{\natexlab{b}}, \bibinfo{journal}{New J. Phys.}
  \textbf{\bibinfo{volume}{6}}, \bibinfo{pages}{45}.

\bibitem[{\citenamefont{Gea-Banacloche}(1998)}]{Gea98}
\bibinfo{author}{\bibnamefont{Gea-Banacloche}, \bibfnamefont{J.}},
  \bibinfo{year}{1998}, \bibinfo{journal}{Phys. Rev. A}
  \textbf{\bibinfo{volume}{58}}, \bibinfo{pages}{4244}.

\bibitem[{\citenamefont{Gea-Banacloche}(2002{\natexlab{a}})}]{Gea02b}
\bibinfo{author}{\bibnamefont{Gea-Banacloche}, \bibfnamefont{J.}},
  \bibinfo{year}{2002}{\natexlab{a}}, e-print quant-ph/0212027.

\bibitem[{\citenamefont{Gea-Banacloche}(2002{\natexlab{b}})}]{Gea02}
\bibinfo{author}{\bibnamefont{Gea-Banacloche}, \bibfnamefont{J.}},
  \bibinfo{year}{2002}{\natexlab{b}}, \bibinfo{journal}{Phys. Rev. A}
  \textbf{\bibinfo{volume}{65}}, \bibinfo{eid}{022308}.

\bibitem[{\citenamefont{Gea-Banacloche and Ozawa}(2005)}]{Gea05}
\bibinfo{author}{\bibnamefont{Gea-Banacloche}, \bibfnamefont{J.}}, and
  \bibinfo{author}{\bibfnamefont{M.} \bibnamefont{Ozawa}},
  \bibinfo{year}{2005}, \bibinfo{journal}{J. Opt. B: Quantum Semiclass.}
  \textbf{\bibinfo{volume}{7}}, \bibinfo{pages}{S326}.

\bibitem[{\citenamefont{Giovannetti}
  \emph{et~al.}(2001)\citenamefont{Giovannetti, Lloyd, and Maccone}}]{Gio01}
\bibinfo{author}{\bibnamefont{Giovannetti}, \bibfnamefont{V.}},
  \bibinfo{author}{\bibfnamefont{S.}~\bibnamefont{Lloyd}}, and
  \bibinfo{author}{\bibfnamefont{L.}~\bibnamefont{Maccone}},
  \bibinfo{year}{2001}, \bibinfo{journal}{Nature (London)}
  \textbf{\bibinfo{volume}{412}}(\bibinfo{number}{26 July 2001}),
  \bibinfo{pages}{417}.

\bibitem[{\citenamefont{Giovannetti}
  \emph{et~al.}(2002)\citenamefont{Giovannetti, Lloyd, and Maccone}}]{Gio02}
\bibinfo{author}{\bibnamefont{Giovannetti}, \bibfnamefont{V.}},
  \bibinfo{author}{\bibfnamefont{S.}~\bibnamefont{Lloyd}}, and
  \bibinfo{author}{\bibfnamefont{L.}~\bibnamefont{Maccone}},
  \bibinfo{year}{2002}, \bibinfo{journal}{Phys. Rev. A}
  \textbf{\bibinfo{volume}{65}}, \bibinfo{eid}{022309}.

\bibitem[{\citenamefont{Giovannetti}
  \emph{et~al.}(2004{\natexlab{a}})\citenamefont{Giovannetti, Lloyd, and
  Maccone}}]{Gio04a}
\bibinfo{author}{\bibnamefont{Giovannetti}, \bibfnamefont{V.}},
  \bibinfo{author}{\bibfnamefont{S.}~\bibnamefont{Lloyd}}, and
  \bibinfo{author}{\bibfnamefont{L.}~\bibnamefont{Maccone}},
  \bibinfo{year}{2004}{\natexlab{a}}, \bibinfo{journal}{Science}
  \textbf{\bibinfo{volume}{306}},
  \bibinfo{pages}{1330}.

\bibitem[{\citenamefont{Giovannetti}
  \emph{et~al.}(2006)\citenamefont{Giovannetti, Lloyd, and Maccone}}]{Gio06}
\bibinfo{author}{\bibnamefont{Giovannetti}, \bibfnamefont{V.}},
  \bibinfo{author}{\bibfnamefont{S.}~\bibnamefont{Lloyd}}, and
  \bibinfo{author}{\bibfnamefont{L.}~\bibnamefont{Maccone}},
  \bibinfo{year}{2006}, \bibinfo{journal}{Phys. Rev. Lett.}
  \textbf{\bibinfo{volume}{96}}, \bibinfo{eid}{010401}.

\bibitem[{\citenamefont{Giovannetti}
  \emph{et~al.}(2004{\natexlab{b}})\citenamefont{Giovannetti, Lloyd, Maccone,
  Shapiro, and Wong}}]{Gio04b}
\bibinfo{author}{\bibnamefont{Giovannetti}, \bibfnamefont{V.}},
  \bibinfo{author}{\bibfnamefont{S.}~\bibnamefont{Lloyd}},
  \bibinfo{author}{\bibfnamefont{L.}~\bibnamefont{Maccone}},
  \bibinfo{author}{\bibfnamefont{J.~H.} \bibnamefont{Shapiro}}, and
  \bibinfo{author}{\bibfnamefont{F.~N.~C.} \bibnamefont{Wong}},
  \bibinfo{year}{2004}{\natexlab{b}}, \bibinfo{journal}{Phys. Rev. A}
  \textbf{\bibinfo{volume}{70}}, \bibinfo{eid}{043808}.

\bibitem[{\citenamefont{Gisin}(2004)}]{Gis04}
\bibinfo{author}{\bibnamefont{Gisin}, \bibfnamefont{N.}}, \bibinfo{year}{2004},
  e-print quant-ph/0408095.

\bibitem[{\citenamefont{Gisin and Popescu}(1999)}]{Gis99}
\bibinfo{author}{\bibnamefont{Gisin}, \bibfnamefont{N.}}, and
  \bibinfo{author}{\bibfnamefont{S.}~\bibnamefont{Popescu}},
  \bibinfo{year}{1999}, \bibinfo{journal}{Phys. Rev. Lett.}
  \textbf{\bibinfo{volume}{83}}, \bibinfo{pages}{432}.

\bibitem[{\citenamefont{Gisin} \emph{et~al.}(2002)\citenamefont{Gisin, Ribordy,
  Tittel, and Zbinden}}]{Gis02}
\bibinfo{author}{\bibnamefont{Gisin}, \bibfnamefont{N.}},
  \bibinfo{author}{\bibfnamefont{G.}~\bibnamefont{Ribordy}},
  \bibinfo{author}{\bibfnamefont{W.}~\bibnamefont{Tittel}}, and
  \bibinfo{author}{\bibfnamefont{H.}~\bibnamefont{Zbinden}},
  \bibinfo{year}{2002}, \bibinfo{journal}{Rev. Mod. Phys.}
  \textbf{\bibinfo{volume}{74}}, \bibinfo{pages}{145}.

\bibitem[{\citenamefont{Giulini}(1996)}]{Giu96}
\bibinfo{author}{\bibnamefont{Giulini}, \bibfnamefont{D.}},
  \bibinfo{year}{1996}, in \emph{\bibinfo{title}{Decoherence and the Appearance of
  a Classical World in Quantum Theory,}} (\bibinfo{publisher}{Springer, Berlin}), chapter~\bibinfo{chapter}{6}, pp.
\bibinfo{pages}{187--222}.

\bibitem[{\citenamefont{Giulini}(2000{\natexlab{a}})}]{Giu00a}
\bibinfo{author}{\bibnamefont{Giulini}, \bibfnamefont{D.}},
  \bibinfo{year}{2000}, \bibinfo{journal}{Lect. Notes Phys.}
  \textbf{\bibinfo{volume}{538}}, \bibinfo{pages}{87}.

\bibitem[{\citenamefont{Giulini}(2000{\natexlab{b}})}]{Giu00b}
\bibinfo{author}{\bibnamefont{Giulini}, \bibfnamefont{D.}},
  \bibinfo{year}{2000}, \bibinfo{journal}{Lect. Notes Phys.}
  \textbf{\bibinfo{volume}{559}}, \bibinfo{pages}{67}.

\bibitem[{\citenamefont{Greenberger}
  \emph{et~al.}(1995)\citenamefont{Greenberger, Horne, and Zeilinger}}]{Gre95}
\bibinfo{author}{\bibnamefont{Greenberger}, \bibfnamefont{D.~M.}},
  \bibinfo{author}{\bibfnamefont{M.~A.} \bibnamefont{Horne}}, and
  \bibinfo{author}{\bibfnamefont{A.}~\bibnamefont{Zeilinger}},
  \bibinfo{year}{1995}, \bibinfo{journal}{Phys. Rev. Lett.}
  \textbf{\bibinfo{volume}{75}}, \bibinfo{pages}{2064}.

\bibitem[{\citenamefont{Haag}(1962)}]{Haa62}
\bibinfo{author}{\bibnamefont{Haag}, \bibfnamefont{R.}}, \bibinfo{year}{1962},
  \bibinfo{journal}{Il Nuovo Cimento} \textbf{\bibinfo{volume}{XXV}},
  \bibinfo{pages}{2695}.

\bibitem[{\citenamefont{Hardy}(1994)}]{Har94}
\bibinfo{author}{\bibnamefont{Hardy}, \bibfnamefont{L.}}, \bibinfo{year}{1994},
  \bibinfo{journal}{Phys. Rev. Lett.}
  \textbf{\bibinfo{volume}{73}}, \bibinfo{pages}{2279}.

\bibitem[{\citenamefont{Hardy}(1995)}]{Har95}
\bibinfo{author}{\bibnamefont{Hardy}, \bibfnamefont{L.}}, \bibinfo{year}{1995},
  \bibinfo{journal}{Phys. Rev. Lett.}
  \textbf{\bibinfo{volume}{75}}, \bibinfo{pages}{2065}.

\bibitem[{\citenamefont{Haroche}(1984)}]{Har84}
\bibinfo{author}{\bibnamefont{Haroche}, \bibfnamefont{S.}},
  \bibinfo{year}{1984}, in \emph{\bibinfo{booktitle}{New Trends in Atomic
  Physics}}, edited by
  \bibinfo{editor}{\bibfnamefont{G.}~\bibnamefont{Grynberg}} and
  \bibinfo{editor}{\bibfnamefont{R.}~\bibnamefont{Stora}}
  (\bibinfo{publisher}{Elsevier, Amsterdam}), Les Houches Session XXXVIII.

\bibitem[{\citenamefont{Harrow} \emph{et~al.}(2006)\citenamefont{Harrow,
  Oliveira, and Terhal}}]{Har06}
\bibinfo{author}{\bibnamefont{Harrow}, \bibfnamefont{A.}},
  \bibinfo{author}{\bibfnamefont{R.}~\bibnamefont{Oliveira}}, and
  \bibinfo{author}{\bibfnamefont{B.~M.} \bibnamefont{Terhal}},
  \bibinfo{year}{2006}, \bibinfo{journal}{Phys. Rev. A}
  \textbf{\bibinfo{volume}{73}}, \bibinfo{eid}{032311}.

\bibitem[{\citenamefont{Hessmo} \emph{et~al.}(2004)\citenamefont{Hessmo,
  Usachev, Heydari, and Bjork}}]{Hes04}
\bibinfo{author}{\bibnamefont{Hessmo}, \bibfnamefont{B.}},
  \bibinfo{author}{\bibfnamefont{P.}~\bibnamefont{Usachev}},
  \bibinfo{author}{\bibfnamefont{H.}~\bibnamefont{Heydari}}, and
  \bibinfo{author}{\bibfnamefont{G.}~\bibnamefont{Bjork}},
  \bibinfo{year}{2004}, \bibinfo{journal}{Phys. Rev. Lett.}
  \textbf{\bibinfo{volume}{92}}, \bibinfo{eid}{180401}.

\bibitem[{\citenamefont{Holevo}(1982)}]{Hol82}
\bibinfo{author}{\bibnamefont{Holevo}, \bibfnamefont{A.~S.}},
  \bibinfo{year}{1982}, \emph{\bibinfo{title}{Probabilistic and Statistical
  Aspects of Quantum Theory}} (\bibinfo{publisher}{North Holland, Amsterdam}).

\bibitem[{\citenamefont{Horodecki} \emph{et~al.}(1998)\citenamefont{Horodecki,
  Horodecki, and Horodecki}}]{Hor98}
\bibinfo{author}{\bibnamefont{Horodecki}, \bibfnamefont{M.}},
  \bibinfo{author}{\bibfnamefont{P.}~\bibnamefont{Horodecki}}, and
  \bibinfo{author}{\bibfnamefont{R.}~\bibnamefont{Horodecki}},
  \bibinfo{year}{1998}, \bibinfo{journal}{Phys. Rev. Lett.}
  \textbf{\bibinfo{volume}{80}}, \bibinfo{pages}{5239}.

\bibitem[{\citenamefont{Horodecki} \emph{et~al.}(1999)\citenamefont{Horodecki,
  Horodecki, and Horodecki}}]{Hor99}
\bibinfo{author}{\bibnamefont{Horodecki}, \bibfnamefont{P.}},
  \bibinfo{author}{\bibfnamefont{M.}~\bibnamefont{Horodecki}}, and
  \bibinfo{author}{\bibfnamefont{R.}~\bibnamefont{Horodecki}},
  \bibinfo{year}{1999}, \bibinfo{journal}{Phys. Rev. Lett.}
  \textbf{\bibinfo{volume}{82}}, \bibinfo{pages}{1056}.

\bibitem[{\citenamefont{Hoston and You}(1996)}]{Hos96}
\bibinfo{author}{\bibnamefont{Hoston}, \bibfnamefont{W.}}, and
  \bibinfo{author}{\bibfnamefont{L.}~\bibnamefont{You}}, \bibinfo{year}{1996},
  \bibinfo{journal}{Phys. Rev. A}
  \textbf{\bibinfo{volume}{53}}, \bibinfo{pages}{4254}.

\bibitem[{\citenamefont{Itano}(2003)}]{Ita03}
\bibinfo{author}{\bibnamefont{Itano}, \bibfnamefont{W.~M.}},
  \bibinfo{year}{2003}, \bibinfo{journal}{Phys. Rev. A}
  \textbf{\bibinfo{volume}{68}}, \bibinfo{eid}{046301}.

\bibitem[{\citenamefont{Janzing and Beth}(2003)}]{Jan03}
\bibinfo{author}{\bibnamefont{Janzing}, \bibfnamefont{D.}}, and
  \bibinfo{author}{\bibfnamefont{T.}~\bibnamefont{Beth}}, \bibinfo{year}{2003},
  \bibinfo{journal}{IEEE Trans. Inf. Th.}
  \textbf{\bibinfo{volume}{49}}, \bibinfo{pages}{230}.

\bibitem[{\citenamefont{Javanainen and Yoo}(1996)}]{Jav96}
\bibinfo{author}{\bibnamefont{Javanainen}, \bibfnamefont{J.}}, and
  \bibinfo{author}{\bibfnamefont{S.~M.} \bibnamefont{Yoo}},
  \bibinfo{year}{1996}, \bibinfo{journal}{Phys. Rev. Lett.}
  \textbf{\bibinfo{volume}{76}}, \bibinfo{pages}{161}.

\bibitem[{\citenamefont{Jeffrey} \emph{et~al.}(2006)\citenamefont{Jeffrey,
  Altepeter, Colci, and Kwiat}}]{Jef06}
\bibinfo{author}{\bibnamefont{Jeffrey}, \bibfnamefont{E.~R.}},
  \bibinfo{author}{\bibfnamefont{J.~B.} \bibnamefont{Altepeter}},
  \bibinfo{author}{\bibfnamefont{M.}~\bibnamefont{Colci}}, and
  \bibinfo{author}{\bibfnamefont{P.~G.} \bibnamefont{Kwiat}},
  \bibinfo{year}{2006}, \bibinfo{journal}{Phys. Rev. Lett.}
  \textbf{\bibinfo{volume}{96}}, \bibinfo{eid}{150503}.

\bibitem[{\citenamefont{Jones} \emph{et~al.}(2005)\citenamefont{Jones, Wiseman,
  and Pope}}]{Jon05}
\bibinfo{author}{\bibnamefont{Jones}, \bibfnamefont{S.~J.}},
  \bibinfo{author}{\bibfnamefont{H.~M.} \bibnamefont{Wiseman}}, and
  \bibinfo{author}{\bibfnamefont{D.~T.} \bibnamefont{Pope}},
  \bibinfo{year}{2005}, \bibinfo{journal}{Phys. Rev. A}
  \textbf{\bibinfo{volume}{72}}, \bibinfo{eid}{022330}.

\bibitem[{\citenamefont{Jones} \emph{et~al.}(2006)\citenamefont{Jones,
  Wiseman, Bartlett, Vaccaro, and Pope}}]{Jon06}
\bibinfo{author}{\bibnamefont{Jones}, \bibfnamefont{S.~J.}},
  \bibinfo{author}{\bibfnamefont{H.~M.}~\bibnamefont{Wiseman}},
  \bibinfo{author}{\bibfnamefont{S.~D.}~\bibnamefont{Bartlett}},
  \bibinfo{author}{\bibfnamefont{J.~A.} \bibnamefont{Vaccaro}}, and
  \bibinfo{author}{\bibfnamefont{D.~T.} \bibnamefont{Pope}},
  \bibinfo{year}{2006}, \bibinfo{journal}{Phys. Rev. A}
  \textbf{\bibinfo{volume}{74}}, \bibinfo{eid}{062313}.

\bibitem[{\citenamefont{Jozsa} \emph{et~al.}(2000)\citenamefont{Jozsa, Abrams,
  Dowling, and Williams}}]{Joz00}
\bibinfo{author}{\bibnamefont{Jozsa}, \bibfnamefont{R.}},
  \bibinfo{author}{\bibfnamefont{D.~S.} \bibnamefont{Abrams}},
  \bibinfo{author}{\bibfnamefont{J.~P.} \bibnamefont{Dowling}}, and
  \bibinfo{author}{\bibfnamefont{C.~P.} \bibnamefont{Williams}},
  \bibinfo{year}{2000}, \bibinfo{journal}{Phys. Rev. Lett.}
  \textbf{\bibinfo{volume}{85}}, \bibinfo{pages}{2010}.

\bibitem[{\citenamefont{Jozsa} \emph{et~al.}(2001)\citenamefont{Jozsa, Abrams,
  Dowling, and Williams}}]{Joz01}
\bibinfo{author}{\bibnamefont{Jozsa}, \bibfnamefont{R.}},
  \bibinfo{author}{\bibfnamefont{D.~S.} \bibnamefont{Abrams}},
  \bibinfo{author}{\bibfnamefont{J.~P.} \bibnamefont{Dowling}}, and
  \bibinfo{author}{\bibfnamefont{C.~P.} \bibnamefont{Williams}},
  \bibinfo{year}{2001}, \bibinfo{journal}{Phys. Rev. Lett.}
  \textbf{\bibinfo{volume}{87}}, \bibinfo{eid}{129802}.

\bibitem[{\citenamefont{Kempe} \emph{et~al.}(2001)\citenamefont{Kempe, Bacon,
  Lidar, and Whaley}}]{Kem01}
\bibinfo{author}{\bibnamefont{Kempe}, \bibfnamefont{J.}},
  \bibinfo{author}{\bibfnamefont{D.}~\bibnamefont{Bacon}},
  \bibinfo{author}{\bibfnamefont{D.~A.} \bibnamefont{Lidar}}, and
  \bibinfo{author}{\bibfnamefont{K.~B.} \bibnamefont{Whaley}},
  \bibinfo{year}{2001}, \bibinfo{journal}{Phys. Rev. A}
  \textbf{\bibinfo{volume}{63}}, \bibinfo{eid}{042307}.

\bibitem[{\citenamefont{Kershaw and Woo}(1974)}]{Ker74}
\bibinfo{author}{\bibnamefont{Kershaw}, \bibfnamefont{D.}}, and
  \bibinfo{author}{\bibfnamefont{C.~H.} \bibnamefont{Woo}},
  \bibinfo{year}{1974}, \bibinfo{journal}{Phys. Rev. Lett.}
  \textbf{\bibinfo{volume}{33}}, \bibinfo{pages}{918}.

\bibitem[{\citenamefont{Kitaev} \emph{et~al.}(2004)\citenamefont{Kitaev,
  Mayers, and Preskill}}]{Kit04}
\bibinfo{author}{\bibnamefont{Kitaev}, \bibfnamefont{A.}},
  \bibinfo{author}{\bibfnamefont{D.}~\bibnamefont{Mayers}}, and
  \bibinfo{author}{\bibfnamefont{J.}~\bibnamefont{Preskill}},
  \bibinfo{year}{2004}, \bibinfo{journal}{Phys. Rev. A}
  \textbf{\bibinfo{volume}{69}}, \bibinfo{eid}{052326}.

\bibitem[{\citenamefont{Knill} \emph{et~al.}(2000)\citenamefont{Knill,
  Laflamme, and Viola}}]{Kni00}
\bibinfo{author}{\bibnamefont{Knill}, \bibfnamefont{E.}},
  \bibinfo{author}{\bibfnamefont{R.}~\bibnamefont{Laflamme}}, and
  \bibinfo{author}{\bibfnamefont{L.}~\bibnamefont{Viola}},
  \bibinfo{year}{2000}, \bibinfo{journal}{Phys. Rev. Lett.}
  \textbf{\bibinfo{volume}{84}}, \bibinfo{pages}{2525}.

\bibitem[{\citenamefont{Kok} \emph{et~al.}(2006)\citenamefont{Kok, Munro,
  Nemoto, Ralph, Dowling, and Milburn}}]{Kok06}
\bibinfo{author}{\bibnamefont{Kok}, \bibfnamefont{P.}},
  \bibinfo{author}{\bibfnamefont{W.}~\bibnamefont{Munro}},
  \bibinfo{author}{\bibfnamefont{K.}~\bibnamefont{Nemoto}},
  \bibinfo{author}{\bibfnamefont{T.}~\bibnamefont{Ralph}},
  \bibinfo{author}{\bibfnamefont{J.~P.} \bibnamefont{Dowling}}, and
  \bibinfo{author}{\bibfnamefont{G.}~\bibnamefont{Milburn}},
  \bibinfo{year}{2006}, \bibinfo{journal}{Rev. Mod. Phys.}
  \textbf{\bibinfo{volume}{79}}, \bibinfo{pages}{135}.

\bibitem[{\citenamefont{Kok} \emph{et~al.}(2005)\citenamefont{Kok, Ralph, and
  Milburn}}]{Kok05}
\bibinfo{author}{\bibnamefont{Kok}, \bibfnamefont{P.}},
  \bibinfo{author}{\bibfnamefont{T.~C.} \bibnamefont{Ralph}}, and
  \bibinfo{author}{\bibfnamefont{G.~J.} \bibnamefont{Milburn}},
  \bibinfo{year}{2005}, \bibinfo{journal}{Quantum Info. Comp.}
  \textbf{\bibinfo{volume}{5}}, \bibinfo{pages}{239}.

\bibitem[{\citenamefont{von Korff and Kempe}(2004)}]{Kor04}
\bibinfo{author}{\bibnamefont{von Korff}, \bibfnamefont{J.}}, and
  \bibinfo{author}{\bibfnamefont{J.}~\bibnamefont{Kempe}},
  \bibinfo{year}{2004}, \bibinfo{journal}{Phys. Rev. Lett.}
  \textbf{\bibinfo{volume}{93}}, \bibinfo{eid}{260502}.

\bibitem[{\citenamefont{Landauer}(1993)}]{Lan93}
\bibinfo{author}{\bibnamefont{Landauer}, \bibfnamefont{R.}},
  \bibinfo{year}{1993}, in \emph{\bibinfo{booktitle}{Proc. Workshop
  on Physics and Computation: PhysComp'92}},
  (\bibinfo{publisher}{IEEE Comp. Sci. Press},
  \bibinfo{address}{Los Alamitos}), pp. \bibinfo{pages}{1--4}.

\bibitem[{\citenamefont{Langford} \emph{et~al.}(2004)\citenamefont{Langford,
  Dalton, Harvey, O'Brien, Pryde, Gilchrist, Bartlett, and White}}]{Lan04}
\bibinfo{author}{\bibnamefont{Langford}, \bibfnamefont{N.~K.}},
  \bibinfo{author}{\bibfnamefont{R.~B.} \bibnamefont{Dalton}},
  \bibinfo{author}{\bibfnamefont{M.~D.} \bibnamefont{Harvey}},
  \bibinfo{author}{\bibfnamefont{J.~L.} \bibnamefont{O'Brien}},
  \bibinfo{author}{\bibfnamefont{G.~J.} \bibnamefont{Pryde}},
  \bibinfo{author}{\bibfnamefont{A.} \bibnamefont{Gilchrist}},
  \bibinfo{author}{\bibfnamefont{S.~D.} \bibnamefont{Bartlett}}, and
  \bibinfo{author}{\bibfnamefont{A.~G.} \bibnamefont{White}},
  \bibinfo{year}{2004}, \bibinfo{journal}{Phys. Rev. Lett.}
  \textbf{\bibinfo{volume}{93}}, \bibinfo{eid}{053601}.

\bibitem[{\citenamefont{Leggett}(2001)}]{Leg01}
\bibinfo{author}{\bibnamefont{Leggett}, \bibfnamefont{A.~J.}},
  \bibinfo{year}{2001}, \bibinfo{journal}{Rev. Mod. Phys.}
  \textbf{\bibinfo{volume}{73}}, \bibinfo{pages}{307}.

\bibitem[{\citenamefont{Lindner} \emph{et~al.}(2003)\citenamefont{Lindner,
  Peres, and Terno}}]{Lin03}
\bibinfo{author}{\bibnamefont{Lindner}, \bibfnamefont{N.~H.}},
  \bibinfo{author}{\bibfnamefont{A.}~\bibnamefont{Peres}}, and
  \bibinfo{author}{\bibfnamefont{D.~R.} \bibnamefont{Terno}},
  \bibinfo{year}{2003}, \bibinfo{journal}{Phys. Rev. A}
  \textbf{\bibinfo{volume}{68}}, \bibinfo{eid}{042308}.

\bibitem[{\citenamefont{Lindner} \emph{et~al.}(2006)\citenamefont{Lindner,
  Scudo, and Bru{\ss}}}]{Lin06}
\bibinfo{author}{\bibnamefont{Lindner}, \bibfnamefont{N.~H.}},
  \bibinfo{author}{\bibfnamefont{P.~F.} \bibnamefont{Scudo}}, and
  \bibinfo{author}{\bibfnamefont{D.}~\bibnamefont{Bru{\ss}}},
  \bibinfo{year}{2006}, \bibinfo{journal}{Int. J. of Quantum Information}
  \textbf{\bibinfo{volume}{4}}, \bibinfo{pages}{131}.

\bibitem[{\citenamefont{Livine and Terno}(2006)}]{Liv06}
\bibinfo{author}{\bibnamefont{Livine}, \bibfnamefont{E.~R.}}, and
  \bibinfo{author}{\bibfnamefont{D.~R.} \bibnamefont{Terno}},
  \bibinfo{year}{2006}, \bibinfo{journal}{Nucl. Phys. B}
  \textbf{\bibinfo{volume}{741}}, \bibinfo{pages}{131}.

\bibitem[{\citenamefont{Lubkin}(1970)}]{Lub70}
\bibinfo{author}{\bibnamefont{Lubkin}, \bibfnamefont{E.}},
  \bibinfo{year}{1970}, \bibinfo{journal}{Ann. Phys.}
  \textbf{\bibinfo{volume}{56}}, \bibinfo{pages}{69}.

\bibitem[{\citenamefont{Mair}
  \emph{et~al.}(2001)\citenamefont{Mair, Vaziri, Weihs, and Zeilinger}}]{Mai01}
\bibinfo{author}{\bibnamefont{Mair}, \bibfnamefont{A.}},
  \bibinfo{author}{\bibfnamefont{A.} \bibnamefont{Vaziri}},
  \bibinfo{author}{\bibfnamefont{G.}~\bibnamefont{Weihs}}, and
  \bibinfo{author}{\bibfnamefont{A.}~\bibnamefont{Zeilinger}},
  \bibinfo{year}{2001}, \bibinfo{journal}{Nature (London)}
  \textbf{\bibinfo{volume}{412}}(\bibinfo{number}{19 July 2001}),
  \bibinfo{pages}{313}.

\bibitem[{\citenamefont{Makhlin} \emph{et~al.}(2001)\citenamefont{Makhlin,
  Sch\"on, and Shnirman}}]{Mak01}
\bibinfo{author}{\bibnamefont{Makhlin}, \bibfnamefont{Y.}},
  \bibinfo{author}{\bibfnamefont{G.}~\bibnamefont{Sch\"on}}, and
  \bibinfo{author}{\bibfnamefont{A.}~\bibnamefont{Shnirman}},
  \bibinfo{year}{2001}, \bibinfo{journal}{Rev. Mod. Phys.}
  \textbf{\bibinfo{volume}{73}}, \bibinfo{pages}{357}.

\bibitem[{\citenamefont{Massar}(2000)}]{Mas00}
\bibinfo{author}{\bibnamefont{Massar}, \bibfnamefont{S.}},
  \bibinfo{year}{2000}, \bibinfo{journal}{Phys. Rev. A}
  \textbf{\bibinfo{volume}{62}}, \bibinfo{pages}{040101}.

\bibitem[{\citenamefont{Massar and Popescu}(1995)}]{Mas95}
\bibinfo{author}{\bibnamefont{Massar}, \bibfnamefont{S.}}, and
  \bibinfo{author}{\bibfnamefont{S.}~\bibnamefont{Popescu}},
  \bibinfo{year}{1995}, \bibinfo{journal}{Phys. Rev. Lett.}
  \textbf{\bibinfo{volume}{74}}, \bibinfo{pages}{1259}.

\bibitem[{\citenamefont{Mayers}(2002)}]{May02}
\bibinfo{author}{\bibnamefont{Mayers}, \bibfnamefont{D.}},
  \bibinfo{year}{2002}, e-print quant-ph/0212159.

\bibitem[{\citenamefont{Milburn and Poulin}(2006)}]{Mil06}
\bibinfo{author}{\bibnamefont{Milburn}, \bibfnamefont{G.~J.}}, and
  \bibinfo{author}{\bibfnamefont{D.}~\bibnamefont{Poulin}},
  \bibinfo{year}{2006}, \bibinfo{journal}{Int. J. of Quantum Information}
  \textbf{\bibinfo{volume}{4}}, \bibinfo{pages}{151}.

\bibitem[{\citenamefont{Mirman}(1969)}]{Mir69}
\bibinfo{author}{\bibnamefont{Mirman}, \bibfnamefont{R.}},
  \bibinfo{year}{1969}, \bibinfo{journal}{Phys. Rev.}
  \textbf{\bibinfo{volume}{186}}, \bibinfo{pages}{1380}.

\bibitem[{\citenamefont{Mirman}(1970)}]{Mir70}
\bibinfo{author}{\bibnamefont{Mirman}, \bibfnamefont{R.}},
  \bibinfo{year}{1970}, \bibinfo{journal}{Phys. Rev. D}
  \textbf{\bibinfo{volume}{1}}, \bibinfo{pages}{3349}.

\bibitem[{\citenamefont{Molmer}(1997)}]{Mol97}
\bibinfo{author}{\bibnamefont{Molmer}, \bibfnamefont{K.}},
  \bibinfo{year}{1997}, \bibinfo{journal}{Phys. Rev. A}
  \textbf{\bibinfo{volume}{55}}, \bibinfo{pages}{3195}.

\bibitem[{\citenamefont{Molmer}(1998)}]{Mol98}
\bibinfo{author}{\bibnamefont{Molmer}, \bibfnamefont{K.}},
  \bibinfo{year}{1998}, \bibinfo{journal}{Phys. Rev. A}
  \textbf{\bibinfo{volume}{58}}, \bibinfo{pages}{4247}.

\bibitem[{\citenamefont{Nemoto and Braunstein}(2002)}]{Nem02}
\bibinfo{author}{\bibnamefont{Nemoto}, \bibfnamefont{K.}}, and
  \bibinfo{author}{\bibfnamefont{S.~L.} \bibnamefont{Braunstein}},
  \bibinfo{year}{2002}, e-print quant-ph/0207135.

\bibitem[{\citenamefont{Nemoto and Braunstein}(2003)}]{Nem03}
\bibinfo{author}{\bibnamefont{Nemoto}, \bibfnamefont{K.}}, and
  \bibinfo{author}{\bibfnamefont{S.~L.} \bibnamefont{Braunstein}},
  \bibinfo{year}{2003}, \bibinfo{journal}{Phys. Rev. A}
  \textbf{\bibinfo{volume}{68}}, \bibinfo{eid}{042326}.

\bibitem[{\citenamefont{Nemoto and Braunstein}(2004)}]{Nem04}
\bibinfo{author}{\bibnamefont{Nemoto}, \bibfnamefont{K.}}, and
  \bibinfo{author}{\bibfnamefont{S.~L.} \bibnamefont{Braunstein}},
  \bibinfo{year}{2004}, \bibinfo{journal}{Phys. Lett. A}
  \textbf{\bibinfo{volume}{333}}, \bibinfo{pages}{378}.

\bibitem[{\citenamefont{Nha and Carmichael}(2005)}]{Nha05}
\bibinfo{author}{\bibnamefont{Nha}, \bibfnamefont{H.}}, and
  \bibinfo{author}{\bibfnamefont{H.~J.} \bibnamefont{Carmichael}},
  \bibinfo{year}{2005}, \bibinfo{journal}{Phys. Rev. A}
  \textbf{\bibinfo{volume}{71}}, \bibinfo{eid}{013805}.

\bibitem[{\citenamefont{Nielsen and Chuang}(2000)}]{Nie00}
\bibinfo{author}{\bibnamefont{Nielsen}, \bibfnamefont{M.~A.}}, and
  \bibinfo{author}{\bibfnamefont{I.~L.} \bibnamefont{Chuang}},
  \bibinfo{year}{2000}, \emph{\bibinfo{title}{Quantum Computation and Quantum
  Information}} (\bibinfo{publisher}{Cambridge University Press, Cambridge}).

\bibitem[{\citenamefont{Nielsen}(2003)}]{Nie03}
\bibinfo{author}{\bibnamefont{Nielsen}, \bibfnamefont{M.~A.}},
  \bibinfo{year}{2003}, unpublished; see
    \texttt{http://www.qinfo.org/talks/2003/repn/}

\bibitem[{\citenamefont{Ou and Mandel}(1988)}]{Ou88}
\bibinfo{author}{\bibnamefont{Ou}, \bibfnamefont{Z.~Y.}}, and
  \bibinfo{author}{\bibfnamefont{L.}~\bibnamefont{Mandel}},
  \bibinfo{year}{1988}, \bibinfo{journal}{Phys. Rev. Lett.}
  \textbf{\bibinfo{volume}{61}}, \bibinfo{pages}{50}.

\bibitem[{\citenamefont{Page and Wootters}(1983)}]{Pag83}
\bibinfo{author}{\bibnamefont{Page}, \bibfnamefont{D.~N.}}, and
  \bibinfo{author}{\bibfnamefont{W.~K.} \bibnamefont{Wootters}},
  \bibinfo{year}{1983}, \bibinfo{journal}{Phys. Rev. D}
  \textbf{\bibinfo{volume}{27}}, \bibinfo{pages}{2885}.

\bibitem[{\citenamefont{Paskauskas and You}(2001)}]{Pas01}
\bibinfo{author}{\bibnamefont{Paskauskas}, \bibfnamefont{R.}}, and
  \bibinfo{author}{\bibfnamefont{L.}~\bibnamefont{You}}, \bibinfo{year}{2001},
  \bibinfo{journal}{Phys. Rev. A}
  \textbf{\bibinfo{volume}{64}}, \bibinfo{eid}{042310}.

\bibitem[{\citenamefont{Peres and Scudo}(2001{\natexlab{a}})}]{Per01a}
\bibinfo{author}{\bibnamefont{Peres}, \bibfnamefont{A.}}, and
  \bibinfo{author}{\bibfnamefont{P.~F.} \bibnamefont{Scudo}},
  \bibinfo{year}{2001}{\natexlab{a}}, \bibinfo{journal}{Phys. Rev. Lett.}
  \textbf{\bibinfo{volume}{86}}, \bibinfo{pages}{4160}.

\bibitem[{\citenamefont{Peres and Scudo}(2001{\natexlab{b}})}]{Per01b}
\bibinfo{author}{\bibnamefont{Peres}, \bibfnamefont{A.}}, and
  \bibinfo{author}{\bibfnamefont{P.~F.} \bibnamefont{Scudo}},
  \bibinfo{year}{2001}{\natexlab{b}}, \bibinfo{journal}{Phys. Rev. Lett.}
  \textbf{\bibinfo{volume}{87}}, \bibinfo{eid}{167901}.

\bibitem[{\citenamefont{Peres and Scudo}(2002{\natexlab{a}})}]{Per02a}
\bibinfo{author}{\bibnamefont{Peres}, \bibfnamefont{A.}}, and
  \bibinfo{author}{\bibfnamefont{P.~F.} \bibnamefont{Scudo}},
  \bibinfo{year}{2002}{\natexlab{a}}, \bibinfo{journal}{J. Mod. Opt.}
  \textbf{\bibinfo{volume}{49}}, \bibinfo{pages}{1235}.

\bibitem[{\citenamefont{Peres and Scudo}(2002{\natexlab{b}})}]{Per02b}
\bibinfo{author}{\bibnamefont{Peres}, \bibfnamefont{A.}}, and
  \bibinfo{author}{\bibfnamefont{P.~F.} \bibnamefont{Scudo}},
  \bibinfo{year}{2002}{\natexlab{b}}, in \emph{\bibinfo{booktitle}{Quantum
  Theory: Reconsideration of Foundations}}, edited by
  \bibinfo{editor}{\bibfnamefont{A.}~\bibnamefont{Khrennikov}}
  (\bibinfo{publisher}{V{\"a}xj{\"o} Univ. Press},
  \bibinfo{address}{V{\"a}xj{\"o}, Sweden}), p.~\bibinfo{pages}{283}, e-print quant-ph/0201017.

\bibitem[{\citenamefont{Peres and Terno}(2004)}]{Per04}
\bibinfo{author}{\bibnamefont{Peres}, \bibfnamefont{A.}}, and
  \bibinfo{author}{\bibfnamefont{D.~R.} \bibnamefont{Terno}},
  \bibinfo{year}{2004}, \bibinfo{journal}{Rev. Mod. Phys.}
  \textbf{\bibinfo{volume}{76}}, \bibinfo{eid}{93}.

\bibitem[{\citenamefont{Petersen} \emph{et~al.}(2005)\citenamefont{Petersen,
  Madsen, and Molmer}}]{Pet05}
\bibinfo{author}{\bibnamefont{Petersen}, \bibfnamefont{V.}},
  \bibinfo{author}{\bibfnamefont{L.~B.} \bibnamefont{Madsen}}, and
  \bibinfo{author}{\bibfnamefont{K.}~\bibnamefont{Molmer}},
  \bibinfo{year}{2005}, \bibinfo{journal}{Phys. Rev. A}
  \textbf{\bibinfo{volume}{71}}, \bibinfo{eid}{012312}.

\bibitem[{\citenamefont{Popescu}(2002)}]{Pop02}
\bibinfo{author}{\bibnamefont{Popescu}, \bibfnamefont{S.}},
  \bibinfo{year}{2002}, unpublished.

\bibitem[{\citenamefont{Poulin}(2006)}]{Pou05}
\bibinfo{author}{\bibnamefont{Poulin}, \bibfnamefont{D.}},
  \bibinfo{year}{2006}, \bibinfo{journal}{Int. J. Theor. Phys.}
  \textbf{\bibinfo{volume}{45}}, \bibinfo{pages}{1189}.

\bibitem[{\citenamefont{Poulin and Yard}(2006)}]{Pou06}
\bibinfo{author}{\bibnamefont{Poulin}, \bibfnamefont{D.}}, and
  \bibinfo{author}{\bibfnamefont{J.}~\bibnamefont{Yard}},
  \bibinfo{year}{2006}, e-print quant-ph/0612126.

\bibitem[{\citenamefont{Preskill}(2000)}]{Pre00}
\bibinfo{author}{\bibnamefont{Preskill}, \bibfnamefont{J.}},
  \bibinfo{year}{2000}, e-print quant-ph/0010098.

\bibitem[{\citenamefont{Pryde} \emph{et~al.}(2005)\citenamefont{Pryde, O'Brien,
  White, and Bartlett}}]{Pry05}
\bibinfo{author}{\bibnamefont{Pryde}, \bibfnamefont{G.~J.}},
  \bibinfo{author}{\bibfnamefont{J.~L.} \bibnamefont{O'Brien}},
  \bibinfo{author}{\bibfnamefont{A.~G.} \bibnamefont{White}}, and
  \bibinfo{author}{\bibfnamefont{S.~D.} \bibnamefont{Bartlett}},
  \bibinfo{year}{2005}, \bibinfo{journal}{Phys. Rev. Lett.}
  \textbf{\bibinfo{volume}{94}}, \bibinfo{eid}{220406}.

\bibitem[{\citenamefont{Rau} \emph{et~al.}(2003)\citenamefont{Rau, Dunningham,
  and Burnett}}]{Rau03}
\bibinfo{author}{\bibnamefont{Rau}, \bibfnamefont{A.~V.}},
  \bibinfo{author}{\bibfnamefont{J.~A.} \bibnamefont{Dunningham}}, and
  \bibinfo{author}{\bibfnamefont{K.}~\bibnamefont{Burnett}},
  \bibinfo{year}{2003}, \bibinfo{journal}{Science}
  \textbf{\bibinfo{volume}{301}}, \bibinfo{pages}{1081}.

\bibitem[{\citenamefont{Rudolph}(1999{\natexlab{a}})}]{Rud99}
\bibinfo{author}{\bibnamefont{Rudolph}, \bibfnamefont{T.}},
  \bibinfo{year}{1999}{\natexlab{a}}, e-print quant-ph/9902010.

\bibitem[{\citenamefont{Rudolph}(1999{\natexlab{b}})}]{Rud99b}
\bibinfo{author}{\bibnamefont{Rudolph}, \bibfnamefont{T.}},
  \bibinfo{year}{1999}{\natexlab{b}}, e-print quant-ph/9904037.

\bibitem[{\citenamefont{Rudolph and Grover}(2003)}]{Rud03}
\bibinfo{author}{\bibnamefont{Rudolph}, \bibfnamefont{T.}}, and
  \bibinfo{author}{\bibfnamefont{L.}~\bibnamefont{Grover}},
  \bibinfo{year}{2003}, \bibinfo{journal}{Phys. Rev. Lett.}
  \textbf{\bibinfo{volume}{91}}, \bibinfo{eid}{217905}.

\bibitem[{\citenamefont{Rudolph and Sanders}(2001{\natexlab{a}})}]{Rud01b}
\bibinfo{author}{\bibnamefont{Rudolph}, \bibfnamefont{T.}}, and
  \bibinfo{author}{\bibfnamefont{B.~C.} \bibnamefont{Sanders}},
  \bibinfo{year}{2001}{\natexlab{a}}, e-print quant-ph/0112020.

\bibitem[{\citenamefont{Rudolph and Sanders}(2001{\natexlab{b}})}]{Rud01a}
\bibinfo{author}{\bibnamefont{Rudolph}, \bibfnamefont{T.}}, and
  \bibinfo{author}{\bibfnamefont{B.~C.} \bibnamefont{Sanders}},
  \bibinfo{year}{2001}{\natexlab{b}}, \bibinfo{journal}{Phys. Rev. Lett.}
  \textbf{\bibinfo{volume}{87}}, \bibinfo{eid}{077903}.

\bibitem[{\citenamefont{Rugar} \emph{et~al.}(2004)\citenamefont{Rugar,
  Budakian, Mamin, and Chui}}]{Rug04}
\bibinfo{author}{\bibnamefont{Rugar}, \bibfnamefont{D.}},
  \bibinfo{author}{\bibfnamefont{R.}~\bibnamefont{Budakian}},
  \bibinfo{author}{\bibfnamefont{H.~J.} \bibnamefont{Mamin}}, and
  \bibinfo{author}{\bibfnamefont{B.~W.} \bibnamefont{Chui}},
  \bibinfo{year}{2004}, \bibinfo{journal}{Nature (London)}
  \textbf{\bibinfo{volume}{430}}, \bibinfo{pages}{329}.

\bibitem[{\citenamefont{Samuelsson}
  \emph{et~al.}(2003)\citenamefont{Samuelsson, Sukhorukov, and
  Buttiker}}]{Sam03}
\bibinfo{author}{\bibnamefont{Samuelsson}, \bibfnamefont{P.}},
  \bibinfo{author}{\bibfnamefont{E.~V.} \bibnamefont{Sukhorukov}}, and
  \bibinfo{author}{\bibfnamefont{M.}~\bibnamefont{Buttiker}},
  \bibinfo{year}{2003}, \bibinfo{journal}{Phys. Rev. Lett.}
  \textbf{\bibinfo{volume}{91}}, \bibinfo{eid}{157002}.

\bibitem[{\citenamefont{Samuelsson}
  \emph{et~al.}(2004)\citenamefont{Samuelsson, Sukhorukov, and
  Buttiker}}]{Sam04}
\bibinfo{author}{\bibnamefont{Samuelsson}, \bibfnamefont{P.}},
  \bibinfo{author}{\bibfnamefont{E.~V.} \bibnamefont{Sukhorukov}}, and
  \bibinfo{author}{\bibfnamefont{M.}~\bibnamefont{Buttiker}},
  \bibinfo{year}{2004}, \bibinfo{journal}{Phys. Rev. Lett.}
  \textbf{\bibinfo{volume}{92}}, \bibinfo{eid}{026805}.

\bibitem[{\citenamefont{Samuelsson}
  \emph{et~al.}(2005)\citenamefont{Samuelsson, Sukhorukov, and
  Buttiker}}]{Sam05}
\bibinfo{author}{\bibnamefont{Samuelsson}, \bibfnamefont{P.}},
  \bibinfo{author}{\bibfnamefont{E.~V.} \bibnamefont{Sukhorukov}}, and
  \bibinfo{author}{\bibfnamefont{M.}~\bibnamefont{Buttiker}},
  \bibinfo{year}{2005}, \bibinfo{journal}{New J. Phys.}
  \textbf{\bibinfo{volume}{7}}, \bibinfo{pages}{176}.

\bibitem[{\citenamefont{Sanders} \emph{et~al.}(2003)\citenamefont{Sanders,
  Bartlett, Rudolph, and Knight}}]{San03}
\bibinfo{author}{\bibnamefont{Sanders}, \bibfnamefont{B.~C.}},
  \bibinfo{author}{\bibfnamefont{S.~D.} \bibnamefont{Bartlett}},
  \bibinfo{author}{\bibfnamefont{T.}~\bibnamefont{Rudolph}}, and
  \bibinfo{author}{\bibfnamefont{P.~L.} \bibnamefont{Knight}},
  \bibinfo{year}{2003}, \bibinfo{journal}{Phys. Rev. A}
  \textbf{\bibinfo{volume}{68}}, \bibinfo{eid}{042329}.

\bibitem[{\citenamefont{Schliemann}
  \emph{et~al.}(2001)\citenamefont{Schliemann, Cirac, Kus, Lewenstein, and
  Loss}}]{Sch01}
\bibinfo{author}{\bibnamefont{Schliemann}, \bibfnamefont{J.}},
  \bibinfo{author}{\bibfnamefont{J.~I.} \bibnamefont{Cirac}},
  \bibinfo{author}{\bibfnamefont{M.}~\bibnamefont{Kus}},
  \bibinfo{author}{\bibfnamefont{M.}~\bibnamefont{Lewenstein}}, and
  \bibinfo{author}{\bibfnamefont{D.}~\bibnamefont{Loss}}, \bibinfo{year}{2001},
  \bibinfo{journal}{Phys. Rev. A}
  \textbf{\bibinfo{volume}{64}}, \bibinfo{eid}{022303}.

\bibitem[{\citenamefont{Schuch}
  \emph{et~al.}(2004{\natexlab{a}})\citenamefont{Schuch, Verstraete, and
  Cirac}}]{Sch04a}
\bibinfo{author}{\bibnamefont{Schuch}, \bibfnamefont{N.}},
  \bibinfo{author}{\bibfnamefont{F.}~\bibnamefont{Verstraete}}, and
  \bibinfo{author}{\bibfnamefont{J.~I.} \bibnamefont{Cirac}},
  \bibinfo{year}{2004}{\natexlab{a}}, \bibinfo{journal}{Phys. Rev. Lett.}
  \textbf{\bibinfo{volume}{92}}, \bibinfo{eid}{087904}.

\bibitem[{\citenamefont{Schuch}
  \emph{et~al.}(2004{\natexlab{b}})\citenamefont{Schuch, Verstraete, and
  Cirac}}]{Sch04b}
\bibinfo{author}{\bibnamefont{Schuch}, \bibfnamefont{N.}},
  \bibinfo{author}{\bibfnamefont{F.}~\bibnamefont{Verstraete}}, and
  \bibinfo{author}{\bibfnamefont{J.~I.} \bibnamefont{Cirac}},
  \bibinfo{year}{2004}{\natexlab{b}}, \bibinfo{journal}{Phys. Rev. A}
  \textbf{\bibinfo{volume}{70}}, \bibinfo{eid}{042310}.

\bibitem[{\citenamefont{Shih and Alley}(1988)}]{Shi88}
\bibinfo{author}{\bibnamefont{Shih}, \bibfnamefont{Y.~H.}}, and
  \bibinfo{author}{\bibfnamefont{C.~O.} \bibnamefont{Alley}},
  \bibinfo{year}{1988}, \bibinfo{journal}{Phys. Rev. Lett.}
  \textbf{\bibinfo{volume}{61}}, \bibinfo{pages}{2921}.

\bibitem[{\citenamefont{Schumacher}(2003)}]{Sch03}
\bibinfo{author}{\bibnamefont{Schumacher}, \bibfnamefont{B.}},
  \bibinfo{year}{2003}, unpublished.

\bibitem[{\citenamefont{Siegman}(1986)}]{Sie86}
\bibinfo{author}{\bibnamefont{Siegman}, \bibfnamefont{A.~E.}},
  \bibinfo{year}{1986}, \emph{\bibinfo{title}{Lasers}}
  (\bibinfo{publisher}{University Science Books, Mill Valley, CA}).

\bibitem[{\citenamefont{Silberfarb and Deutsch}(2003)}]{Sil03}
\bibinfo{author}{\bibnamefont{Silberfarb}, \bibfnamefont{A.}}, and
  \bibinfo{author}{\bibfnamefont{I.~H.} \bibnamefont{Deutsch}},
  \bibinfo{year}{2003}, \bibinfo{journal}{Phys. Rev. A}
  \textbf{\bibinfo{volume}{68}}, \bibinfo{eid}{013817}.

\bibitem[{\citenamefont{Smolin}(2004)}]{Smo04}
\bibinfo{author}{\bibnamefont{Smolin}, \bibfnamefont{J.~A.}},
  \bibinfo{year}{2004}, e-print quant-ph/0407009.

\bibitem[{\citenamefont{Spedalieri}(2004)}]{Spe04}
\bibinfo{author}{\bibnamefont{Spedalieri}, \bibfnamefont{F.~M.}}, \bibinfo{year}{2004},
  e-print quant-ph/0409057.

\bibitem[{\citenamefont{Spekkens and Rudolph}(2001)}]{SR01}
\bibinfo{author}{\bibnamefont{Spekkens}, \bibfnamefont{R.~W.}}, and
  \bibinfo{author}{\bibfnamefont{T.}~\bibnamefont{Rudolph}},
  \bibinfo{year}{2001}, \bibinfo{journal}{Phys. Rev. A}
  \textbf{\bibinfo{volume}{65}}, \bibinfo{pages}{012310}.

\bibitem[{\citenamefont{Spekkens and Sipe}(2003)}]{SS03}
\bibinfo{author}{\bibnamefont{Spekkens}, \bibfnamefont{R.~W.}}, and
  \bibinfo{author}{\bibfnamefont{J.~E.}~\bibnamefont{Sipe}},
  \bibinfo{year}{2003}, in \emph{\bibinfo{title}{Coherence and Quantum Optics VIII,}}
   (\bibinfo{publisher}{Kluwer Academic, New York}),
   p.~\bibinfo{pages}{465}.

\bibitem[{\citenamefont{Sternberg}(1994)}]{Sternberg}
\bibinfo{author}{\bibnamefont{Sternberg}, \bibfnamefont{S.}},
  \bibinfo{year}{1994}, \emph{\bibinfo{title}{Group theory and physics}}
  (\bibinfo{publisher}{Cambridge University Press}).

\bibitem[{\citenamefont{Strocchi and Wightman}(1974)}]{Str74}
\bibinfo{author}{\bibnamefont{Strocchi}, \bibfnamefont{F.}}, and
  \bibinfo{author}{\bibfnamefont{A.~S.} \bibnamefont{Wightman}},
  \bibinfo{year}{1974}, \bibinfo{journal}{J. Math. Phys.}
  \textbf{\bibinfo{volume}{15}}, \bibinfo{pages}{2198}.

\bibitem[{\citenamefont{Tan} \emph{et~al.}(1991)\citenamefont{Tan, Walls, and
  Collett}}]{Tan91}
\bibinfo{author}{\bibnamefont{Tan}, \bibfnamefont{S.~M.}},
  \bibinfo{author}{\bibfnamefont{D.~F.} \bibnamefont{Walls}}, and
  \bibinfo{author}{\bibfnamefont{M.~J.} \bibnamefont{Collett}},
  \bibinfo{year}{1991}, \bibinfo{journal}{Phys. Rev. Lett.}
  \textbf{\bibinfo{volume}{66}}, \bibinfo{pages}{252}.

\bibitem[{\citenamefont{Terhal} \emph{et~al.}(2001)\citenamefont{Terhal,
  DiVincenzo, and Leung}}]{Ter01}
\bibinfo{author}{\bibnamefont{Terhal}, \bibfnamefont{B.~M.}},
  \bibinfo{author}{\bibfnamefont{D.~P.} \bibnamefont{DiVincenzo}}, and
  \bibinfo{author}{\bibfnamefont{D.~W.} \bibnamefont{Leung}},
  \bibinfo{year}{2001}, \bibinfo{journal}{Phys. Rev. Lett.}
  \textbf{\bibinfo{volume}{86}}, \bibinfo{pages}{5807}.

\bibitem[{\citenamefont{Vaccaro} \emph{et~al.}(2003)\citenamefont{Vaccaro,
  Anselmi, and Wiseman}}]{Vac03}
\bibinfo{author}{\bibnamefont{Vaccaro}, \bibfnamefont{J.~A.}},
  \bibinfo{author}{\bibfnamefont{F.}~\bibnamefont{Anselmi}}, and
  \bibinfo{author}{\bibfnamefont{H.~M.} \bibnamefont{Wiseman}},
  \bibinfo{year}{2003}, \bibinfo{journal}{Int. J. of Quantum Information}
  \textbf{\bibinfo{volume}{1}}, \bibinfo{pages}{427}.

\bibitem[{\citenamefont{Vaccaro} \emph{et~al.}(2005)\citenamefont{Vaccaro,
  Anselmi, Wiseman, and Jacobs}}]{Vac05}
\bibinfo{author}{\bibnamefont{Vaccaro}, \bibfnamefont{J.~A.}},
  \bibinfo{author}{\bibfnamefont{F.}~\bibnamefont{Anselmi}},
  \bibinfo{author}{\bibfnamefont{H.~M.} \bibnamefont{Wiseman}}, and
  \bibinfo{author}{\bibfnamefont{K.}~\bibnamefont{Jacobs}},
  \bibinfo{year}{2005}, e-print quant-ph/0501121.

\bibitem[{\citenamefont{Vaziri} \emph{et~al.}(2003)\citenamefont{Vaziri,
  Pan, Jennewein, Weihs, and Zeilinger}}]{Vaz03}
\bibinfo{author}{\bibnamefont{Vaziri}, \bibfnamefont{A.}},
  \bibinfo{author}{\bibfnamefont{J.-W.} \bibnamefont{Pan}},
  \bibinfo{author}{\bibfnamefont{T.} \bibnamefont{Jennewein}},
  \bibinfo{author}{\bibfnamefont{G.} \bibnamefont{Weihs}}, and
  \bibinfo{author}{\bibfnamefont{A.} \bibnamefont{Zeilinger}},
  \bibinfo{year}{2003}, \bibinfo{journal}{Phys. Rev. Lett.}
  \textbf{\bibinfo{volume}{91}}, \bibinfo{eid}{227902}.

\bibitem[{\citenamefont{Verstraete and Cirac}(2003)}]{Ver03}
\bibinfo{author}{\bibnamefont{Verstraete}, \bibfnamefont{F.}}, and
  \bibinfo{author}{\bibfnamefont{J.~I.} \bibnamefont{Cirac}},
  \bibinfo{year}{2003}, \bibinfo{journal}{Phys. Rev. Lett.}
  \textbf{\bibinfo{volume}{91}}, \bibinfo{eid}{010404}.

\bibitem[{\citenamefont{Walton} \emph{et~al.}(2003)\citenamefont{Walton,
  Abouraddy, Sergienko, Saleh, and Teich}}]{Wal03}
\bibinfo{author}{\bibnamefont{Walton}, \bibfnamefont{Z.~D.}},
  \bibinfo{author}{\bibfnamefont{A.~F.} \bibnamefont{Abouraddy}},
  \bibinfo{author}{\bibfnamefont{A.~V.} \bibnamefont{Sergienko}},
  \bibinfo{author}{\bibfnamefont{B.~E.~A.} \bibnamefont{Saleh}}, and
  \bibinfo{author}{\bibfnamefont{M.~C.} \bibnamefont{Teich}},
  \bibinfo{year}{2003}, \bibinfo{journal}{Phys. Rev. Lett.}
  \textbf{\bibinfo{volume}{91}}, \bibinfo{eid}{087901}.

\bibitem[{\citenamefont{Watrous}(2004)}]{Wat04}
\bibinfo{author}{\bibnamefont{Watrous}, \bibfnamefont{J.}},
  \bibinfo{year}{2004}, \bibinfo{journal}{Phys. Rev. Lett.}
  \textbf{\bibinfo{volume}{93}}, \bibinfo{eid}{010502}.

\bibitem[{\citenamefont{Wigner}(1952)}]{Wig52}
\bibinfo{author}{\bibnamefont{Wigner}, \bibfnamefont{E.~P.}},
  \bibinfo{year}{1952}, \bibinfo{journal}{Z. Phys.}
  \textbf{\bibinfo{volume}{133}}, \bibinfo{pages}{101}.

\bibitem[{\citenamefont{Wick} \emph{et~al.}(1952)\citenamefont{Wick, Wightman,
  and Wigner}}]{Wic52}
\bibinfo{author}{\bibnamefont{Wick}, \bibfnamefont{G.~C.}},
  \bibinfo{author}{\bibfnamefont{A.~S.} \bibnamefont{Wightman}}, and
  \bibinfo{author}{\bibfnamefont{E.~P.} \bibnamefont{Wigner}},
  \bibinfo{year}{1952}, \bibinfo{journal}{Phys. Rev.}
  \textbf{\bibinfo{volume}{88}}, \bibinfo{pages}{101}.

\bibitem[{\citenamefont{Wick} \emph{et~al.}(1970)\citenamefont{Wick, Wightman,
  and Wigner}}]{WWW70}
\bibinfo{author}{\bibnamefont{Wick}, \bibfnamefont{G.~C.}},
  \bibinfo{author}{\bibfnamefont{A.~S.} \bibnamefont{Wightman}}, and
  \bibinfo{author}{\bibfnamefont{E.~P.} \bibnamefont{Wigner}},
  \bibinfo{year}{1970}, \bibinfo{journal}{Phys. Rev. D}
  \textbf{\bibinfo{volume}{1}}, \bibinfo{pages}{3267}.

\bibitem[{\citenamefont{Wiseman}(2003)}]{Wis03}
\bibinfo{author}{\bibnamefont{Wiseman}, \bibfnamefont{H.~M.}},
  \bibinfo{year}{2003}, in \emph{\bibinfo{booktitle}{Proceedings of SPIE Vol.
  5111, Fluctuations and Noise in Photonics and Quantum Optics}}, edited by
  \bibinfo{editor}{\bibfnamefont{D.}~\bibnamefont{Abbott}},
  \bibinfo{editor}{\bibfnamefont{J.~H.} \bibnamefont{Shapiro}}, and
  \bibinfo{editor}{\bibfnamefont{Y.}~\bibnamefont{Yamamoto}}
  (\bibinfo{publisher}{SPIE}, \bibinfo{address}{Bellingham, WA}), pp.
  \bibinfo{pages}{78--91}, e-print quant-ph/0303116.

\bibitem[{\citenamefont{Wiseman}(2004)}]{Wis04}
\bibinfo{author}{\bibnamefont{Wiseman}, \bibfnamefont{H.~M.}},
  \bibinfo{year}{2004}, \bibinfo{journal}{J. Opt. B: Quantum Semiclass.}
  \textbf{\bibinfo{volume}{6}}, \bibinfo{pages}{S849}.

\bibitem[{\citenamefont{Wiseman and Vaccaro}(2003)}]{Wis03b}
\bibinfo{author}{\bibnamefont{Wiseman}, \bibfnamefont{H.~M.}}, and
  \bibinfo{author}{\bibfnamefont{J.~A.} \bibnamefont{Vaccaro}},
  \bibinfo{year}{2003}, \bibinfo{journal}{Phys. Rev. Lett.}
  \textbf{\bibinfo{volume}{91}}, \bibinfo{eid}{097902}.

\bibitem[{\citenamefont{Yoo} \emph{et~al.}(1997)\citenamefont{Yoo, Ruostekoski,
  and Javanainen}}]{Yoo97}
\bibinfo{author}{\bibnamefont{Yoo}, \bibfnamefont{S.~M.}},
  \bibinfo{author}{\bibfnamefont{J.}~\bibnamefont{Ruostekoski}}, and
  \bibinfo{author}{\bibfnamefont{J.}~\bibnamefont{Javanainen}},
  \bibinfo{year}{1997}, \bibinfo{journal}{J. Mod. Opt.}
  \textbf{\bibinfo{volume}{44}}, \bibinfo{pages}{1763}.

\bibitem[{\citenamefont{Yurtsever and Dowling}(2002)}]{Yur02}
\bibinfo{author}{\bibnamefont{Yurtsever}, \bibfnamefont{U.}}, and
  \bibinfo{author}{\bibfnamefont{J.~P.}~\bibnamefont{Dowling}},
  \bibinfo{year}{2002}, \bibinfo{journal}{Phys. Rev. A}
  \textbf{\bibinfo{volume}{65}}, \bibinfo{eid}{052317}.

\bibitem[{\citenamefont{Zanardi}(2001)}]{Zan01b}
\bibinfo{author}{\bibnamefont{Zanardi}, \bibfnamefont{P.}},
  \bibinfo{year}{2001}, \bibinfo{journal}{Phys. Rev. Lett.}
  \textbf{\bibinfo{volume}{87}}, \bibinfo{eid}{077901}.

\bibitem[{\citenamefont{Zanardi and Rasetti}(1997)}]{Zan97}
\bibinfo{author}{\bibnamefont{Zanardi}, \bibfnamefont{P.}}, and
  \bibinfo{author}{\bibfnamefont{M.}~\bibnamefont{Rasetti}},
  \bibinfo{year}{1997}, \bibinfo{journal}{Phys. Rev. Lett.}
  \textbf{\bibinfo{volume}{79}}, \bibinfo{pages}{3306}.

\bibitem[{\citenamefont{Zou} \emph{et~al.}(2006)\citenamefont{Zou, Shu, and
  Guo}}]{Zou06}
\bibinfo{author}{\bibnamefont{Zou}, \bibfnamefont{X.}},
  \bibinfo{author}{\bibfnamefont{J.}~\bibnamefont{Shu}}, and
  \bibinfo{author}{\bibfnamefont{G.}~\bibnamefont{Guo}}, \bibinfo{year}{2006},
  \bibinfo{journal}{Phys. Rev. A}
  \textbf{\bibinfo{volume}{73}}, \bibinfo{eid}{054301}.

\end{thebibliography}
\end{document}